\documentclass[twocolumn,amsmath,amssymb,10pt,floatfix,superscriptaddress,a4paper,letterpaper,fleqn,nofootinbib]{revtex4-1}
\usepackage{amssymb}
\usepackage{graphicx}
\usepackage{epsfig}
\usepackage{dcolumn}
\usepackage{array}
\usepackage{bm}
\usepackage{fancyhdr}
\usepackage{floatflt}
\usepackage{longtable}
\usepackage{threeparttable}
\usepackage{multirow}
\usepackage{float}
\usepackage{afterpage}
\usepackage{color}
\usepackage{textcomp}
\usepackage{multirow}
\usepackage{mathdesign}
\usepackage{booktabs}
\usepackage{paralist}

\newcommand{\nuc}[2]{$^{{\mathrm{#1}}}${#2}}
\newcommand{\za}[2]{{$^{#2}$}{#1}}

\newcommand{\kadonis}{KADoNiS}

\usepackage[warnundef]{jabbrv}
\DefineJournalAbbreviation{Test}{Tes} 
\setlongtables
\definecolor{Darkgreen}{rgb}{0,0.4,0}
\usepackage[breaklinks=true,linkbordercolor={1 1 1}]{hyperref}
\hypersetup{
colorlinks = true,    
linktoc    = all,     
linkcolor  = blue,    
citecolor  = Darkgreen
}
\def\etal{\emph{et al.}\,}

\begin{document}
\setcounter{page}{1}
\title{
 \qquad \\ \qquad \\ \qquad \\  \qquad \\ \qquad \\ \qquad \\
Newly Evaluated Neutron Reaction Data on Chromium Isotopes
}

\author {G.P.A.~Nobre}
  	\email[Corresponding author: ]{gnobre@bnl.gov}
  	\affiliation{National Nuclear Data Center, Brookhaven National Laboratory, Upton, NY 11973, USA}
\author {M.T.~Pigni}
  	\affiliation{Oak Ridge National Laboratory, Oak Ridge, TN 37831, USA}
\author {D.A.~Brown}
  	\affiliation{National Nuclear Data Center, Brookhaven National Laboratory, Upton, NY 11973, USA}
\author {R.~Capote}
  	\affiliation{NAPC--Nuclear Data Section, International Atomic Energy Agency, A-1040 Vienna, Austria}
\author {A.~Trkov}
        \affiliation{Jozef Stefan Institute, Jamova 39, 1000 Ljubljana, Slovenia}
\author{K.H.~Guber}
	\affiliation{Oak Ridge National Laboratory, Oak Ridge, TN 37831, USA}
\author {R.~Arcilla}
  \affiliation{National Nuclear Data Center, Brookhaven National Laboratory, Upton, NY 11973, USA}
\author{J.~Gutierrez}
	\affiliation{Gonzaga University, Spokane, WA, USA}
\author{A.~Cuadra}
        \affiliation{National Nuclear Data Center, Brookhaven National Laboratory, Upton, NY 11973, USA}
  	\affiliation{Nuclear Systems \& Structural Analysis Group, Brookhaven National Laboratory, Upton, NY 11973, USA}
\author {G.~Arbanas}
  	\affiliation{Oak Ridge National Laboratory, Oak Ridge, TN 37831, USA}
\author {B.~Kos}
        \affiliation{Jozef Stefan Institute, Jamova 39, 1000 Ljubljana, Slovenia}
\author{D.~Bernard}
	\affiliation{CEA/DEN/CAD/DER/SPRC/Laboratoire d'\'Etudes de Physique Centre de Cadarache, B\^{a}timent 230, F-13108 Saint-Paul-Lez-Durance, France}
\author{P.~Leconte}
	\affiliation{CEA/DEN/CAD/DER/SPRC/Laboratoire d'\'Etudes de Physique Centre de Cadarache, B\^{a}timent 230, F-13108 Saint-Paul-Lez-Durance, France}

\date{\today}

\begin{abstract}{
Neutron reaction data for the set of major chromium isotopes were reevaluated
from the thermal energy range up to 20~MeV. In the low energy region,
updates to the thermal values together with an improved $R$-matrix
analysis of the resonance parameters characterizing the cluster of large $s$-wave resonances for
\nuc{50,53}{Cr} isotopes were performed. In the intermediate and high energy
range up to 20~MeV, the evaluation methodology used statistical nuclear reaction
models implemented in the EMPIRE code within the Hauser-Feshbach framework to evaluate the reaction
cross sections and angular distributions. Exceptionally, experimental data were used to
evaluate relevant cross sections above the resonance region up to 5 MeV in the major \nuc{52}{Cr} isotope.
Evaluations were benchmarked with Monte Carlo simulations of a small suite of critical assemblies highly sensitive to Chromium data, and with the Oktavian shielding benchmark to judge deep penetration performance with a 14-MeV D-T neutron source.
A significant improvement in performance is demonstrated compared to existing evaluations.

}
\end{abstract}
\maketitle

\lhead{Newly Evaluated Neutron $\dots$}     
\chead{NUCLEAR DATA SHEETS}                  
\rhead{G.P.A. Nobre \textit{et al.}}        
\lfoot{}
\rfoot{}
\renewcommand{\footrulewidth}{0.4pt}
\tableofcontents{}


\section{Introduction}
\label{Sec:Introduction}

Chromium is a tarnish-resistant metal with a high melting point
commonly alloyed with iron and other materials to produce stainless
steel. In nuclear fission and fusion applications, the need to
increase thermal efficiency led to the development of martensitic
steels that, containing small amounts of Cr~(10-16\%) and Mo~(<1\%),
are characterized by an excellent resistance to thermal
stresses~\cite{Klueh:2001}. Tempered martensite gives steel good hardness
and high toughness.
Chromium has also been recently proposed as a coating on zircaloy to improve the oxidation resistance of the cladding, and as dopant in UO$_2$ to improve the fuel’s corrosion and fragmentation characteristics~\cite{Delafoy:2018}.

Naturally occurring chromium \textsuperscript{nat}Cr consists of three
stable isotopes \nuc{52,53,54}{Cr} plus an extremely long-lived
isotope \nuc{50}{Cr} (half-life >$1.8\times10^{17}$~y). Having proton number
$Z$=24, these isotopes are very close to the $N$=28 neutron shell
closure, the major isotope \nuc{52}{Cr} being ``magic'', with $N$=28 exactly.
Therefore, like other nuclei in the same mass region,
chromium isotopes have very low level densities and a propensity for
low-lying unnatural parity states. The low level density manifests itself as
non-statistical fluctuations of neutron cross sections at low energies,
enhancing the role of the neutron capture by the first resonance levels of the odd isotopes
\cite{Pronyaev:2013} but also causing the resonance fluctuations to
extend to higher energies ($\gtrsim$ 1~MeV), well into the energy range typical of
fission neutron spectra. As expected, these fluctuations can be impactful on
the neutron flux in many applications and also affect the neutron leakage.

The evaluated data for the chromium isotopes released in the
\mbox{ENDF/B-VIII.0} library followed recent measurements~\cite{Guber:2011}
and evaluation efforts~\cite{jkps.59.1644} in support of the
U.S. Department of Energy (DOE) Nuclear Criticality Safety Program
(NCSP) with particular focus in the resolved resonance region. In the
intermediate and high energy region, although the recent work by
Pereslavtsev~\cite{jkps.59.931} is currently included in the \mbox{JEFF-3.3}
library~\cite{Plompen2020}, \mbox{ENDF/B-VIII.0} library \cite{Brown:2018} still relies on the work performed in
1997 by Chiba and released in the \mbox{ENDF/B-VI.1}
library~\cite{LA150,ENDF-VI.8}, which was extended to 200~MeV.
The current evaluation is limited to 20~MeV.
Besides the dated and somewhat limited
evaluation work on the chromium isotopes, the most evident problem
common to all released nuclear data libraries is related to the poor description of the
cluster of large $s$-wave resonances that primarily characterize the
neutron capture on the \nuc{53}{Cr} isotope, which dominates the neutron capture on
natural chromium near 5~keV of the neutron incident energy.


A special set of criticality experiments ($k_\infty$), highly sensitive to structural materials,
was carried out at IPPE (Institute of Physics and Power Engineering), Obninsk, Russia and are designed as KBR assemblies. 
Those experiments were testing zirconium, molybdenum, chromium, and stainless steel in fast neutron flux assemblies driven by
highly enriched uranium (HEU). Large discrepancies of several percent (1~\% is equal to 1000~pcm)  between calculated and measured reactivity were found when the \mbox{ENDF/B-VII.1} (or \mbox{ENDF/B-VIII.0}) library was used. The largest disagreement of 11~\% in the most
chromium-sensitive experiment, KBR-15, (HEU-MET INTER-015 in ICBESP notation \cite{ICSBEP}) was traced to
heavily underestimated neutron capture on natural chromium in the keV region \cite{Koscheev:2017}.
Those authors attempted to address the issue \cite{Koscheev:2017} within the BROND-3.1 library~\cite{BROND-3.1} by fitting the
\nuc{53}{Cr}$(n,\gamma)$ cross sections published by Guber~\cite{Guber:2011} at the NDST~2010 conference. The
digitized capture yield data in EXFOR were interpreted as cross section data, therefore, neglecting
important corrections such as multiple scattering and self-shielding effects, which may reach 80\% in the keV region.
Data were fitted by adjusting the bound resonance levels and the widths of a few low-lying resonances but, in doing that,
the value of the measured thermal capture cross section was greatly overestimated.
Despite the incorrect interpretation of the measured capture yield as cross sections, the BROND-3.1 evaluation showed a significantly improved
performance for the KBR cases, particularly
due to the large increase of capture cross sections in the energy region of the cluster of $s$-wave resonances between 1--10~keV.
However, this improvement in integral performance could not be understood because the digitized data
taken from~\cite{Guber:2011} were, later on, found to be affected by a mistake in calculating the density of the
Cr\textsubscript{2}O\textsubscript{3} sample used in the measurement. In fact, after the correction of Guber \etal, the
\nuc{53}{Cr}$(n,\gamma)$ capture yield available in the EXFOR library (EXFOR subentry 14324006) is decreased by a factor of two with respect to
the data published in Ref.~\cite{Guber:2011}.
Unfortunately, these corrected Guber \etal data are very discrepant from earlier comprehensive Stieglitz \etal measurements on
enriched chromium isotopes undertaken at RPI in the 1970's and reported in the EXFOR library, especially regarding the normalization. Reasons for such differences are still not understood.

\begin{figure}[htbp]
\begin{center}
\includegraphics[width=0.49\textwidth]{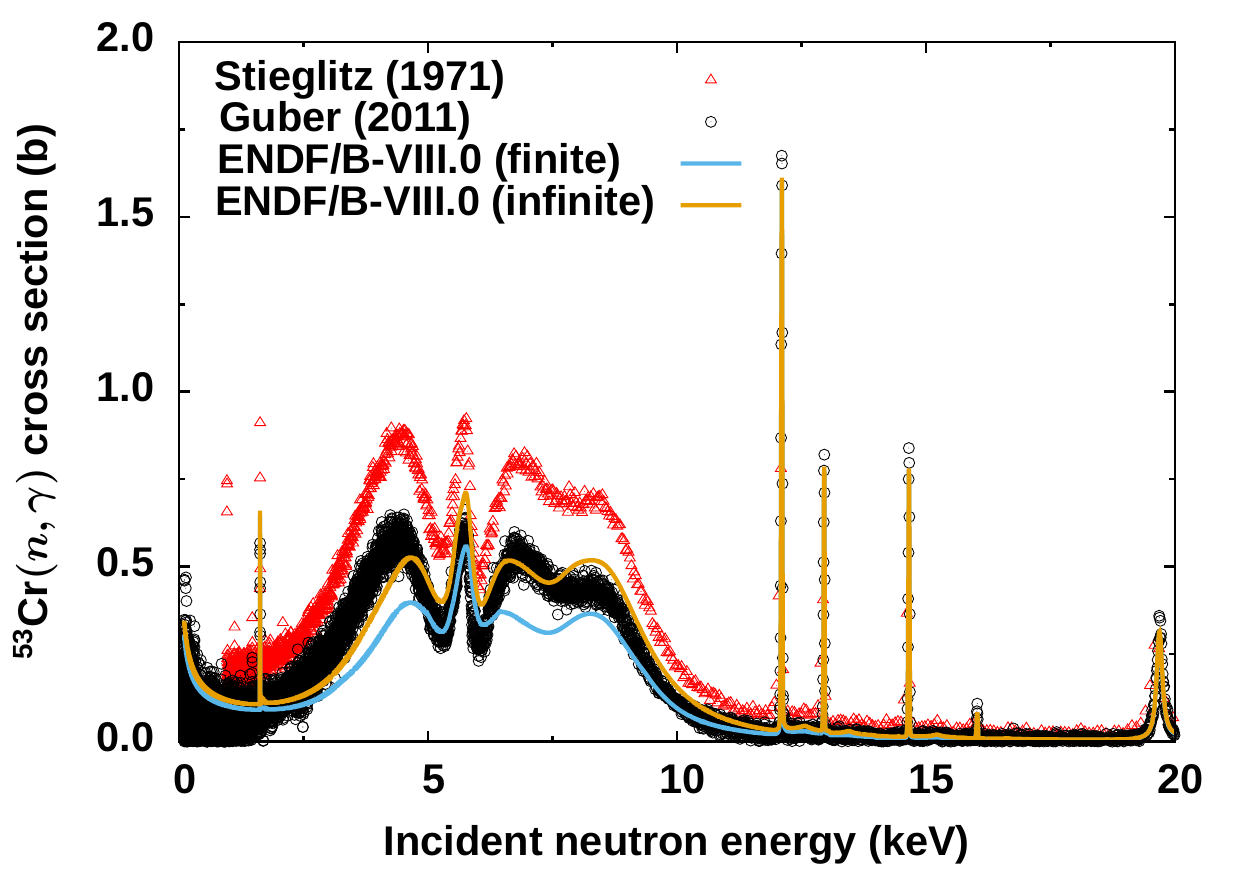}
\caption{(color online) \nuc{53}{Cr}$(n,\gamma)$ cross
  sections from Guber \etal~\cite{Guber:2011}  and Stieglitz \etal~\cite{Stieglitz:1971}
 as reported in the EXFOR library and (in blue and yellow
  solid lines) the cross
  sections reconstructed from ENDF/B-VIII.0 resonance
  parameters including multiple-scattering and self-shielding
  corrections based on the experimental configuration of  Guber \etal}
\label{fig:53cr-guber-stieg-diff}
\end{center}
\end{figure}

Although there are differences in the enrichment and geometry of the
oxide sample between the experiments of Stieglitz \etal and Guber \etal, these are
not sufficient to explain the large discrepancies of the reported
\nuc{53}{Cr}$(n,\gamma)$ cross sections as shown in
Fig.~\ref{fig:53cr-guber-stieg-diff}. Here, the cross sections
reconstructed from ENDF/B-VIII.0 resonance parameters and including
multiple-scattering and self-shielding corrections based on the experimental configuration of Guber \etal, {\it i.e.}, thickness, isotopic abundances, and
sample geometry, are also shown in blue for the finite size case and
in yellow for the infinite-slab approximation case. Differently from
the infinite slab approximation, when finite-sized corrections are
included, the geometric properties of the sample are used to calculate
the multiple scattering corrections. Quantities such as the
estimated mean-free-path for a given incident neutron energy or Monte Carlo simulations for
given sample properties can be used as independent ways to determine the accuracy of
the multiple scattering corrections calculated in the finite or infinite
approximation. It is also visible in
Fig.~\ref{fig:53cr-guber-stieg-diff}, in the energy
below 2.5~keV, resonance-like data points in the of Stieglitz \etal that are not
visible in the data of  Guber \etal 
This discrepancy was not investigated in the
present work.

The present work is an attempt to, first, resolve the discrepancy
between the measured capture data, second, understand the large
discrepancy in quantifying multiple scattering corrections in finite
and infinite approximation and, finally, derive a consistent set of
resonance parameters (energy, capture, and neutron widths) based on a
fitting procedure including appropriate experimental corrections. The
ultimate goal is to improve the agreement with the KBR integral
test cases and other criticality and shielding experiments
highly-sensitive to chromium.

\begin{figure}[htbp]
\begin{center}
\includegraphics[width=0.45\textwidth]{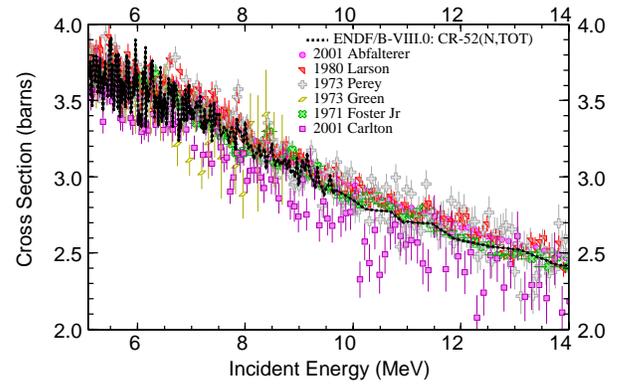}
\caption{(color online) Normalization problems in \nuc{52}{Cr} total cross section data by Carlton \textit{et al.} data \cite{Carlton:2000}. The other datasets are taken from EXFOR.}
\label{fig:Carlton-wrong}
\end{center}
\end{figure}

At higher energies, the \nuc{52}{Cr} total cross section from ENDF/B-VIII.0 is larger than Carlton \textit{et al.} data in EXFOR 13840002 subentry \cite{Carlton:2000} above 4 MeV. 
However, Carlton data are discrepant in normalization\footnote{Carlton data also show a discontinuity around 4 MeV as discussed in Sec.~\ref{Sec:fluctuating_data}.} with all other datasets in the energy region from $\sim$5 up to 10 MeV as shown in Fig.~\ref{fig:Carlton-wrong}.
Higher-energy fluctuations extend up to around 10 MeV in the total and may extend up to 5 MeV in elastic and inelastic cross sections, angular distributions, and partial gamma production cross sections. Those fluctuations have to be considered in the evaluations at least up to 6 MeV based on measured data. Additionally, Hauser-Feshbach calculations have to agree with averaged cross-section data. Finally we note that the angular distributions in all Cr isotopes are a copy of the ENDF/B-V \nuc{nat}{Cr} angular distribution \cite{Hetrick:1991lf}, which were fitted to available experimental data.

As mentioned above, just as chromium can mask problems with iron, iron can also mask problems with chromium.  It behooves us to understand (and address) the shortcomings in iron.  We first note the biggest \nuc{56}{Fe} shortcoming, already noted in the CIELO iron paper \cite{CIELO-Fe}: the evaluated inelastic cross section based on available experimental data is too high, leading to an elastic scattering cross section that is too low.  This causes the fast neutron leakage spectrum from 2 up to 10 MeV to be too low in the integral testing used for CIELO.  A recent experiment by Pirovano \textit{et al.} \cite{PhysRevC.99.024601} further supports this observation.  Next we note that deep penetration (leakage) experiments indicate that the cross-section ``valleys'' in between resonances are too low and that direct capture needs to be included in the evaluation \cite{CIELO-Fe}.  Revising the direct capture may also help resolve the discrepancy with Firestone \textit{et al.}'s evaluation of the thermal capture cross section \cite{PhysRevC.95.014328}.  Finally we note that at high energies ($>40$ MeV) shielding experiments reported by Konno \textit{et al.} indicate that further corrections are needed \cite{KWON20182,KONNO20152178}.

Despite chromium's ubiquity in steel alloys, there is only one integral criticality test that we are aware of that explicitly targets chromium: the KBR-15 assembly discussed above. There are of course several other criticality tests with high stainless steel content and a sizable sensitivity to chromium including the PU-MET-INT-002 test noted in the iron CIELO paper \cite{CIELO-Fe}.  CEA/DEN/CAD/DER/SPRC/Laboratoire d'\`Etudes de Physique in Cadarache, France also included the study of chromium within their comprehensive experimental program on structural, moderating, and absorbing materials entitled MAESTRO~\cite{jef-1849:2017}, particularly in MAESTRO phase III (2013/14). The MAESTRO programme relies on pile-oscillation measurements in the MINERVE reactor and a very careful modeling of the experimental setup allowing for benchmark quality of derived results.

The organization of this paper is as follows. First we describe improvements to the resonance evaluations in all of the chromium isotopes.  Following this, we detail the determination of the total cross section and validation of the optical model used in the evaluation.  With this in hand, we turn to the main business of evaluating the fast regions of the chromium isotopes.  This discussion will include a careful consideration of the level densities involved and the detailed tuning of the Hauser-Feshbach calculations to experimental data. A comprehensive comparison of available experimental data including cross sections, angular distributions, neutron and gamma emission spectra, and double differential cross sections are presented. We conclude with a discussion of the validation of the evaluations using both criticality and shielding benchmarks, as well as pile-oscillation measurements in the MINERVE reactor.


\section{Resolved Resonance Region}\label{Sec:rrr}

\subsection{Background}\label{Sec:background}

\begin{figure}[htbp]
\begin{center}
\includegraphics[width=0.49\textwidth]{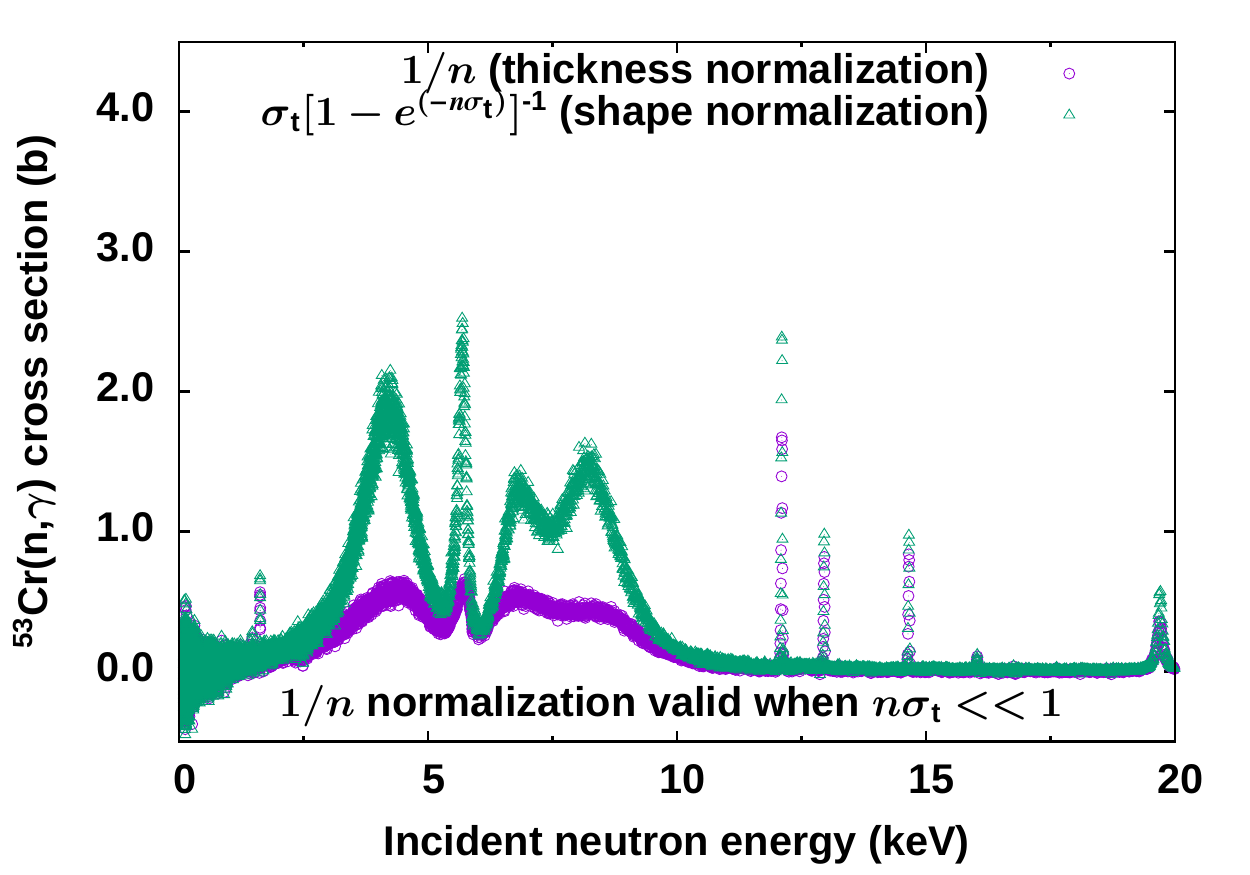}
\caption{(color online) \nuc{53}{Cr}$(n,\gamma)$ cross sections from Guber \etal~\cite{Guber:2011} obtained by
  normalizing the neutron capture yield $Y$ by the inverse of the sample
  thickness (in purple) and a shape normalization (in green).}
\label{fig:53cr-norm-guber}
\end{center}
\end{figure}

\begin{figure}[htbp]
\begin{center}
\includegraphics[width=0.49\textwidth]{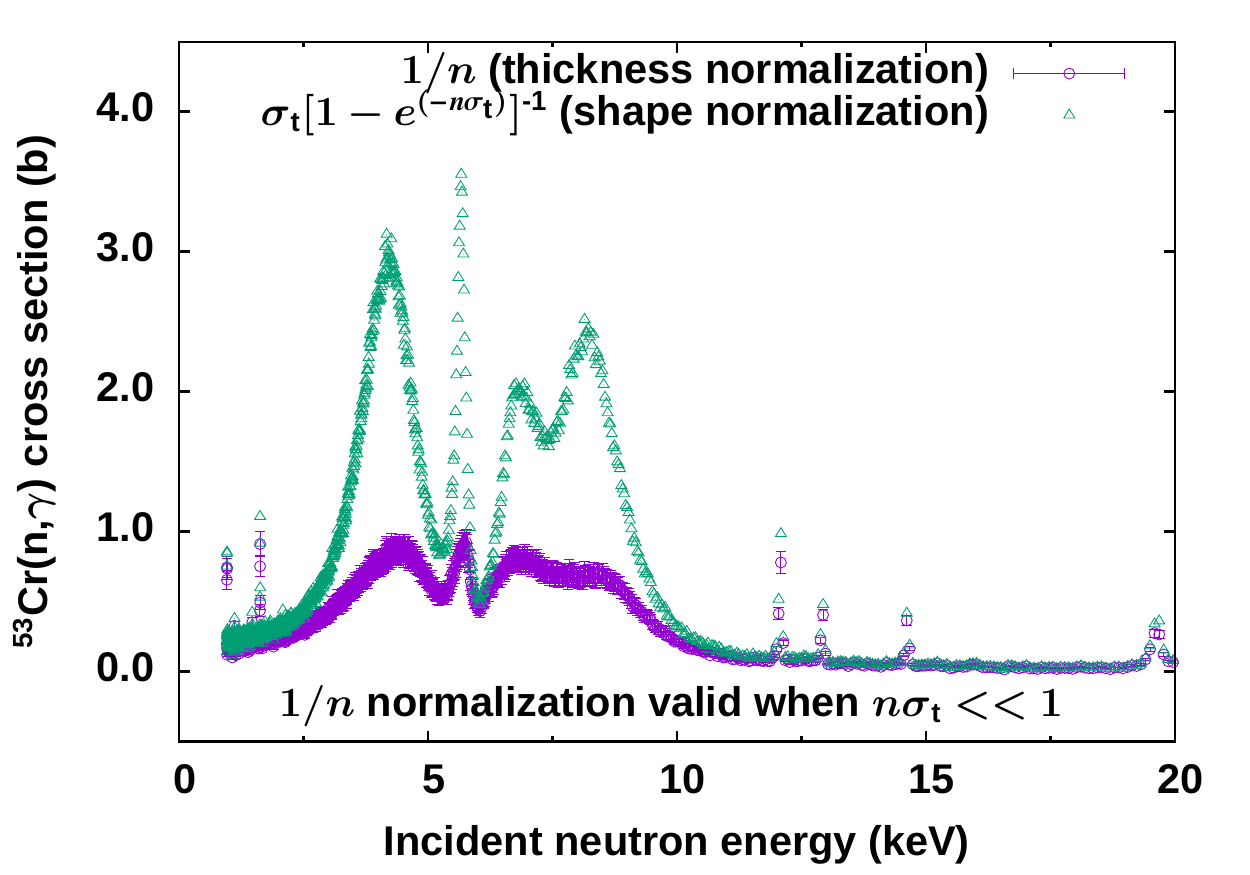}
\caption{(color online) \nuc{53}{Cr}$(n,\gamma)$ cross sections from Stieglitz \etal~\cite{Stieglitz:1971} obtained by
  normalizing the neutron capture yield $Y$ by the inverse of the sample
  thickness (in purple) and a shape normalization (in green).}
\label{fig:53cr-norm-stieg}
\end{center}
\end{figure}

In the low-energy neutron range from a few keVs up to a few tens of
keV, structural materials such as chromium and vanadium are
characterized by clusters of $s$-wave resonances with large neutron and
capture widths. Due to the finite size of sample thickness used in
capture yield measurements, the measured cluster of $s$-wave
resonances is associated with strong multiple scattering interactions
as well as self-shielding and neutron sensitivity effects. In the
specific case of \nuc{53}{Cr}$(n,\gamma)$, the four energy levels
between $4$ and $10$~keV are clearly an example for which the
multiple-scattering interaction of the incident neutrons has the
effect of greatly enhancing the number of measured counts. Although the
self-shielding effects tend to reduce the measured counts, the overall
result is that the observed capture widths are largely
overestimated. Therefore, to properly evaluate the resonance capture
widths for these levels, it is of fundamental importance to account
for these experimental effects in the fitting evaluation
procedure. Most often used in the fit of measured capture yield data,
methods of calculation to correct the theoretical cross sections or
yield for both self-shielding and multiple scattering effects are
implemented in the $R$-matrix code SAMMY. Other input options of the
code are the geometric characteristic of the sample such as thickness,
given in atom/barn (a/b) or in cm, and its height, given in cm, as well as the type of
normalization of the fitted data. The latter represents another
important point of discussion in the fitting procedure of the measured
capture data aimed to derive the resonance parameters such as energies
and related widths. In this regard, the measured capture cross
sections (in barns) are often reported in literature as the neutron
capture yield (a dimensionless quantity) multiplied by the inverse of the 
sample thickness. This type of normalization can be understood by the
relationship between the (self-shielded) neutron capture yield and the
total and capture cross section,
\begin{equation}
Y=(1-e^{-n\sigma_{\textrm{t}}})\frac{\sigma_{\gamma}}{\sigma_{\textrm{t}}}\approx n~\sigma_{\gamma}\,,
\end{equation}
in the limit the sample thickness $n$ is very thin such as the
quantity $n\sigma_{\textrm{t}}<<1$. Under this assumption, the capture
cross sections are determined in shape by the measured
energy-dependent neutron yield data and in magnitude by the sample
thickness. However, for the measured \nuc{53}{Cr}$(n,\gamma)$ cross
sections, the thickness normalization as reported by Stieglitz~\cite{Stieglitz:1971} and
Guber~\cite{Guber:2011} seems to be the incorrect choice. In fact, a simple comparison
of the capture cross sections normalized by the constant value $1/n$
and by the shape normalization
$\sigma_{\textrm{t}}(1-e^{-n\sigma_{\textrm{t}}})^{-1}$ defined by the energy dependent total
cross section as well as the thickness, shows
remarkable differences as displayed in
Figs.~\ref{fig:53cr-norm-guber}--\ref{fig:53cr-norm-stieg}. This can
be explained by the fact, although the thickness for both
experiments is relatively thin, the \nuc{53}{Cr} total cross sections
are such that the condition $n\sigma_{\textrm{t}}<<1$ is not
met. For this reason, the  \nuc{53}{Cr}$(n,\gamma)$
cross sections from Stieglitz \etal and Guber \etal were not used in the fitting procedure. Instead, the
related neutron capture yields were used and the set of resonance
parameters was derived by the simultaneous fit of the neutron capture
yields and transmission data in the same energy range.

In addition to the normalization of the capture cross sections,
another relevant point of discussion is the discrepancy between measured data from 
Stieglitz \etal and Guber \etal as clearly shown in
Figs.~\ref{fig:53cr-norm-guber}--\ref{fig:53cr-norm-stieg}. With a
slight deviation in shape below 1~keV, the two measured capture yield
data sets differ by a constant value of about 50\% in the energy
region of the cluster of $s$-wave resonances between
1--10~keV. Although a re-analysis of the thickness value of data from Guber \etal 
 showed the reported value of the thickness, $n$=0.0137~a/b, should
be reduced by 3\%, for an updated value of $n$=0.0133~a/b, there is no
clear explanation of the large difference between the two data
sets. An indirect proof that  \nuc{53}{Cr} yield data from Stieglitz \etal are
correct is their consistency with capture yield data measured for a
natural sample. The capture widths obtained by the fit of the natural
yield data consistently reproduced the yield data from Stieglitz \etal for the
enriched \nuc{53}{Cr} sample. Moreover, mainly due to the large
capture width of the resonance level at 4.1~keV, it is clear that,
although the cluster of $s$-wave resonances is in the keV region, its
contribution to the evaluation of the capture cross sections in
thermal energy range is not negligible.

In the next subsection the results of the evaluation procedure of the two
isotopes \nuc{50,53}{Cr} will be discussed in details.

\subsection{Estimation of Multiple-Scattering Corrections}

\subsubsection{Finite size and infinite-slab approximation}\label{subsub:fin_int_approx}

\begin{figure}[htbp]
\begin{center}
\includegraphics[width=0.49\textwidth]{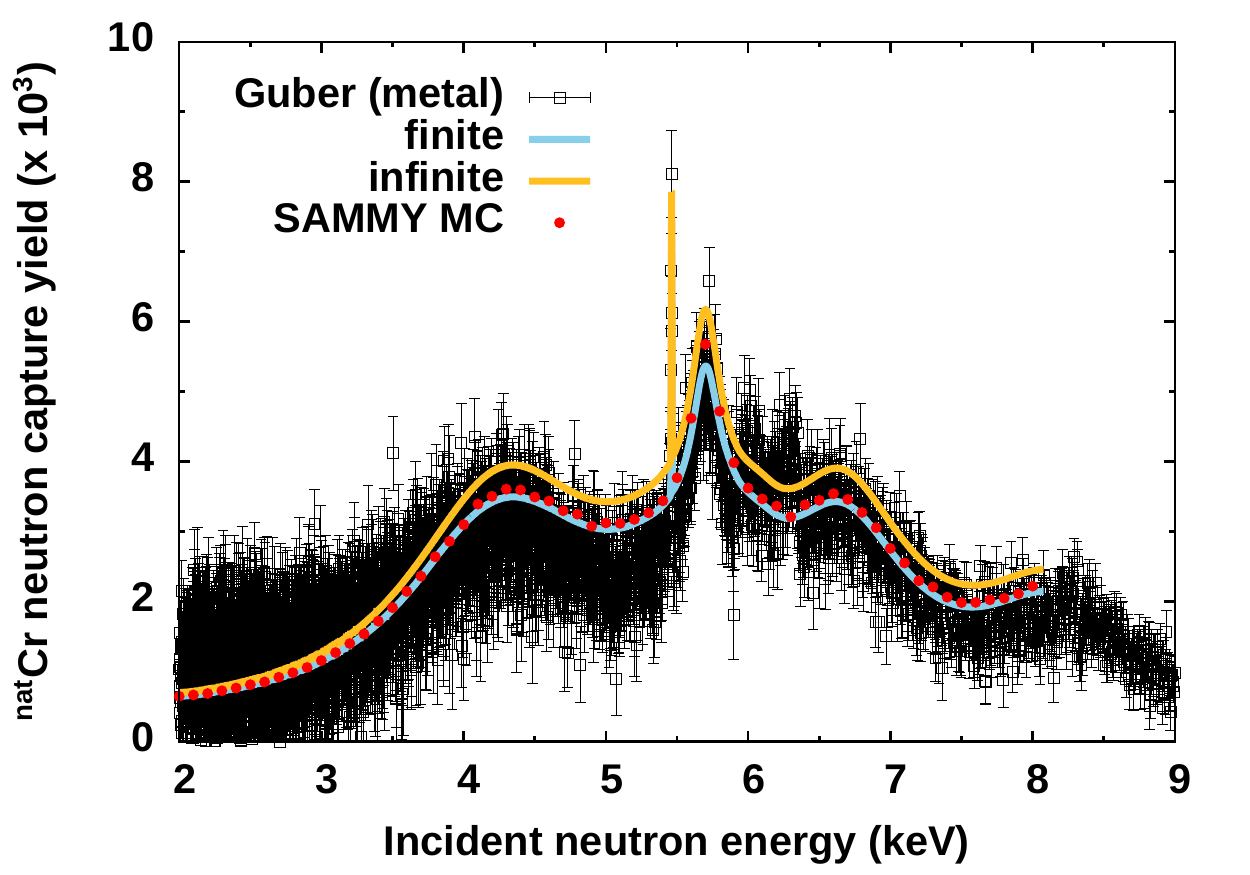}
\caption{(color online) \nuc{nat}{Cr} neutron capture yield calculated for a
  finite size sample (in light blue) and in infinite-slab
  approximation (in yellow) compared to Monte Carlo
  simulations (in red) performed by SAMMY. The calculations used the sample
  configuration of  the metal natural sample of Guber \etal}
\label{fig:guber_metal_mltsc}
\end{center}
\end{figure}

\begin{figure}[htbp]
\begin{center}
\includegraphics[width=0.49\textwidth]{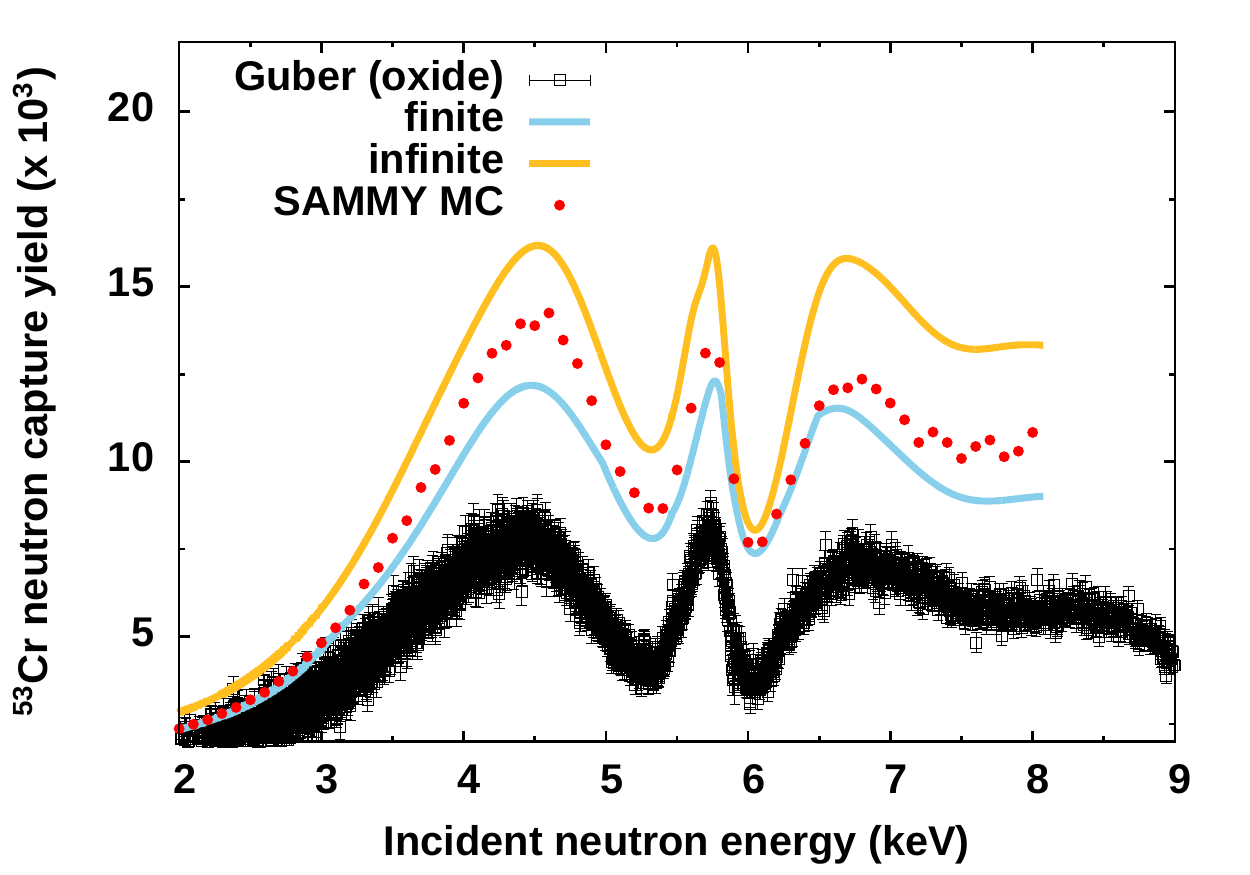}
\caption{(color online) \nuc{53}{Cr} neutron capture yield calculated for a
  finite size sample (in light blue) and in infinite-slab
  approximation (in yellow) compared to Monte Carlo
  simulations (in red) performed by SAMMY. The calculations used the sample
  configuration of the oxide enriched sample from  Guber \etal}
\label{fig:guber_oxide_mltsc}
\end{center}
\end{figure}

\begin{figure}[htbp]
\begin{center}
\includegraphics[width=0.49\textwidth]{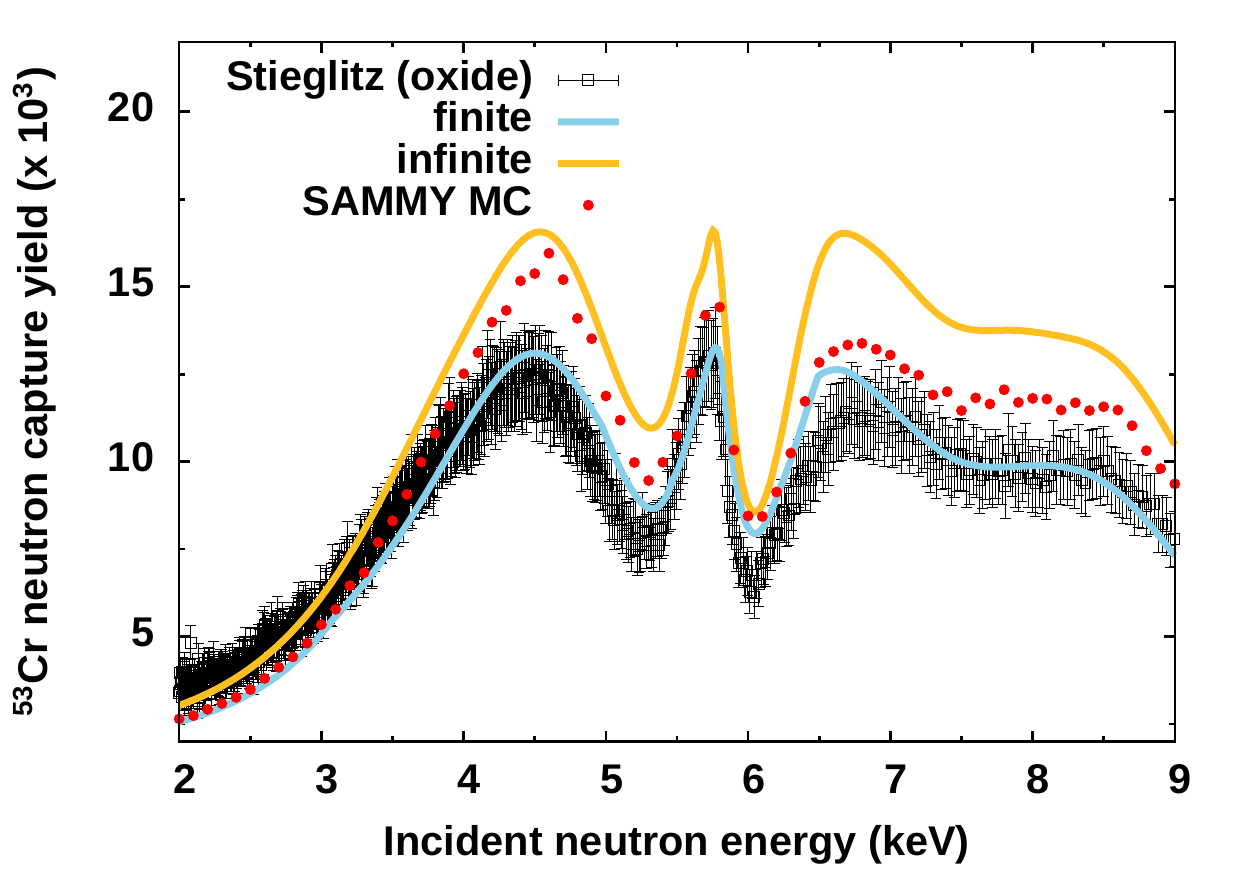}
\caption{(color online) \nuc{53}{Cr} neutron capture yield calculated for a
  finite size sample (in light blue) and in infinite-slab
  approximation (in yellow) compared to Monte Carlo
  simulations (in red) performed by SAMMY. The calculations used the sample
  configuration of the oxide enriched sample from Stieglitz \etal}
\label{fig:stieglitz_oxide_mltsc}
\end{center}
\end{figure}

In addition to intrinsic properties of the sample (density and
weight), the sample's geometric configuration (thickness, height or
radius) in relation to the beam size plays an important role in the
quantification of the multiple scattering corrections. In fact, the use
of either the finite size sample or infinite-slab approximation can lead to
large differences in the calculated neutron capture yield as shown in
Figs~\ref{fig:guber_metal_mltsc}--\ref{fig:stieglitz_oxide_mltsc}. These
discrepancies can be more or less remarkable depending on the
experimental configuration of the sample. In
Fig.~\ref{fig:guber_metal_mltsc} the discrepancy for the measurement on the metal
natural sample is of the order of 15\% differently
from the other two measurements for oxide enriched sample shown in
Figs~\ref{fig:guber_oxide_mltsc}--\ref{fig:stieglitz_oxide_mltsc} for
which the discrepancy is about 30\%. As expected, the impact of the finite size or
infinite slab approximation on the metal natural sample calculations
is relatively small compared to the oxide cases since its a thinner sample configuration.
Another way to calculate the neutron
capture yield is to perform Monte Carlo simulations including the same
experimental configurations for a defined set of resonance
parameters. Plotted in red dots, these simulations agree nicely with
the finite size approximation in the case of the metal natural sample
as shown in Fig~\ref{fig:guber_metal_mltsc}. However, for both oxide
enriched samples, it is clearly difficult to determine if either the finite size
sample or infinite slab approximation are the suitable option. For this
reason, the data set measured on the natural metal sample was selected and
used in the fit with finite size sample option. The sample
characteristic used in the Monte Carlo simulations are reported in
Table~\ref{table:ExperimentalSetupValues} in which the beam size is
assumed for all measurements equals to the size of the sample. Due to
edge effects this parameter can have a big impact on the Monte Carlo calculations.
Moreover, the plotted yields calculated in finite size or infinite slab approximation as
well as the Monte Carlo simulations were performed by using the set of
resonance parameters obtained from the fit of the natural data as shown in Fig.~\ref{fig:guber_metal_mltsc}.
The comparison of the calculated yields with measured data from Guber \etal and Stieglitz \etal
 displayed in
Figs~\ref{fig:guber_oxide_mltsc}--\ref{fig:stieglitz_oxide_mltsc},
respectively, shows that the magnitude of  data from Stieglitz \etal is
definitely more consistent with the measurements on the metal
sample. Although additional investigation on the agreement between
Monte Carlo and analytical calculations for the multiple scattering
corrections is needed, this analysis suggests that enriched data
measured on oxide sample by Guber \etal are affected by a normalization issue.
\begin{table}
\caption{\label{table:ExperimentalSetupValues} \nuc{50,53}{Cr} sample characteristics
  for both natural metal and oxide enriched samples used in the
  calculations and Monte Carlo simulations.}

\begin{tabular}{ccccc}
\toprule\toprule
Sample & \multicolumn{2}{c}{Thickness} & Radius & Beam height \\
       & a/b  & cm     & cm     & cm \\
\midrule
 \nuc{50}{Cr}~\cite{Stieglitz:1971} & 0.0088 &0.914 & 3.17 & 3.17\\
 \nuc{nat}{Cr}~\cite{Guber:2011} & 0.0263 &0.318& 1.27 & 1.27 \\
 \nuc{53}{Cr}~\cite{Guber:2011} & 0.0133 &0.978& 2.03$^{(a)}$ & 2.03\\
 \nuc{53}{Cr}~\cite{Stieglitz:1971} & 0.0142 &1.234 & 3.17 & 3.17\\
\bottomrule
\bottomrule
\end{tabular}
\begin{tablenotes}
\item ${}^{(a)}$ This is the effective radius calculated from the area
  $A=2\times 6.844$ cm$^{2}$ of the  $2.54\times 5.08$~cm
  (or $1\times2$~inches) rectangular sample formed by two 1-inch
  square samples placed one over the other.
\end{tablenotes}
\end{table}

\begin{figure}[htbp]
\begin{center}
\includegraphics[width=0.49\textwidth]{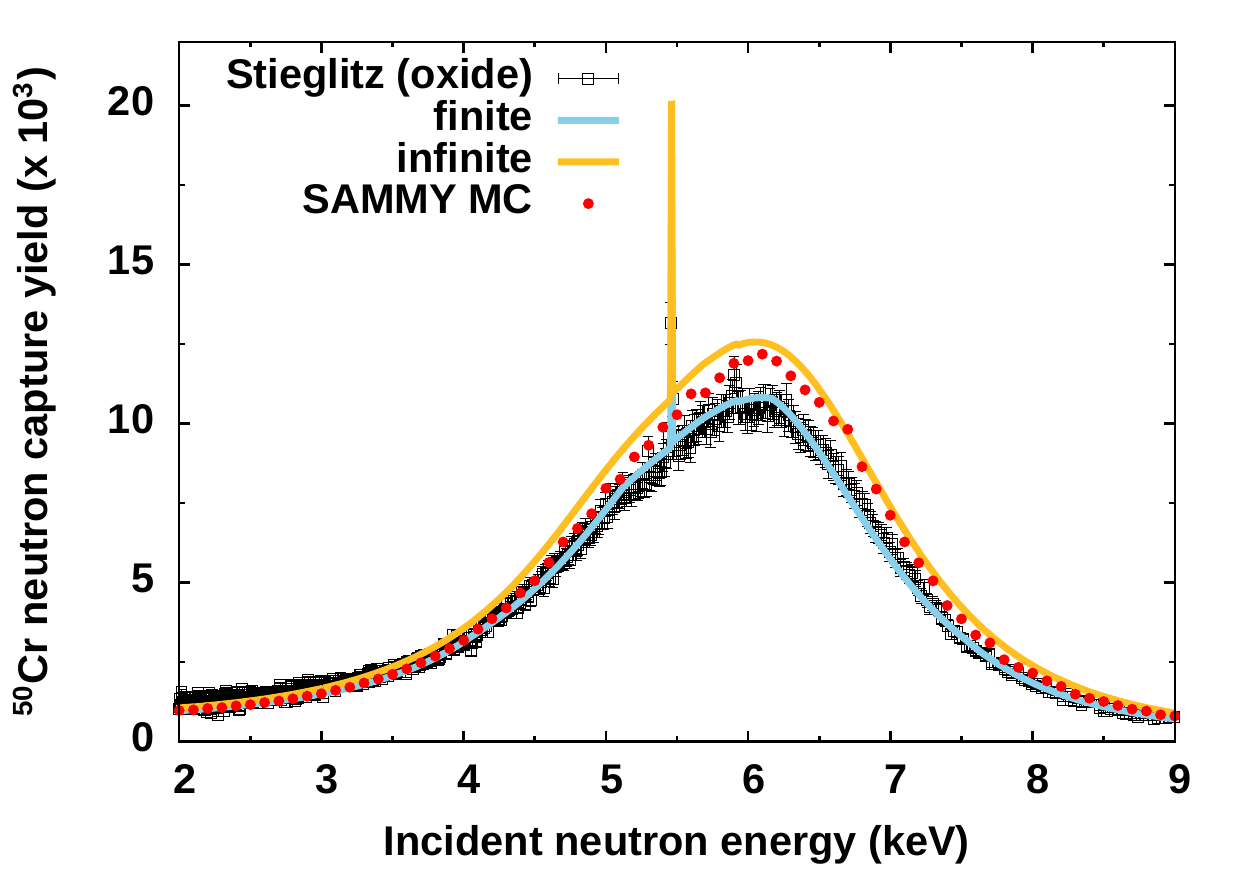}
\caption{(color online) \nuc{50}{Cr} neutron capture yield calculated for a
  finite size sample (in light blue) and in infinite-slab
  approximation (in yellow) compared to Monte Carlo
  simulations (in red) performed by SAMMY. The calculations used the sample
  configuration of the oxide enriched sample from  Stieglitz \etal and scaled by a
  factor 0.92.}
\label{fig:stieglitz_oxide_mltsc_50}
\end{center}
\end{figure}

Similarly to the oxide configurations related to the \nuc{53}{Cr}
isotope, Figure~\ref{fig:stieglitz_oxide_mltsc_50} shows the
Monte Carlo simulations for the \nuc{50}{Cr}, which are found in the middle of the finite size
and infinite slab approximation calculations. Since the thickness is
slightly thinner than the \nuc{53}{Cr} oxide sample, the Monte Carlo
results tend to be close to the infinite slab approximation. The fit
for this set of data was performed with finite size approximation and,
for future updates, the plan is, first, to resolve the discrepancy between
Monte Carlo simulations and the analytical treatment of the multiple
scattering corrections, and second, to derive the proper set of parameters.

\subsubsection{Experimental corrections estimated by the MCNP}

Independently from the efforts to understand the Guber and Stieglitz data, simplified MCNP6.2 \cite{MCNP6,MCNP6.2} models of the \nuc{53}{Cr}  capture experiments were developed to assess the multiple scattering cross sections and potentially compare MCNP and SAMMY implementations. Note that the models were not intended to fully represent the experiments or replicate their results, rather to understand the magnitude of the multiple scattering correction and its dependence on energy and sample thickness for geometries representative of both experiments. The results obtained did not appear sensitive to sample geometry other than thickness, and the discussion and results shown below focus on the results of the simplified ORELA model used in Guber et al experiment. The model included a disc-shaped uniform source sufficiently large ($r$=6cm) to fully illuminate the enriched chromium sample described in \cite{Guber:2011}. To study the dependence of the multiple scattering correction on sample thickness, the mono-energetic source neutrons were assumed to have energies uniformly distributed between 4 and 10 keV. In both cases, the neutrons were assumed to be mono-directional and perpendicular to the sample, which allowed to limit the size of the model. Note that this assumption is a good approximation of the typical experimental situation as the source of neutrons is more than 40 meters away from the sample in the closest setup. The rectangular sample size was, as described in \cite{Guber:2011}, 2.54 cm by 5.08 cm long (1" by 2") with nominal thickness $t_n$ = 0.98 cm. The studied thickness varied from 0.01 up to 3 cm following the estimated mean-free path of 5~keV neutrons in the chromium oxide. 
No attempt was made to model the C6D6 detector, rather appropriately tagged photon current tallies were implemented on various relevant surfaces.

\begin{figure}[htbp]
\begin{center}
\includegraphics[width=0.47\textwidth,clip,trim = 38mm 85mm 45mm 88mm]{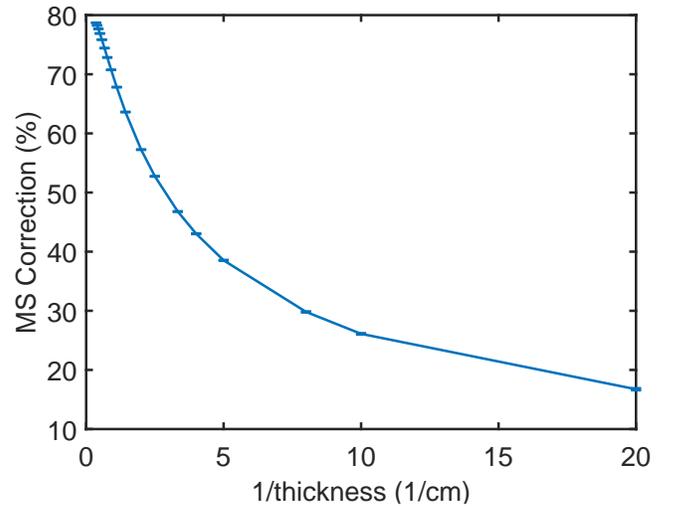}
\caption{(color online) Multiple scattering correction dependence on sample thickness from simplified MCNP6 model.}
\label{fig:MS-thickness}
\end{center}
\end{figure}

\begin{figure}[htbp]
\begin{center}
\includegraphics[width=0.48\textwidth,clip,trim = 12mm 63mm 3mm 70mm]{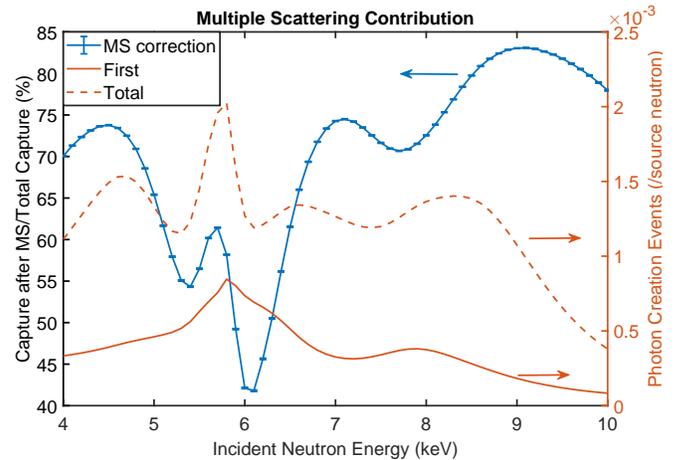}
\caption{(color online) Multiple scattering correction dependency on energy from simplified MCNP6 model.}
\label{fig:MSvsE}
\end{center}
\end{figure}

\begin{figure}[htbp]
\begin{center}
\includegraphics[width=0.47\textwidth,clip,trim = 35mm 81mm 42mm 85mm]{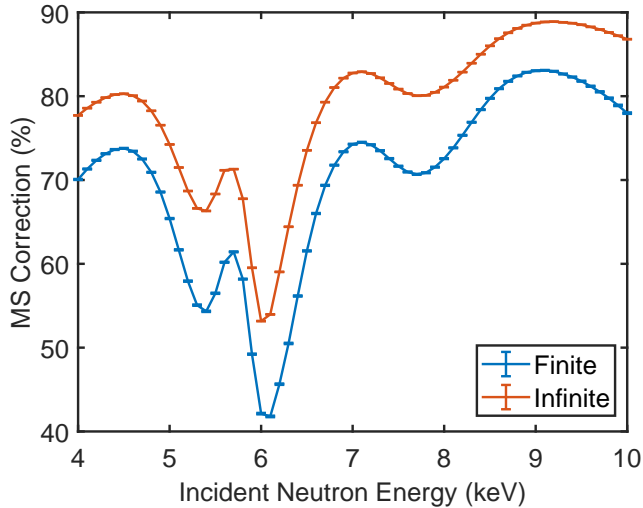}
\caption{(color online) Multiple scattering correction comparison, finite vs. infinite sample sensitivity.}
\label{fig:MS-finite-infinite}
\end{center}
\end{figure}

The results are shown in Fig.~\ref{fig:MS-thickness} for different sample thicknesses. For the nominal sample thickness, the multiple scattering correction, defined as the number of photons having undergone more than one collision (including the collisions of the photon’s parent neutron), divided by the total number of photons, is 69\%; for thicker samples, 
the multiple scattering correction can reach up to 80\%; as expected, the correction asymptotically approaches 0 for extremely thin samples (for 0.01 cm, the correction is ~2\%). It should be noted that the narrow uncertainty bars shown on the plot correspond to the statistical uncertainty only, and do not account for other sources of systematic uncertainties in the estimation of the effect of multiple scattering, such as the inclusion of non-capture photons, background, etc.
Those effects, however, are expected to be orders of magnitude lower than the multiple scattering and capture phenomena under study.  The energy dependence of the correction is shown in Fig.~\ref{fig:MSvsE} (blue line). The correction exhibits peaks at energies slightly shifted to the right from the resonance energies shown in Table~\ref{table:ResonanceWidths}, because of the interplay of the contributions shown in orange (MS correction = 100$\times$(Total--First)/Total). Note that the number of first collision photon creation events (labeled ``First'') only reflects the largest resonance at approximately 5.7~keV, due to the energy grid chosen for the study and the narrowness of the resonances, while the total number of photon creation events (labeled ``Total'') shows all four resonances, slightly shifted to higher energies, since the neutrons emitted from the source have the opportunity to downscatter into the resonances. An additional sensitivity was performed to the finite vs infinite sample approximations discussed in the previous section. As seen in Fig.~\ref{fig:MS-finite-infinite}, the multiple scattering corrections for the finite and infinite sample approximations differ basically by a normalization factor, with a stronger correction resulting for the infinite sample.

%

\subsection{Results}

The evaluation of the total and capture cross sections of the
\nuc{50,53}{Cr} isotopes was performed with particular focus on the
cluster of $s$-wave resonances as well as the related thermal values
in the attempt to address the inconsistencies of the measured data
sets and the large discrepancies with the evaluated data reported in the
ENDF/B-VIII.0 library.

The simultaneous fit of neutron capture yields and transmission data
showed that, in addition to the large discrepancy between  measured capture yield data  by Stieglitz \etal 
and Guber \etal, as explained in
section~\ref{Sec:background}, there is no perfect compatibility among
 available transmission data in the alignment of the energy levels measured in
the $s$-wave cluster energy region for the \nuc{53}{Cr}
isotope. Although a shift in the alignment between the energy levels
for transmission and capture yield data is expected because of the
effect of the strong multiple scattering and self-shielding corrections in
the capture yield data (clearly visible in
Figs~\ref{fig:53cr-norm-guber}--\ref{fig:53cr-norm-stieg}), this
misalignment seems too remarked and an
optimal fit between transmission and capture data could not be
achieved.

\begin{figure}[htbp]
\begin{center}
\includegraphics[width=0.49\textwidth]{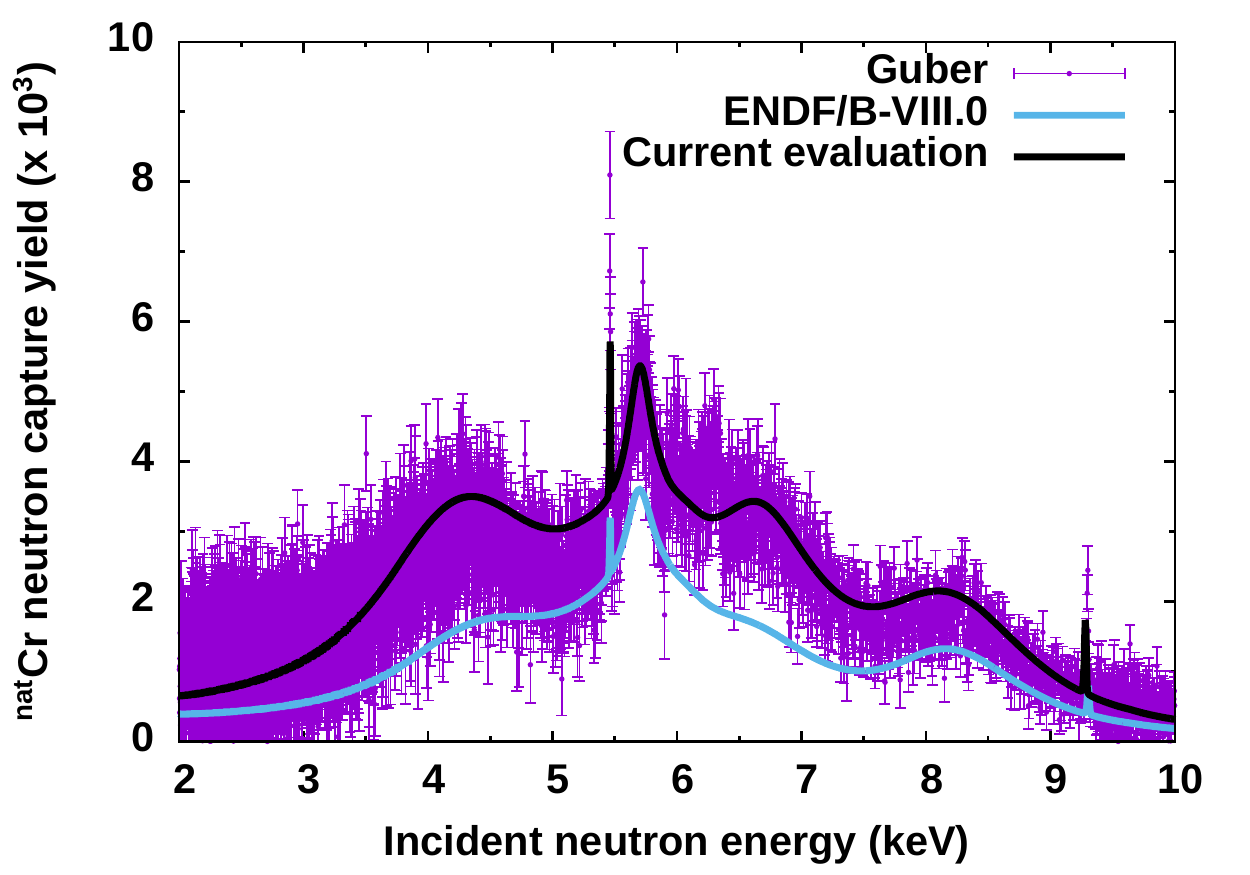}
\caption{(color online) Measured~\cite{Guber:2011} and calculated \nuc{nat}{Cr} neutron capture yield
  data. The neutron capture yield shown for all cases was calculated
  in finite size approximation.}
\label{fig:natcr-guber}
\end{center}
\end{figure}

\begin{figure}[htbp]
\begin{center}
\includegraphics[width=0.49\textwidth]{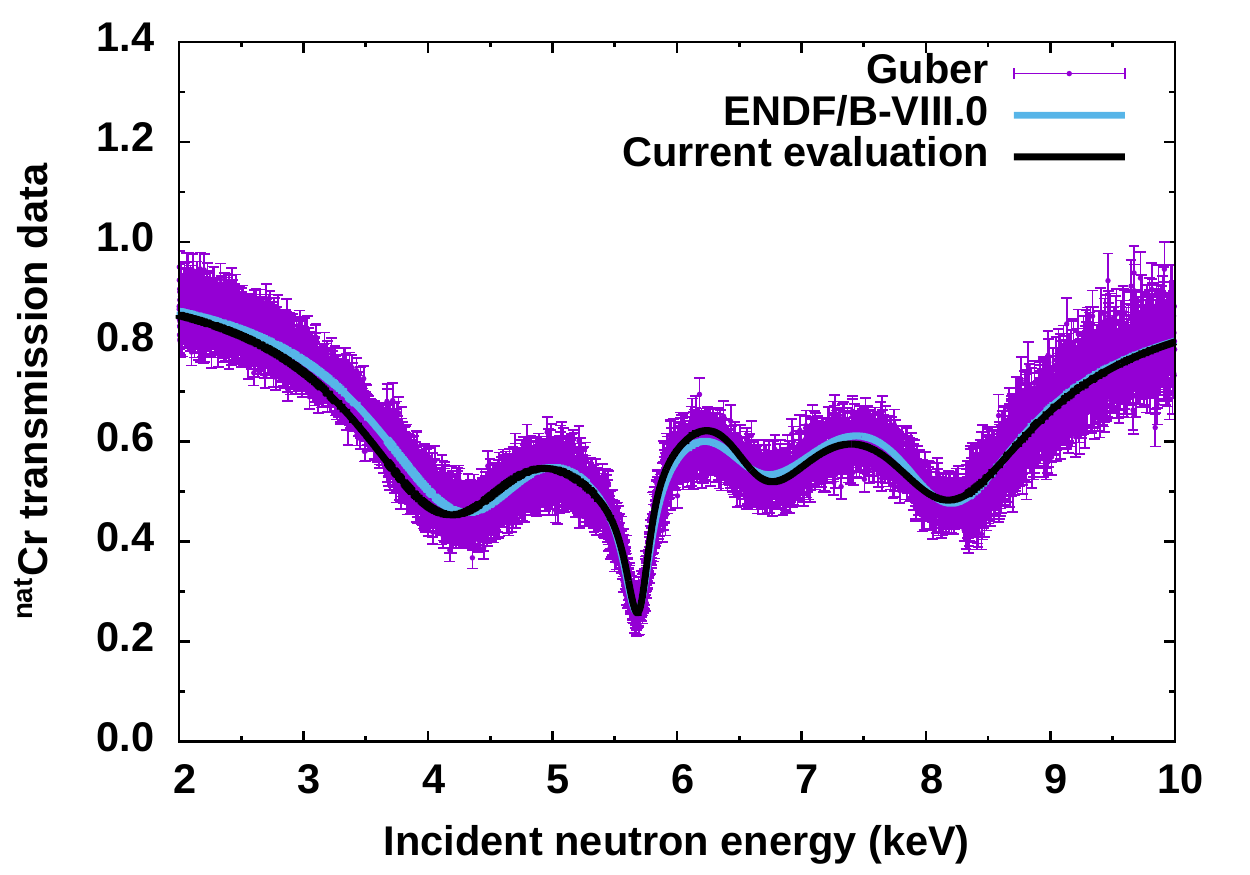}
\caption{(color online) Measured~\cite{Guber:2011} and calculated \nuc{nat}{Cr} thin-target transmission
  data.}
\label{fig:natcr-thin-guber}
\end{center}
\end{figure}

\begin{figure}[htbp]
\begin{center}
\includegraphics[width=0.49\textwidth]{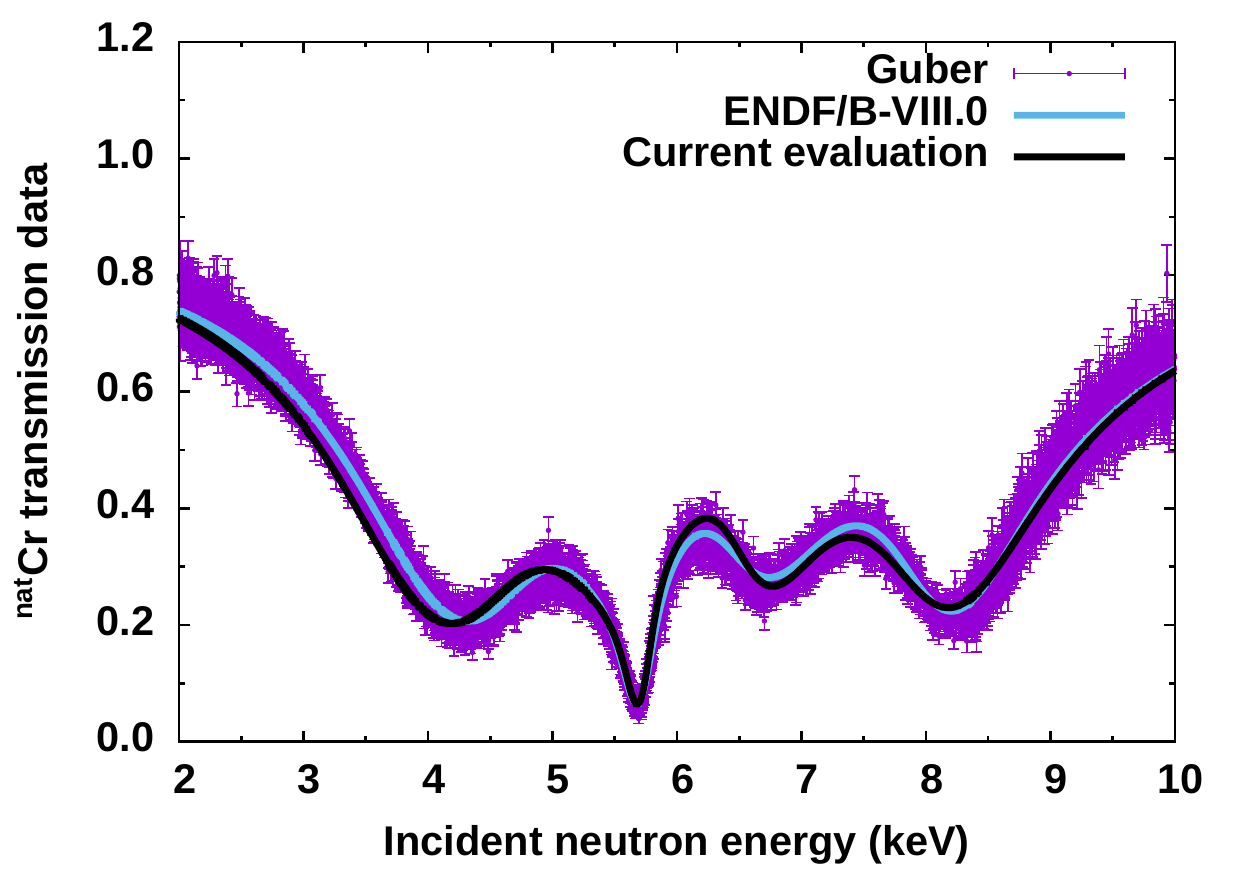}
\caption{(color online) Measured~\cite{Guber:2011} and calculated \nuc{nat}{Cr} thick-target
  transmission data.}
\label{fig:natcr-thick-guber}
\end{center}
\end{figure}

Following the analysis described in \ref{subsub:fin_int_approx}, the
strategy adopted in the evaluation procedure was the simultaneous fit
of transmission and capture yield data measured on the natural sample
performed at ORELA~\cite{Guber:2011}. The capture resonance widths
fitted to neutron capture yield data on natural sample indirectly
inferred the proper magnitude of the \nuc{53}{Cr}$(n,\gamma)$ cross
sections. This was possible also because the cluster of $s$-wave
resonances is populated only by levels of the two analyzed isotopes,
\nuc{50,53}{Cr}. On the other side, the simultaneous fit together with
thin-
and thick-target transmission data measured on natural sample still
revealed the issue with the level misalignment most evident for the
resonance energy level at 4.1~keV as shown in
Figs~\ref{fig:natcr-guber}--\ref{fig:natcr-thick-guber}. Here, in
addition to the current evaluation (in black solid line), the yield
and transmission data calculated from the resonance parameters
reported in the ENDF/B-VIII.0 library are shown by
light blue lines.


In the case of neutron capture yield calculations as well as the related
fitting procedure, the finite size option in the SAMMY code was adopted.
By using the set of resonance parameters fitted to the natural
measured data, it could be possible to compare the calculated yield data to the measurements of 
Stieglitz \etal  and Guber \etal performed with enriched
\nuc{53}{Cr} sample. However, due to the discrepancies for both oxide
sample configurations between Monte Carlo simulations and analytical
approximations, the related measured data were excluded from the fit
of the resonance widths.

Table~\ref{table:ResonanceWidths} reports the n+\nuc{53}{Cr} energy
levels and related capture and neutron widths for selected nuclear
data libraries and Reich-Moore $R$-matrix analyses. As expected, ENDF/B-VIII.0
values of the resonance capture widths are systematically smaller than
other sets. As shown in Fig.\ref{fig:53cr-guber-stieg-diff}, we can
only guess these values were fitted in infinite slab approximation to
data from Guber \etal measured on the enriched oxide sample although the
agreement is not perfect. Although the interplay with the neutron
widths is an important factor,  capture widths from Stieglitz \etal are
consistent with the current evaluation for the 8~keV level, however,
they seem to be overestimated for the 5 and 6~keV levels. The
differences between the current evaluation and the widths reported
 by Stieglitz \etal can be understood in differences in both the quantification of
the multiple scattering corrections and the possible slight
disagreement between  oxide enriched data  from Stieglitz \etal and  metal
data from Guber \etal as displayed in Figs.~\ref{fig:guber_metal_mltsc}
and~\ref{fig:stieglitz_oxide_mltsc}.

\begin{table}
\caption{\label{table:ResonanceWidths} \nuc{53}{Cr} energy levels and related
  capture and neutron widths reported in selected nuclear data libraries and
  $R$-matrix analyses.}

\begin{tabular}{cccc}
\toprule\toprule
& $E$~(keV) & $\Gamma_{\gamma}$~(eV) & $\Gamma_{n}$~(eV)  \\
\midrule
\multirow{4}{*}{ENDF/B-VIII.0}                  & 4.25 & 2.07 & 1360 \\
                                                & 5.67 & 0.62 & 231  \\
                                                & 6.79 & 1.72 & 1183 \\
                                                & 8.21 & 1.91 & 1139 \\
\hline
\multirow{4}{*}{Stieglitz~\cite{Stieglitz:1971}}& 4.18 & 3.23 & 1520 \\
                                                & 5.67 & 1.23 &  220 \\
                                                & 6.74 & 5.28 & 1200 \\
                                                & 8.18 & 3.25 & 1030 \\
\hline
\multirow{4}{*}{Current                        }& 4.31 & 5.04 & 1408 \\
                                                & 5.68 & 0.79 &  202 \\
                                                & 6.78 & 4.14 &  956 \\
                                                & 8.19 & 3.35 & 1313 \\
\bottomrule
\bottomrule
\end{tabular}
\end{table}

In
Tabs.~\ref{table:natCrIntegralValues}--~\ref{table:54CrIntegralValues}
the thermal constant values and resonance integrals (RI) of the
natural Chromium and \nuc{50-54}{Cr} isotopes are reported for several evaluated and
measured data. As expected, the major updates are related to the
\nuc{50,53}{Cr} isotopes. The thermal capture on natural chromium increased by 0.11 barn, 
a 0.3\% increase driven by changes in \nuc{53}{Cr} still within the uncertainties of the 
Mughabghab Atlas 2017 value. However, the capture RI of the natural chromium increased 
by 9\% to 1.67 b, but it is still low compared to the recommended Atlas value of $2.1\pm0.2$ barn.

Although the analysis for \nuc{50}{Cr} needs to be updated due to the multiple scattering 
corrections, the current analysis shows a relatively large increase in the scattering cross
section and a slight decrease in the capture cross sections keeping
the total thermal value almost unchanged. However, the \nuc{50}{Cr} RI increased to 7.73~b
by 7\% compared to the Atlas value of 7.2~b in better agreement with 
Kayzero data-base value~\cite{kayzero}.  

The \nuc{53}{Cr} isotope features an increase of around 7\% (13\%) for the thermal
capture cross section (RI) with respect to ENDF/B-VIII.0 library and the Atlas
compilation. The values of the BROND-3.1 library~\cite{BROND-3.1} are clearly
overestimated due to the assumption used in the fitting procedure of
the capture cross sections.

\begin{table}
\caption{\label{table:natCrIntegralValues} \nuc{nat}{Cr} resonance region integral metrics. Atlas results marked with ``exp'' refer to the Mughabghab's evaluation based on the experimental data, while  Atlas values derived from calculations are marked with ``calc''.  `` * '' indicates the measurement is actually a measurement of (n,abs) which must equal (n,$\gamma$) at thermal energies.
}
\begin{tabular}{r|ccc}
\toprule \toprule
$\sigma_{\textrm{therm}}$ (b)					  & $(n,$tot)	    & $(n,$el)	      & $(n,\gamma)$	   \\
\midrule
This work                                         & 6.71            & 3.48            & 3.23 \\
ENDF/B-VIII.0 \cite{Brown:2018}					  & 6.58            & 3.46            & 3.12 \\
Atlas (2018) \cite{Atlas2018}					  &                 & 3.38 $\pm$ 0.01 & 3.17 $\pm$ 0.08 \\
BROND-3.1~\cite{BROND-3.1}					      & 6.83            & 3.36            & 3.47       \\
\midrule
(1935) Dunning \cite{PhysRev.48.265}              & 4.9             &                 &  \\
(1937) Goldhaber \cite{Goldhaber1937}             &                 & 3.6             &  \\
(1946) Coltman \cite{PhysRev.69.411}              &                 &                 & 2.5	$\pm$ 0.05  \\
(1950) Harris* \cite{PhysRev.80.342}              &                 &                 & 3.05 $\pm$ 0.15 \\
(1950) Colmer* \cite{Colmer1950}                  &                 &                 & 3.1 $\pm$ 0.2   \\
(1951) Grimeland \cite{Grimeland1951}             &                 &                 & 2.88 $\pm$ 0.15 \\
(1951) Pomerance \cite{PhysRev.83.641}            &	                &                 & 2.83 $\pm$ 0.14 \\
(1953) Melkonian \cite{PhysRev.92.702}            & 7.3             &                 & \\
(1956) Prokhorov \cite{Prokhorov1956}             &                 &                 & 3.2	$\pm$ 0.1   \\
(1973) Salama \cite{Salama1973}                   & 6.47 $\pm 0.33$ &                 & \\
(1976) Kropff \cite{Kropff1976}                   & 6.50 $\pm$ 0.14 &                 & 3.25 $\pm$ 0.15  \\
\bottomrule
\bottomrule
\end{tabular}
\vspace{0.25cm}

\begin{tabular}{r|ccc}
\toprule
\toprule
RI (b)						                & $(n,$tot)	      & $(n,$el)	    & $(n,\gamma)$	  \\
\midrule
This work                                   & 90.79           & 86.05           & 1.67      \\
ENDF/B-VIII.0 \cite{Brown:2018}	            & 89.64           & 84.96           & 1.53  \\
Atlas (2018) \cite{Atlas2018}	            & 			      &			        & 2.1 $\pm$ 0.2   \\
BROND-3.1~\cite{BROND-3.1}					& 89.65           & 84.85           & 2.02       \\
\midrule
(1955) Macklin \cite{NSR1955MAZZ}	        &	              &                 & 1.9   \\
(1956) Prokhorov \cite{Prokhorov1956}       &                 &                 & 1.9 $\pm$ 0.4 \\
(1957) Klimentov* \cite{SJA.3.1387.1957}    &                 &                 & 2.6 $\pm$ 1.1 \\
(1964) Kapchigashev \cite{NSR1964KA33}      &                 &                 & 1.5 $\pm$ 0.1 \\
\bottomrule
\bottomrule
\end{tabular}
\end{table}


\begin{table}
\caption{\label{table:50CrIntegralValues} $^{50}$Cr resonance region integral metrics.
Atlas results marked with ``exp'' refer to the Mughabghab's evaluation based on the experimental data, while  Atlas values derived from calculations are marked with ``calc''.
All major libraries except JENDL-4.0 used the same ORNL RRR from Ref.~\cite{jkps.59.1644}.
}
\begin{tabular}{r|ccc}
\toprule
\toprule
$\sigma_{\textrm{therm}}$ (b)					  & $(n,$tot)	    & $(n,$el)	     & $(n,\gamma)$	   \\
\midrule
This work                                         & 18.06           & 2.85           & 15.21  \\
ENDF/B-VIII.0 \cite{Brown:2018}					  & 17.81           & 2.40           & 15.42  \\
BROND-3.1~\cite{BROND-3.1}					      & 17.81           & 2.40           & 15.42    \\
Atlas (2018) \cite{Atlas2018}					  & 17.1 $\pm$ 0.4  &	 	         & 14.7 $\pm$ 0.4  \\
k0 (2019) \cite{kayzero}                          &                 &                & 15.1 $\pm$ 0.8 \\
\midrule
(1947) Seren \cite{PhysRev.72.888}                &	                &                & 11.0 $\pm$ 4.4 \\
(1952) Pomerance \cite{PhysRev.88.412}            &	                &                & 16.3 $\pm$ 1.3 \\
(1960) Lyon \cite{Lyon1960}                       &                 &                & 16.6 $\pm$ 1.7 \\
(1968) Sims \cite{SIMS1968349}	                  &	                &                & 15.9 $\pm$ 0.1 \\
(1975) Gleason \cite{Gleason1975}                 &	                &                & 15.8 $\pm$ 0.2 \\
(1978) Heft	\cite{1978HEZL}                       &	                &                & 16.1 $\pm$ 0.1 \\
(1978) Koester \cite{Koester1978}                 &                 & 2.41 $\pm$ 0.06& \\
(1984) Simonits \cite{Simonits1984}               &                 &                & 15.2	$\pm$ 0.2\\
(1988) De Corte \cite{DeCorte1988}                &                 &                & 14.9 $\pm$ 0.36\\
(1997) Venturini \cite{VENTURINI1997493}          &                 &                & 14.2 $\pm$ 0.5 \\
(2014) Arbocco \cite{Arbocco2014}                 &                 &                & 14.41 $\pm$ 0.13\\
\bottomrule
\bottomrule
\end{tabular}
\vspace{0.25cm}

\begin{tabular}{r|ccc}
\toprule
\toprule
RI (b)						                & $(n,$tot)	      & $(n,$el)	    & $(n,\gamma)$	\\
\midrule
This work                                   & 273.13          & 261.97          & 7.73          \\
ENDF/B-VIII.0 \cite{Brown:2018}	            & 259.73          & 249.02          & 7.24          \\
BROND-3.1~\cite{BROND-3.1}				    & 259.73          & 249.02          & 7.24          \\
Atlas (2018) \cite{Atlas2018}	            & 			      &			        & 7.2 $\pm$ 0.2 \\
k0 (2019) \cite{kayzero}                    &                 &                 & 8 $\pm$ 2.5   \\
\midrule
(1968) Sims \cite{SIMS1968349}	            &	              &                 & 10.4 $\pm$ 0.4  \\
(1971) De Corte \cite{DeCorte1971}          &                 &                 & 8.3 \\
(1972) Van Der Linden \cite{VanderLinden1972}&                &                 & 7.8 $\pm$ 0.4   \\
(1975) Gleason \cite{Gleason1975}           &	              &                 & 9.0 $\pm$ 0.5   \\
(1976) Sage \cite{Sage1976}                 &	              &                 & 12.5 $\pm$ 3.24 \\
(1978) Heft	\cite{1978HEZL}                 &	              &                 & 10.5 $\pm$ 0.4  \\
\bottomrule
\bottomrule
\end{tabular}
\vspace{0.25cm}

\begin{tabular}{l|ccc}
\toprule
\toprule
MACS(30 keV) (mb) 								& $(n,\gamma)$   \\
\midrule
This work                                       & 45.02          \\
ENDF/B-VIII.0 \cite{Brown:2018}  				& 38.25  \\
JENDL-4.0 \cite{JENDL4.0}						& 36.95  \\
BROND-3.1~\cite{BROND-3.1}					        & 38.25      \\
Atlas (exp) (2018) \cite{Atlas2018}				& 40.3 $\pm$ 4.0 \\
Atlas (calc) (2018) \cite{Atlas2018}			& 37.3 $\pm$ 2.7 \\
\kadonis-0.3  \cite{Kadonis0.3} 			    & 49.0 $\pm$ 13.0 \\
\midrule
(1971) Allen \cite{AGM71}                       & 31 $\pm$ 4 \\
(1974) Beer \cite{J.AAA.37.197.1974}, (1975) Beer \cite{Beer:1975lf} & 50 $\pm$ 15 \\
(1977) Kenny \cite{NSR1977KEZF}                 & 52 $\pm$ 13 \\
\bottomrule
\bottomrule
\end{tabular}
\end{table}



\begin{table}
\caption{\label{table:integralMetrics52Cr} \nuc{52}{Cr} resonance region integral metrics.
Atlas results marked with ``exp'' refer to the Mughabghab's evaluation based on the experimental data, while  Atlas values derived from calculations are marked with ``calc''.
All major libraries used the same ORNL RRR from  Ref.~\cite{jkps.59.1644}.
}
\begin{tabular}{r|ccc}
\toprule
\toprule
$\sigma_{\textrm{therm}}$ (b)					  & $(n,$tot)	    & $(n,$el)	      & $(n,\gamma)$	\\
\midrule
This work                                         & 3.905           & 3.047           & 0.857             \\
ENDF/B-VIII.0 \cite{Brown:2018}					  & 3.905           & 3.047           & 0.857           \\
BROND-3.1~\cite{BROND-3.1}					          & 3.687           & 2.926           & 0.761      \\
Atlas (2018) \cite{Atlas2018}					  & 3.64 $\pm$ 0.03 & 2.78 $\pm$ 0.02 & 0.86 $\pm$ 0.02 \\
\midrule
(1952) Pomerance \cite{PhysRev.88.412}            &                 &                 & 0.73 $\pm$ 0.06 \\
(1997) Venturini \cite{VENTURINI1997493}          &                 &                 & 0.86 $\pm$ 0.03 \\
\bottomrule
\bottomrule
\end{tabular}
\vspace{0.25cm}

\begin{tabular}{r|ccc}
\toprule
\toprule
RI (b)						                & $(n,$tot)	      & $(n,$el)	    & $(n,\gamma)$	 \\
\midrule
This work                                   & 56.05           & 52.52           & 0.494       \\
ENDF/B-VIII.0 \cite{Brown:2018}             & 56.05           & 52.52           & 0.494          \\
BROND-3.1~\cite{BROND-3.1}					    & 55.90           & 52.53           & 0.761       \\
Atlas (2018) \cite{Atlas2018}	            & 			      &			        & 0.66           \\
\bottomrule
\bottomrule
\end{tabular}
\vspace{0.25cm}

\begin{tabular}{l|ccc}
\toprule
\toprule
MACS(30 keV) (mb) 								& $(n,\gamma)$    \\
\midrule
This work                                       & 7.991       \\
ENDF/B-VIII.0 \cite{Brown:2018}					& 7.991           \\
BROND-3.1~\cite{BROND-3.1}					        & 8.976           \\
Atlas (exp) (2018) \cite{Atlas2018}				& 9.1 $\pm$ 2.1   \\
Atlas (calc) (2018) \cite{Atlas2018}			& 9.1 $\pm$ 0.6   \\
\kadonis-0.3  \cite{Kadonis0.3} 			    & 8.8 $\pm$ 2.3   \\
\midrule
(1971) Allen \cite{AGM71}                       & 3.8 $\pm$ 1.0 \\ 
(1974) Beer \cite{J.AAA.37.197.1974}, (1975) Beer \cite{Beer:1975lf}  & 8.3 $\pm$ 3.0 \\
(1977) Kenny  \cite{NSR1977KEZF}                & 9.8 $\pm$ 2.3   \\
(1989) Rohr \cite{Rohr:1989}              & 8.79 $\pm$ 2.30 \\
\bottomrule
\bottomrule
\end{tabular}
\end{table}


\begin{table}
\caption{\label{table:53CrIntegralValues} \nuc{53}{Cr} resonance region integral metrics.
Atlas results marked with ``exp'' refer to the Mughabghab's evaluation based on the experimental data, while  Atlas values derived from calculations are marked with ``calc''.
All major libraries used the same ORNL RRR from  Ref.~\cite{jkps.59.1644}.
}
\begin{tabular}{r|ccc}
\toprule
\toprule
$\sigma_{\textrm{therm}}$ (b)					  & $(n,$tot)	    & $(n,$el)	     & $(n,\gamma)$   \\
\midrule
This work                                         & 27.17           & 7.78           & 19.40           \\
ENDF/B-VIII.0 \cite{Brown:2018}					  & 25.94           & 7.82           & 18.11  \\
BROND-3.1~\cite{BROND-3.1}					          & 30.51           & 7.82           & 22.69     \\
Atlas (2018) \cite{Atlas2018}					  & 25.9 $\pm$ 0.7  & 7.3 $\pm$ 0.2  & 18.6 $\pm$ 0.6 \\
\midrule
(1952) Pomerance \cite{PhysRev.88.412}            &	                &                & 17.5 $\pm$ 1.4 \\
(1997) Venturini \cite{VENTURINI1997493}          &                 &                & 18.6 $\pm$ 0.6 \\
\bottomrule
\bottomrule
\end{tabular}
\vspace{0.25cm}

\begin{tabular}{r|ccc}
\toprule
\toprule
RI (b)						                & $(n,$tot)	      & $(n,$el)	    & $(n,\gamma)$	  \\
\midrule
This work                                   & 323.71          & 311.01          & 9.59   \\
ENDF/B-VIII.0 \cite{Brown:2018}	            & 317.76          & 305.52          & 8.43    \\
BROND-3.1~\cite{BROND-3.1}					    & 319.27          & 304.29          & 11.19       \\
Atlas (2018) \cite{Atlas2018}	            & 			      &			        & 8.66 $\pm$ 0.40 \\
\midrule
(1950) Harris \cite{PhysRev.79.11}          &	              & 78.               &  \\
\bottomrule
\bottomrule
\end{tabular}
\vspace{0.25cm}

\begin{tabular}{l|ccc}
\toprule
\toprule
MACS(30 keV) (mb) 								& $(n,\gamma)$    \\
\midrule
This work                                       & 51.73     \\
ENDF/B-VIII.0 \cite{Brown:2018}					& 25.94   \\
JENDL-4.0 \cite{JENDL4.0}						& 26.17   \\
BROND-3.1~\cite{BROND-3.1}					        & 48.99    \\
Atlas (exp) (2018) \cite{Atlas2018}				& 54.0 $\pm$ 9.4  \\
Atlas (calc) (2018) \cite{Atlas2018}			& 36.0 $\pm$ 5.0  \\
\kadonis-0.3  \cite{Kadonis0.3} 			    & 58.0 $\pm$ 10.0 \\
\midrule
(1971) Allen  \cite{AGM71}                      & 40 $\pm$ 5  \\
(1974) Beer \cite{J.AAA.37.197.1974}, (1975) Beer \cite{Beer:1975lf}  & 61 $\pm$ 24  \\
(1977) Kenny \cite{NSR1977KEZF}                 & 58 $\pm$ 10  \\
\bottomrule
\bottomrule
\end{tabular}
\end{table}



\begin{table}
\caption{\label{table:54CrIntegralValues} \nuc{54}{Cr} resonance region integral metrics.
Atlas results marked with ``exp'' refer to the Mughabghab's evaluation based on the experimental data, while
Atlas values derived from calculations are marked with ``calc''.
All major libraries except JENDL-4.0 used the same ORNL RRR from  Ref.~\cite{jkps.59.1644}.
}
\begin{tabular}{r|ccc}
\toprule
\toprule
$\sigma_{\textrm{therm}}$ (b)					  & $(n,$tot)	    & $(n,$el)	      & $(n,\gamma)$    \\
\midrule
This work                                         & 2.94            & 2.52            & 0.41         \\
ENDF/B-VIII.0 \cite{Brown:2018}					  & 2.93            & 2.52            & 0.41            \\
BROND-3.1~\cite{BROND-3.1}					          & 2.93            & 2.52            & 0.41          \\
Atlas (2018) \cite{Atlas2018}					  & 2.95 $\pm$ 0.11 & 2.54 $\pm$ 0.20 & 0.41 $\pm$ 0.04 \\
\midrule
(1952) Pomerance \cite{PhysRev.88.412}            &	                &                 & $<$ 0.3 \\
(1954) Bazorgan \cite{PhysRev.95.781}             &                 &                 & 0.36 $\pm$ 0.04 \\
(1972) White \cite{White1972}                     &                 &                 & 0.34 $\pm$ 0.04 \\
\bottomrule
\bottomrule
\end{tabular}
\vspace{0.25cm}

\begin{tabular}{r|ccc}
\toprule
\toprule
RI (b)						                & $(n,$tot)	      & $(n,$el)	    & $(n,\gamma)$	  \\
\midrule
This work                                   & 50.98           & 47.24           & 0.21  \\
ENDF/B-VIII.0 \cite{Brown:2018}	            & 50.50           & 46.70           & 0.21            \\
BROND-3.1~\cite{BROND-3.1}					    & 50.50           & 46.70           & 0.21         \\
Atlas (2018) \cite{Atlas2018}	            & 			      &			        & 0.20 $\pm$ 0.04 \\
\bottomrule
\bottomrule
\end{tabular}
\vspace{0.25cm}

\begin{tabular}{l|ccc}
\toprule
\toprule
MACS(30 keV) (mb) 								& $(n,\gamma)$   \\
\midrule
This work                                       & 4.78          \\
ENDF/B-VIII.0 	\cite{Brown:2018}				& 4.78           \\
JENDL-4.0 \cite{JENDL4.0}						& 4.69   \\
BROND-3.1~\cite{BROND-3.1}					        & 4.78         \\
Atlas (exp) (2018) \cite{Atlas2018}				& 6.2 $\pm$ 1.5  \\
Atlas (calc) (2018) \cite{Atlas2018}			& 5.3 $\pm$ 1.0  \\
\kadonis-0.3  \cite{Kadonis0.3} 			    & 6.7 $\pm$ 1.6  \\
\midrule
(1971) Allen  \cite{AGM71}                      & 4.5 \\
(1977) Kenny \cite{NSR1977KEZF}                 & 6.7 $\pm$ 1.6  \\
\bottomrule
\bottomrule
\end{tabular}
\end{table}

\subsection{Direct-Semidirect (DSD) neutron capture cross sections}
\label{sec:DSD}

Direct-semidirect (DSD) neutron capture cross sections on \nuc{50,52,53,54}{Cr} were computed using a single-particle direct-semidirect capture model implemented in computer code CUPIDO that was originally developed to calculate fluctuation effects in radiative capture in the giant dipole resonance energy region \cite{PARKER1995}.  The CUPIDO code has also been used for computation of direct-semidirect capture  in the thermal neutron and resolved resonance energy ranges \cite{ARBANAS2005}, in conjunction with R-matrix nuclear data evaluations  using the code SAMMY \cite{LARSON2008}.

The original calculations of DSD capture used for this evaluation were performed in 2008 using then available version of CUPIDO, which did not yet have the energy dependence of the Koning-Delaroche optical model potential (OMP) parameters programmed into it. Consequently, the Koning-Delaroche OMP was first translated into a linear energy-dependent OMP (with 1 keV incident neutron energy as a reference point), and this OMP was then used as input for DSD calculations by CUPIDO.  CUPIDO calculations using this OMP yielded DSD thermal neutron capture cross sections that were anomalously large. In fact, they were found to be much larger than the measured total thermal neutron capture of chromium isotopes.  This anomalously large DSD capture was interpreted as a manifestation of a potential scattering divergence caused by the $s-$wave ``resonance'' effect that is inherent to single-particle potential models \cite{MEYER1976}.   In the mass range spanned by the four stable chromium isotopes, the said $s$-wave resonance is due to the energy of the $3s$ single-particle neutron state in the nuclear potential being nearly equal to the neuron binding energy.

\begin{figure}[htbp]
\begin{center}
\includegraphics[width=0.50\textwidth,clip,trim = 10mm 5mm 8mm 31mm]{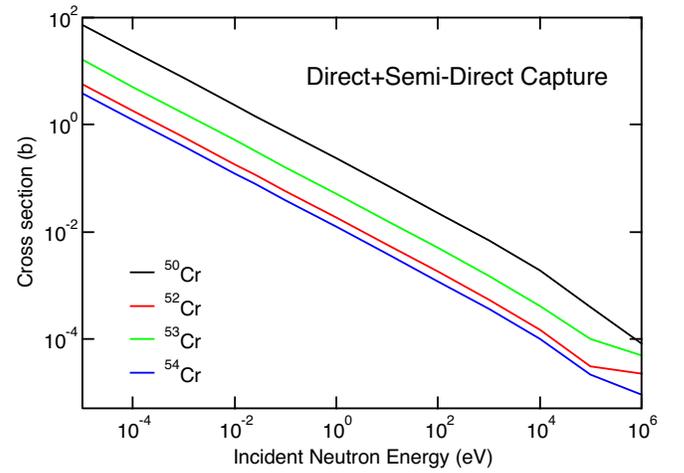}
\caption{(color online) Direct-semidirect (DSD) capture cross sections for all stable chromium isotopes used in this evaluation.}
\label{fig:DSD}
\end{center}
\end{figure}

These anomalously large DSD thermal neutron cross sections were regularized by adjusting the single-particle Woods-Saxon optical potential well depth, so that the potential elastic scattering length ($R^{\prime}$) of the adjusted well depth would fall within the uncertainty range of the values reported in the Mughabghab's Atlas \cite{Atlas2018}.  In particular, we have adjusted the depth of the real volume part of the Woods-Saxon potential for each chromium isotope.  Modified parameter values used to calculate the reported DSD are shown in Table \ref{table:V0Rprime}.  These adjusted values of the real volume potential depth were then used for calculations of DSD capture by the CUPIDO code, and the calculated thermal neutron DSD capture was then found to be reasonable and generally smaller than the measured total thermal capture cross sections of chromium isotopes.  DSD neutron capture cross sections were computed in this way using the code CUPIDO between 0.01 meV and 1 MeV incoming neutron energy, and are shown in Fig.~\ref{fig:DSD}.

For calculations of captured neutron's final bound state wave functions, the nuclear potential used by CUPIDO for all of the reported calculations in this work is the one of Bear and Hodgson \cite{Bear_1978}.  In our single particle model of DSD capture into bound states of known binding energies, we have adopted a conventional approximation in which the bound state potential depth is varied to reproduce the binding energy of each bound state, e.g. \cite{Krausmann1995}\footnote{An alternative convention is to use a bound state potential consistent with the OMP parameterization \cite{Xu2012}.}. Consequently, in this work the potential well depth of this potential was fitted by CUPIDO to reproduce the reported bound state energy of each bound state, which were supplied to CUPIDO as input parameters.  Finally, amplitudes of DSD capture are computed as expectation values of an electromagnetic operator sandwiched between the incoming neutron wave function in the continuum and the captured neutron bound state.  All DSD calculations reported in this work were performed using a current form of EM operator \cite{PARKER1995}.

The dominant contribution to the DSD capture for low energy neutrons is due to electric dipole (E1) transitions of neutrons in the incoming $s$-wave function.  Therefore the most important capturing bound states in the corresponding (A+1) chromium isotopes are those that could be reached by s-wave neutrons via an E1 transition.  The spin-coupling algebra of quantum mechanics then implies that the bound states contributing the majority of DSD capture are those with orbital angular momentum value of 1, i.e. the bound $p$-levels.  Consequently, we have collected published spectroscopic data of bound $p$-levels of the respective (A+1) chromium isotopes, as deduced from (d,p) reactions, or from other neutron-depositing transfer reactions.  (For $^{52,53}$Cr, $l=$ 0 and 2 bound states were used in addition to $l=1$ bound states.)  Neutron single-particle spectroscopic factors of $^{51,53,54,55}$Cr were taken from Refs.~\cite{DAVID1969, XIAOLONG2006}, \cite{KOPECKY1980, BEENE1978}, \cite{BROWN1970, JUNDE2006}, and \cite{MACGREGOR1972, JUNDE2008}, respectively.

The DSD capture cross sections computed in this way were then used as input to the SAMMY code, where the (measured) total capture cross section was fitted as the sum of this DSD capture cross section and the resonant capture cross section computed by SAMMY using the Reich-Moore approximation.  The inherent relative uncertainties of the DSD capture cross sections due to its model defect, spectroscopic factor uncertainties, etc., are approximately 20\%.

\begin{figure}[htbp]
\begin{center}
\includegraphics[width=0.50\textwidth,clip,trim = 10mm 5mm 8mm 31mm]{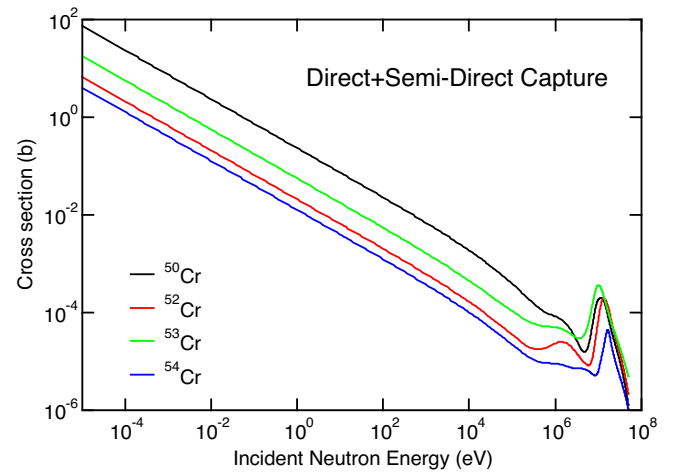}
\caption{(color online) Direct-semidirect (DSD) cross sections for all stable chromium isotopes computed using the latest version of the CUPIDO code using a genuine Koning-Delaroche OMP parameterization is found to be well within the experimental and the model uncertainties of the original calculations plotted in Fig.~\ref{fig:DSD}, and consequently the original DSD calculations were retained for the current evaluations.  The new DSD calculations, including the large semidirect contribution due to giant dipole resonance in the 15 MeV range will be utilized when more accurate capture data become available in the future.}
\label{fig:DSD_new}
\end{center}
\end{figure}

Calculations of DSD capture were repeated recently using the newer version of the CUPIDO code that has a genuine Koning-Delaroche OMP parameterization built into it. These recent DSD calculations were found to be at most few percent apart from the original CUPIDO calculations, i.e., well within the experimental uncertainties, and therefore we have simply retained the original calculations for this evaluation.  For completeness, the DSD capture computed using the Koning-Delaroche parameterization in the latest version of the CUPIDO code is plotted in Fig.~\ref{fig:DSD_new}, over a broader energy range showing the semidirect capture via giant dipole resonance in the 15 MeV range.

\begin{table}
\caption{\label{table:V0Rprime}  The strength of real volume part of the effective Woods-Saxon potential (at low energy) listed in column $V_V$, were adjusted to a corresponding value in column $V'_V$ in order to make the potential scattering length $R'$ of the modified potential fall within the uncertainty reported in Atlas in column $R'_{\textrm{Atlas}}$ \cite{Atlas2018}. (No such modifications were needed for calculations of DSD capture on $^{50}$Cr.)}
\begin{tabular}{c|cccc}
\toprule
\toprule
Cr mass number					  &    $V_V$[MeV] 	    & $V'_V$[MeV] 	      & $R'$[fm]   &  $R'_{\textrm{Atlas}}$[fm] \\
\midrule
52 & 52.5 & 53.2 & 5.23 & 5.4 $\pm$ 0.3 \\
53 & 52.2 & 52.7 & 5.41 & 5.64 $\pm$ 0.3 \\
54 & 51.9 & 52.0 & 5.35 & 5.5 $\pm$ 0.3 \\
\bottomrule
\bottomrule
\end{tabular}
\vspace{0.25cm}
\end{table}

%
%
%
%

\subsection{Angular distributions in the RRR}
\label{sec:ang_dist_RR}

In the resolved resonance region, the angular distributions for all chromium isotopes were reconstructed from resonance parameters using the Blatt-Biedenharn formalism.  This formalism uniquely reconstructs the angular distributions for a given scattering angle from the assigned spins of the resonances, and produces fluctuations in the angular distribution on energy scales somewhat smaller than the energy grid used in the cross section reconstruction and occasionally finer than what can be accommodated by the ENDF 11-column floating point number format.  Therefore the angular distributions were minimally smoothed so that the energy dependence of the angular distributions could be represented as faithfully as possible given the limitations of the ENDF format.

\begin{figure}[htbp]
\begin{center}
\includegraphics[scale=0.41,angle = -90,clip,trim = 31mm 28mm 17mm 46mm]{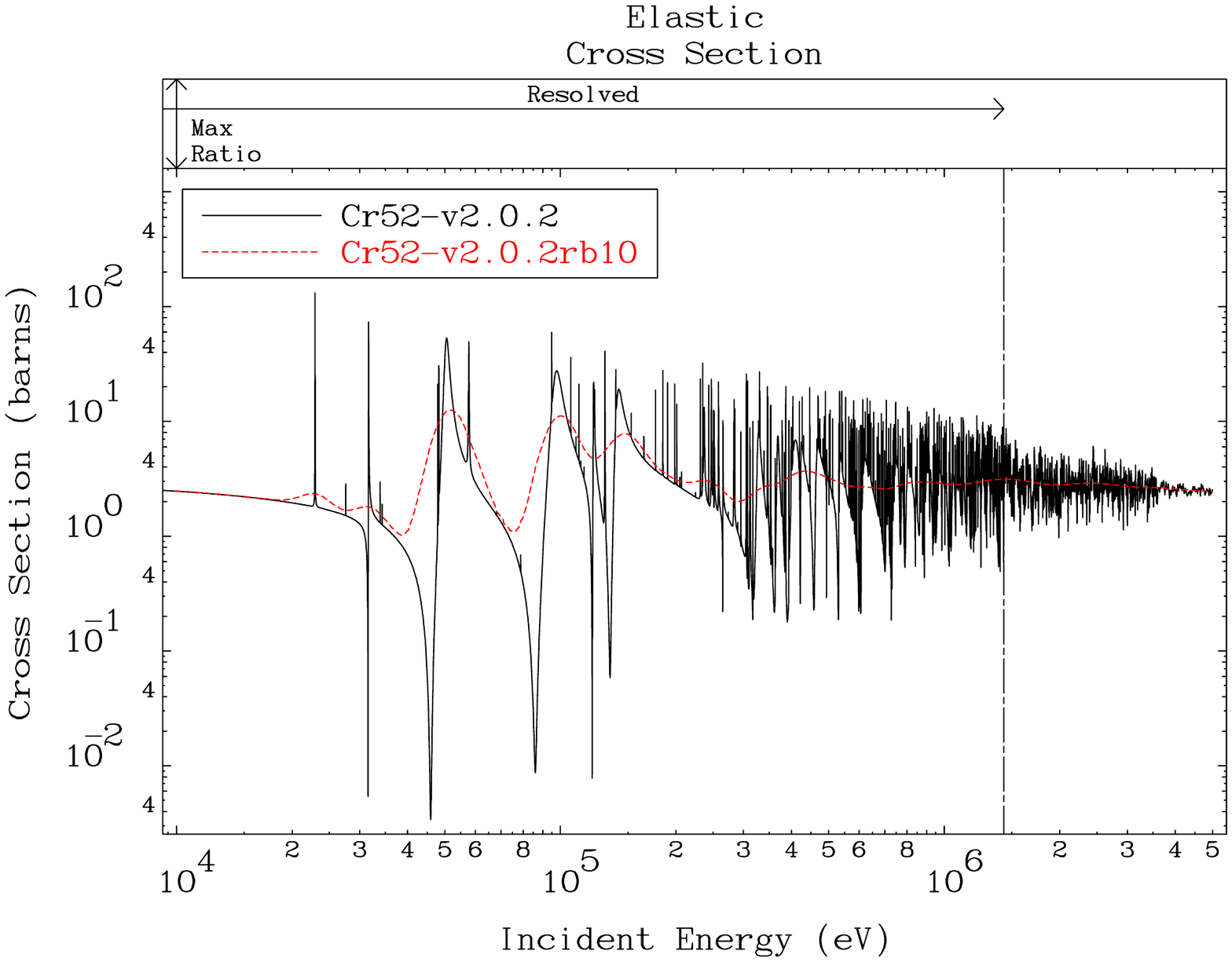} \\ \vspace{0mm}
\includegraphics[scale=0.41,angle = -90,clip,trim = 31mm 28mm 17mm 46mm]{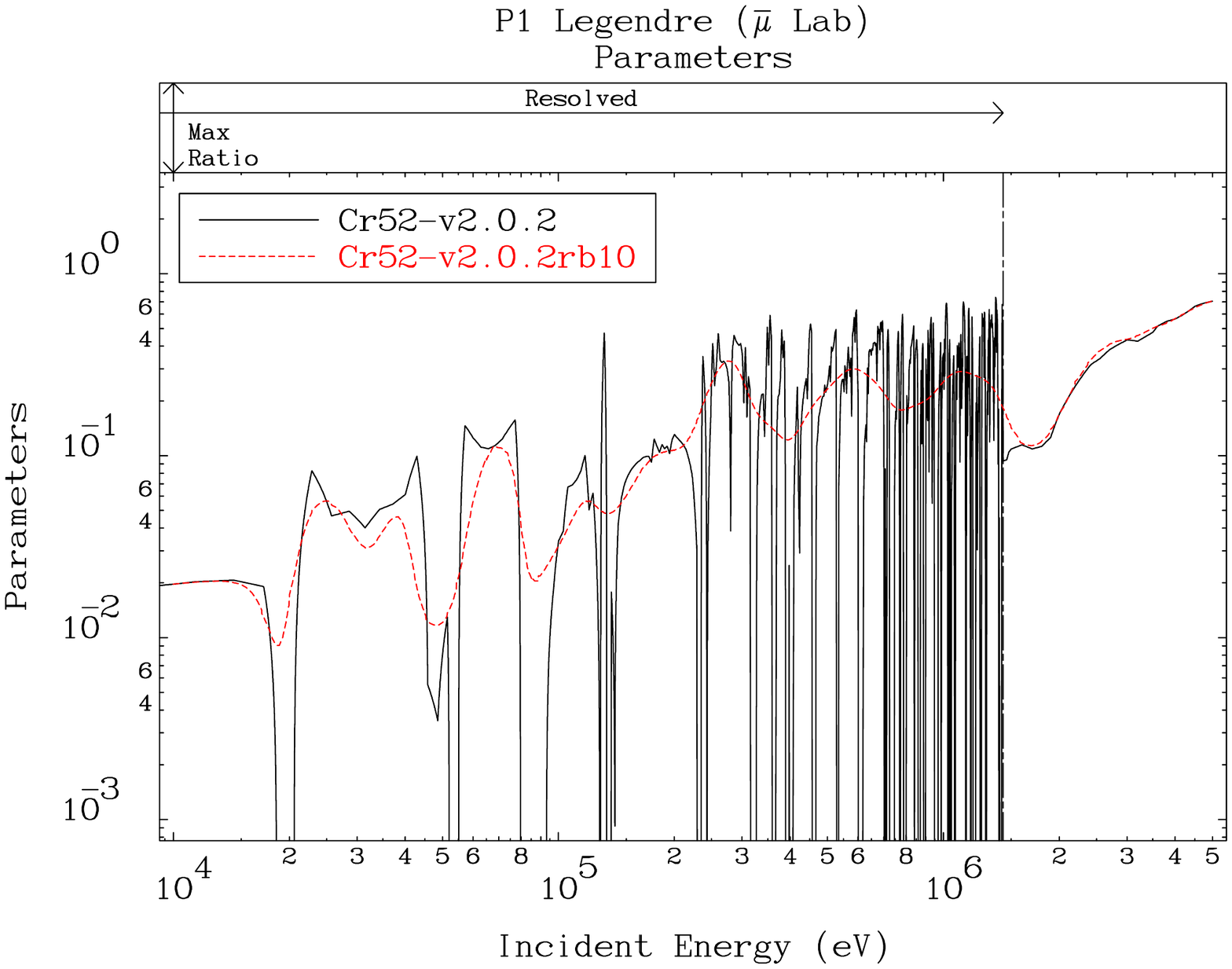} \\ \vspace{0mm}
\includegraphics[scale=0.41,angle = -90,clip,trim = 31mm 28mm 17mm 46mm]{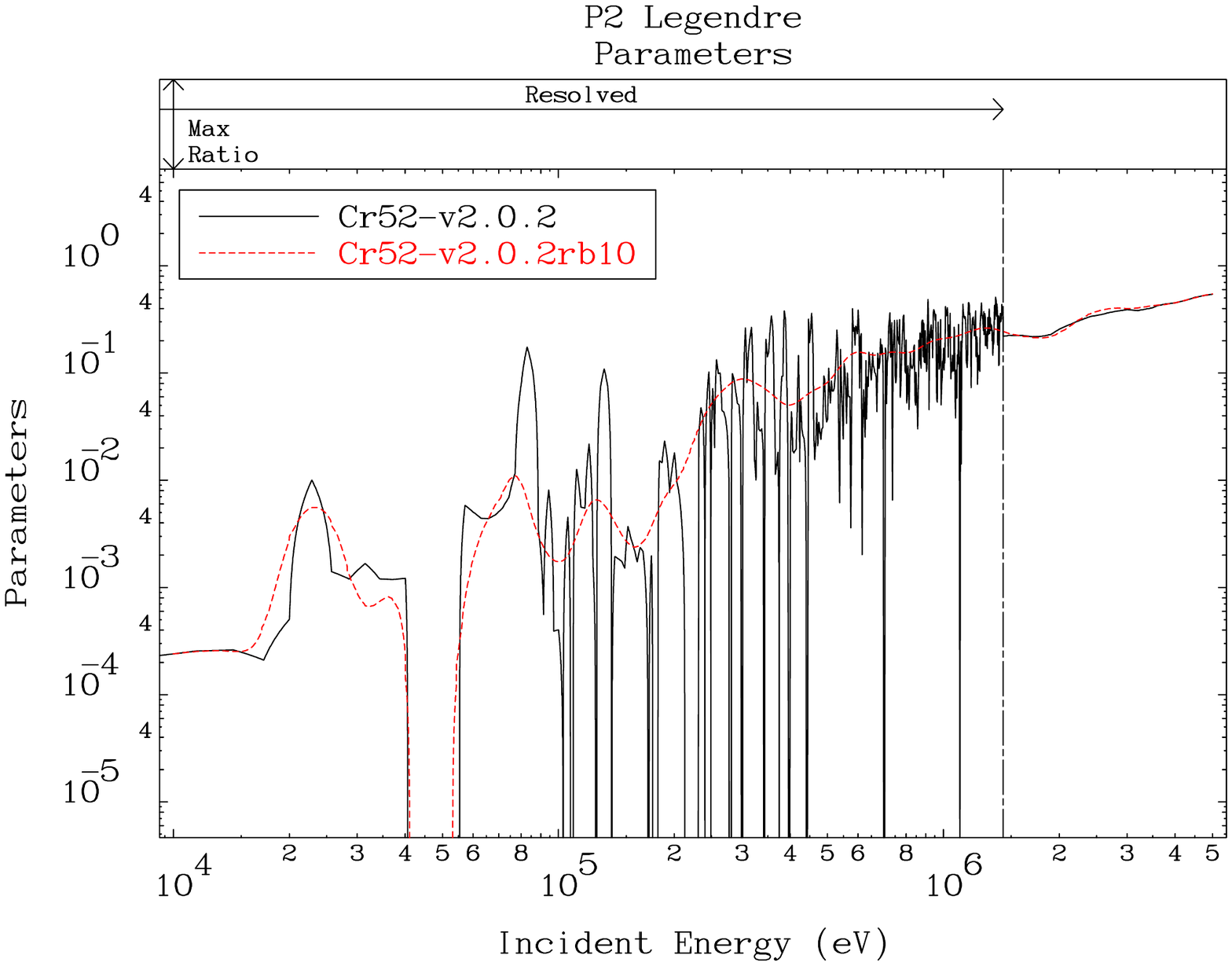}
\caption{(color online) Elastic cross section (top panel) and the $P_1\equiv\bar{\mu}(E)$ coefficient (middle panel) and $P_2\equiv\xi$(E) coefficient (bottom panel) of the Legendre polynomial expansion of the elastic scattering distributions as a function of the incident neutron energy. Angular distributuons on file are shown as black curves, while 10\%--resolution broadened curves are shown in red.}
\label{fig:Resolution_broadening}
\end{center}
\end{figure}

The resulting resonant behavior as a function of the incident energy can be seen in the black curves in Fig.~\ref{fig:Resolution_broadening}, where the elastic cross section (top panel) and the first two coefficients ($P_1\equiv\bar{\mu}(E)$ and $P_2\equiv\xi$(E)) of the Legendre polynomial expansion of the elastic scattering distributions are plotted as a function of the incident neutron energy. Note that the plotted black curves were already smoothed to reduce the size of the file. However, additional resolution broadening of about 10\% should be applied to compare with measured elastic angular distributions in the resonance region. The resolution broadened elastic cross sections and $\bar{\mu}(E)$ and $\xi(E)$ Legendre coefficients are shown as a function of the neutron incident energy in Fig.~\ref{fig:Resolution_broadening} with red lines. Note that significant fluctuations remain in the whole RRR (roughly below 2 MeV) despite the resolution broadening.

\begin{figure*}[p]
\begin{center}
\includegraphics[scale=0.40,clip,trim = 0mm 0mm 0mm 0mm]{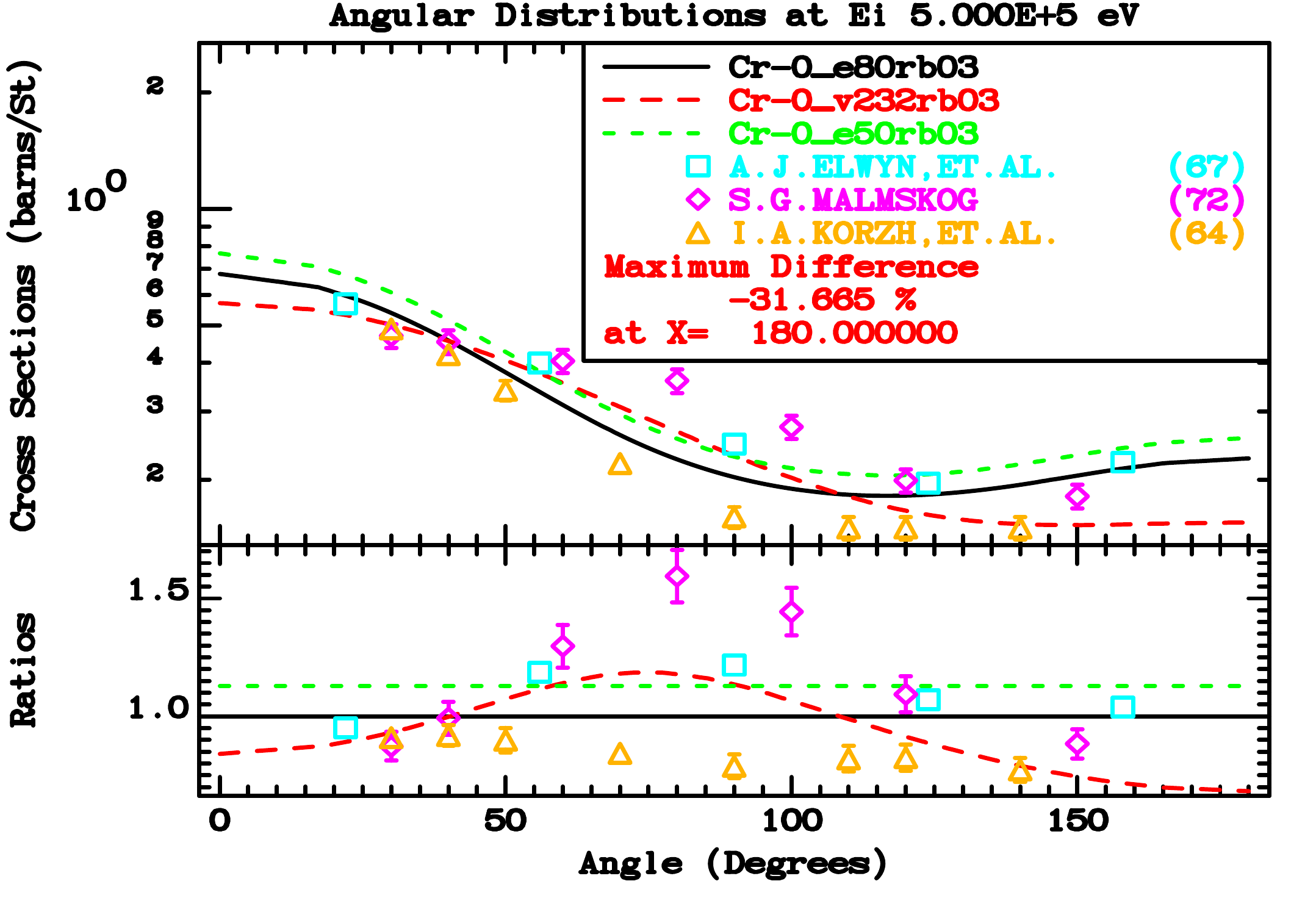}
\includegraphics[scale=0.40,clip,trim = 0mm 0mm 0mm 0mm]{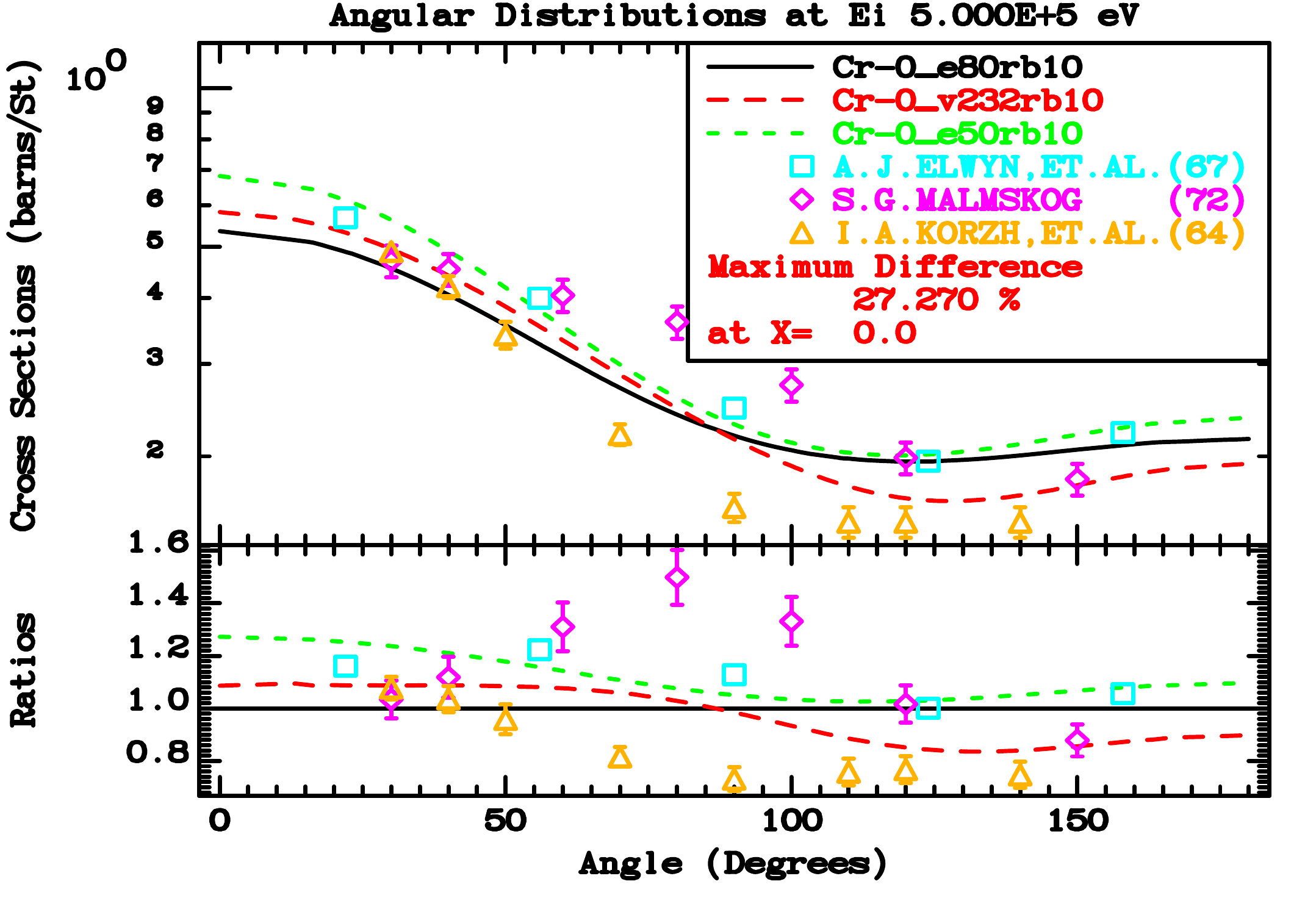} 
\\
\includegraphics[scale=0.40,clip,trim = 0mm 0mm 0mm 0mm]{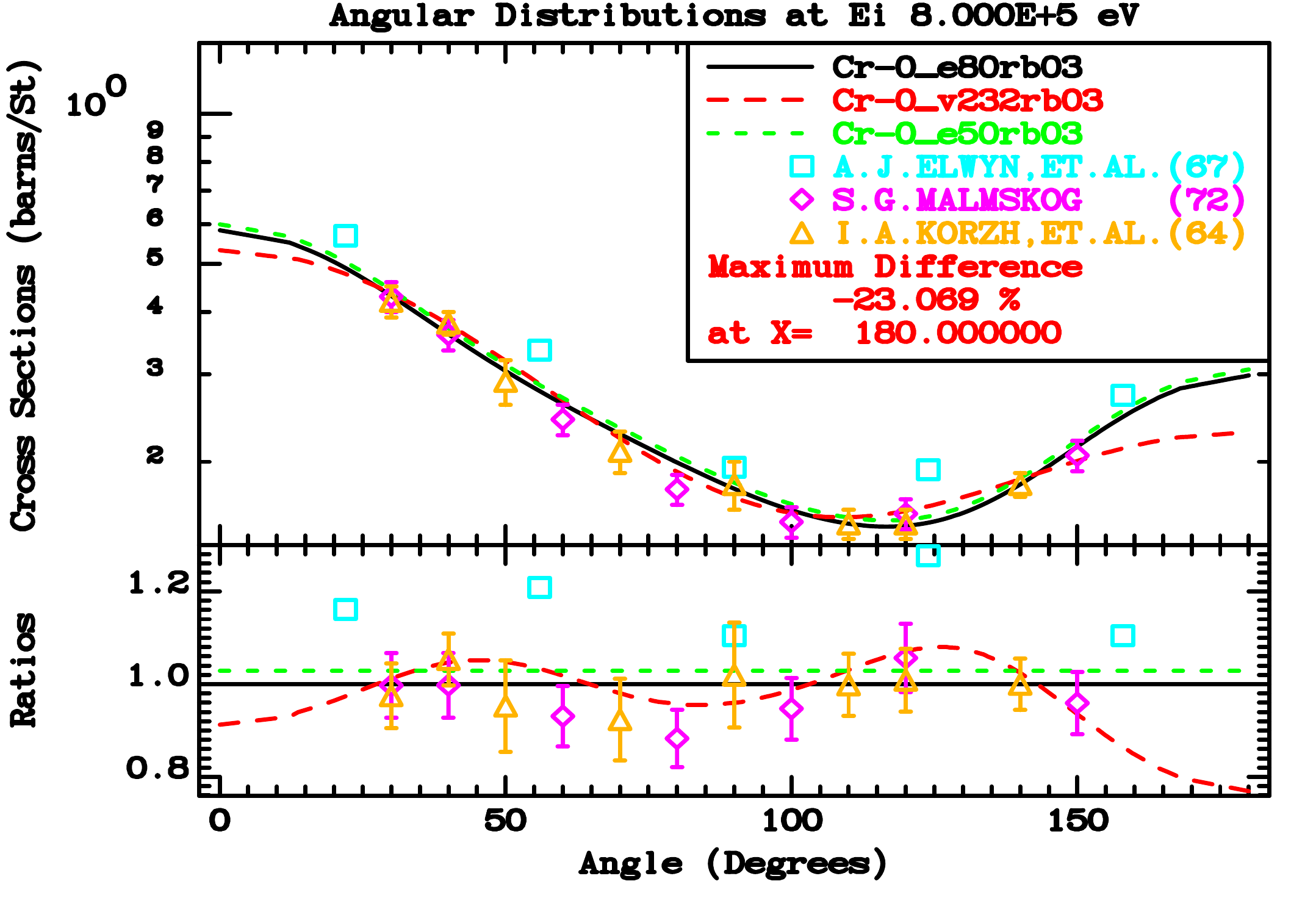}
\includegraphics[scale=0.40,clip,trim = 0mm 0mm 0mm 0mm]{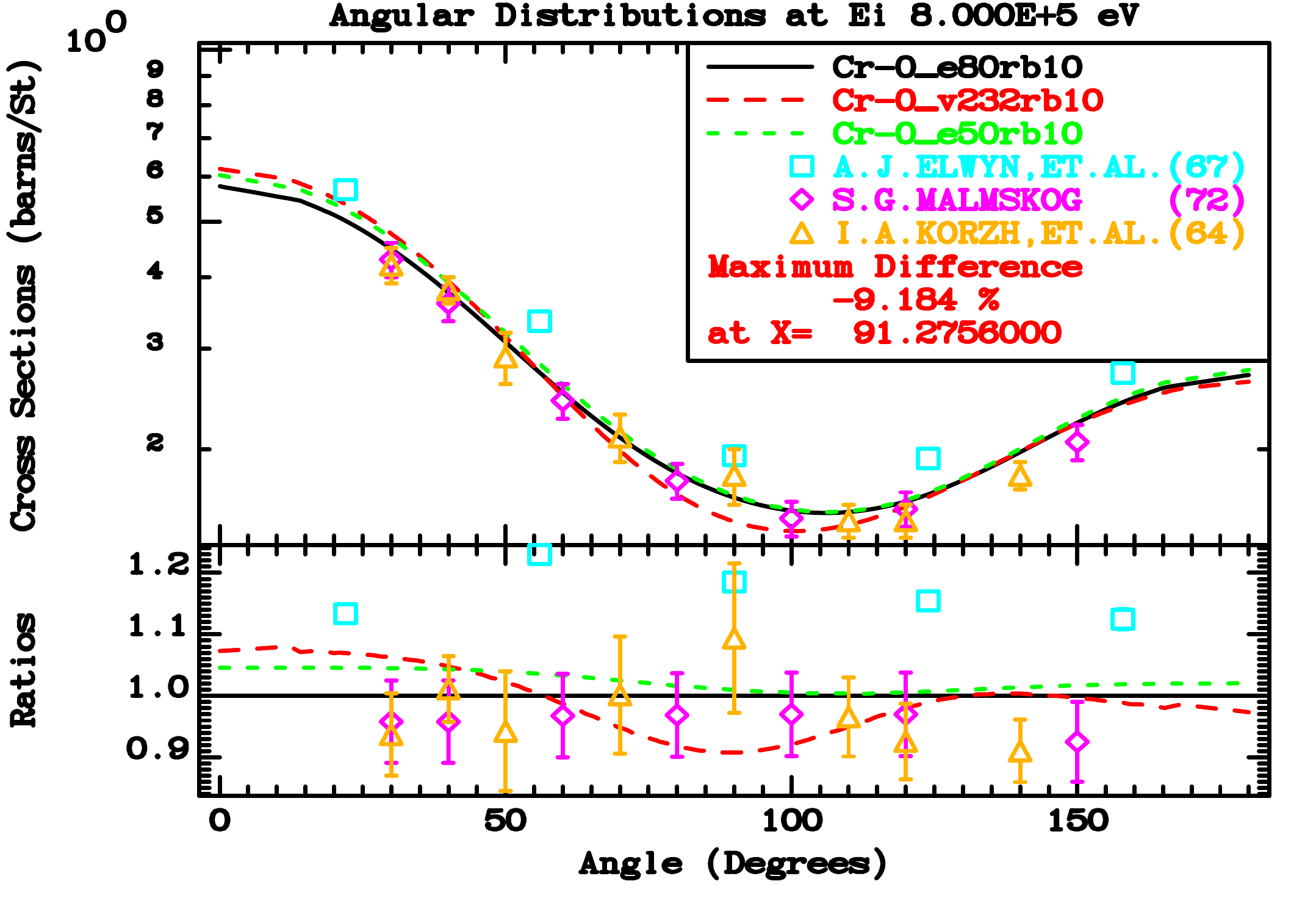} 
\\
\includegraphics[scale=0.40,clip,trim = 0mm 0mm 0mm 0mm]{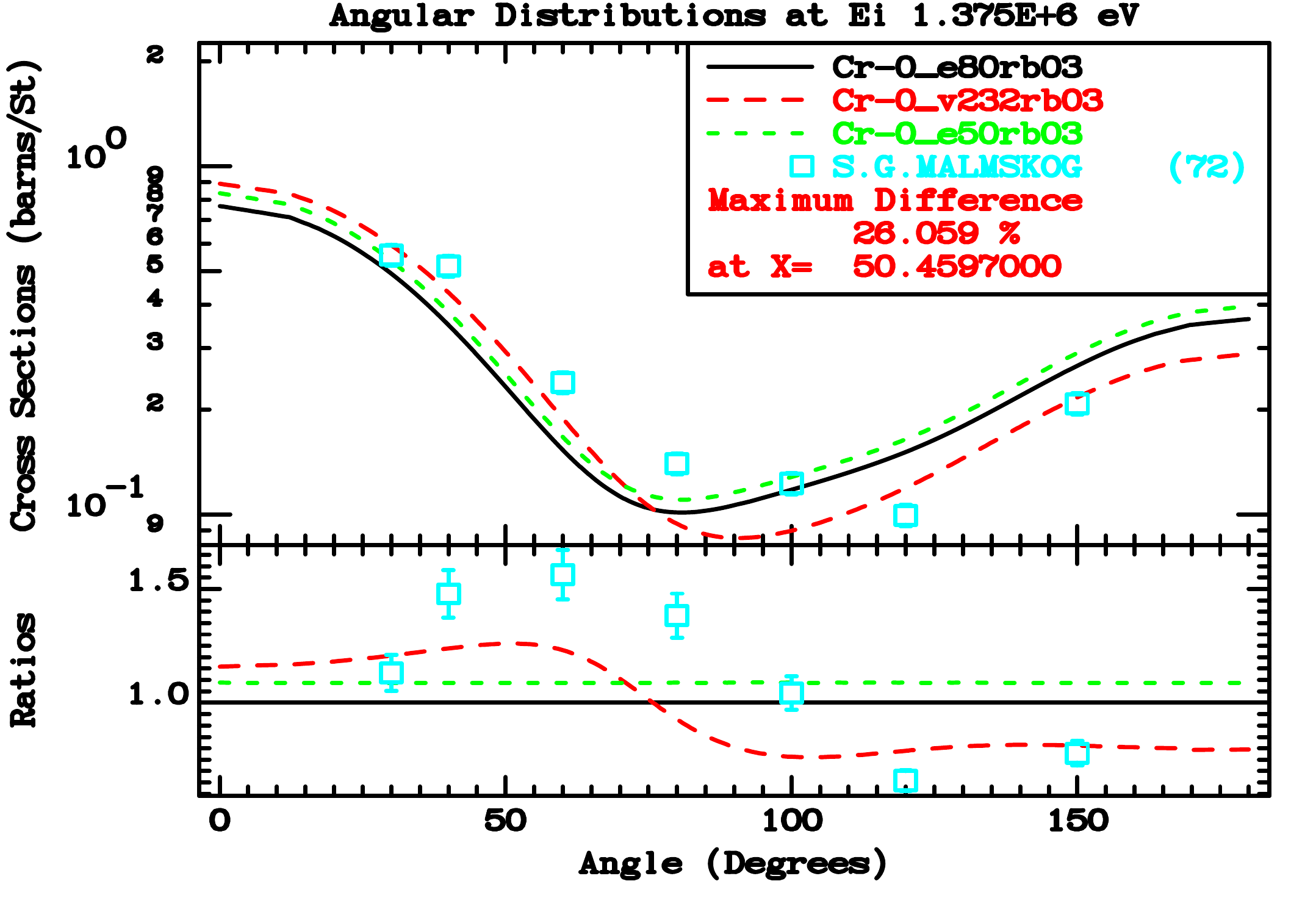}
\includegraphics[scale=0.40,clip,trim = 0mm 0mm 0mm 0mm]{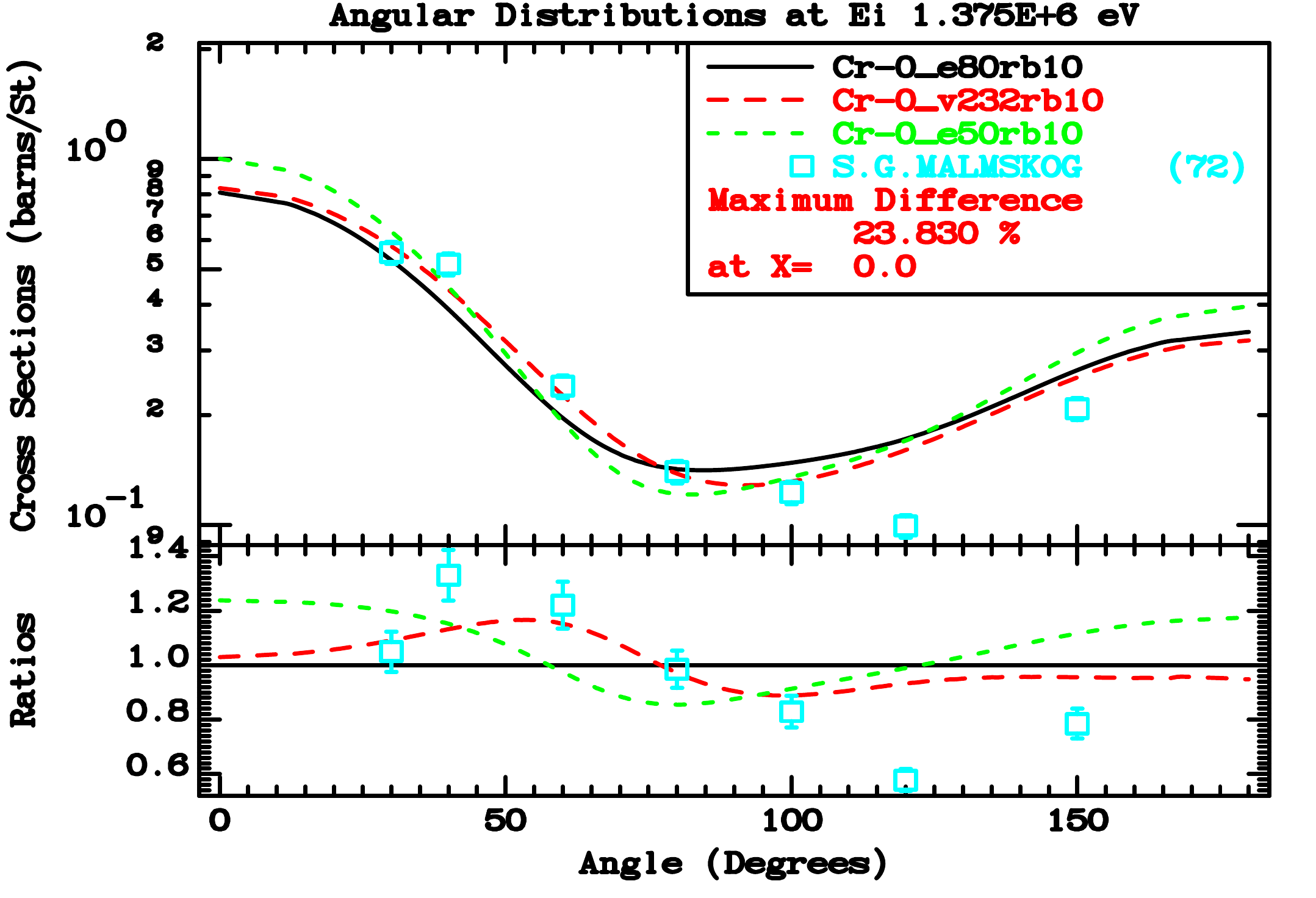} 
\\
\caption{(color online) Angular distributions of elastic scattered neutrons on natural Cr targets at three incident neutron energies of 500 (top), 800 (middle),
and 1375~keV (bottom rows). The left column compares the ENDF/B-V evaluation (black), ENDF/B-VIII.0 evaluation (red dashed) and the current work (green dashed) with 3\% resolution broadening; The right column compares the ENDF/B-VIII.0 evaluation (black) and the current work (red dashed) with 10\% resolution broadening.}
\label{fig:ang_dist_RRR-NatCr}
\end{center}
\end{figure*}

\begin{figure*}[p]
\begin{center}
\includegraphics[scale=0.40,clip,trim = 0mm 0mm 0mm 0mm]{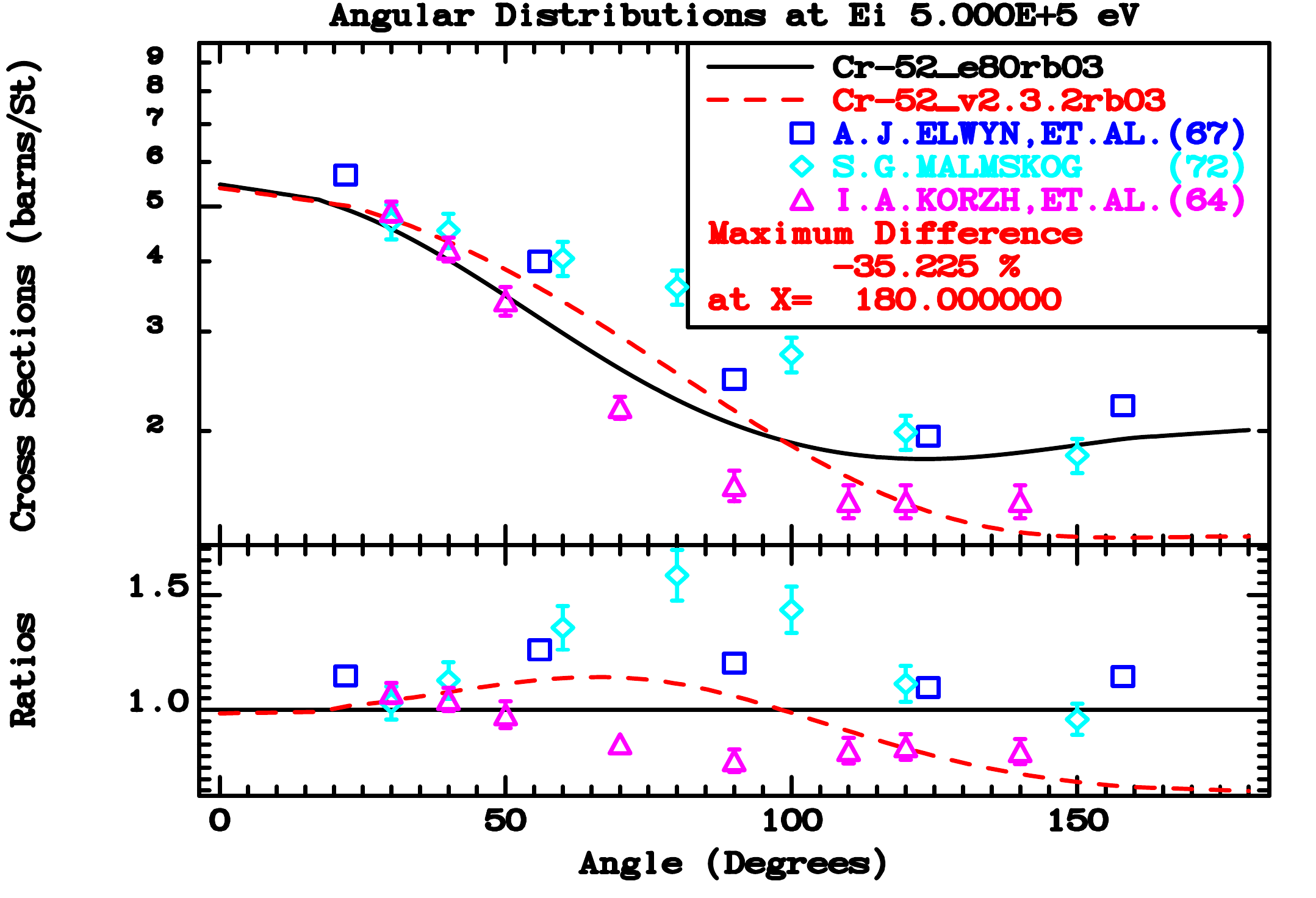}
\includegraphics[scale=0.40,clip,trim = 0mm 0mm 0mm 0mm]{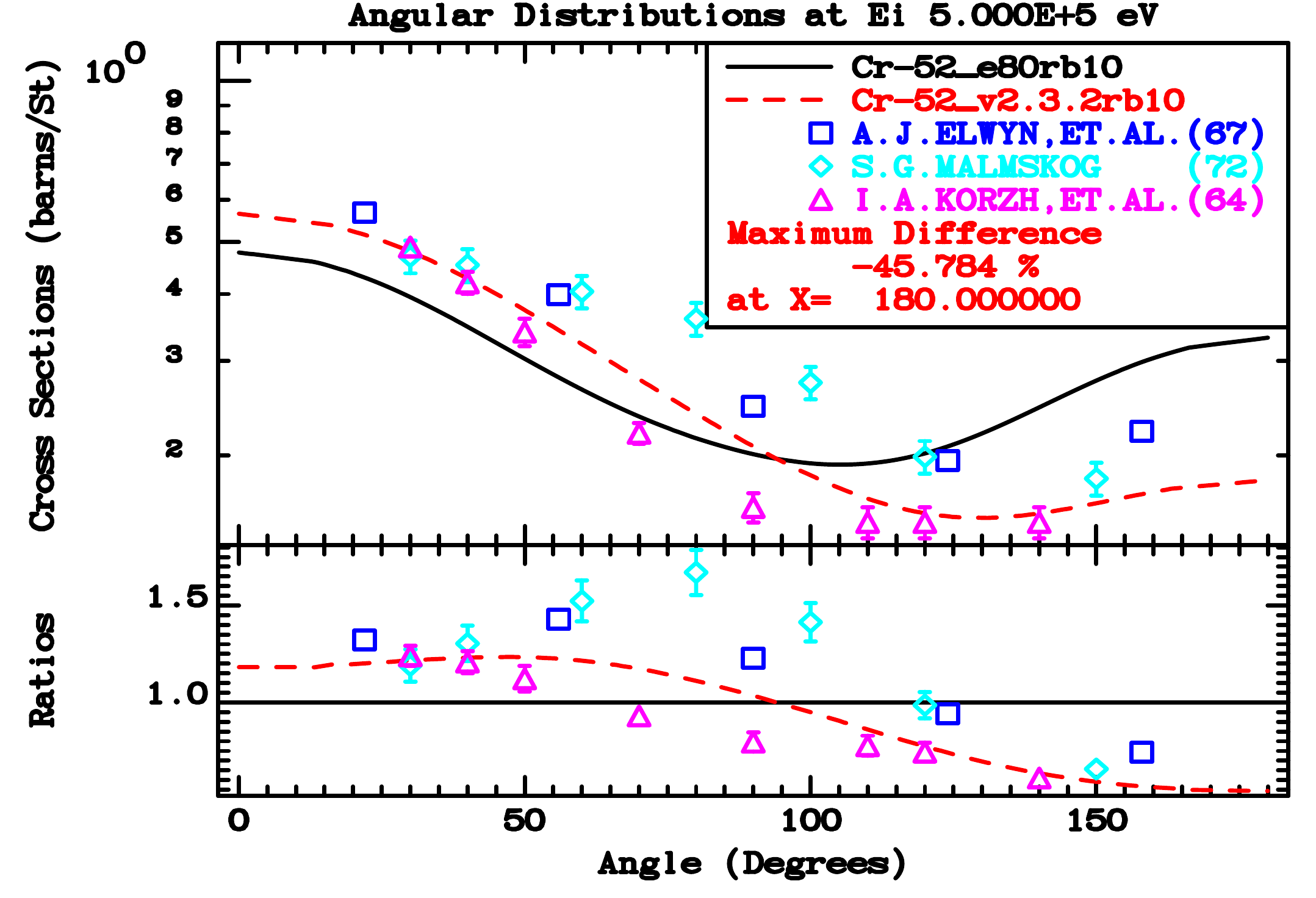} 
\\
\includegraphics[scale=0.40,clip,trim = 0mm 0mm 0mm 0mm]{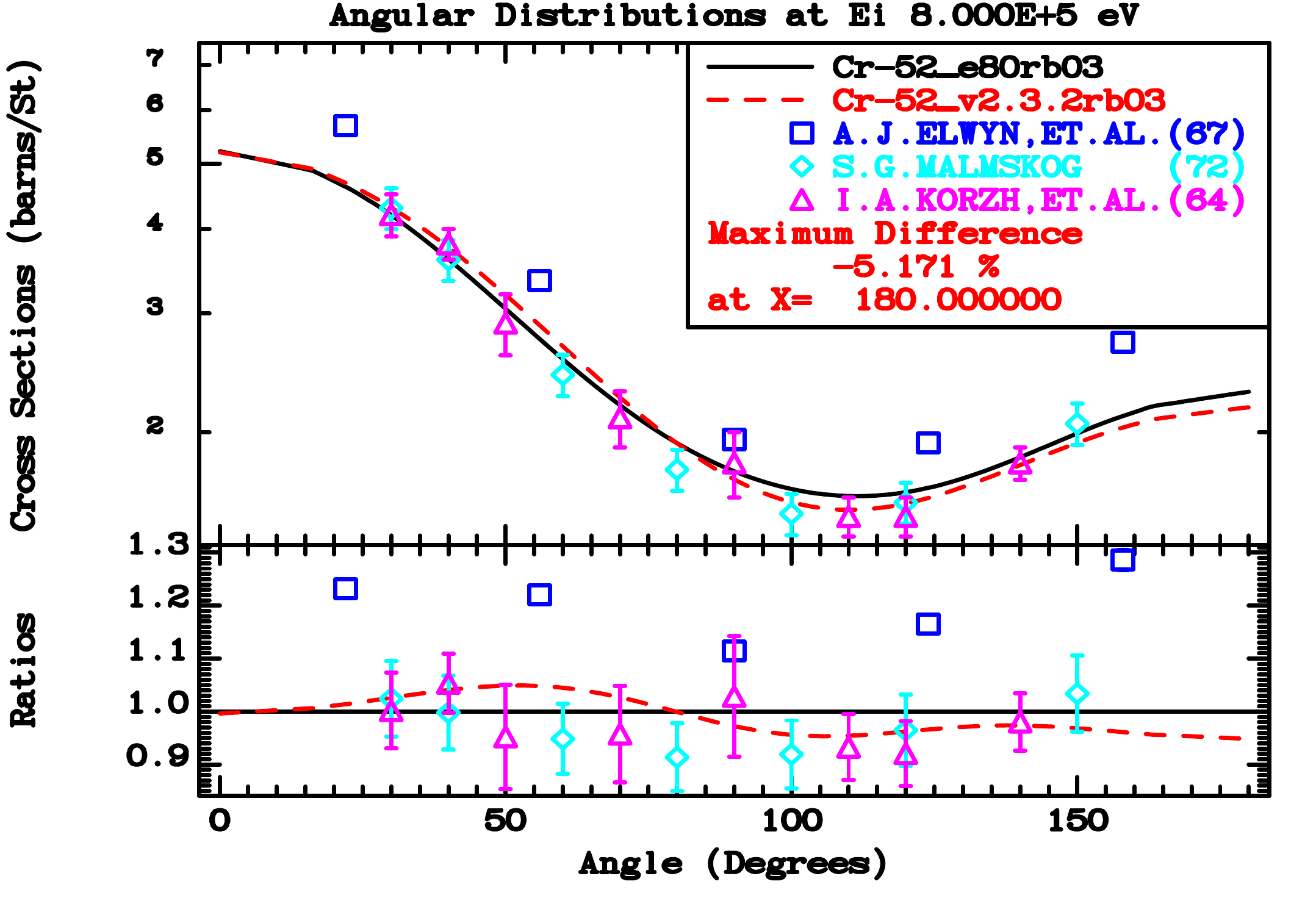}
\includegraphics[scale=0.40,clip,trim = 0mm 0mm 0mm 0mm]{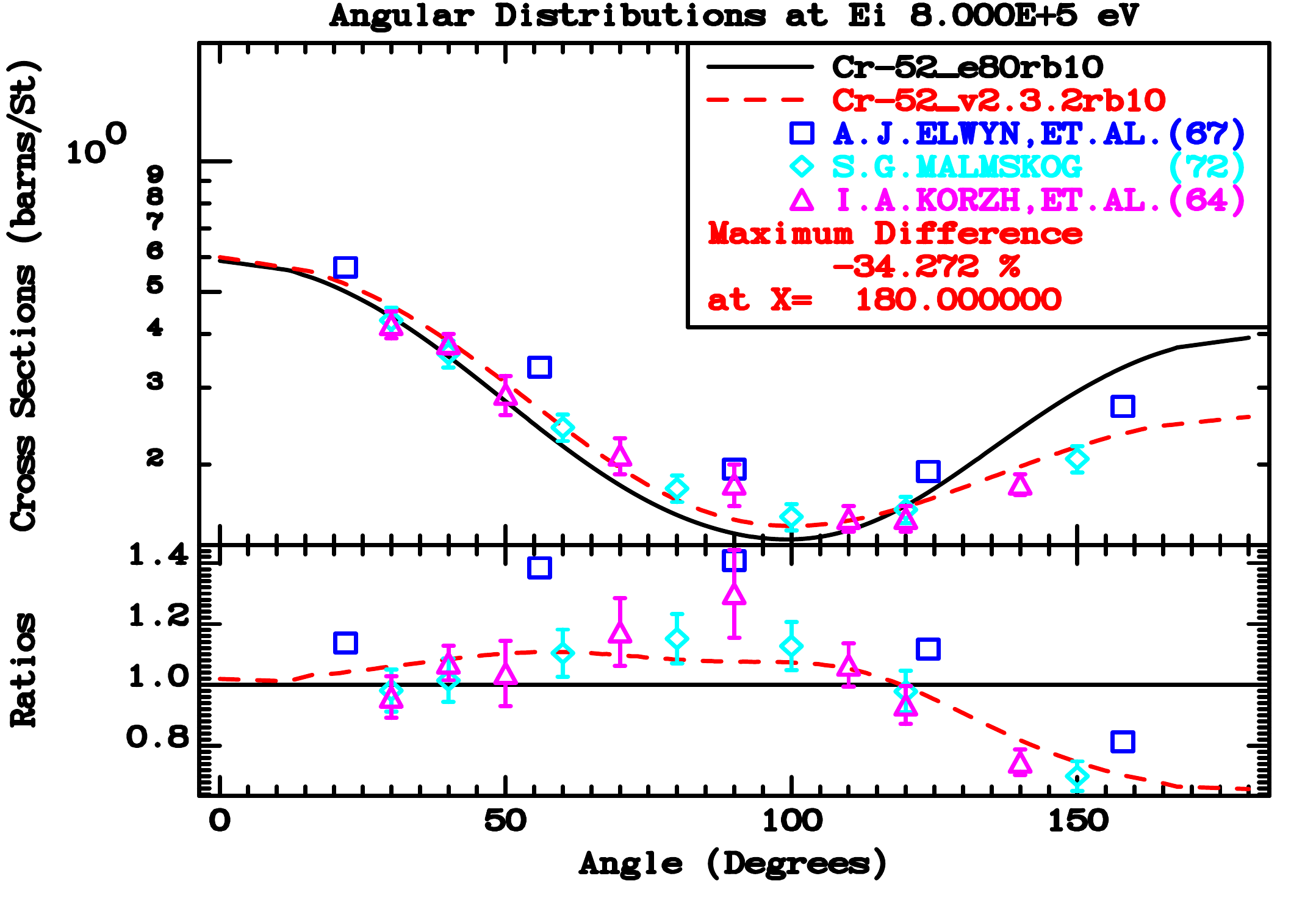} 
\\
\includegraphics[scale=0.40,clip,trim = 0mm 0mm 0mm 0mm]{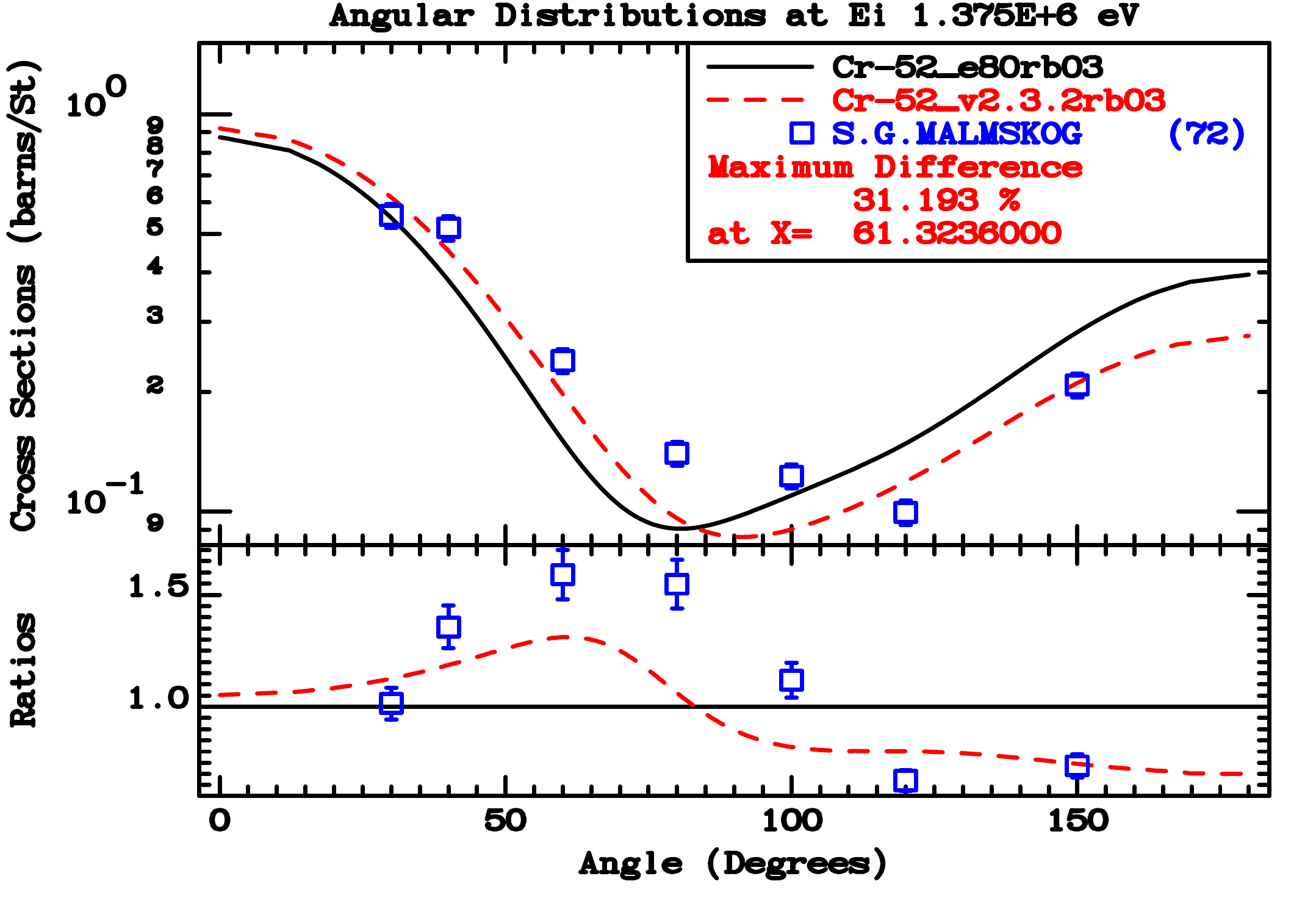}
\includegraphics[scale=0.40,clip,trim = 0mm 0mm 0mm 0mm]{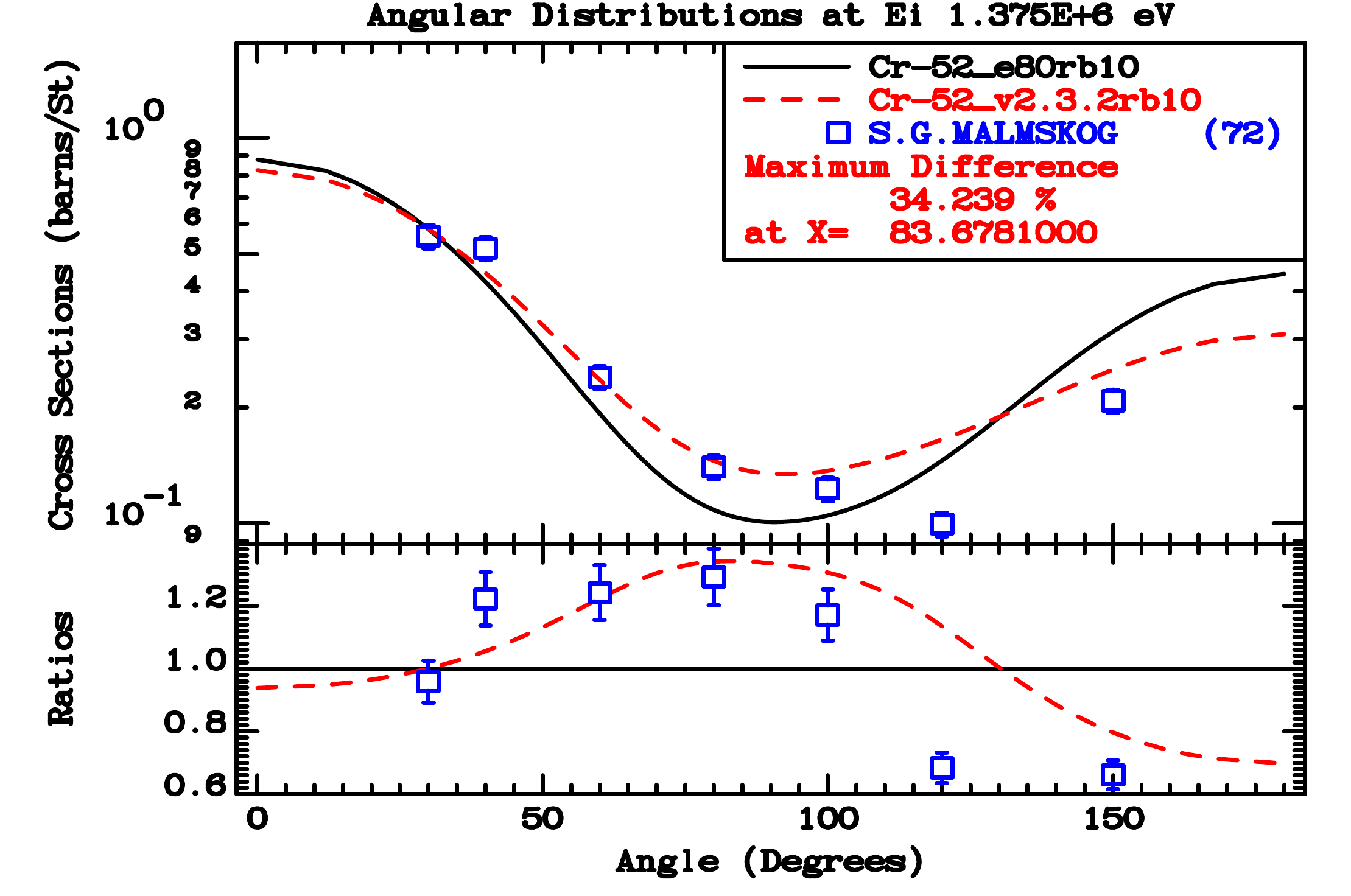} 
\\
\caption{(color online) Angular distributions of elastic scattered neutrons on \nuc{52}{Cr} targets at three incident neutron energies of 500 (top), 800 (middle),
and 1375~keV (bottom rows). The left column compares the ENDF/B-V evaluation (black), ENDF/B-VIII.0 evaluation (red dashed) and the current work (green dashed) with 3\% resolution broadening; The right column compares the ENDF/B-VIII.0 evaluation (black) and the current work (red dashed) with 10\% resolution broadening.}
\label{fig:ang_dist_RRR-52Cr}
\end{center}
\end{figure*}

A comparison of resolution broadened angular distributions in the resonance region with ENDF/B-V, ENDF/B-VIII.0, and the current evaluation is shown in Figs.~\ref{fig:ang_dist_RRR-NatCr} and \ref{fig:ang_dist_RRR-52Cr} for scattering on natural chromium and on \nuc{52}{Cr} major isotope, respectively. Note that the ENDF/B-V evaluation of angular distributions shown in Fig.~\ref{fig:ang_dist_RRR-NatCr} was based on low-resolution experimental data, while the ENDF/B-VIII.0 angular distributions were adopted from the ENDF/B-V evaluation with renormalized elastic cross sections\footnote{Therefore, the shapes are exactly the same, shifted up or down according to the elastic cross section difference between the two evaluations.}. The left column in Figs.~\ref{fig:ang_dist_RRR-NatCr} and \ref{fig:ang_dist_RRR-52Cr} contains the 3\% resolution-broadened curves, while the right column shows the 10\% resolution-broadened curves; three different incident energies of 500, 800, and 1375~keV are shown in the upper, middle, and bottom rows for both cases. All these energies correspond to the RRR of the major \nuc{52}{Cr} isotope. The main point being stressed here is that the comparison of evaluations with low-resolution measured data requires strong smoothing to reproduce the experimental resolution.  It should serve as a warning to avoid comparing low-resolution measured data with reconstructed angular distribution from the resonance parameters without applying resolution broadening to the reconstructed cross sections. 

An improved description of experimental data can be clearly seen in Fig.~\ref{fig:ang_dist_RRR-52Cr} for the 10\% curve (dashed red curve in the right column) compared to the 3\% resolution broadened curve (dashed red curve in the left column) especially when the shape of the ratios is compared at all three analysed energies. Note also that the observed agreement with data both in normalization and shape is good in the whole energy range, and better than what was achieved in previous evaluations (see the bottom row of Fig.~\ref{fig:ang_dist_RRR-52Cr}), especially at energies between 800 and 1375~keV, which correspond to the maximum probability of the fission neutrons.

\begin{figure}[htbp]
\begin{center}
\includegraphics[scale=0.40,angle=-90,clip,trim = 37mm 42mm 35mm 26mm]{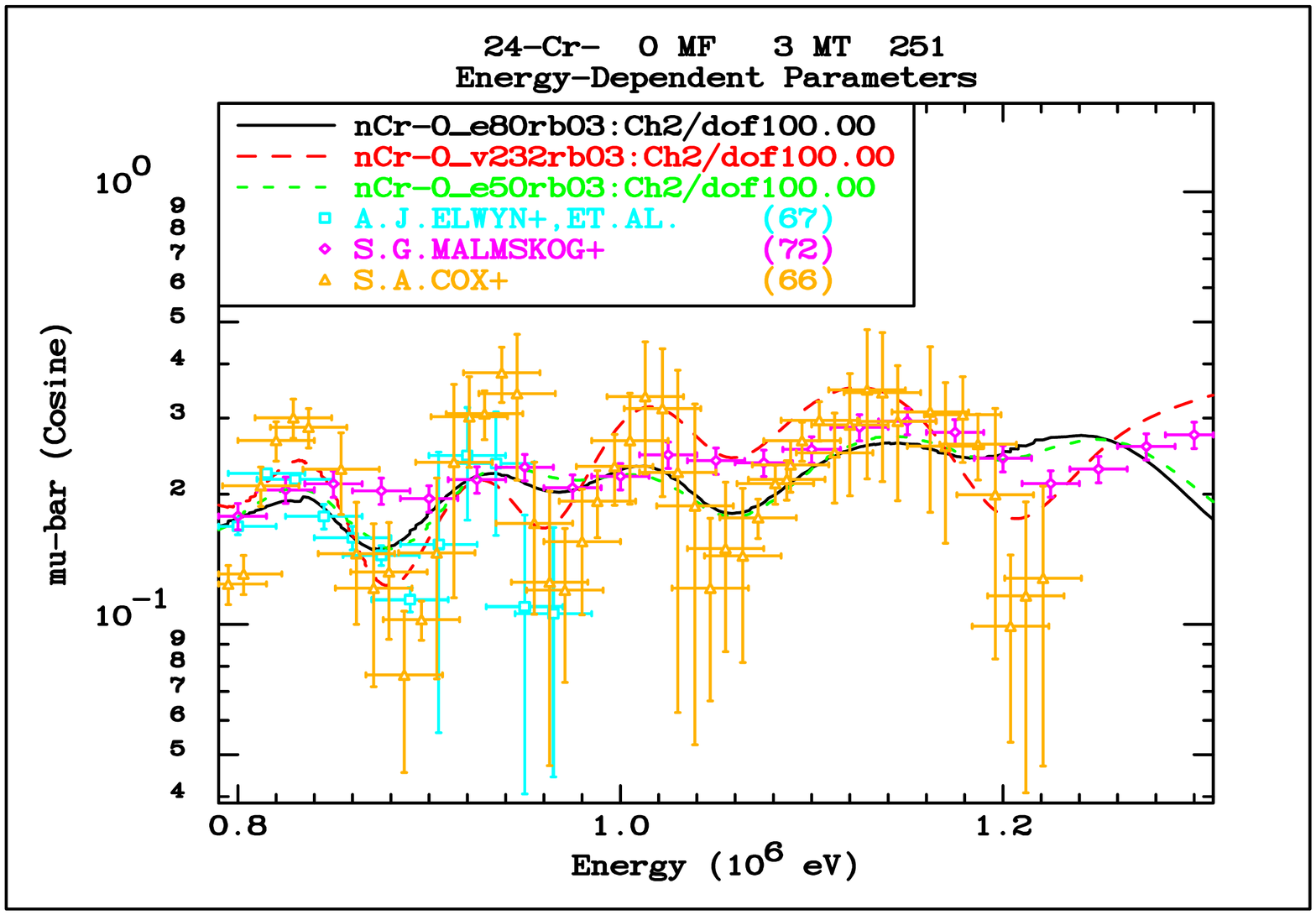} \\
\includegraphics[scale=0.40,angle=-90,clip,trim = 37mm 42mm 35mm 26mm]{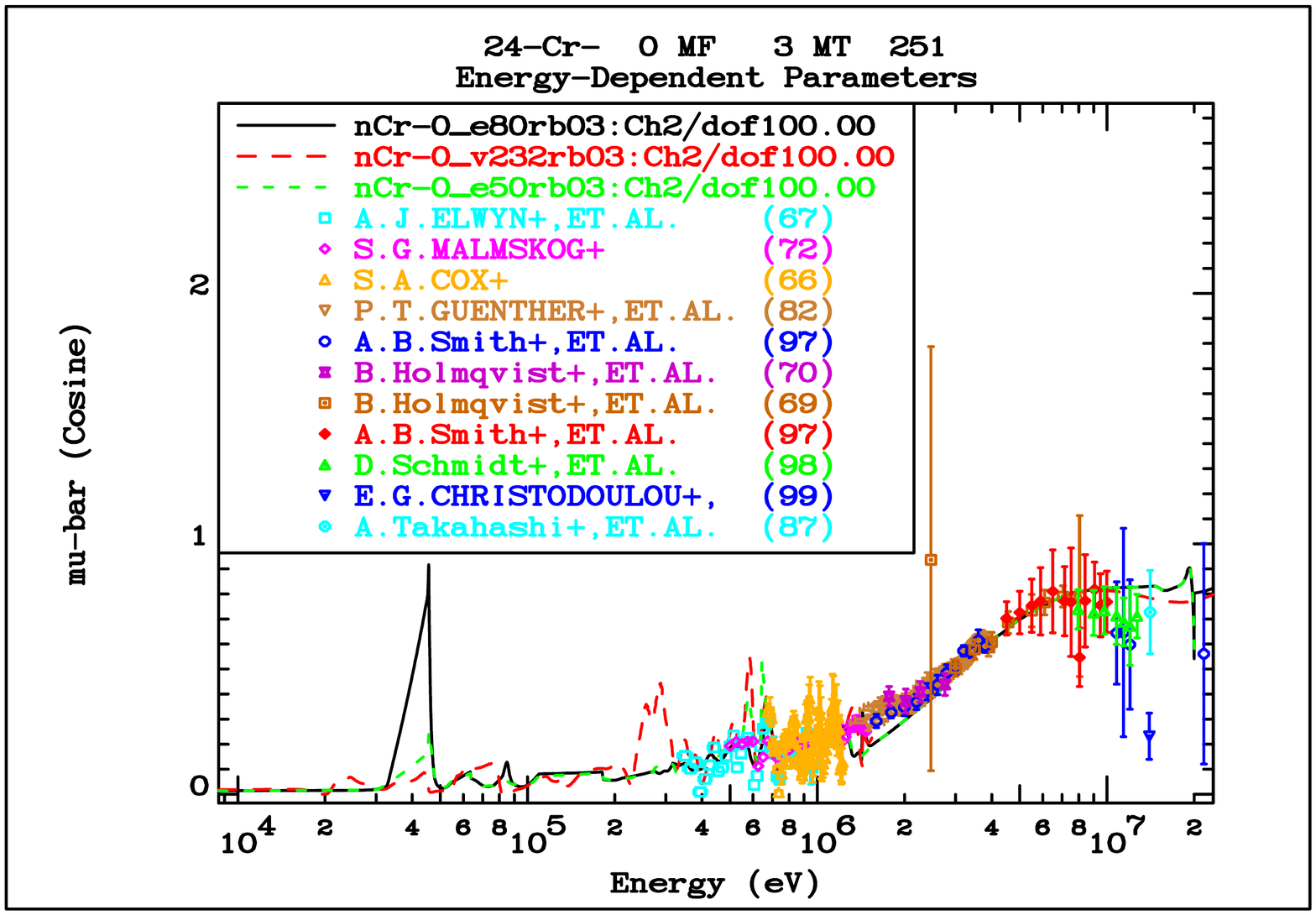}
\caption{(color online) Energy dependence of the $P_1$ coefficient ($\bar{\mu}$) of the Legendre polynomial expansion in the LAB system. The energy region from 10~keV up to 20 MeV is shown in the bottom panel, while the important region from 0.8 up to 1.3~MeV is zoomed on the upper panel. Black solid curves and red dashed curves correspond to 3\% resolution-broadened ENDF/B-VIII.0 and current evaluation, respectively.}
\label{fig:mubar}
\end{center}
\end{figure}

A different comparison to experimental data, which is related to the calculated angular distributions, is shown in Fig.~\ref{fig:mubar}, where the 3\% resolution broadened $P_1$ Legendre coefficient ($\bar{\mu}$) of the angular distribution in the laboratory system is shown as a function of the incident neutron energy. The energy range from 100 keV up to about 8~MeV corresponds to the region of relatively monotonic increase of the $\bar{\mu}$. There is a marked improvement in the description of measured data from 1 to 2 MeV as seen in the lower panel. A clear improvement of the description of measured data can be seen from 0.8 up to 1.3~MeV, near the end of the RRR. Fluctuations in data, specially in higher resolution data from Cox \etal, are much better captured by the new evaluation.

\section{Experimental Constraints on Fast-region Fluctuations}
\label{Sec:fluctuating_data}

In the energy range above the typical RRR ($\sim 0.5$ MeV) to around 10 MeV, nearly all chromium isotopes exhibit fluctuations in their total cross section.  This region of course is very important in reactors as it covers the majority of fission spectrum neutrons. This poses two problems:
\begin{enumerate}
\item In order to properly model leakage in nuclear systems, we require a detailed representation of the total and elastic cross sections as well as the elastic angular distribution.  Because of the model lack of capability to predict observed fluctuations, this can only be done by directly fitting high-resolution experimental data.
\item Hauser-Feshbach theory combined with coupled channel calculations using an optical model potential can give us the best estimates of \textit{energy averaged} cross sections and angular distributions.  However to either demonstrate the efficacy of these calculations or to tune the model parameters, one requires energy averaged experimental results.  Ideally, the optical model should reproduce the average cross section down into the resolved resonance region.
\end{enumerate}
In either case, one relies on accurate, high resolution, experimental values.   Table \ref{table:totalData} summarizes all total cross section measurements on all chromium isotopes (and elemental chromium samples) in the energy range 100 keV-10 MeV.  In this table, we highlight several measurements with a single asterisk to denote that the experiment in question either has a fine enough energy resolution ($\Delta E < 250$ keV) over a large energy range to resolve fluctuations in the cross section or has a well enough determined average cross section to be helpful in baselining the optical model calculations.  There are two sets \cite{Guber:2011,Carlton:2006} which were only available as transmission data.  Although these were used in the resonance region evaluations, they have not been considered here.  In the future we may consider converting these data to cross section data for the cross section fluctuation analysis.

We now discuss the sets highlighted in Table \ref{table:totalData} with the single asterisk.

\paragraph*{(2001) W.P.Abfalterer, F.B.Bateman, \etal} -
EXFOR entry 13753 \cite{Abfalterer:2001} contains data from another ``factory scale'' systematic study of (n,tot) cross sections for 31 materials including $^{nat}$Cr,
in support of the Accelerator Production of Tritium project \cite{1999ABZZ,1998ABZY}.
These experiments were carried out at the LANSCE WNR target area 4 at Los Alamos National Laboratory.
The goal of these measurements were a 1 \% statistical accuracy in 1 \% energy bins with systematic errors less than 1 \% and this goal
was achieved for the chromium target used.
Indeed, the authors maintained such good control over the systematic uncertainties that the cross section covariance matrix is essentially diagonal in energy.
The only potential source of error is that on the areal density of the chromium target which the authors claim is known to better than 0.5\%.
The target was a natural chromium powder encapsulated in a aluminum can and as metallic powder targets are prone to oxidation,
there is a possibility of contamination. The authors did not note any structure from oxygen in the reconstructed cross sections
given the statistical noise in the measurement.

\paragraph*{(1973) F.G.Perey, T.A.Love, \etal} -
EXFOR entry  10342 contains data from ORELA measurements that were transmitted in a private communication \cite{EXFOR.10342:Ref.2} to Brookhaven, with very little information given.
What we do know is that the authors performed transmission measurements on two separate elemental chromium targets with areal densities of 0.21060 a/b and 1.0490 a/b
using the 47.350 m ORELA flight path with a burst width of 5 nsec.  These experiments piggybacked off the measurements described in Ref.~\cite{Perey:1973}.
A careful reading of the EXFOR entry reveals that the targets were compressed chromium metal powder in a lucite binder
and were formed into a right cylinder approximately 10 cm in diameter.
This suggests that oxidation could be a problem with these samples, but we have no simple way to check.
As the authors are deceased and ORELA is decommissioned, it is unlikely we will ever be able to find out more information about these experiments.

Three of the most interesting data sets are \za{Cr}{52}(n,tot) from Agrawal \etal, \za{Cr}{54}(n,tot) from Agrawal \etal \cite{Agrawal:1984} and \za{Cr}{52}(n,tot) from Carlton \etal \cite{Carlton:2000}.  All are high resolution total cross section sets with detailed fluctuating structure above the first inelastic state threshold ($\sim 600$ keV) that may help understand the total cross sections of \za{Cr}{52-54}.  
Both the Agrawal sets and the Carlton set are believed to be chromium(III) oxide (Cr$_2$O$_3$) powder.  Carlton \etal explicitly state that their targets are Cr$_2$O$_3$ powder enriched with \za{Cr}{52}.  For their transmission experiments they state that the target has an inverse thickness of 17.0 b/atom.  Given that Agrawal \etal's targets have inverse thicknesses of 25.33 b/atom (\za{Cr}{52} target) and 18.48 b/atom (\za{Cr}{54} target), we strongly suspect that the Agrawal targets are also powders in target containers of comparable dimension to those used by Carlton \etal  However, since both sets of experiments use powder and powder has a tendency to absorb water, the actual amount of H and O atoms in the targets may not be as well known as planned.  This fact is openly acknowledged in the Carlton \etal paper \cite{Carlton:2000}.  This, coupled with many other problems, limit the usefulness of the Agrawal and Carlton datasets.


\begin{widetext}
\begin{center}
\begin{footnotesize}
\begin{longtable}{lllllllll}
\caption{\label{table:totalData} {\small Collection of (n,tot) data on chromium isotopes and natural chromium in the energy range 100 keV-10 MeV.  Interesting data sets are flagged with an asterisk (*) next to the EXFOR entry number~\cite{ZERKIN201831}.  These sets are discussed in the text.  Data sets marked with the double asterisk (**) are transmission measurement data and have not been converted to cross section data.  We note that there are no usable sets for \za{Cr}{50}.} }\\
\toprule
\toprule
\textbf{EXFOR} & \textbf{Material} & \textbf{Ref.}         &\textbf{Num.} & \textbf{$E_{min}$} & \textbf{$E_{max}$} & \textbf{Facility} & \textbf{Year/}  \\
\textbf{Entry} &                   &                       &\textbf{Pts.} &     \textbf{(MeV)} &     \textbf{(MeV)} &                   & \textbf{Authors}  \\
\midrule
11674.002      & natCr             & \cite{Hibdon:1957,EXFOR.11674:Ref.2}  & 472          & 0.00303            & 0.4085             &  ANL              & (1957) C.T.Hibdon  \\
11540.003      & natCr             & \cite{Whalen:1966}    & 545          & 0.1007             & 0.6487             &  ANL              & (1966) J.F.Whalen, J.W.Meadows    \\
20482.002      & natCr             & \cite{Cabe:1967,EXFOR.20482:Ref.2,EXFOR.20482:Ref.3,EXFOR.20482:Ref.5}
                                                            & 164          & 0.5                & 1.206              &  CEA              & (1967) J.Cabe, M.Laurat, \etal   \\
20012.003      & natCr             & \cite{Cierjacks:1968,EXFOR.20012:Ref.2,EXFOR.20012:Ref.3,EXFOR.20012:Ref.4}
                                                            & 3883         & 0.5001             & 4.398              &  KfK              & (1968) S.Cierjacks, P.Forti, \etal  \\
20012.005      & natCr             & \cite{Cierjacks:1968,EXFOR.20012:Ref.2,EXFOR.20012:Ref.3,EXFOR.20012:Ref.4}
                                                            & 1231         & 4.4069             & 31.94              &  KfK              & (1968) S.Cierjacks, P.Forti, \etal  \\
10047.028*     & natCr             & \cite{FosterJr:1971}  & 240          & 2.339              & 14.888             &  PNNL             & (1971) D.G.Foster Jr, D.W.Glasgow  \\
10225.021      & natCr             & \cite{Green:1973,EXFOR.10225:Ref.2}  & 218          & 1.001              & 8.696              &  BAPL             & (1973) L.Green, J.A.Mitchell   \\
10225.022      & natCr             & \cite{Green:1973,EXFOR.10225:Ref.2}  & 214          & 1.011              & 8.39               &  BAPL             & (1973) L.Green, J.A.Mitchell    \\
10342.004*     & natCr             & \cite{Perey:1973,EXFOR.10342:Ref.2}  & 3658         & 0.18505            & 29.49684           & ORNL              & (1973) F.G.Perey, T.A.Love, \etal    \\
10342.005*     & natCr             & \cite{Perey:1973,EXFOR.10342:Ref.2}  & 3658         & 0.18502            & 29.41653           & ORNL              & (1973) F.G.Perey, T.A.Love, \etal    \\
12882.008      & natCr             & \cite{Larson:1980,EXFOR.12882:Ref.2}  & 685          & 1.9994             & 80.62              &  ORNL             & (1980) D.C.Larson, J.A.Harvey, \etal   \\
13753.017*     & natCr             & \cite{Abfalterer:2001} & 467          & 5.293              & 559.1              &  LANL             & (2001) W.P.Abfalterer, F.B.Bateman, \etal  \\
14324.002**    & natCr             & \cite{Guber:2011}      & 30663        & 1.03e-5            & 0.631               & ORNL  & (2011) K. Guber, P. Koehler \etal \\

\midrule

11601.004      & $^{50}$Cr         & \cite{Farrell:1966}    & 1082         & 0.08               & 0.6205             &   Duke            & (1966) J.A.Farrell, E.G.Bilpuch, \etal  \\
\noindent 20435.002      & $^{50}$Cr         & \cite{KfK1517}         & 5181         & 0.026              & 0.285              &   KfK             & (1972) R.R.Spencer, H.Beer, \etal   \\

\midrule

10047.029*     & $^{52}$Cr         & \cite{FosterJr:1971}   & 240          & 2.339              & 14.888             &  PNNL             & (1971) D.G.Foster Jr, D.W.Glasgow  \\
20435.003      & $^{52}$Cr         & \cite{KfK1517}         & 5281         & 0.021              & 0.285              &  KfK              & (1972) R.R.Spencer, H.Beer, \etal   \\
22131.005      & $^{52}$Cr         & \cite{Rohr:1989,EXFOR.22131:Ref.2}    & 8137         & 0.24001            & 1.2923             &  Geel             & (1989) G.Rohr, R.Shelley, \etal    \\
12830.008*     & $^{52}$Cr         & \cite{Agrawal:1984}    & 8310         & 0.0058679          & 4.9335             &  ORNL             & (1984) H.M.Agrawal, J.B.Garg, \etal    \\
13840.002*     & $^{52}$Cr         & \cite{Carlton:2000}    & 13620        & 0.050185           & 31.162             &  ORNL             & (2000) R.F.Carlton, J.A.Harvey, \etal   \\
13840.005*     & $^{52}$Cr         & \cite{Carlton:2000}    & 12213        & 0.066633           & 69.131             &  ORNL             & (2000) R.F.Carlton, J.A.Harvey, \etal  \\

\midrule

10047.030*     & $^{53}$Cr         & \cite{FosterJr:1971}   & 248          & 2.262              & 14.828              &  PNNL & (1971) D.G.Foster Jr, D.W.Glasgow  \\
20155.012      & $^{53}$Cr         & \cite{Muller:1971,EXFOR.20155:Ref.2,EXFOR.20155:Ref.3,EXFOR.20155:Ref.4}
                                                            & 1657         & 0.016879           & 0.36003             &  KfK & (1971) K.-N.Mueller, G.Rohr, \etal    \\
14324.004**    & $^{53}$Cr        & \cite{Guber:2011}       & 30663        & 1.03e-5            & 0.631               & ORNL  & (2011) K. Guber, P. Koehler \etal \\

\midrule

11601.005      & $^{54}$Cr         & \cite{Farrell:1966}    & 992          & 0.1                & 0.5955              &  Duke  & (1966) J.A.Farrell, E.G.Bilpuch, \etal   \\
12830.009*     & $^{54}$Cr         & \cite{Agrawal:1984}    & 8207         & 0.0053161          & 4.9277              &  ORNL  & (1984) H.M.Agrawal, J.B.Garg, \etal   \\
14114.004**    & $^{54}$Cr         & \cite{Carlton:2006}    & 5169         & 0.0235             & 2.88                &  ORNL  & (2006) R.F.Carlton, C.Baker, J.A.Harvey   \\
\bottomrule
\bottomrule
\end{longtable}
\end{footnotesize}
\end{center}
\end{widetext}

\paragraph*{(1984) H.M.Agrawal, J.B.Garg, \etal} -
EXFOR entry 12830 contains neutron transmission data for  \za{Cr}{52}(n,tot) (Run ID \# 8914) and \za{Cr}{54}(n,tot) (Run ID \# 8910) from the paper of Agrawal, \etal \cite{Agrawal:1984}.  These measurements were undertaken at the ORELA facility at Oak Ridge National Laboratory on the 78.203 m flightpath.  Table I of this reference provides the isotopic composition of the targets.  Neither ref. \cite{Agrawal:1984} nor ORELA log books \cite{LogBooks} indicate the nature of the sample holder.
Both targets were run for about a week.  Agrawal \etal\  state that they corrected their data for the oxide contribution using a background correction of
\begin{equation}
	\sigma^{corr}(E)=\sigma^{meas}-1.5\times\left[3.6-(2\times 10^{-6})E\right]
\end{equation}
where $E$ is given in eV and $\sigma$ in barns.  While the prefactor of 1.5 arises from stoichiometry of a Cr$_2$O$_3$ unit cell, the additional 3.6 barns is a (poor) estimate of the zero energy intercept of the total $^{16}$O cross section.  Furthermore, this factor only appears that it was used in the R-matrix analysis they performed using their measurements.  At higher energies there is an obvious contribution from oxygen resonances.  We attempted to correct for the presence of both oxygen and hydrogen (supposing that the powder sample absorbed water), but were unable to fully correct because of systematic energy shifts in both Agrawal \etal data.  It appears that the ORELA resolution function, including flight path corrections due to rescattering in the ORELA water moderator \cite{Coceva:1983}, was not properly accounted for.  Nevertheless, correcting for the rescattering and our best notion of the isotopic composition of the Agrawal \etal targets,
we extracted a \za{Cr}{52}(n,tot) cross section that is only in modest agreement with the data of Foster Jr. \etal \cite{FosterJr:1971}.  In the end, the Agrawal data could not be used.

\begin{table}
\caption{\label{table:AgrawalIsotopics}Table of isotopic composition of the targets used by Agrawal \etal \cite{Agrawal:1984}.  Taken from Table I of Ref. \cite{Agrawal:1984}.}
\begin{tabular}{lcc}
\toprule
\toprule
\multirow{2}{*}{\textbf{Sample} }   & \multirow{2}{*}{\textbf{Isotope}} & \textbf{Atomic}\\
                   &                  & \textbf{Fraction (\%)}\\
\midrule
$^{52}$Cr$_2$O$_3$ & 50               & 0.01	\\
                   & 52               & 99.87 $\pm$ 0.02\\
                   & 53               & 0.12 $\pm$ 0.02\\
                   & 54               & 0.01 \\
\midrule
$^{54}$Cr$_2$O$_3$ & 50               & 0.18 $\pm$ 0.05\\
                   & 52               & 3.09 $\pm$ 0.10\\
                   & 53               & 1.33 $\pm$ 0.10\\
                   & 54               & 95.40 $\pm$ 0.10\\
\bottomrule
\bottomrule
\end{tabular}
\end{table}

\begin{table}
\caption{\label{table:CarltonIsotopics}Table of isotopic composition of the targets used by Carlton \etal \cite{Carlton:2000}.  Given the nearly identical isotopic makeup of this target and the $^{52}$Cr$_2$O$_3$ target of Agrawal \etal, it is likely that the same target was used in both experiments.}
\begin{tabular}{lcc}
\toprule
\toprule
\multirow{2}{*}{\textbf{Sample} }   & \multirow{2}{*}{\textbf{Isotope}} & \textbf{Atomic}\\
                   &                  & \textbf{Fraction (\%)}\\
\midrule
$^{52}$Cr$_2$O$_3$ & 50               & 0.01	\\
                   & 52               & 99.87\\
                   & 53               & 0.12\\
                   & 54               & 0.01 \\
\bottomrule
\bottomrule
\end{tabular}
\end{table}

\begin{figure*}
\includegraphics[width=0.80\textwidth]{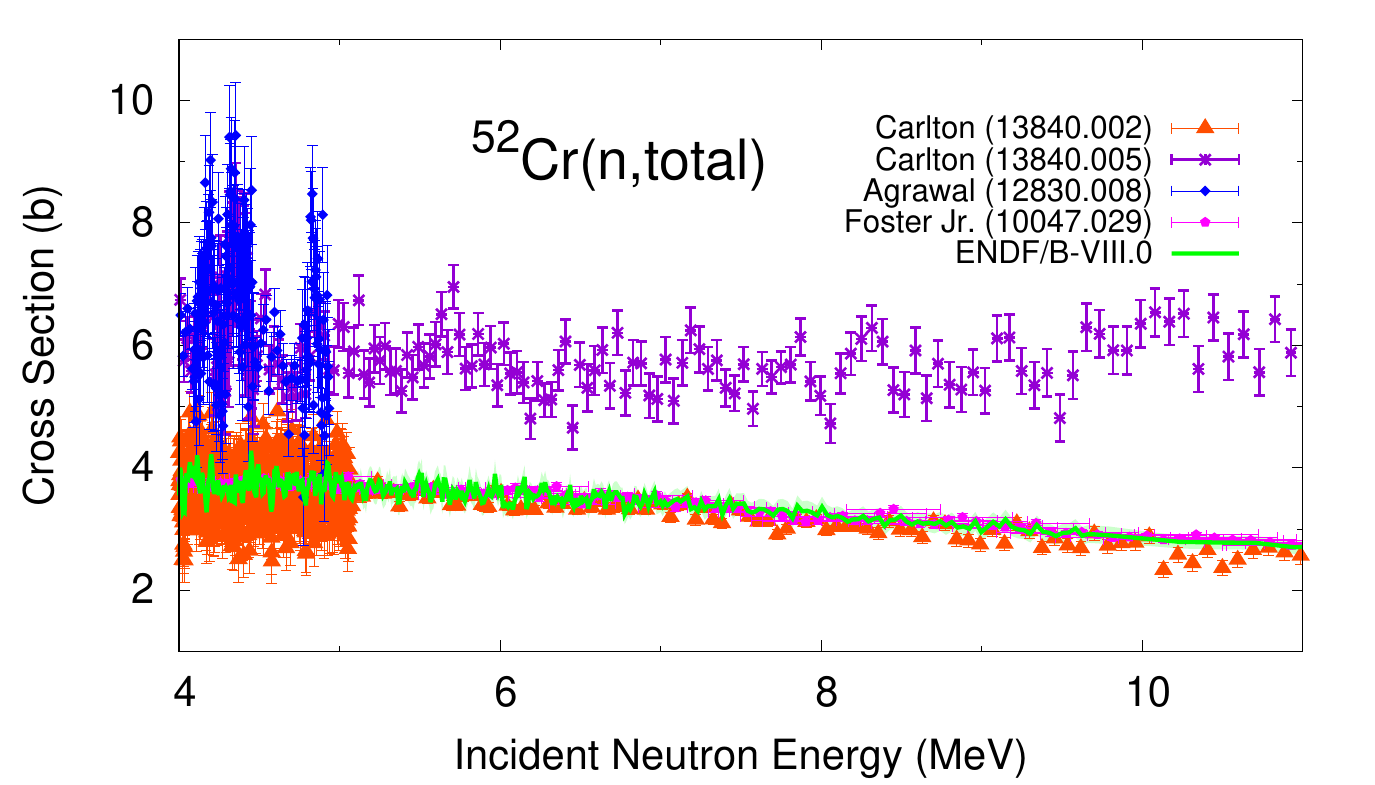}
\caption{(color online) Total cross-section data available for \nuc{52}{Cr} in EXFOR in the 4-20~MeV region, with ENDF/B-VIII.0 plotted as reference. Numbers in parentheses indicate EXFOR entries and subentries~\cite{ZERKIN201831}. References for the experimental data can be found in Table~\ref{table:totalData}. The reasons for the limited usability of the experimental data sets is discussed in the text.}
\label{fig:cr52-total-badstuff}
\end{figure*}

\paragraph*{(2000) R.F.Carlton, J.A.Harvey, \etal} -
EXFOR entry 13840 contains neutron transmission data for \za{Cr}{52}(n,tot) from the paper of Carlton, \etal \cite{Carlton:2000}.  This measurement was taken at the ORELA facility at Oak Ridge National Laboratory on the $201.575\pm0.005$ m flightpath.  The longer flight path provide superb energy resolution even at high energies, eliminating the timing and energy resolution problems in the Agrawal \etal sets.
The isotopic composition of the target used by Carlton \etal is shown in Table \ref{table:CarltonIsotopics}.  The composition is nearly identical to Agrawal \etal's \za{Cr}{52}(n,tot) target, leading us to suspect that both Carlton \etal and Agrawal \etal used the same sample to produce their targets.  That said, Carlton \etal took a more ingenious approach to correcting for the oxygen content of the target.  Instead of a pure $^{52}$Cr$_2$O$_3$ target, Carlton \etal used a $^{52}$Cr$_2$O$_3$+Be target and cycled it with a second BeO target.  The thickness of the BeO target was designed so that the number density of oxygen scatterers in the BeO target was identical to that in the $^{52}$Cr$_2$O$_3$+Be target.  The amount of added Be was chosen to be identical to the number of Be scatterers in the BeO target.  Together, it was hoped that the oxygen and Be content could be ``subtracted'' away \cite{Carlton:2000}.
At first glance, the correction appears to have worked. In Fig. \ref{fig:cr52-total-badstuff} we show the \za{Cr}{52} total cross section in the range 4-20 MeV.  The light blue set that is above all of the others is Carlton \etal's original, uncorrected set.  There are two things to note about this set.  First, it appears to have not been corrected for the other isotopes in the target.  We subtracted our best estimate of the contributions from other isotopes, using ENDF/B-VIII.0 cross sections for the other chromium isotopes, oxygen, hydrogen and beryllium, resulting in the red points with the error band.  This corrected the data below 5 MeV, but was unable to explain the portion above 5 MeV.  This brings us to the second issue, this set appears to be two sets merged in an undocumented manner.  Carlton \etal corrected this set, resulting in the other set in the red triangles.  The nature of the correction is not documented in Ref. \cite{Carlton:2000} or in the ORELA logbooks \cite{LogBooks}.
We note that these corrected points are systematically below both the ENDF/B-VIII.0 evaluation and the data of Foster and Glasgow \cite{FosterJr:1971}.

\section{Evaluated Results in the Fast Range}
\label{Sec:eval}
Multi-channel evaluation in the fast region, regarded here to be where incident neutron energies are higher than the first inelastic of each isotope, were determined by the EMPIRE nuclear reaction code \cite{empire}, unless stated otherwise. Selection of reaction models was based on quality of overall agreement of calculated cross sections with experimental data. Due to low-energy level densities being strongly parity asymmetric, we adopted RIPL-3 HFB level densities and tuned using the approach of Ref.~\cite{Nobre:2020} for \nuc{52}{Cr}. For the minor chromium isotopes we simply adopted the level-density model of Gilbert-Cameron~\cite{Gilbert}. Width fluctuation corrections were applied up to 9.9 MeV and pre-equilibrium was determined using the exciton model~\cite{exciton} as implemented in the code PCROSS~\cite{PCROSS}. 
Full  $\gamma$-ray cascades for every excited nucleus were calculated.  We employed  $\gamma$-strength functions from the RIPL3 MLO1 formulation by Plujko for E1 transitions initiated from the continuum, for all nuclei. M1, E1 and E2 electromagnetic transitions
were considered.  Compound-nucleus decay and direct cross sections were added incoherently and compound-nucleus anisotropy was calculated using Blatt-Biedenharn coefficients.

In Table~\ref{Tab:omp-ripl} we identify the optical model potentials (OMP) used for the different reaction channels. In particular, for the incoming channel of \nuc{50,52,54}{Cr} targets we used the recently-developed, even-even chromium-specific, dispersive soft-rotor of Li et al.~\cite{Li:2013} (index \#616 in RIPL) which was fitted as described in Ref.~\cite{Li:2013} using both neutron and proton scattering data and a soft-rotor model of nuclear structure that typically coupled the first five or six discrete levels of each target nucleus. For \nuc{53}{Cr} we used the same OMP,  indexed in RIPL as \#617, which is equivalent to \#616 with rigid rotor couplings. The soft-rotor structure model provides an excellent description of the nuclear structure of quasi-spherical and very soft to deformation near-magic even-even chromium isotopes.

\begin{table}[!htp]
\caption{Optical model potentials used in the EMPIRE calculations.}
\begin{tabular}{lccc}
\toprule \toprule
 Ejectile        & Type   & RIPL \#                    & Reference \\ 
 \midrule
 \multirow{2}{*}{$n$ (direct)}  & CC & \multirow{2}{*}{ 616       }                 &  \multirow{2}{*}{Li+~\cite{Li:2013} }    \\
                                              & (soft rotor)                &                                &                 \\
 \multirow{2}{*}{$n$ (direct)}  & CC & \multirow{2}{*}{617 for \nuc{53}{Cr}      }                 &  \multirow{2}{*}{Li+~\cite{Li:2013} }    \\
                                              & (rigid rotor)                &                                &                 \\
 $n$ (compound)  & Spher. &  2405                      &  Koning+~\cite{Koning:2003} \\
 $p$             & Spher. &  5405                      &  Koning+~\cite{Koning:2003} \\
 $\alpha$        & Spher. &  9600                      &  Avrigeanu+~\cite{Avrigeanu:1994}  \\
 $d$             & Spher. &  6200                      &  Haixia+~\cite{Haixia:2006}  \\
 $t$             & Spher. &  7100                      &  Becchetti+~\cite{Becchetti-Greenlees}  \\
 \nuc{3}{He}     & Spher. &  8100                      &  Becchetti+~\cite{Becchetti-Greenlees} \\  
 \bottomrule \bottomrule
\end{tabular}
\label{Tab:omp-ripl}
\end{table}

Model parameters were adjusted using the KALMAN code~\cite{Kawano:1997yi} using selected data sets as described in the sections below. Once optimal agreement through fitting was achieved, localized minor tuning was done for specific reactions to maximize agreement with certain data sets.

\subsection{Cross Sections}
   \label{subSec:x-sec}
In this section we discuss cross sections for the major neutron-induced reactions on isotopes of chromium in the fast region.
For the reactions that are not explicitly mentioned we accepted  EMPIRE results  without modifications. In general, in such cases, there were no significant (or any) data to compare with, or the calculated results were close to those measured.

\subsubsection{Total Cross Sections}



In ENDF/B-VIII.0, total cross sections in the fast range for chromium isotopes date back to the ENDF/B-VI MOD1 evaluation of Larson et al. from 1989 and were based on calculations with the spherical potential of Wilmore-Hodgson \cite{Wilmore-OMP} using elemental transmission data. We provide significant improvements by using an isotopic-dependent coupled-channel approach with modern dispersive soft- and rigid-rotor optical model potentials. In our present evaluation we defined the total cross sections in the fast region as follows: \\

\paragraph*{\nuc{52}{Cr}($n$,total) - } For \nuc{52}{Cr} we made minor parameter adjustments to optical model parameters (changes smaller than 1\%) to optimize agreement with smoothed \nuc{nat}{Cr} data of Green et al.~\cite{Green:1973}  and Foster Jr.~\cite{FosterJr:1971}. The optical model potential employed fails to describe the quick decrease of the total cross section data below 6~MeV which is a typical behavior for Cr/Ni/Fe targets.
For this reason we replaced total cross sections between 3.6 and 6 MeV by the \nuc{nat}{Cr} experimental data of Perey~\cite{Perey:1973}, and, between the inelastic threshold and 3.6 MeV by the \nuc{52}{Cr} total data of Carlton et al.~\cite{Carlton:2000}, which in this energy range is compatible with Perey, but of higher resolution. The higher resolution of Carlton et al data is critical at energies lower than 3 MeV for the correct calculation of self-shielding factors, which are extremely sensitive to fluctuating data.

\begin{figure}
\includegraphics[scale=0.70,keepaspectratio=true,clip=true,trim=0mm 0mm 0mm 0mm]{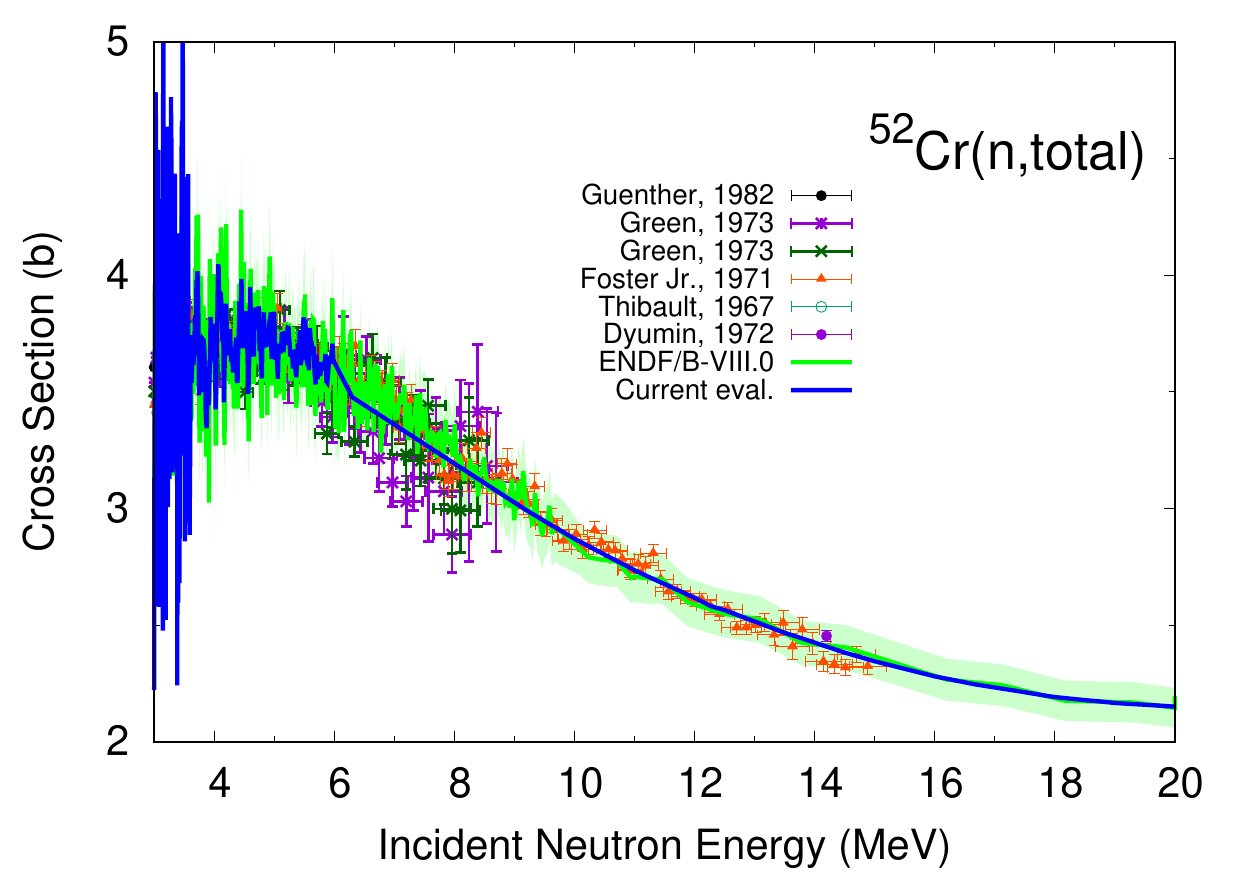}
\caption{(color online) Total cross section in the fast range region for \nuc{52}{Cr}. Data taken from Refs.~\cite{Guenther:1982,Green:1973,EXFOR.10225:Ref.2,FosterJr:1971,Thibault:1967,Dyumin:1972}.}
\label{fig:cr52-total}
\end{figure}

Fig.~\ref{fig:cr52-total} compares the total cross section from the current work and ENDF/B-VIII.0 and relevant experimental data.

\begin{figure}
\includegraphics[scale=0.70,keepaspectratio=true,clip=true,trim=0mm 0mm 0mm 0mm]{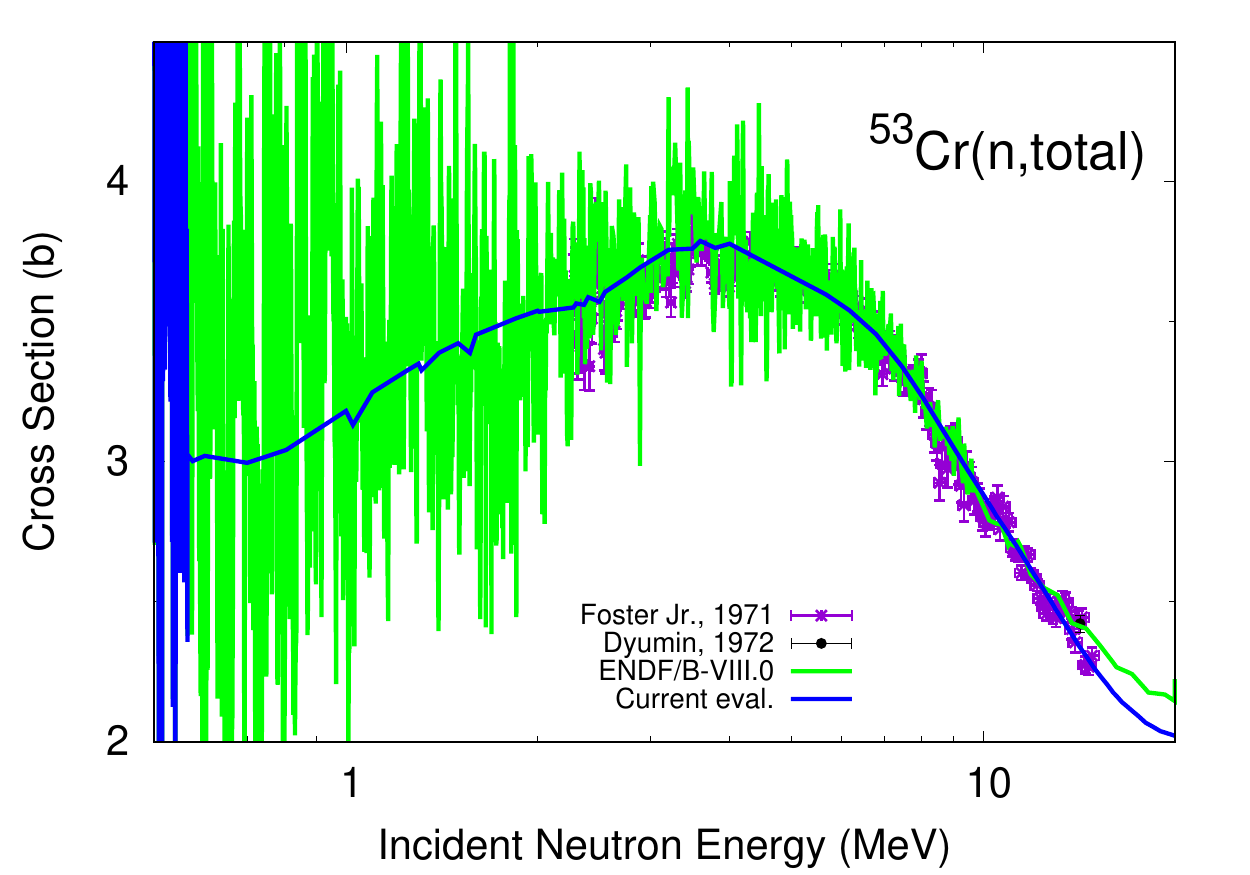}
\caption{(color online) Total cross section in the fast range region for \nuc{53}{Cr}. Data taken from Refs.~\cite{FosterJr:1971,Dyumin:1972}.}
\label{fig:cr53-total}
\end{figure}

\paragraph*{\nuc{53}{Cr}($n$,total) - } Due to the lack of higher resolution data specific for \nuc{53}{Cr} we simply fitted to experimental of Foster Jr.~\cite{FosterJr:1971}. Fig.~\ref{fig:cr53-total} shows our evaluated \nuc{53}{Cr}(n,total) cross sections compared to ENDF/B-VIII.0 and experimental data. It is important to note that the fluctuations seen in ENDF/B curve in Fig.~\ref{fig:cr53-total} were taken from elemental chromium data and therefore is not supported by any \nuc{53}{Cr}-specific data. Such fluctuations are already accounted for in our evaluation for \nuc{52}{Cr} total cross section.

\paragraph*{\nuc{50}{Cr}($n$,total) - } There were no experimental data available in the fast region, except a single point at around 14 MeV from Dyumin et al.~\cite{Dyumin:1972}. Therefore, our recommendation follows EMPIRE calculations. Both ENDF/B-VIII.0 and current evaluation agree with this single datum within 3\%.

\paragraph*{\nuc{54}{Cr}($n$,total) - } Unfortunately, we were not able to use the data of Agrawal et al.~\cite{Agrawal:1984}, as described in Section~\ref{Sec:fluctuating_data}. Since there was no usable data in fast region, apart from a single point from Dyumin et al.~\cite{Dyumin:1972} at $\sim$ 14 MeV, we again followed EMPIRE calculations.  Our evaluation agree with this single data point within $\sim$2\%.

\subsubsection{Elastic Cross Sections}
The elastic cross sections are generally calculated from the optical model above 5 MeV. In the region of fluctuations they are defined as the difference between the total and the sum of the remaining partial cross sections.

\begin{figure}
\includegraphics[scale=0.70,keepaspectratio=true,clip=true,trim=0mm 0mm 0mm 0mm]{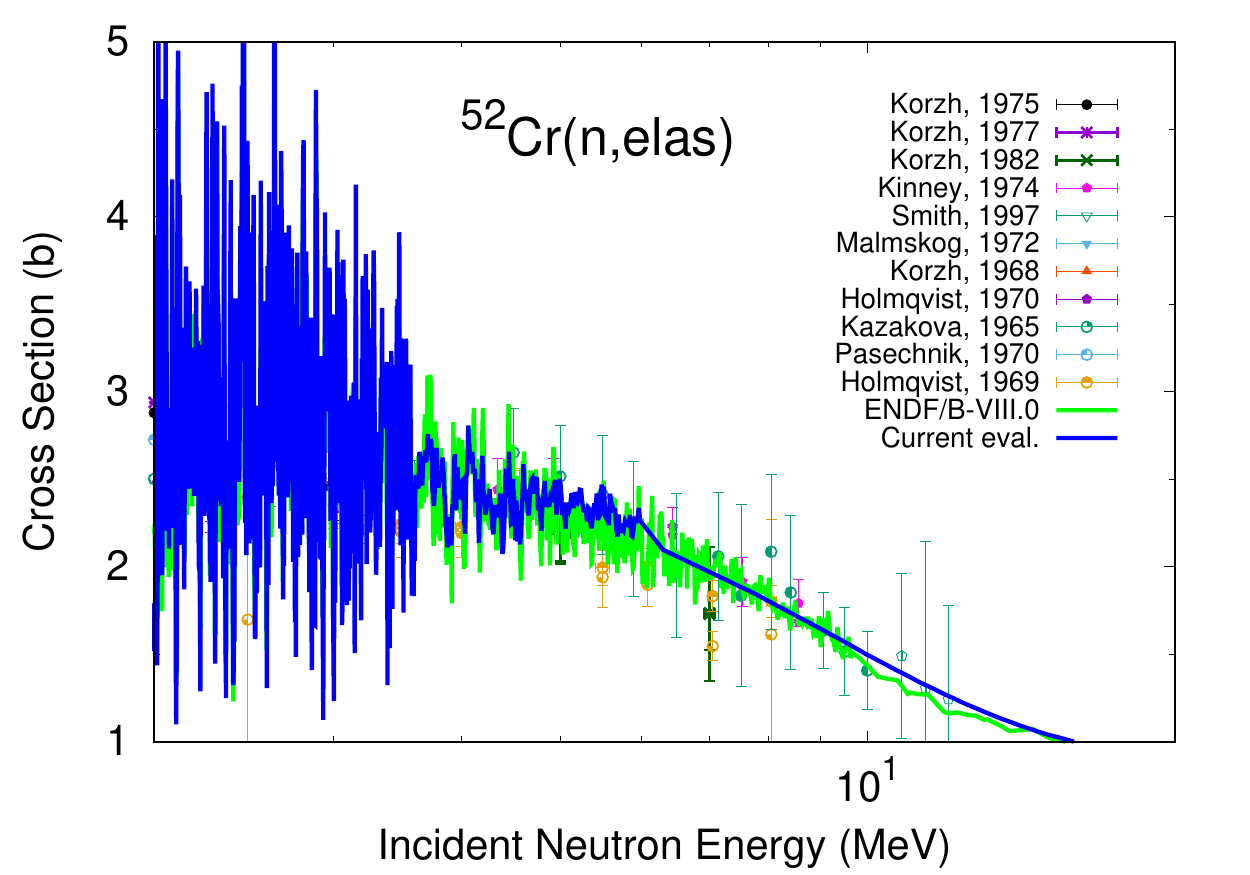}
\caption{(color online) Elastic cross section in the fast range region for \nuc{52}{Cr}. Data taken from Refs.~\cite{Korzh:1975,Korzh:1977,Korzh:1982,Kinney:1974,Smith:1997,Malmskog:1972,Korzh:1968,Holmqvist:1970,Kazakova:1965,Pasechnik:1970,Holmqvist:1969}.}
\label{fig:cr52-elastic}
\end{figure}

\paragraph*{\nuc{52}{Cr}($n$,elas) - } Even though the elastic cross sections were indirectly defined, through the difference between total and all other reactions, we can see in Fig.~\ref{fig:cr52-elastic} that we achieve  good agreement with experimental data, especially  above fluctuations.

\paragraph*{\nuc{50,53,54}{Cr}($n$,elas) - } There are no available elastic experimental data for \nuc{53}{Cr} and the few data points available for \nuc{50}{Cr} and \nuc{54}{Cr} have large error bars and are not constraining enough. We therefore adopted EMPIRE calculations for elastic cross section of the minor chromium isotopes.




\subsubsection{Inelastic Cross Sections}
\label{sec:inel}

In the fast neutron range, defined in the present evaluation as the region above the first inelastic threshold for each chromium isotope, the inelastic channel increasingly becomes the most important constraint, apart from total and elastic cross sections, at least until the opening of the (n,2n) channel.

The inelastic thresholds for chromium isotopes are 564.03 keV, 783.32 keV, 834.855 keV and 1.4341 MeV for \nuc{53}{Cr},  \nuc{50}{Cr}, \nuc{54}{Cr}, and \nuc{52}{Cr}, respectively. Due to their relative abundance, the elemental inelastic cross section is strongly defined, above 1.4341 MeV, by the \nuc{52}{Cr}, hence its careful discussion below. The main constraint for the inelastic channel were the \nuc{52}{Cr}(n,n$^\prime \gamma$) measurements of Mihailescu et al.~\cite{Mihailescu:2007} even though they in some cases differ strongly from previous measurements of Voss \etal For the minor isotopes, the most recent inelastic data dates back to the 1978 experiment of Karatzas \etal~\cite{Karatzas:1978}. We were able to make the data set of Karatzas useful to our minor evaluation after realizing that their values reported in EXFOR did not correspond to the isotopic cross sections, but rather isotopic inelastic cross sections on a natural target, thus needing to be renormalized by their abundance. After this, we realized that the \nuc{52}{Cr} Karatzas data were in fact consistent with smoothed Milhailescu, bringing confidence to use it as constraint for the minor isotopes.

\paragraph*{\nuc{52}{Cr}($n$,inelastic) - } Like many near-closed shell nuclides, chromium isotopes present strong fluctuations up to relatively high neutron incident energies. In this evaluation we make use of measurements of Mihailescu et al.~\cite{Mihailescu:2007}, which were obtained through a 200 meter time-of-flight experiment using the white neutron spectrum of GELINA  on a Cr$_2$O$_3$ sample.  This thorough and detailed measurement was not taken into account in previous ENDF/B evaluations, even though JEFF-3.3 \cite{Plompen2020} has incorporated them in some form, at least for the neutron level cross sections.

\begin{figure*}
\includegraphics[scale=0.50,keepaspectratio=true,clip=true,trim=0mm 7mm 0mm 0mm]{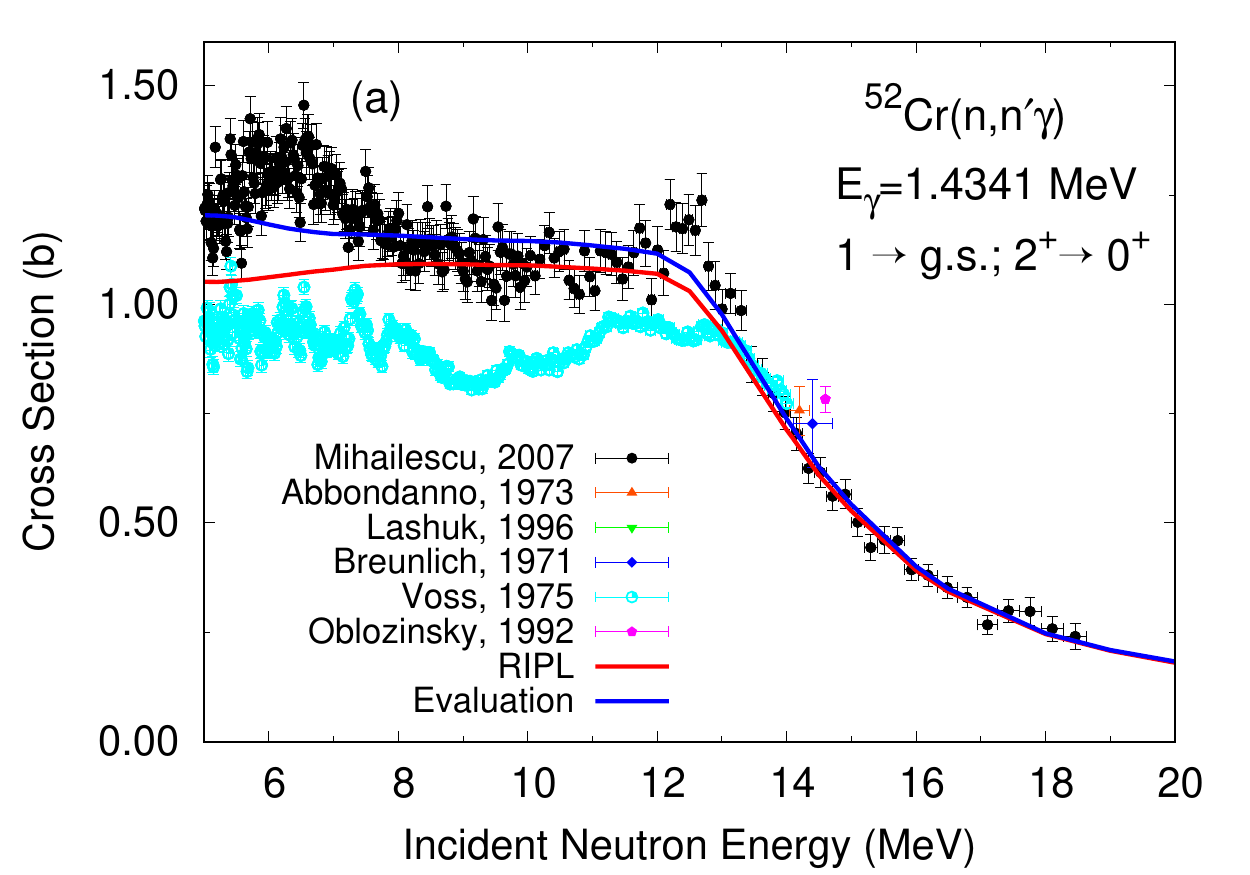} \hspace{-4mm}
\includegraphics[scale=0.50,keepaspectratio=true,clip=true,trim=9mm 7mm 0mm 0mm]{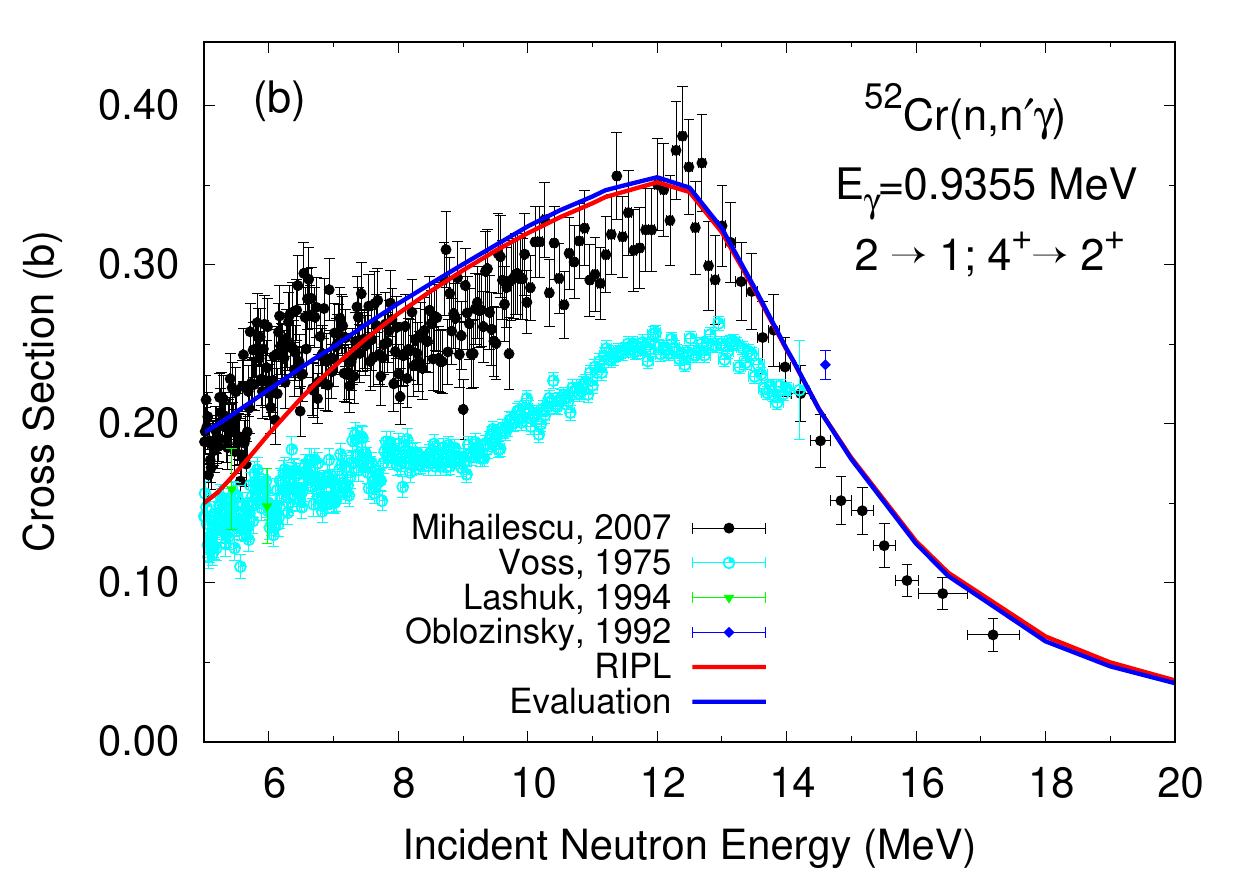} \hspace{-4mm}
\includegraphics[scale=0.50,keepaspectratio=true,clip=true,trim=9mm 7mm 0mm 0mm]{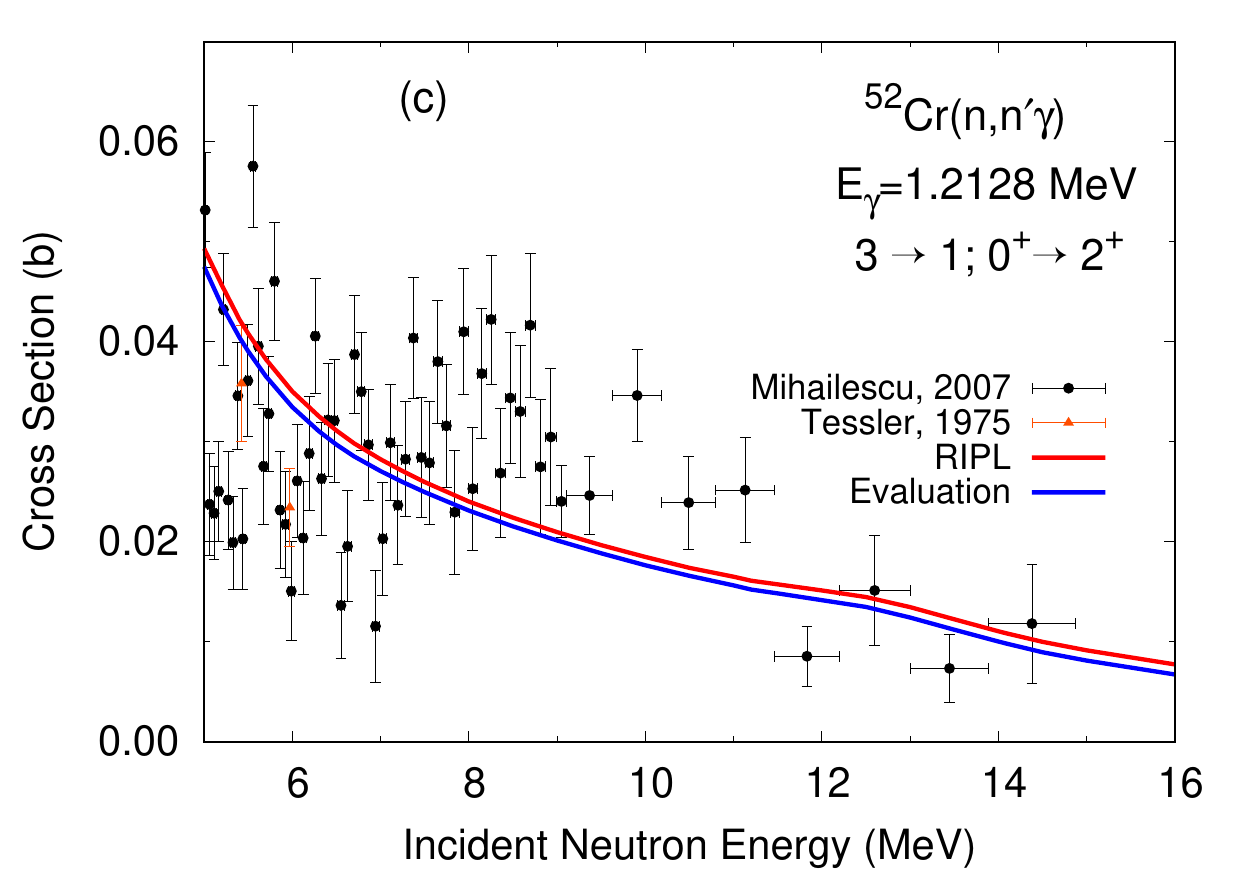} \\
\includegraphics[scale=0.50,keepaspectratio=true,clip=true,trim=0mm 7mm 0mm 0mm]{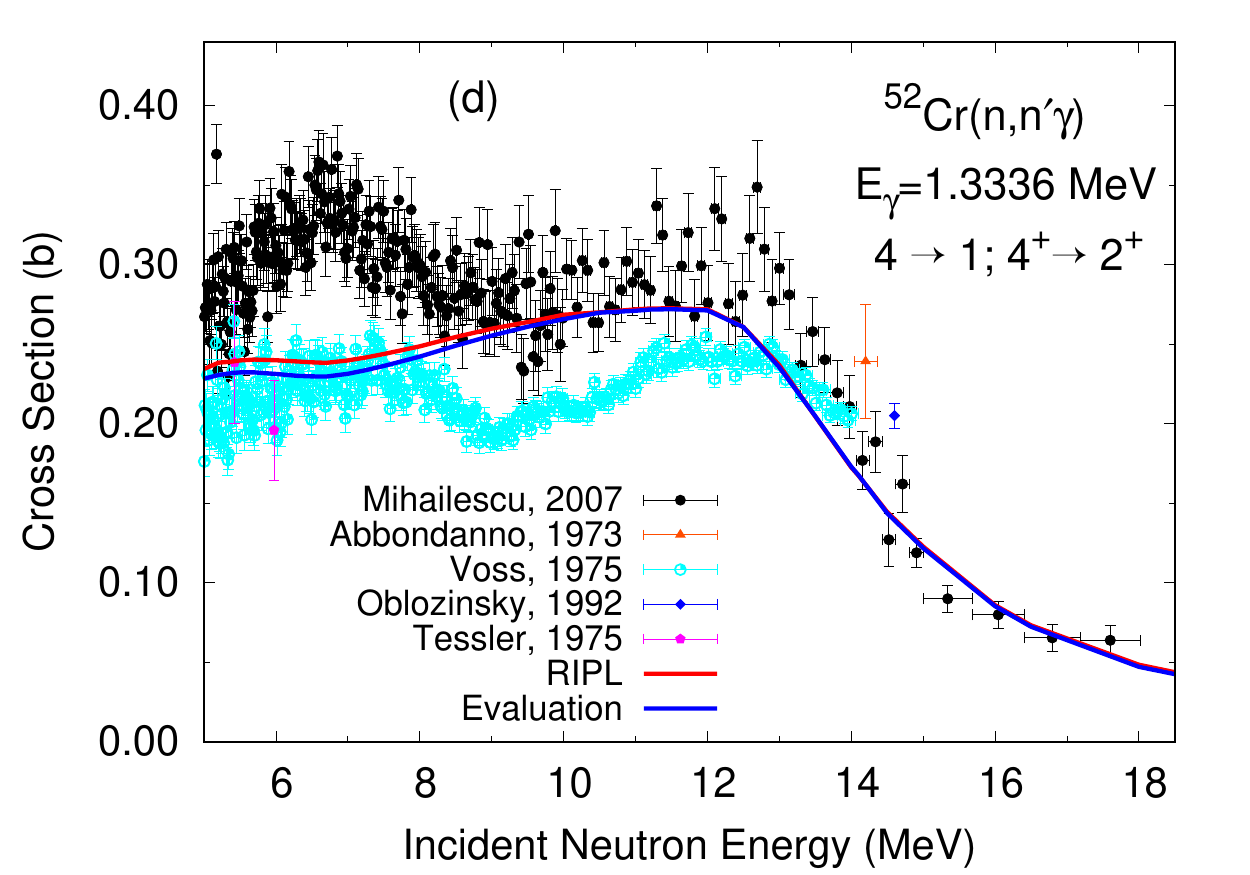} \hspace{-4mm}
\includegraphics[scale=0.50,keepaspectratio=true,clip=true,trim=9mm 7mm 0mm 0mm]{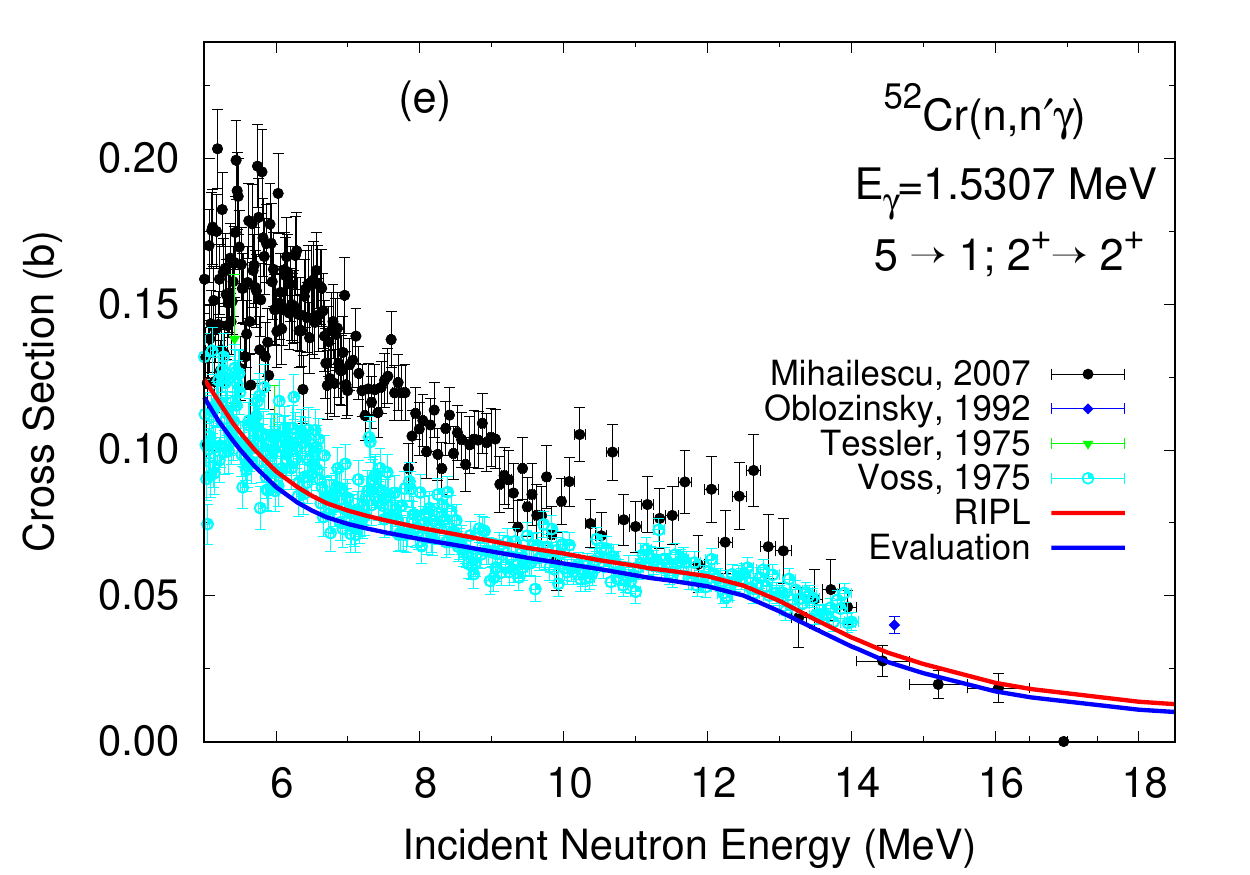} \hspace{-4mm}
\includegraphics[scale=0.50,keepaspectratio=true,clip=true,trim=9mm 7mm 0mm 0mm]{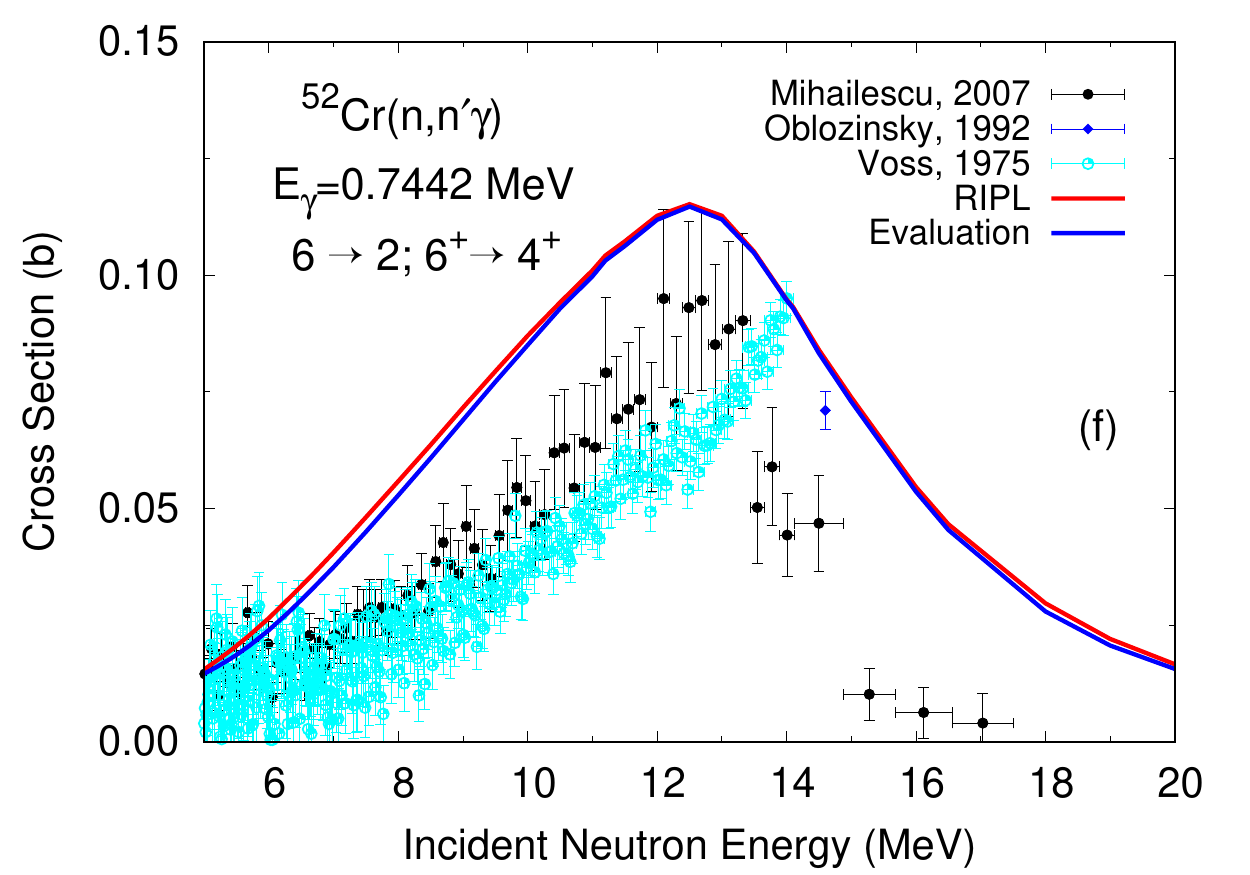} \\
\includegraphics[scale=0.50,keepaspectratio=true,clip=true,trim=0mm 7mm 0mm 0mm]{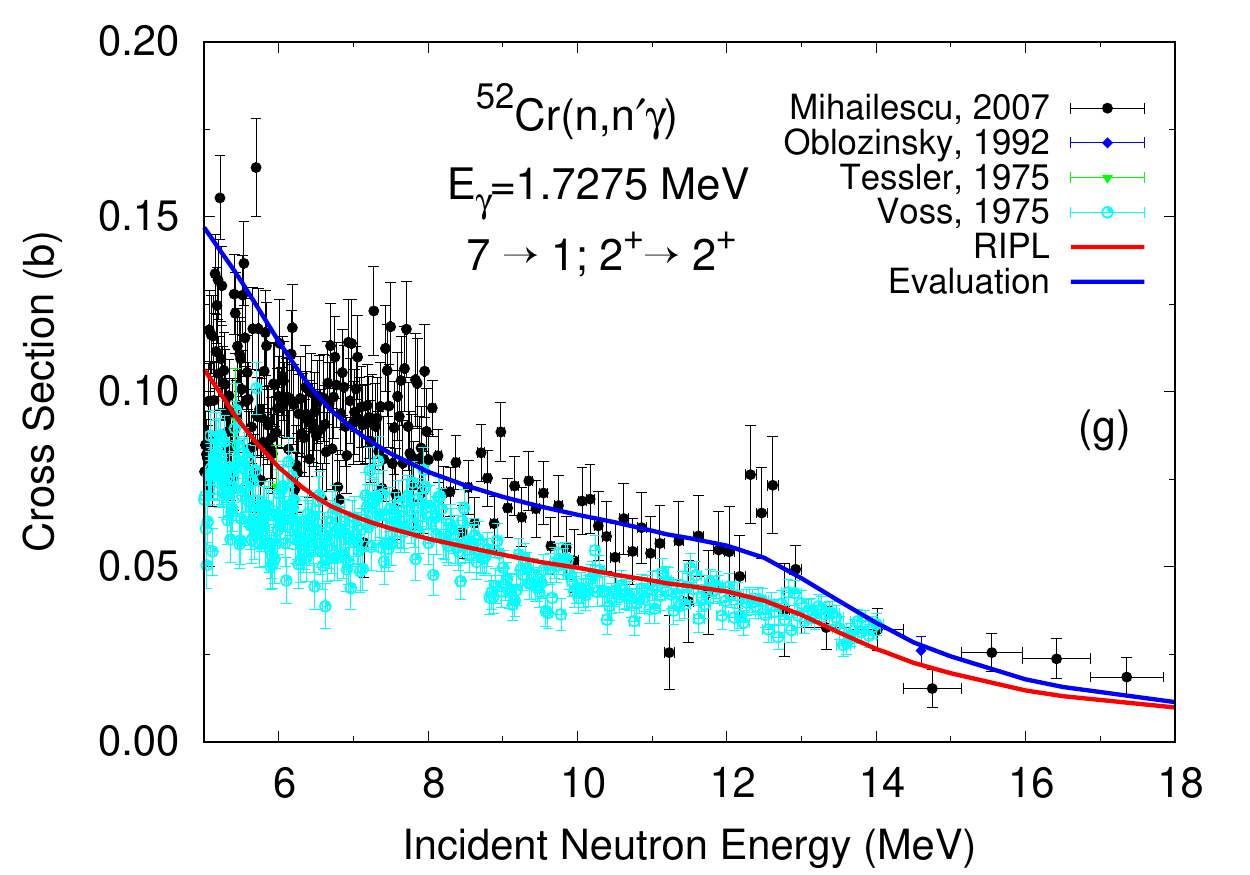} \hspace{-4mm}
\includegraphics[scale=0.50,keepaspectratio=true,clip=true,trim=9mm 7mm 0mm 0mm]{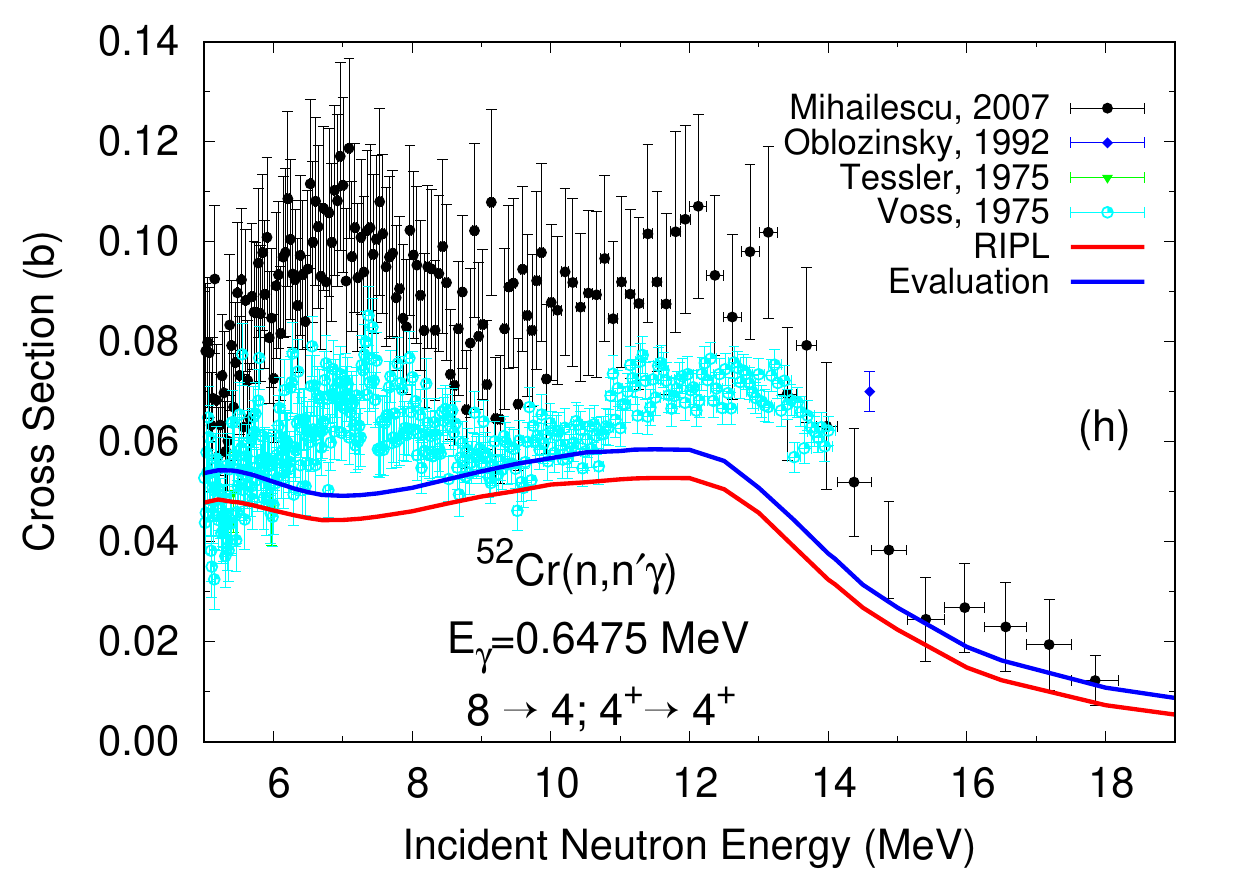} \hspace{-4mm}
\includegraphics[scale=0.50,keepaspectratio=true,clip=true,trim=9mm 7mm 0mm 0mm]{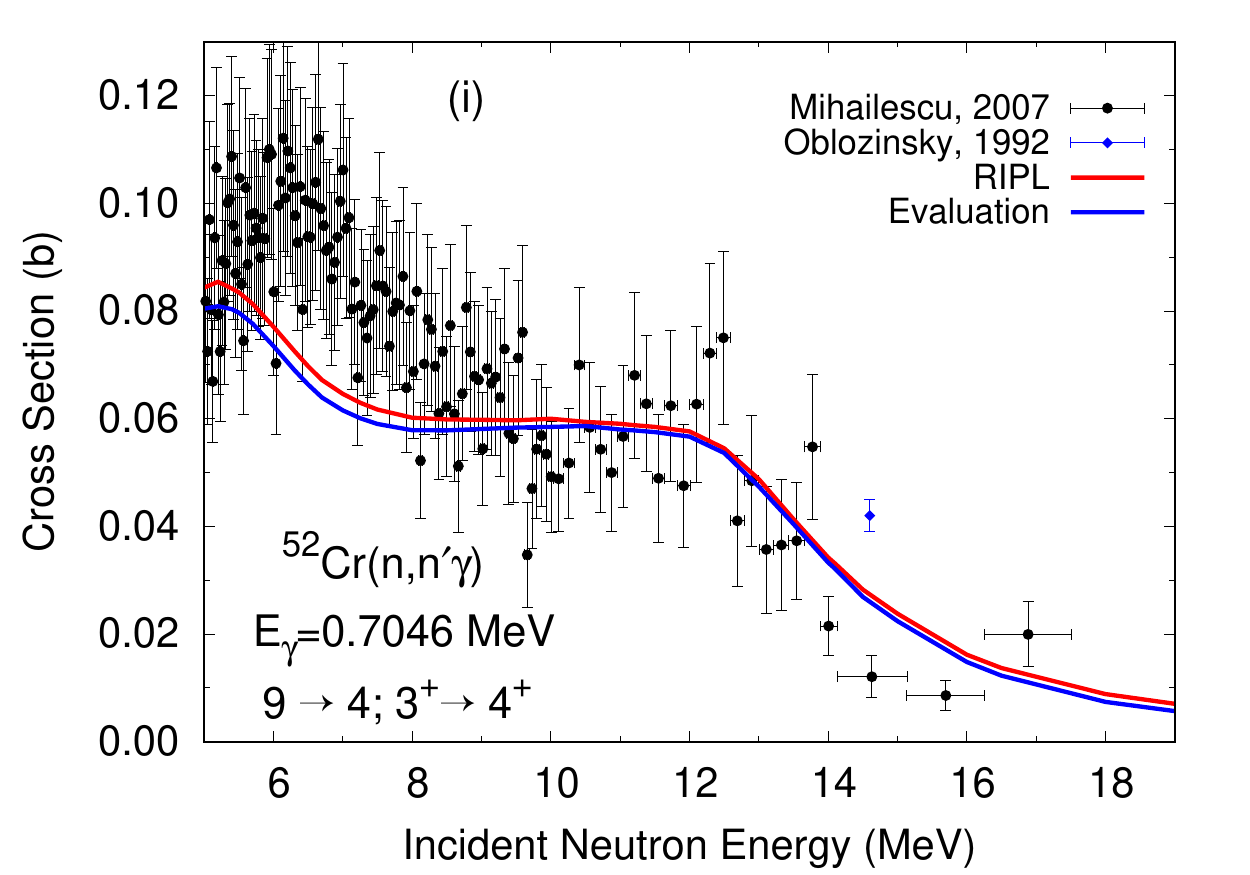}\\
\includegraphics[scale=0.50,keepaspectratio=true,clip=true,trim=0mm 0mm 0mm 0mm]{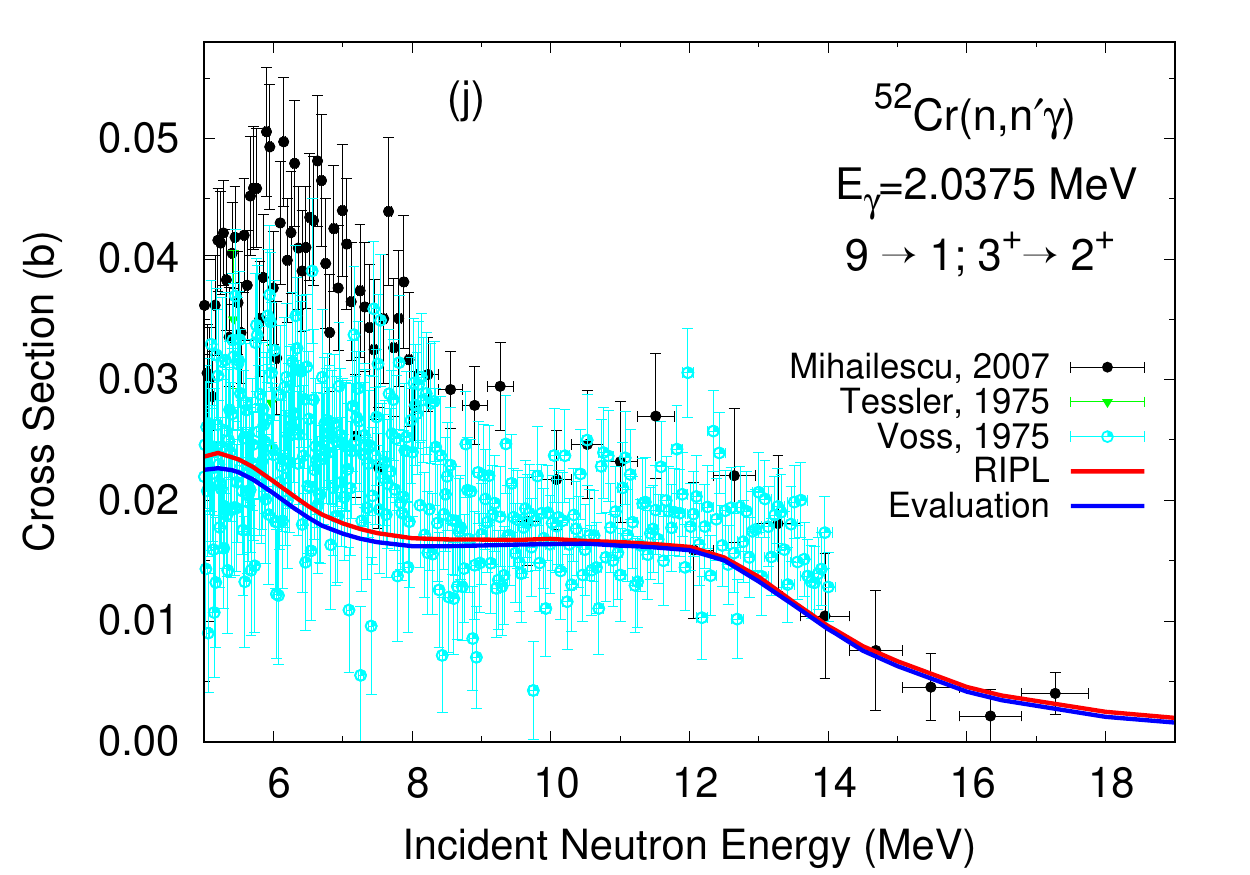} \hspace{-4mm}
\includegraphics[scale=0.50,keepaspectratio=true,clip=true,trim=9mm 0mm 0mm 0mm]{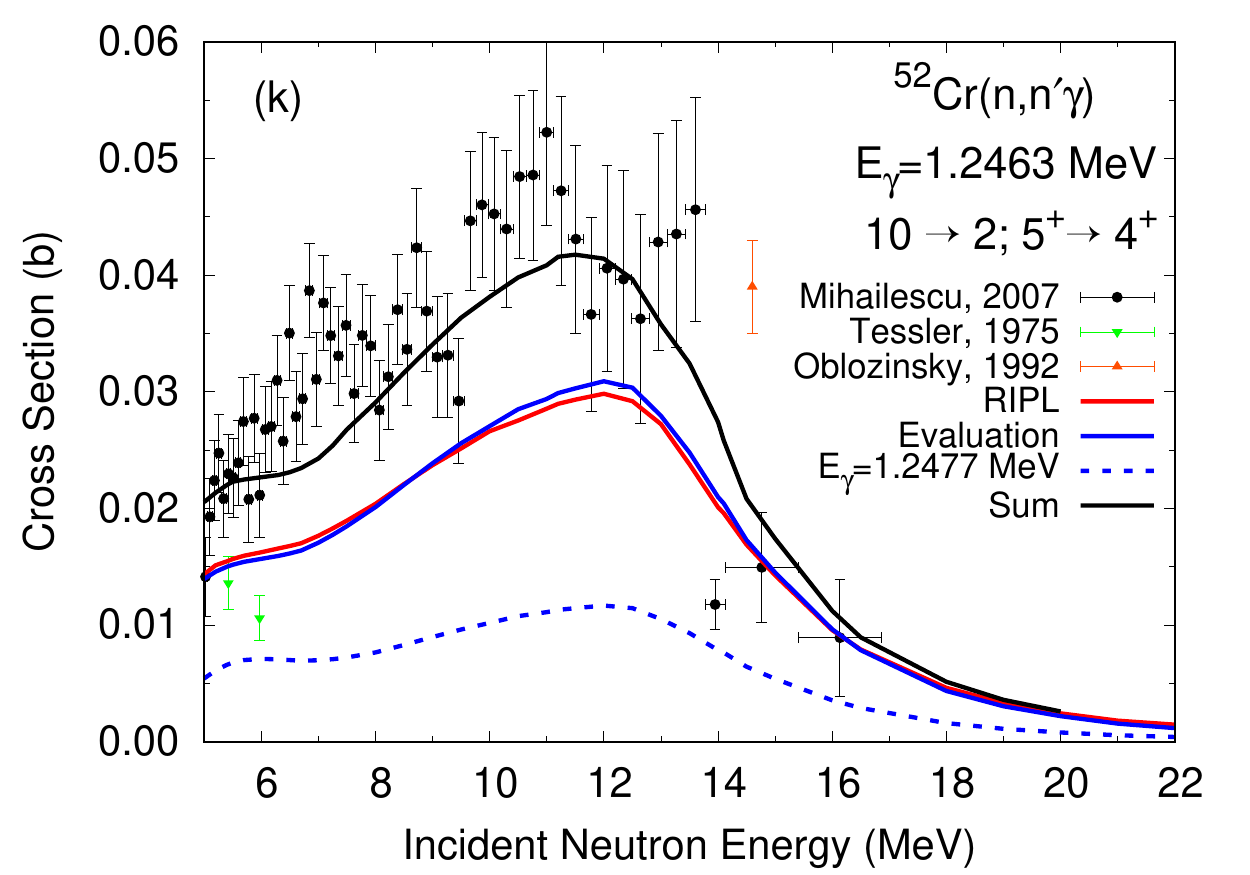} \hspace{-4mm}
\includegraphics[scale=0.50,keepaspectratio=true,clip=true,trim=9mm 0mm 0mm 0mm]{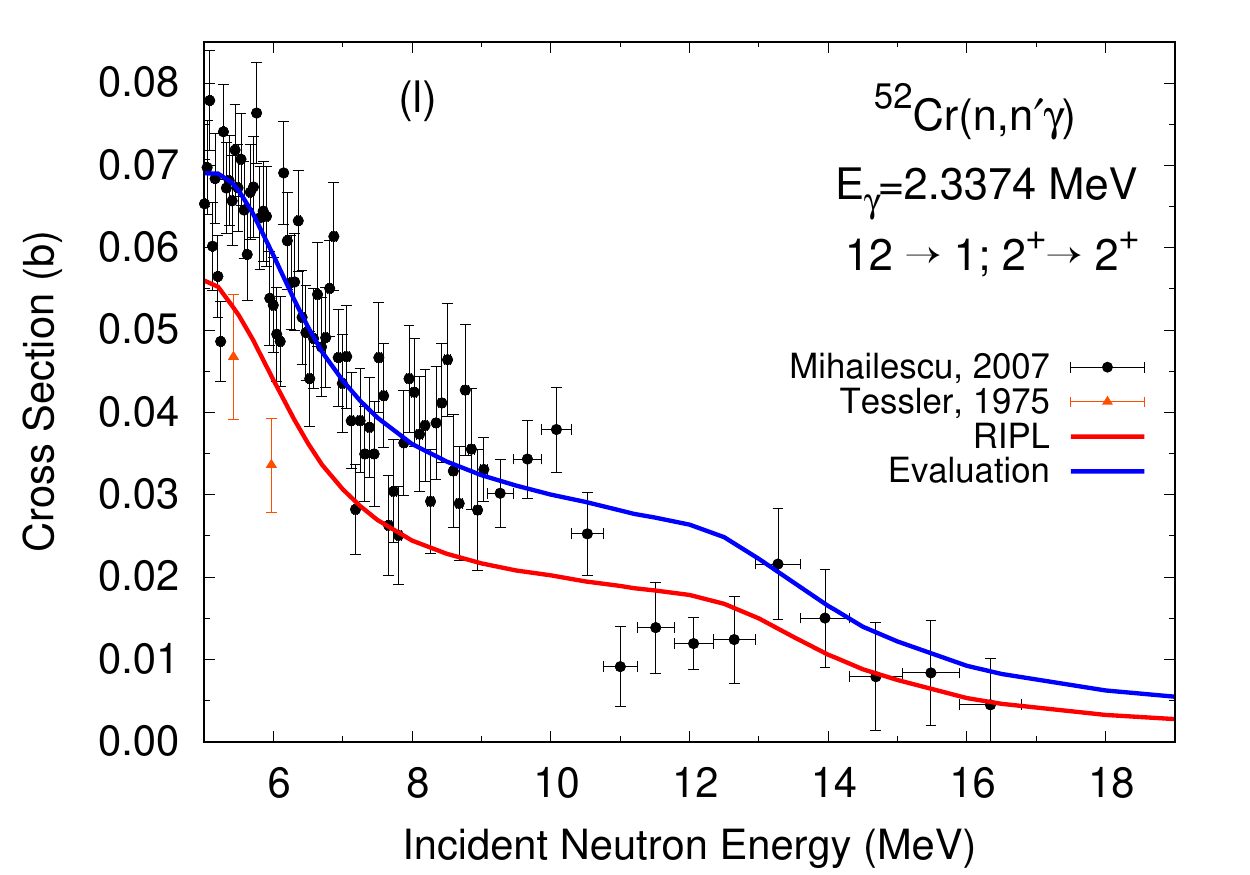} \\
\caption{(color online) Inelastic gamma cross section measurements for \nuc{52}{Cr}. The energy of the transition $\gamma$ is indicated in each panel, as well as the index, spins, and parities of the levels involved in the transition. Results are shown only above neutron incident energy of 5~MeV as model calculations were replaced by experimental data for the first neutron inelastic level cross section below 5~MeV and second to fifth levels below 4~MeV. Data taken from Refs.~\cite{Mihailescu:2007,Abbondanno:1973ws,Lashuk:1996,J.ZN_A.26.451.7103,Voss:1975,Oblozinsky:1992,Tessler:1975}.}
\label{fig:inel_gammas}
\end{figure*}

 Maximum consistency between inelastic neutron data and inelastic gamma cross-section measurements were attempted. For that, missing experimental unambiguous information related to branching ratios and high-level spin and parity in RIPL were complemented through constraints from gamma cross-section data.
 As examples of such changes, the level with excitation energy 3.9512 MeV was missing in RIPL, even though it can be found in ENSDF, and was added to the evaluation level scheme; levels with excitation energy 3.7396, 3.9475, 3.9512, and 4.1000~MeV had incomplete or non-existent decay information in RIPL and/or ENSDF.
 In Fig.~\ref{fig:inel_gammas} we show the predicted EMPIRE calculations for the gamma cross sections using decay scheme from RIPL, where missing branching ratios decay directly to ground state, as the red curve, and the current evaluation (blue curve), compared with Mihailescu data. In some few cases the agreement with data is not optimal: e.g., panels (f) and (h) of Fig.~\ref{fig:inel_gammas} are either consistently lower or higher than measured data;  or in one case the evaluation agrees well with data of Voss et al. \cite{Voss:1975} but not as well with Mihailescu. However, generally, the agreement of the present evaluation with Mihailescu data is very good, in particular for neutron incident energies above $\sim$ 5 MeV. We draw the attention for the good description of the main $\gamma$-emission cross section, i.e., the transition from the first excited state ($E_1$ = 1.4341 MeV) to the ground state, shown in Fig.~\ref{fig:inel_gammas}  (a). Since a large majority of the level excitations in \nuc{52}{Cr} will eventually decay to the first excited level before reaching the ground state, this $\gamma$ transition is a good approximation of total neutron inelastic, up to certain energy. We also point out to the measured transition of $E_\gamma$= 1.2463 MeV, shown in Fig.~\ref{fig:inel_gammas} (k). Since the experimental resolution of the Mihailescu measurement (1.1 keV at 1 MeV \cite{Mihailescu:2007}) was not enough to separate the transitions $E_\gamma=$ 1246.278 MeV (10$^{\mathrm{th}}$ excited level to the 2$^{\mathrm{nd}}$) from the $E_\gamma= $1247.88  MeV (15$^{\mathrm{th}}$ excited level to the 4$^{\mathrm{th}}$). When the evaluated cross section for both transitions is summed, a good agreement with measured data is reached.

\begin{figure}
\includegraphics[scale=0.70,keepaspectratio=true,clip=true,trim=0mm 0mm 0mm 0mm]{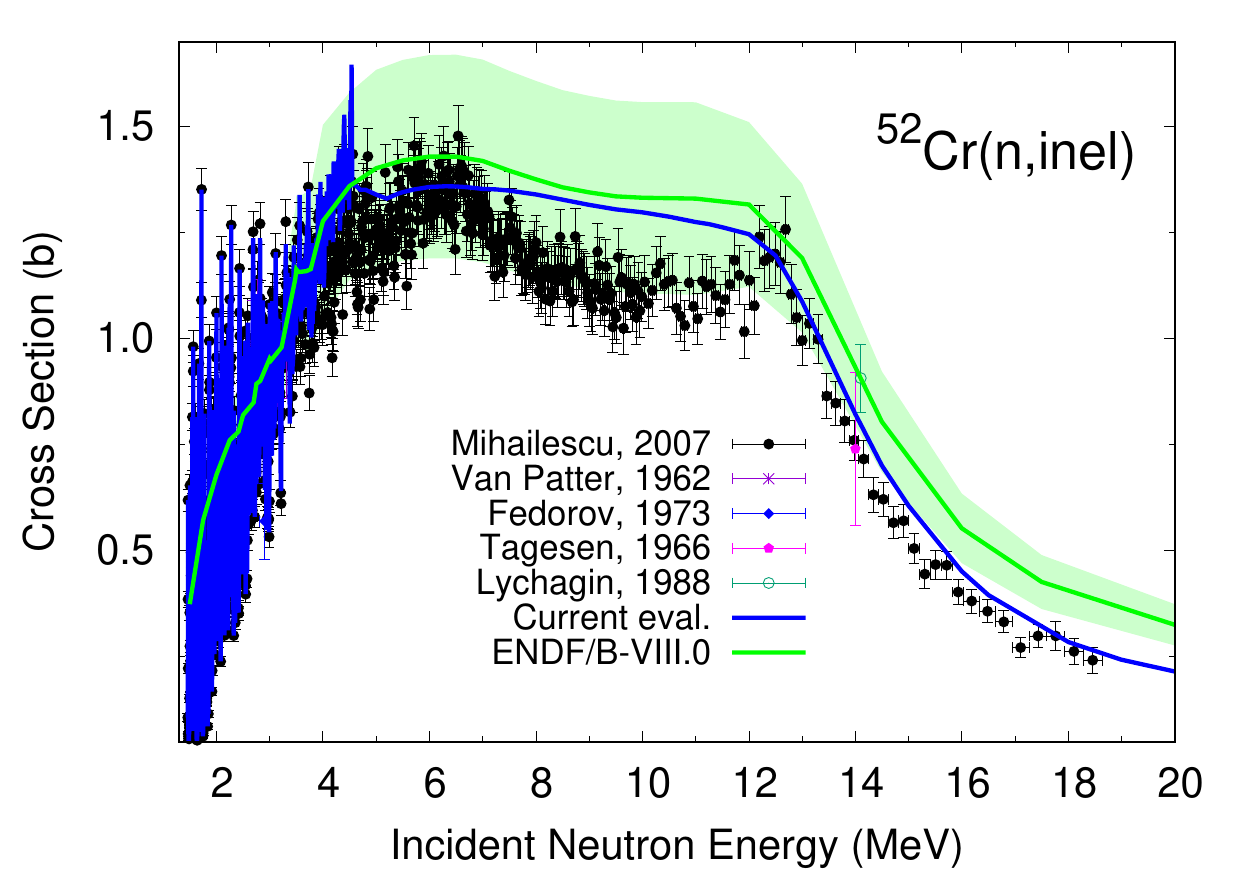}
\caption{(color online) Total inelastic cross section measurements for \nuc{52}{Cr}. Data taken from Refs.~\cite{Mihailescu:2007,VanPatter:1962,Fedorov:1973,Tagesen:1966,Lychagin:1988}.}
\label{fig:cr52-MT4}
\end{figure}

 We smoothed the total inelastic neutron cross section, as well as partial inelastic neutron cross sections so we could better assess the quality of our model-parameter fits. Because we adopted a realistic coupled-channels optical model potential, we were able to develop a consistent framework between Mihailescu data for inelastic neutrons, as well as for inelastic gamma cross sections, which are the actual observables of the experiment. In Fig.~\ref{fig:cr52-MT4} we compare the total inelastic from the current evaluation with ENDF/B-VIII.0 and experimental data, in particular the ones from Mihailescu et al. We clearly identified the need to lower total inelastic cross section above 5 MeV, when compared to ENDF/B-VIII.0. In the decreasing part of total inelastic (above $\sim$ 13 MeV), current evaluation is significantly lower than ENDF/B-VIII.0, in agreement with the upper bound of Mihailescu data. Attempts to lower it further to improve agreement with data would implicate poor description of other threshold reaction channels open at this energy such as (n,p) and (n,2n). However, considering the good agreement with Mihailescu data seen in Fig.~\ref{fig:inel_gammas}(a) in this same energy region suggests that the total inelastic neutron data may be slightly underestimated at such higher energies.

\begin{figure}
\includegraphics[scale=0.60,keepaspectratio=true,clip=true,trim=0mm 7mm 0mm 0mm]{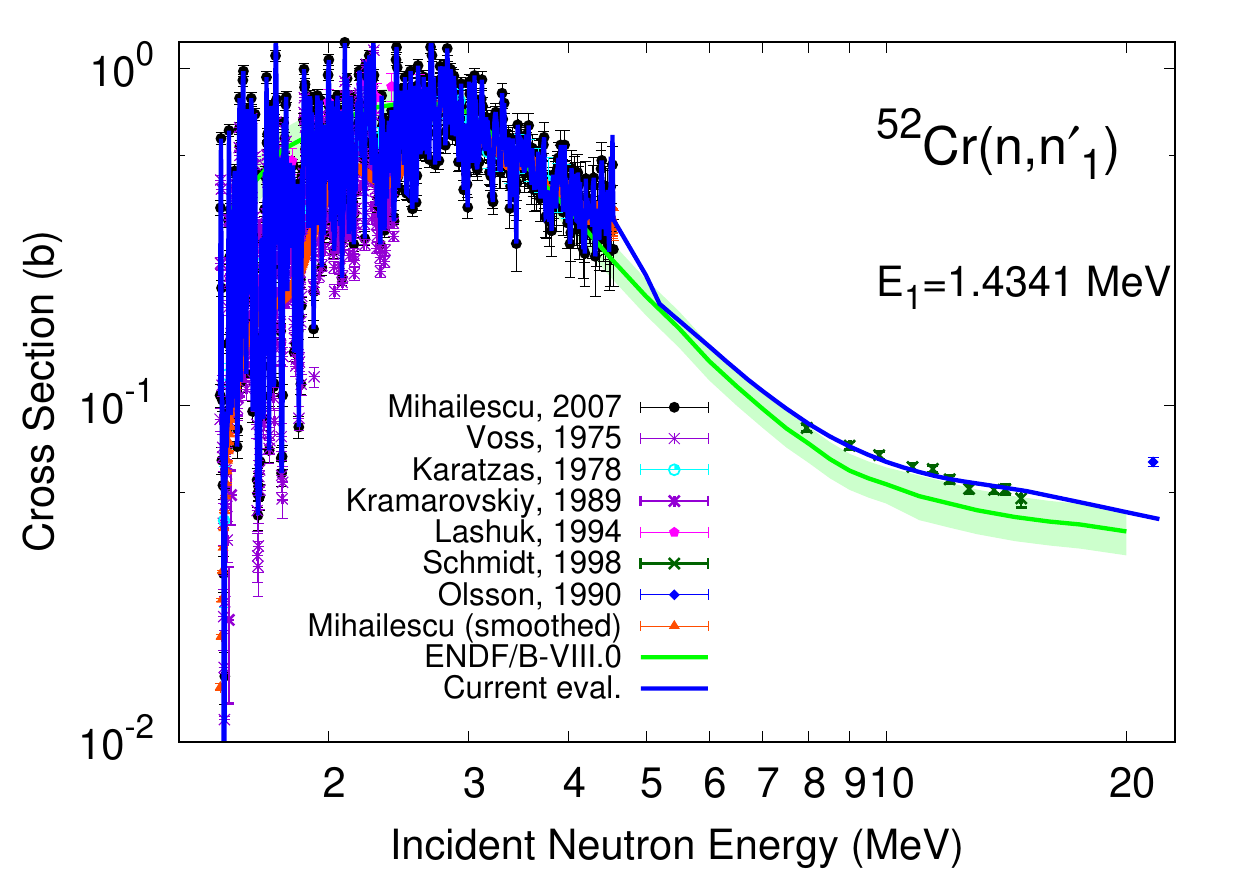}
\includegraphics[scale=0.60,keepaspectratio=true,clip=true,trim=0mm 7mm 0mm 0mm]{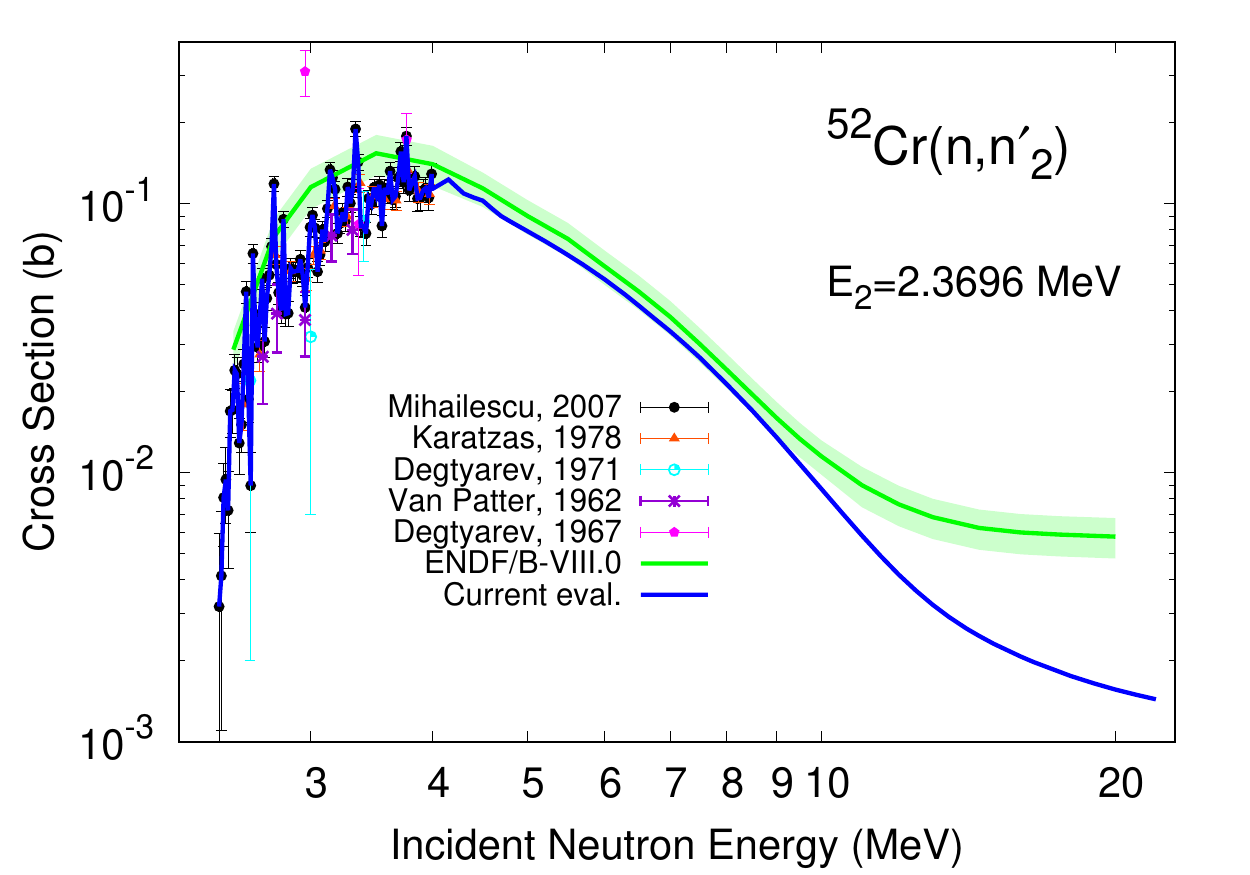}
\includegraphics[scale=0.60,keepaspectratio=true,clip=true,trim=0mm 7mm 0mm 0mm]{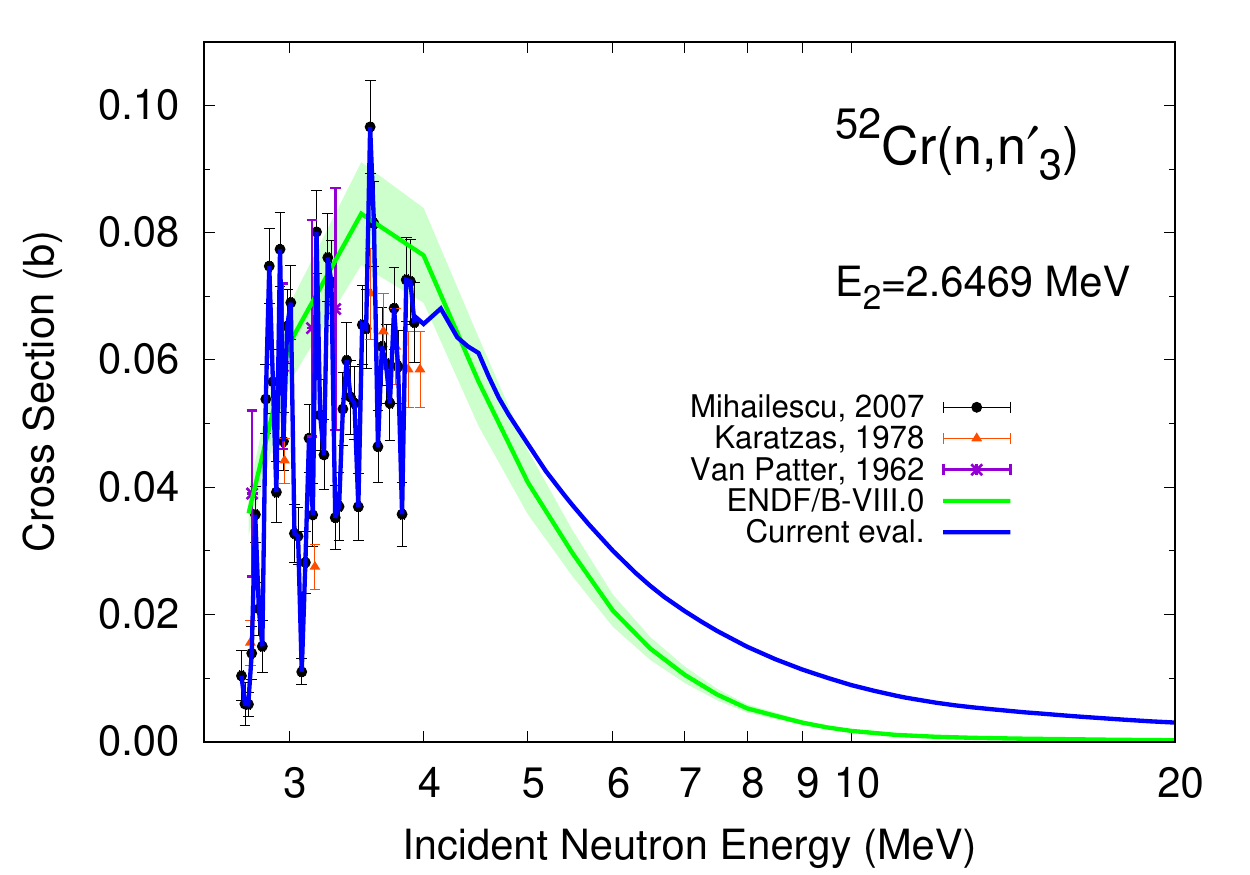}
\includegraphics[scale=0.60,keepaspectratio=true,clip=true,trim=0mm 0mm 0mm 0mm]{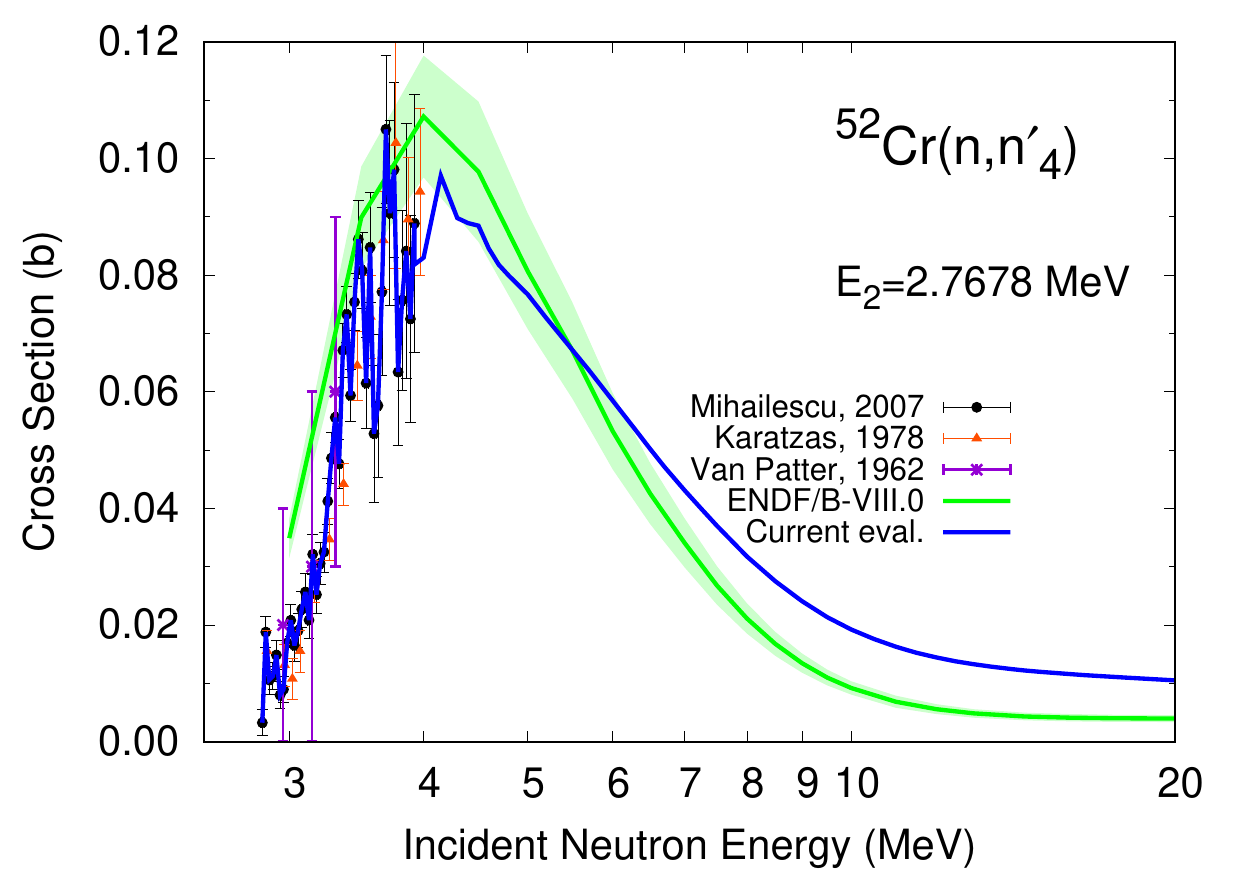}
\caption{(color online) Partial inelastic cross section measurements for \nuc{52}{Cr}.  Data taken from Refs.~\cite{Mihailescu:2007,Voss:1975,Karatzas:1978,Kramarovskiy:1989,Lashuk:1996,Schmidt:1998,Olsson:1990vy,Degtyarev:1971,VanPatter:1962,Degtjarev:1967}.
}
\label{fig:cr52-MT51-54}
\end{figure}

 Due to the expected diminishing of predictive quality of the optical model approach at energies below 5 MeV, we decided to simply adopt Mihailescu data for the first 5 excited level cross sections below 5~MeV (first inelastic) and 4~MeV (second to fifth inelastic channel), constrained by total cross-section data and with the difference put into the elastic channel.  Fig.~\ref{fig:cr52-MT51-54} shows the partial inelastic cross sections for the first four excited levels. Noteworthy is the fact that, for the first excited level, our evaluation also agrees well with the higher-energy set of Schmidt~\cite{C.97TRIEST.1.505.1997}, bringing to consistency inelastic gamma, partial level, and total neutron inelastic cross-section data.

\begin{figure}
\includegraphics[scale=0.70,keepaspectratio=true,clip=true,trim=0mm 0mm 0mm 0mm]{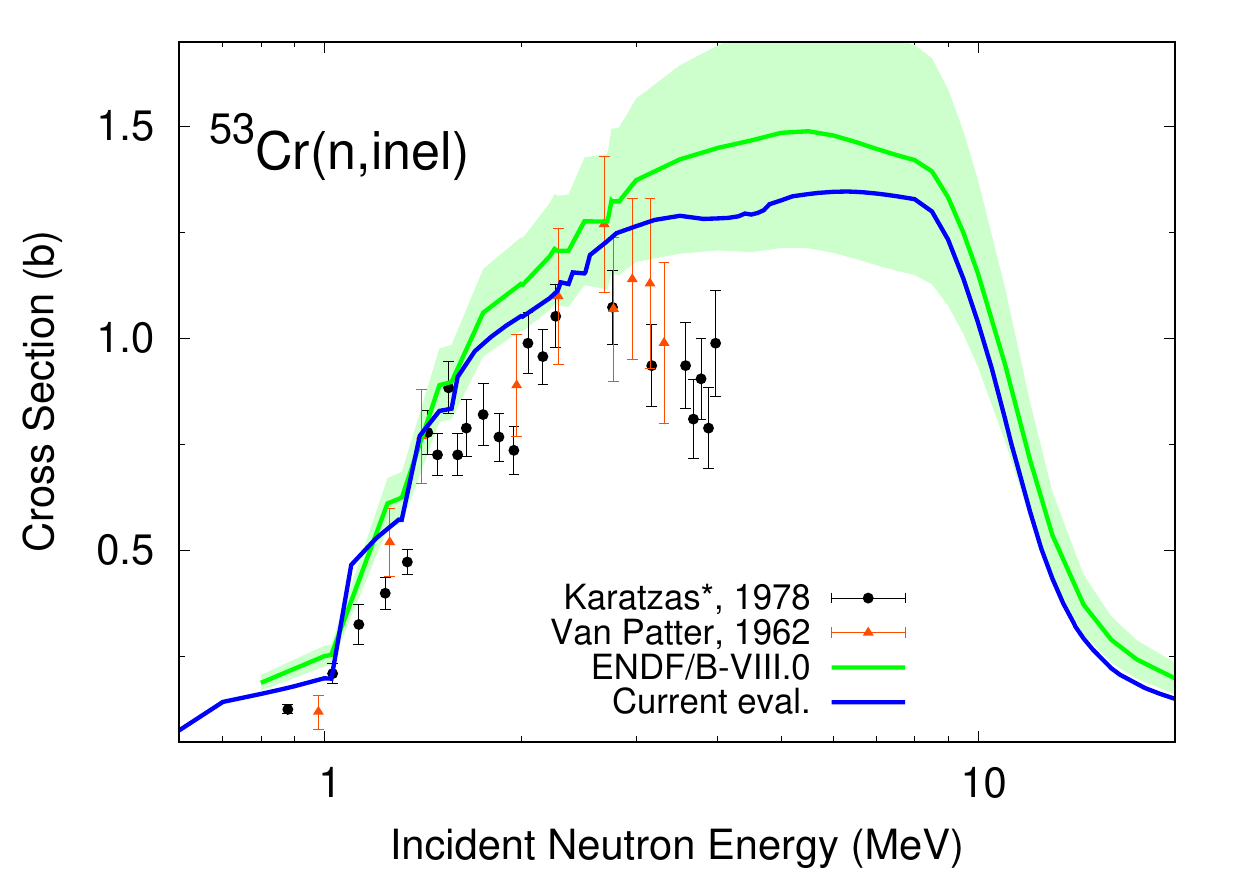}
\includegraphics[scale=0.70,keepaspectratio=true,clip=true,trim=0mm 0mm 0mm 0mm]{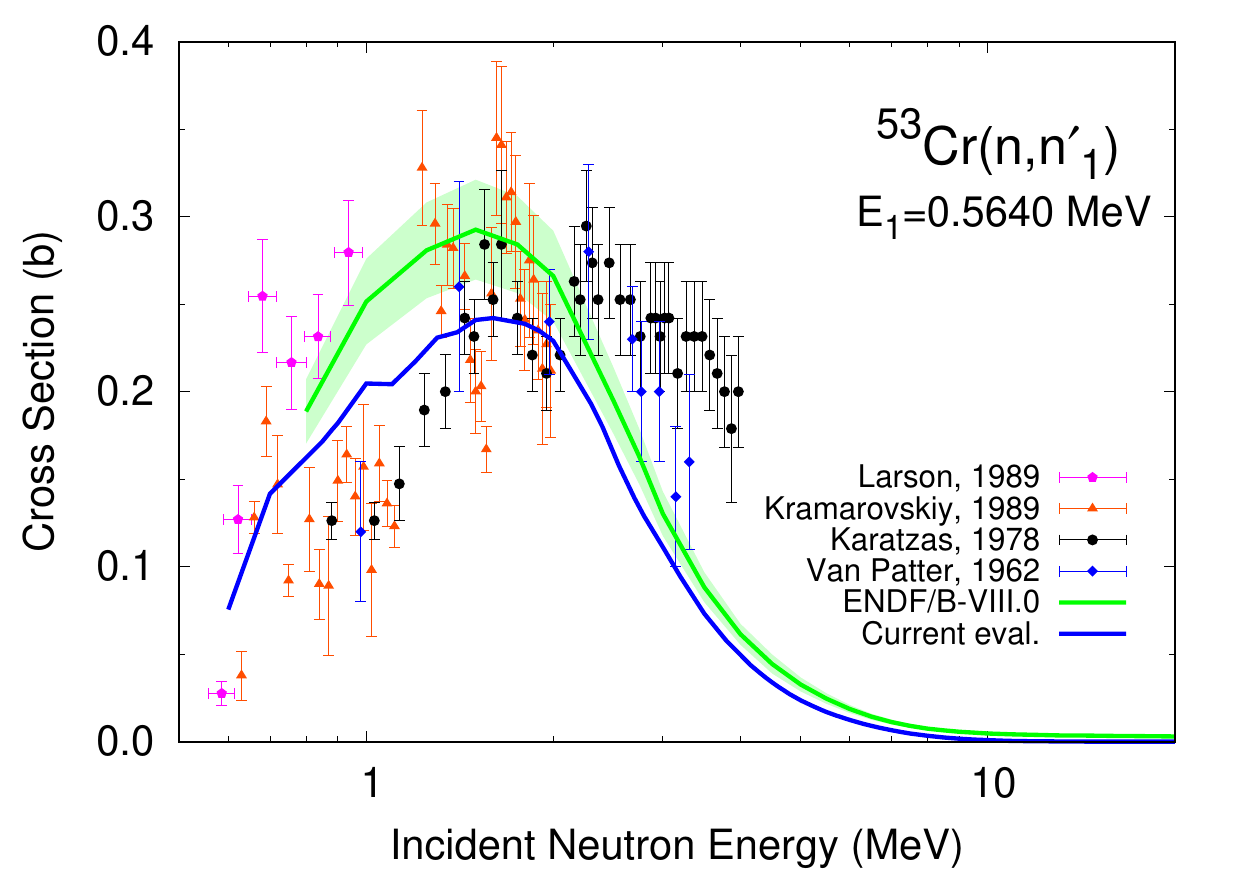}
\caption{(color online) Total inelastic cross section for \nuc{53}{Cr} (top panel) and for the first excited level, which has excitation energy of 0.5640 MeV.  Data taken from Refs.~\cite{Karatzas:1978,VanPatter:1962,Larson:1989,Kramarovskiy:1989}.}
\label{fig:cr53-inelastic}
\end{figure}

  \paragraph*{\nuc{53}{Cr}($n$,inelastic) - } For \nuc{53}{Cr} we fitted to total inelastic data of Karatzas~\etal~\cite{Karatzas:1978}, reaching a good agreement. The current evaluation lowers total inelastic significantly above $\sim$2 MeV in comparison with ENDF/B-VIII.0, as can be seen in Fig.~\ref{fig:cr53-inelastic} (top panel). In the bottom panel of Fig.~\ref{fig:cr53-inelastic} we show a comparison for the inelastic cross section of the first level. The agreement with data is not as good but these data carry additional measurement difficulties due to the low excitation energy.

\begin{figure*}
\includegraphics[scale=0.67,keepaspectratio=true,clip=true,trim=  0mm 7mm 4mm 3mm]{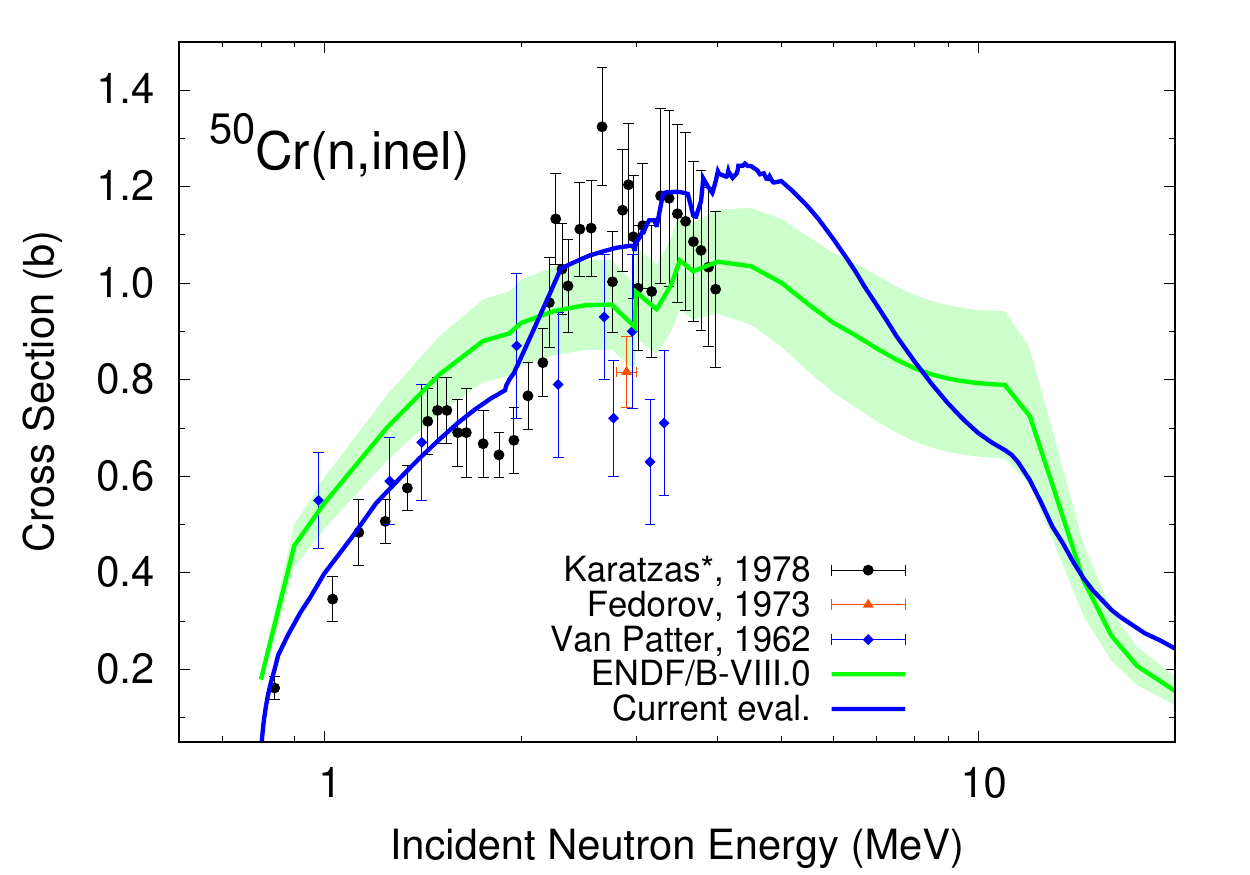}
\includegraphics[scale=0.67,keepaspectratio=true,clip=true,trim=18mm 7mm 4mm 3mm]{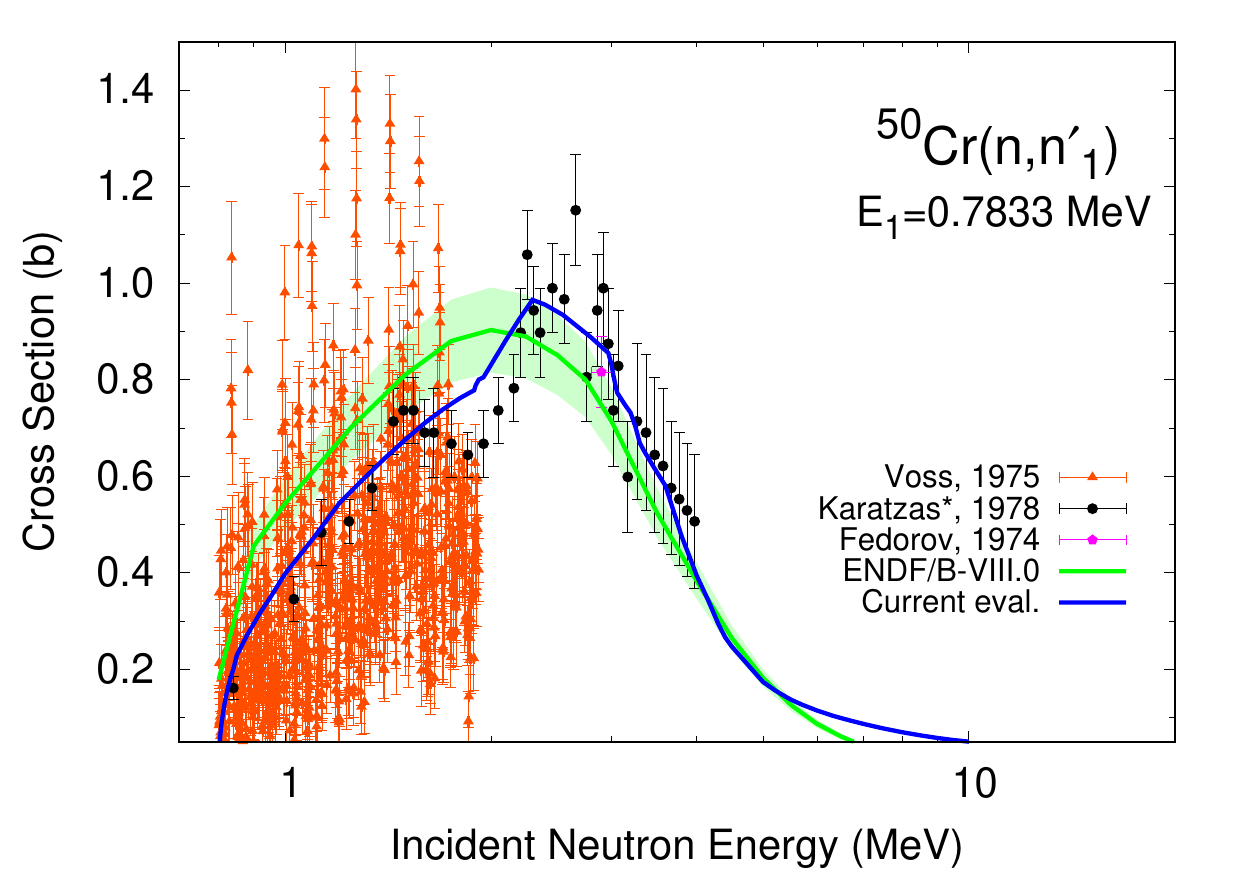} \\
\includegraphics[scale=0.67,keepaspectratio=true,clip=true,trim=  0mm 0mm 4mm 3mm]{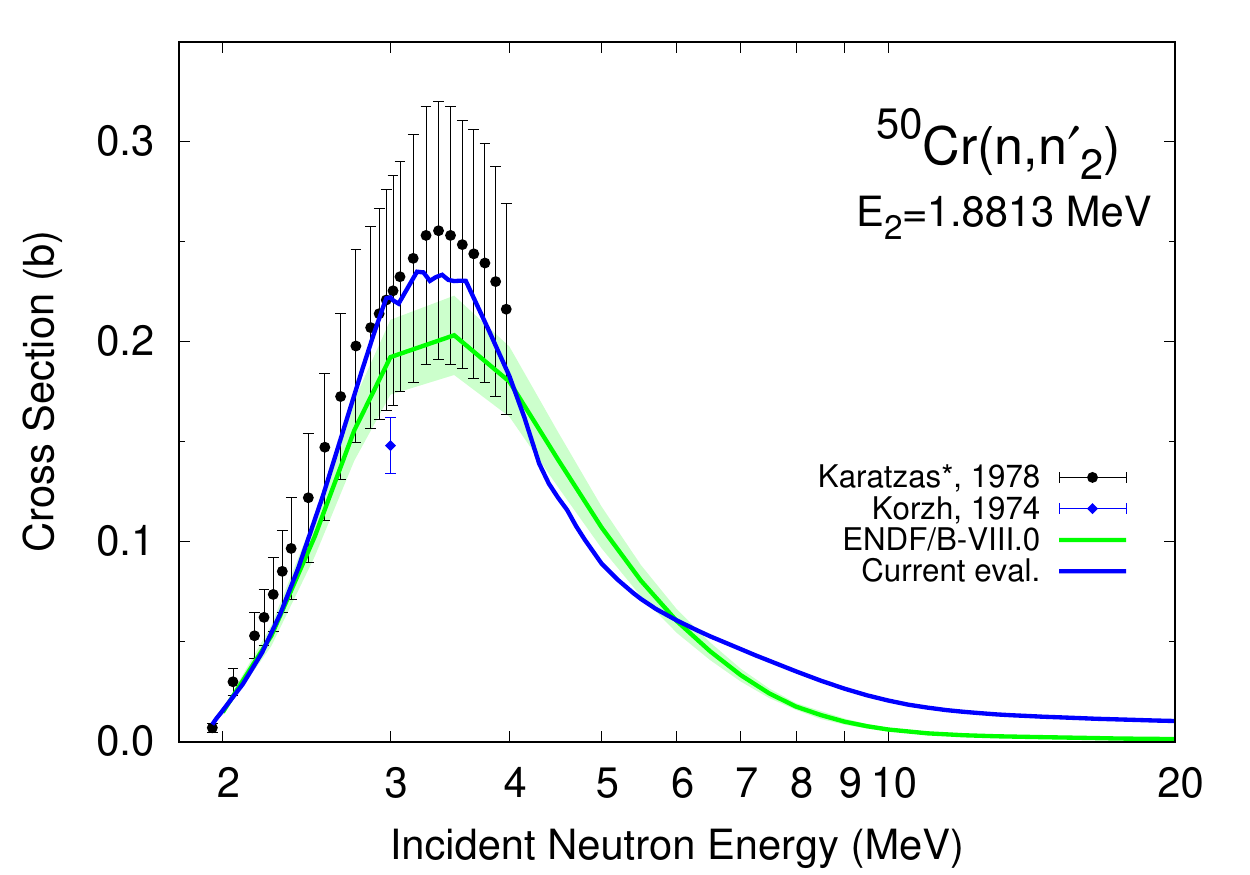}
\includegraphics[scale=0.67,keepaspectratio=true,clip=true,trim=18mm 0mm 4mm 3mm]{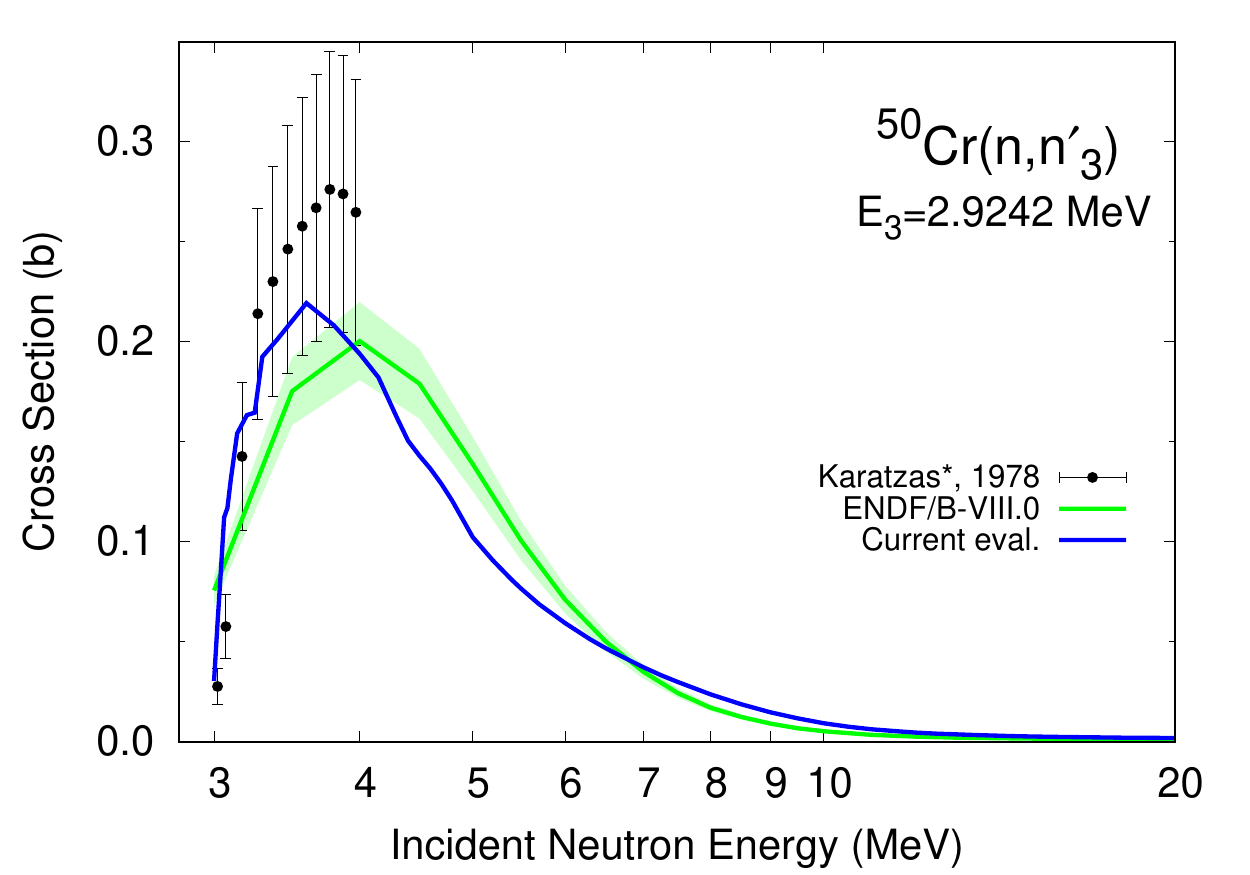}
\caption{(color online) \nuc{50}{Cr} inelastic cross sections for the sum of all levels (top left) and for the first, second, and third excited levels, which have excitation energy of 0.7833 MeV, 1.8813 MeV, and 2.9242 MeV, respectively.  Data taken from Refs.~\cite{Karatzas:1978,Fedorov:1973,VanPatter:1962,Voss:1975,Korzh:1975}.}
\label{fig:cr50-inelastic}
\end{figure*}

 \paragraph*{\nuc{50}{Cr}($n$,inelastic) - } Fig.~\ref{fig:cr50-inelastic} shows the \nuc{50}{Cr} total inelastic cross section, as well as the partial cross sections for the first three excited levels. As mentioned above we fitted to data of Karatzas~\etal~\cite{Karatzas:1978}, obtaining a good agreement with experimental data, and generally better than ENDF/B-VIII.0.

\subsubsection{Capture Cross Sections}
\label{sec:capture_fast}

\begin{figure}
\includegraphics[scale=0.70,keepaspectratio=true,clip=true,trim=0mm 0mm 0mm 0mm]{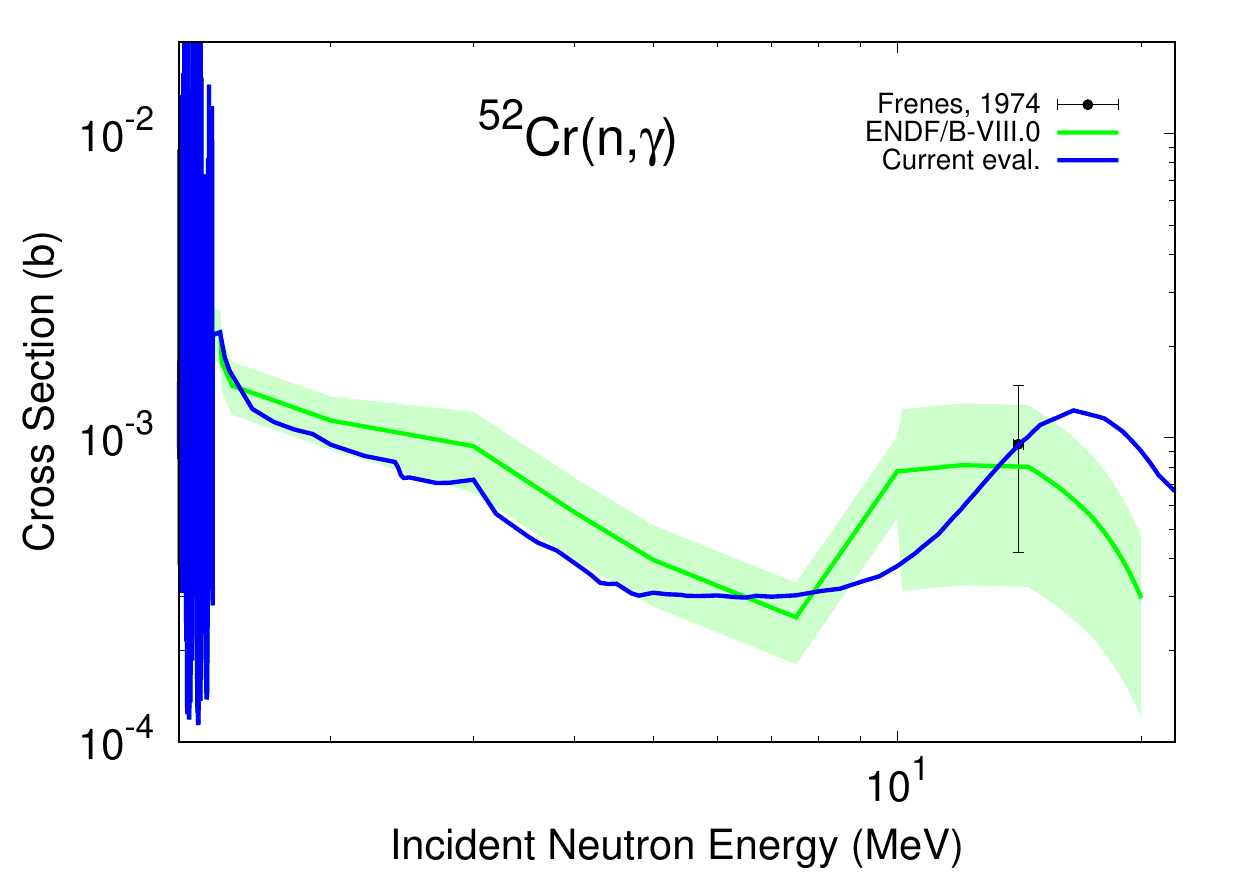}
\caption{(color online) Capture cross section for \nuc{52}{Cr}  in the fast neutron range. Data taken from Refs.~\cite{NSR1974FR15}.}
\label{fig:cr52-MT102}
\end{figure}

\paragraph*{\nuc{52}{Cr}($n,\gamma$) - } For a magic nucleus like \nuc{52}{Cr} capture cross sections in the fast region are small. Availability of data for this nucleus is also poor as the resulting compound nucleus is stable as can be seen in Fig.~\ref{fig:cr52-MT102}, where we compare the present evaluation with ENDF/B-VIII.0 and the single datum available, from Frenes et al. Our evaluation is consistently lower than ENDF/B-VIII.0 from threshold up to around 14 MeV, but there are no data available to constrain them in this region. Our evaluation does however reproduce better the cross-section cusps due the opening of inelastic channels, as well as the 14 MeV datum.

\begin{figure}
\includegraphics[scale=0.70,keepaspectratio=true,clip=true,trim=0mm 0mm 0mm 0mm]{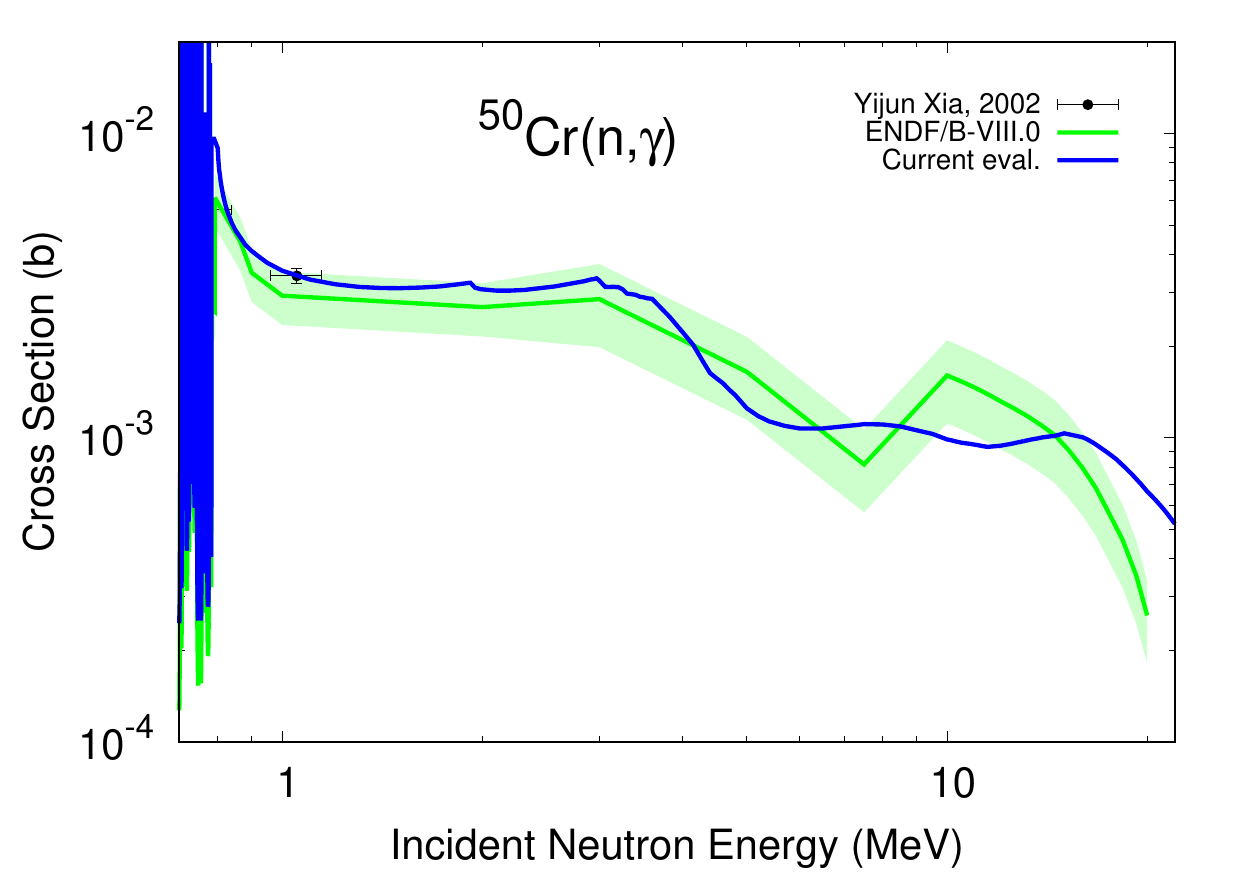}
\caption{(color online) Capture cross section for \nuc{50}{Cr}  in the fast neutron range. Data taken from Refs.~\cite{NSR2002XIZZ}.}
\label{fig:cr50-MT102}
\end{figure}

\paragraph*{\nuc{50}{Cr}($n,\gamma$) - } The only other chromium isotope to have capture data in the fast region is \nuc{50}{Cr}, even though, like \nuc{52}{Cr} it is just a single point. Nevertheless, this point of Yijun Xia \etal \cite{NSR2002XIZZ} helped us constrain the capture normalization, as seen in Fig.~\ref{fig:cr50-MT102}.

\subsubsection{($n,2n$) Cross Sections}

\paragraph*{\nuc{52}{Cr}($n,2n$) -} The existing ENDF/B evaluation for \nuc{52}{Cr}($n$,$2n$)\nuc{51}{Cr} cross sections dates back from ENDF/B-VI-MOD1 \cite{ENDF-VI.8}, done by D. Larson (ORNL) in 1989. In that evaluation, the cross sections were taken from the activation file for \nuc{52}Cr but were adjusted to be smaller below the energy of 15.5 MeV to match the data of Wagner et al.~\cite{Wagner:1989} which agreed well with other data from up to the same period in this energy region.

\begin{figure}
\includegraphics[scale=0.70,keepaspectratio=true,clip=true,trim=0mm 0mm 0mm 0mm]{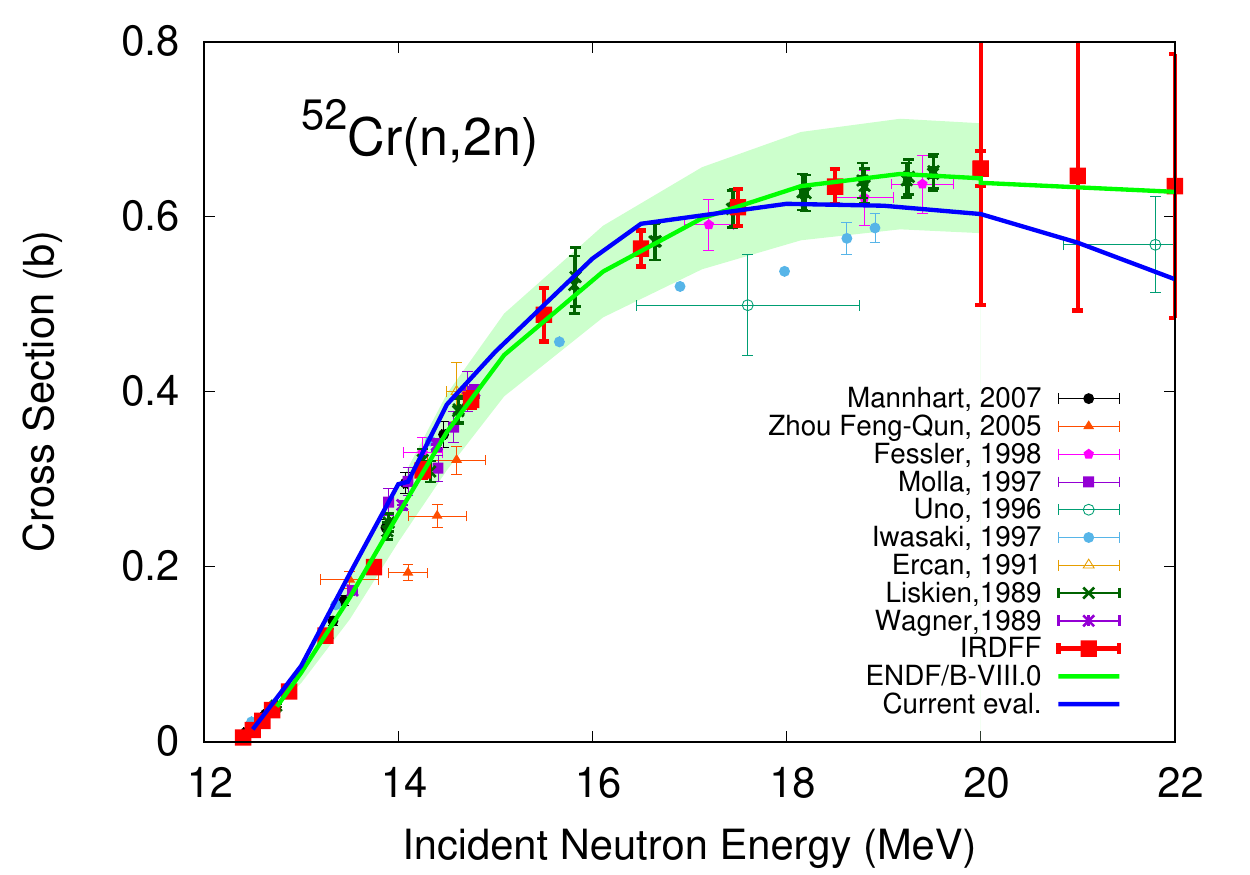}
\caption{(color online) Evaluated \nuc{52}{Cr}(n,2n) cross section compared to ENDF/B-VIII.0, IRDFF~\cite{IRDFF1,IRDFF2} and experimental data~\cite{Mannhart:2007,Zhou:2005,Fessler:1998,Molla:1997,Uno:1996,Iwasaki:1997,Ercan:1991,Liskien:1989,Wagner:1989}.}
\label{fig:cr52-MT16}
\end{figure}

For our new evaluation, in addition of the data of Wagner~\cite{Wagner:1989}, we considered \nuc{52}{Cr}($n$,$2n$) experimental sets equally or more recent than the previous 1989 evaluation (Refs.~\cite{Mannhart:2007,Zhou:2005,Fessler:1998,Molla:1997,Uno:1996,Iwasaki:1997,Ercan:1991,Liskien:1989}). Ultimately, we fitted to the dosimetry file IRDFF v1.05~\cite{IRDFF1,IRDFF2} below 20 MeV, which is compatible with the data sets listed. Since the IRDFF-1.05 above 20 MeV corresponds to a renormalized TENDL library not fitted to data, we decided to fit to Uno data above 20 MeV. Fig.~\ref{fig:cr52-MT16} shows the \nuc{52}{Cr}(n,2n) cross section from our evaluation compared to ENDF/B-VIII.0, IRDFF-1.05 and experimental data. It should be noted that our evaluation is slightly higher than the IRDFF file below 20 MeV and the shape between 16 and 20 MeV disagree with data. The IRDFF file is recommended for neutron dosimetry applications.

\subsubsection{($n,p$) Cross Sections}

\begin{figure}
\includegraphics[scale=0.70,keepaspectratio=true,clip=true,trim=0mm 0mm 0mm 0mm]{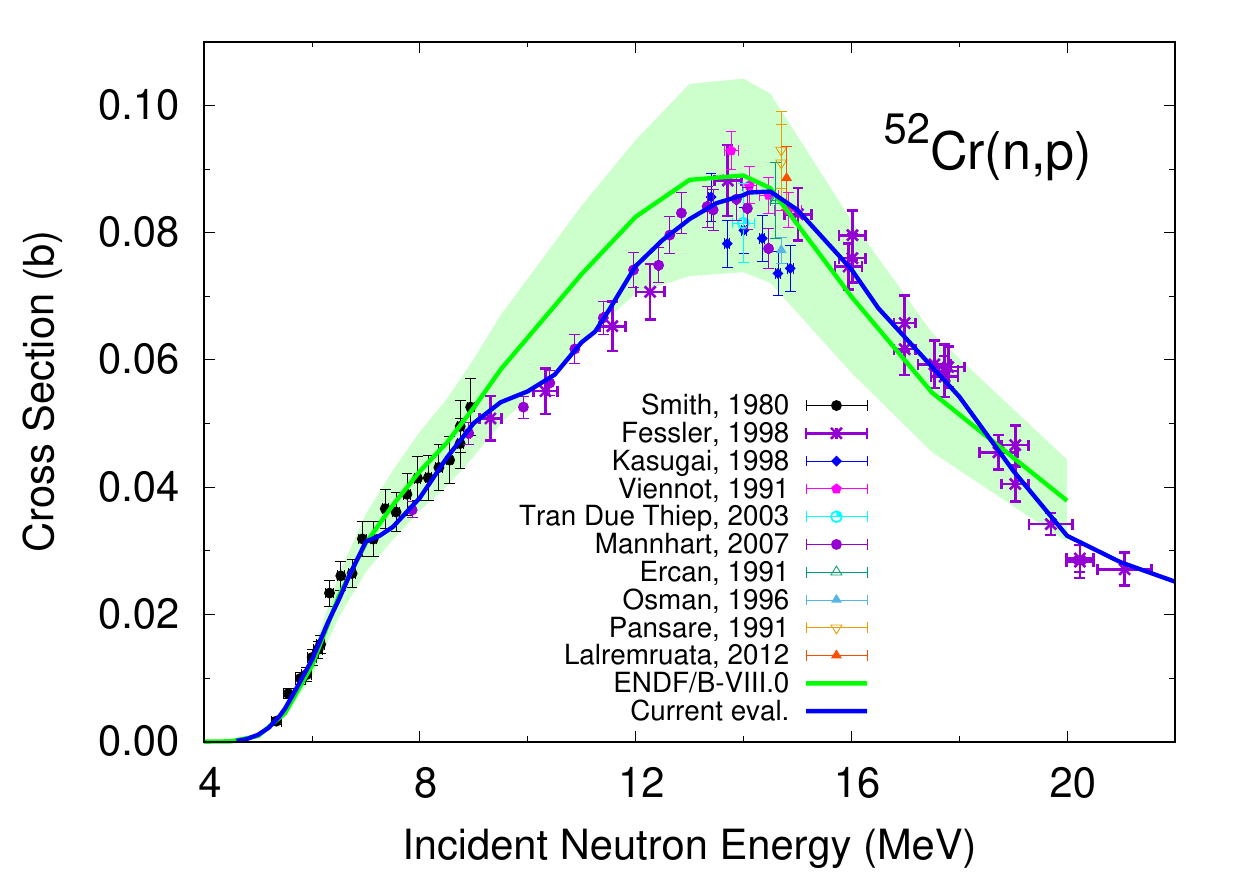}
\caption{(color online) Evaluated \nuc{52}{Cr}(n,p) cross section compared with ENDF/B-VIII.0 and with relevant experimental data \cite{Sm80,Gu85,Ar80,Mo77,Fessler:1998,Mannhart2007,Lalremruata2012,TranDucThiep2003,Kasugai1998,Osman1996,Viennot1991,C.91JUELIC..376.199105,Pansare1991}.}
\label{fig:cr52-MT103}
\end{figure}

\paragraph*{\nuc{52}{Cr}($n,p$) -} The previous \nuc{52}{Cr}($n,p$)\nuc{52}{V} dates back from ENDF/B-VI-MOD1 \cite{ENDF-VI.8}, done by D. Larson (ORNL) in 1989. The ($n$,$p$) cross sections had been calculated by the TNG code \cite{Fu80-BNL,Fu88-1,Fu88-2,Sh86} but modified between 7-16 MeV for better fit of the experimental data up to 9 MeV of Smith et al. \cite{Sm80} and  measurements, by then recent, at 14.5 MeV \cite{Gu85,Ar80,Mo77}. There are many experimental data sets which are newer than the previous evaluation, even though some display inconsistencies among them. The data of Fessler et al. \cite{Fessler:1998} begins at 9.31 MeV, which is above all Smith data used in the previous evaluation,  and goes up to 21.1 MeV. Data from Mannhart et al. \cite{Mannhart2007} overlaps with Smith at 7.9 MeV and goes up to 14.5 MeV. Both sets of Fessler~\cite{Fessler:1998} and Mannhart~\cite{Mannhart2007} play an important complementary role relative to the previous evaluation. In addition to these sets, the new (n,p) data considered in this evaluation were Refs.~\cite{Lalremruata2012,TranDucThiep2003,Kasugai1998,Osman1996,Viennot1991,C.91JUELIC..376.199105,Pansare1991}.
Older sets or data sets with clear conflicting points were excluded from fits and tuning.
Fig.~\ref{fig:cr52-MT103} shows a comparison between the present work and ENDF/B-VIII.0 and the experimental data mentioned above. A clear improvement in the agreement with data, especially above around 9 MeV can be observed.




\begin{figure}
\includegraphics[scale=0.70,keepaspectratio=true,clip=true,trim=0mm 0mm 0mm 0mm]{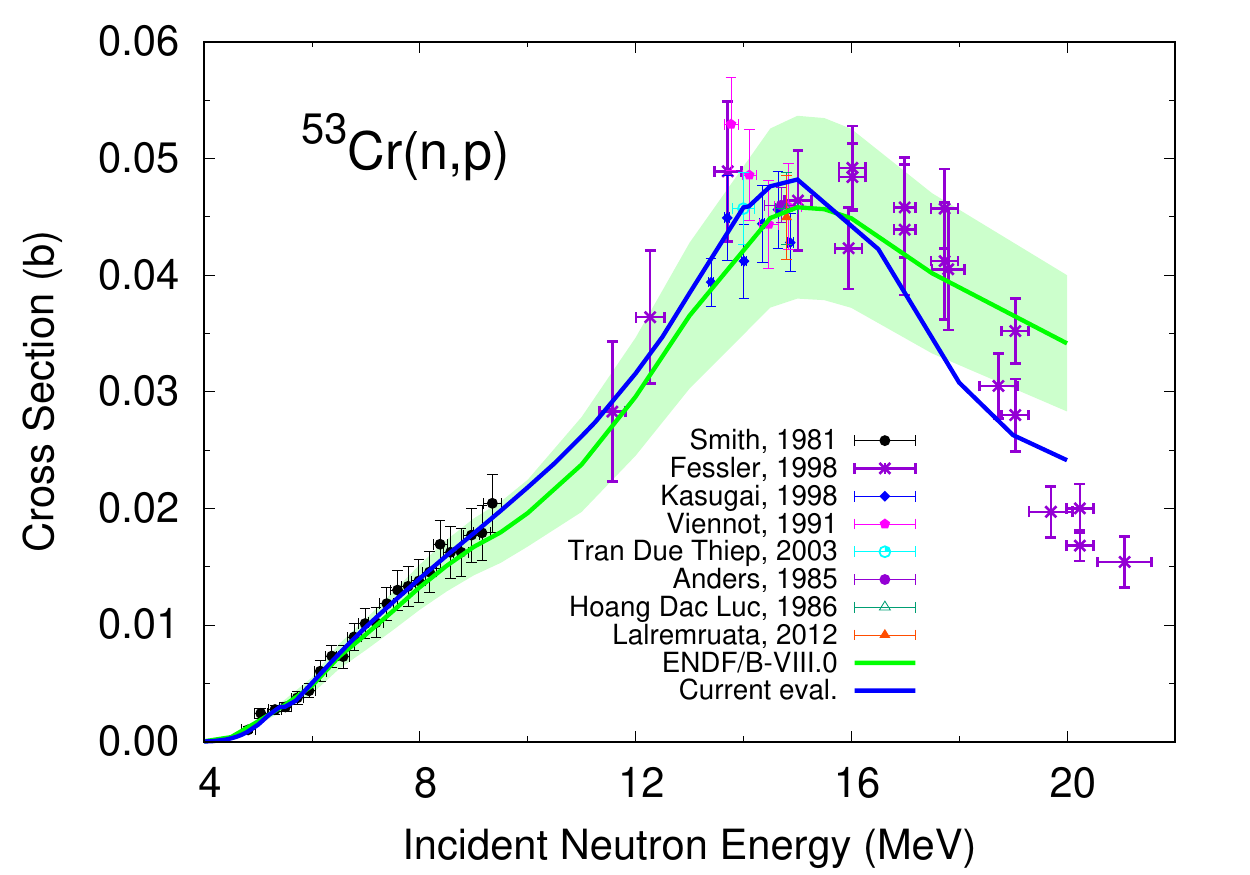}
\caption{(color online) Evaluated \nuc{53}{Cr}(n,p) cross section compared with ENDF/B-VIII.0 and with relevant experimental data \cite{Smith:1981,Fessler:1998,Kasugai1998,Viennot1991,TranDucThiep2003,Anders:1985,Luc:1986,Lalremruata2012}.}
\label{fig:cr53-MT103}
\end{figure}

\paragraph*{\nuc{50}{Cr}($n,p$) -} Fig.~\ref{fig:cr53-MT103} shows a comparison between the present evaluation and ENDF/B-VIII.0 with experimental data. It can be seen that both describe equally well the data from Smith et al. \cite{Sm80} at lower range and the peak of (n,p) cross section, but the current evaluation describe better the high-energy data of Fessler et al.~\cite{Fessler:1998}.

\begin{figure}
\includegraphics[scale=0.70,keepaspectratio=true,clip=true,trim=0mm 0mm 0mm 0mm]{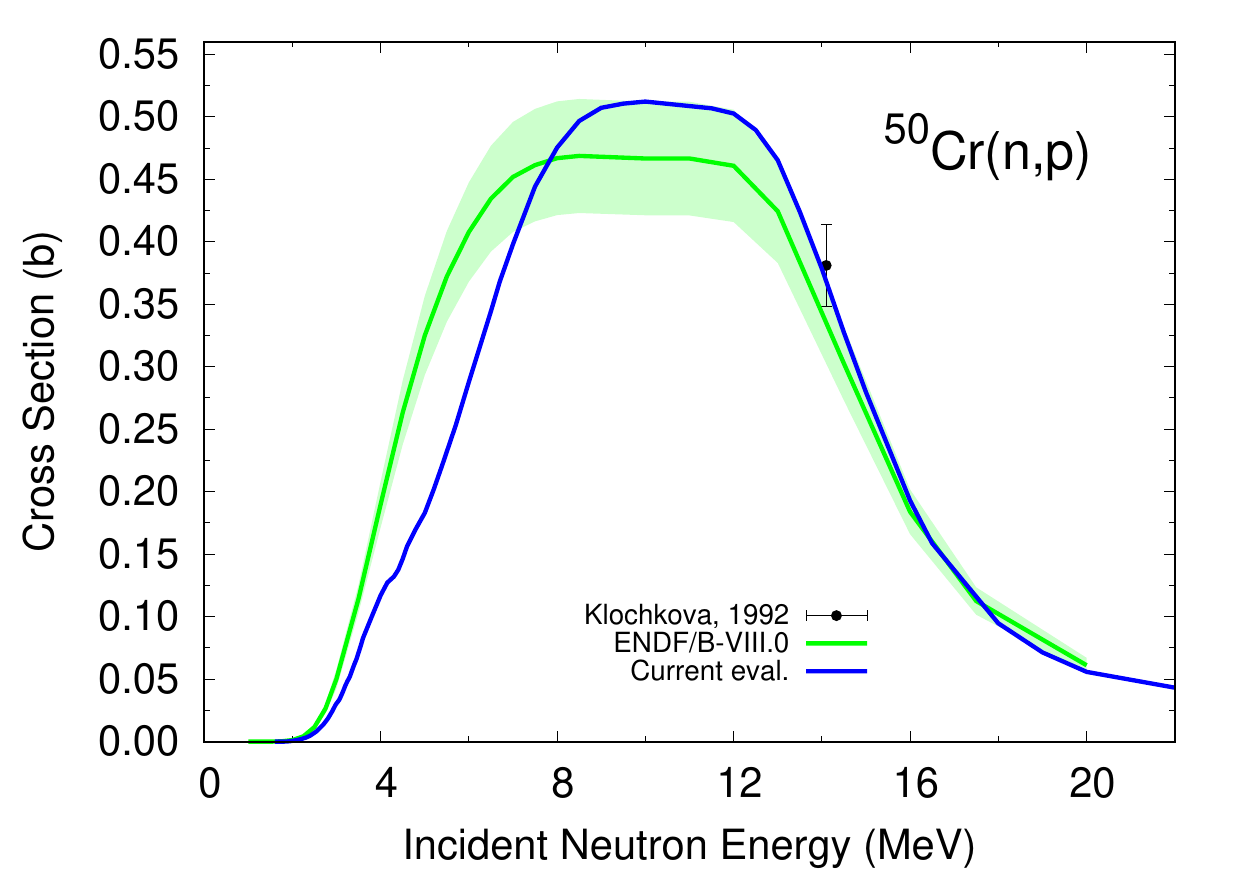}
\caption{(color online) Evaluated \nuc{50}{Cr}(n,p) cross section compared with ENDF/B-VIII.0 and with relevant experimental data~\cite{Klochkova:1992}.}
\label{fig:cr50-MT103}
\end{figure}

\paragraph*{\nuc{50}{Cr}($n,p$) -} As seen in Fig.~\ref{fig:cr50-MT103}, there is only a single datum for the \nuc{50}{Cr}($n,p$), which is from Klochkova \etal from 1992~\cite{Klochkova:1992}. In our fits we ensure that the evaluation go through that experimental point.

\begin{figure}
\includegraphics[scale=0.70,keepaspectratio=true,clip=true,trim=0mm 0mm 0mm 0mm]{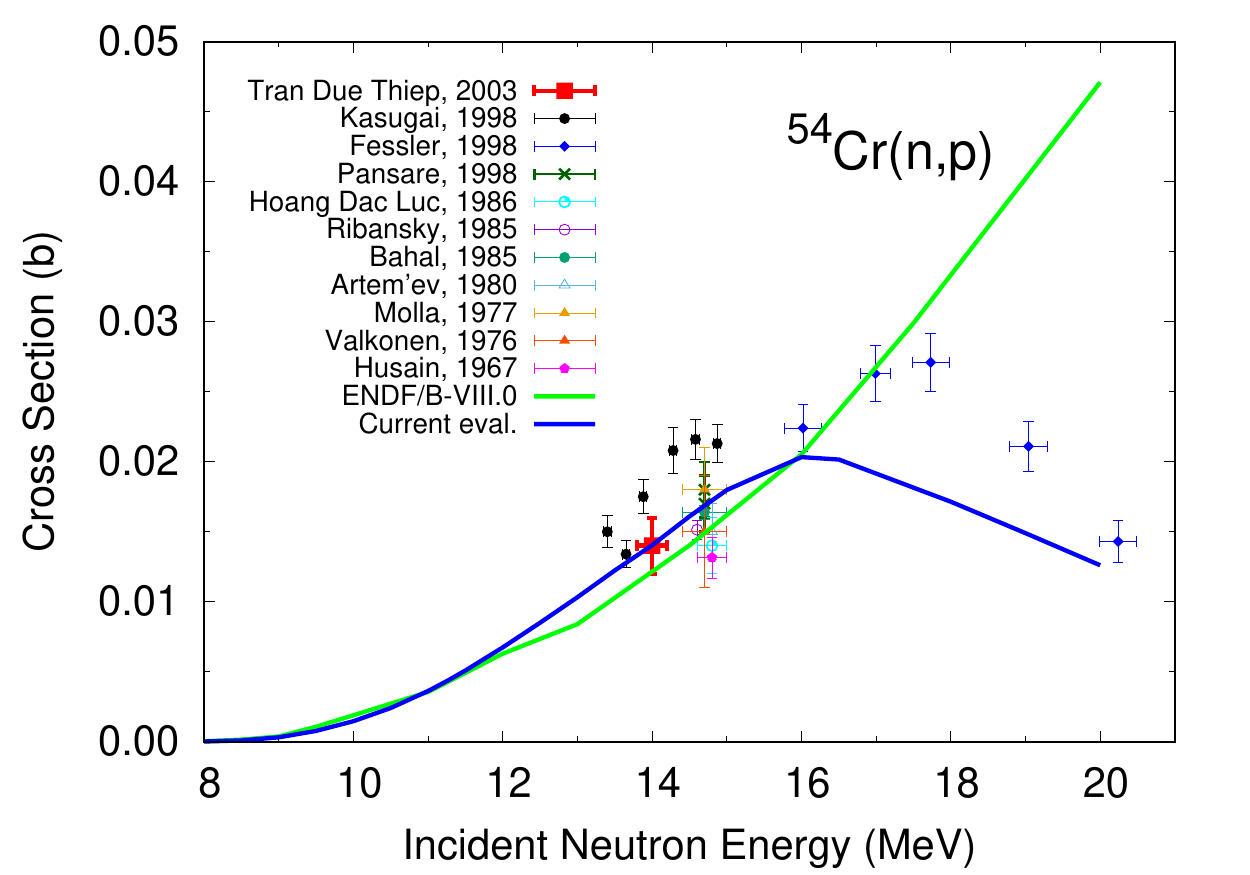}
\caption{(color online) Evaluated \nuc{54}{Cr}(n,p) cross section compared with ENDF/B-VIII.0 and with relevant experimental data~\cite{TranDucThiep2003,Kasugai1998,Fessler:1998,Pansare1991,Luc:1986,Ribansky:1985,Bahal:1985,Ar80,Molla:1977hi,Valkonen:1976,Husain:1967}.}
\label{fig:cr54-MT103}
\end{figure}

\paragraph*{\nuc{54}{Cr}($n,p$) -} In the case of \nuc{54}{Cr}, the (n,p) reaction can be constrained by some experiments that overlap those of \nuc{52}{Cr} (Fig.~\ref{fig:cr54-MT103}). Some of those are newer than the previous ENDF/B evaluation. We performed fits of model parameters taking in consideration the sets shown in Fig.~\ref{fig:cr54-MT103}, reaching a good agreement in the rising part of the cross section, in particular with the most recent experiment of Tran Due Thiep \etal~\cite{TranDucThiep2003}. Even though the agreement with the set of Fessler \etal~\cite{Fessler:1998} in the decreasing part of the cross section is far from optimal, it is in a much better qualitative agreement with data than ENDF/B-VIII.0, for which the (n,p) cross section continues to increase after 16 MeV and beyond 20 MeV.

\subsection{Elastic  and Inelastic Angular Distributions}
   \label{subSec:da}

In the fast region, the  \nuc{52}{Cr} elastic angular distributions between 1.6 and 2.8 Mev were  taken from a fit of experimental data of Guenther et al. \cite{Guenther:1982}, after small tuning to increase forward scattering by half the uncertainty. Above 4.5 MeV, angular distributions were taken directly from EMPIRE calculation.

For \nuc{53}{Cr}, we adopted elastic angular distributions from ENDF/B-VIII.0 between 600 keV and 3 MeV. Above that, angular distributions were provided by EMPIRE calculations. For \nuc{50,54}{Cr}, all angular distributions in the fast region were taken from EMPIRE, as well as all inelastic angular distributions for all isotopes.

\begin{figure}
\includegraphics[scale=0.70,keepaspectratio=true,clip=true,trim=0mm 13mm 0mm 0mm]{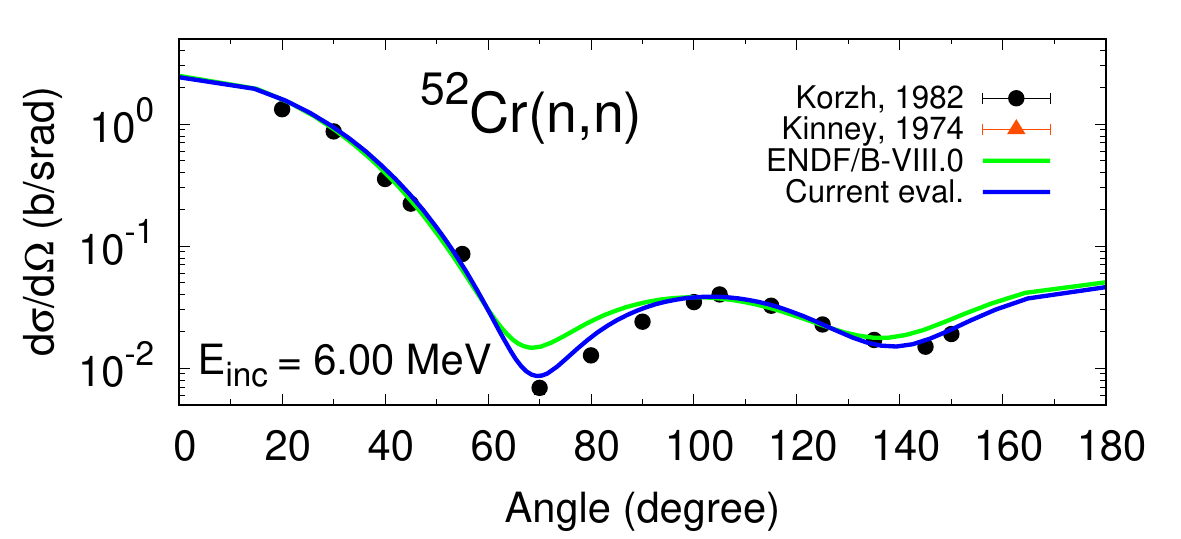} \\ \vspace{-3.4mm}
\includegraphics[scale=0.70,keepaspectratio=true,clip=true,trim=0mm 13mm 0mm 0mm]{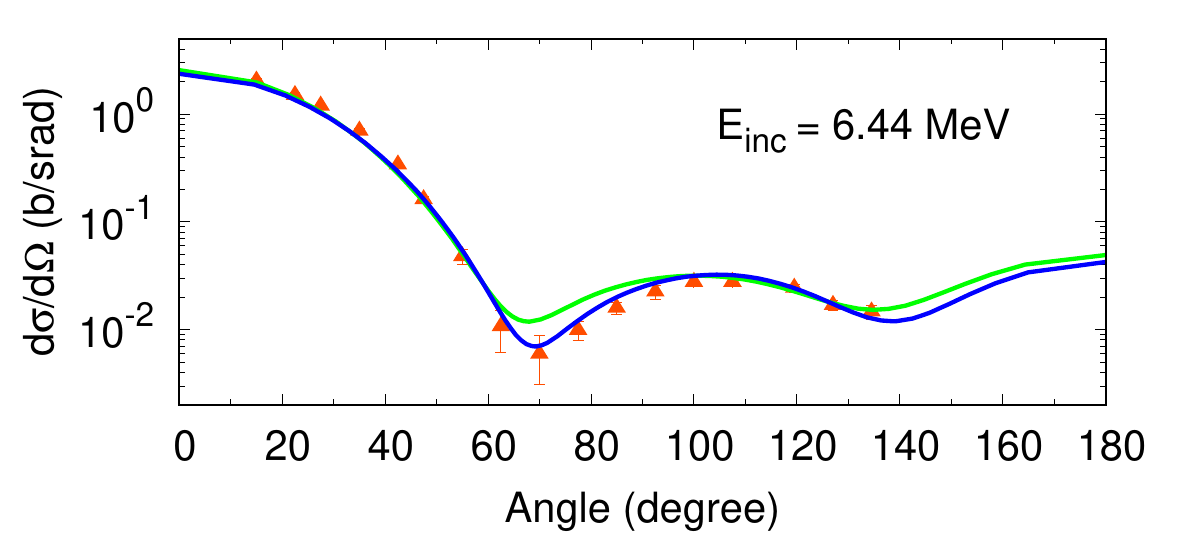} \\ \vspace{-3.4mm}
\includegraphics[scale=0.70,keepaspectratio=true,clip=true,trim=0mm 13mm 0mm 0mm]{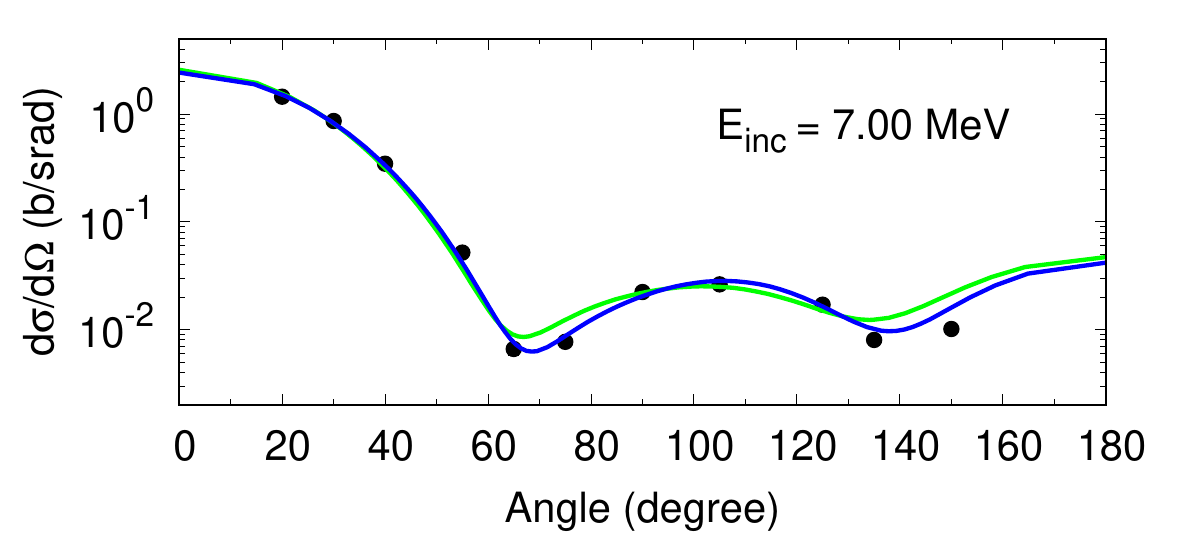} \\ \vspace{-3.4mm}
\includegraphics[scale=0.70,keepaspectratio=true,clip=true,trim=0mm 13mm 0mm 0mm]{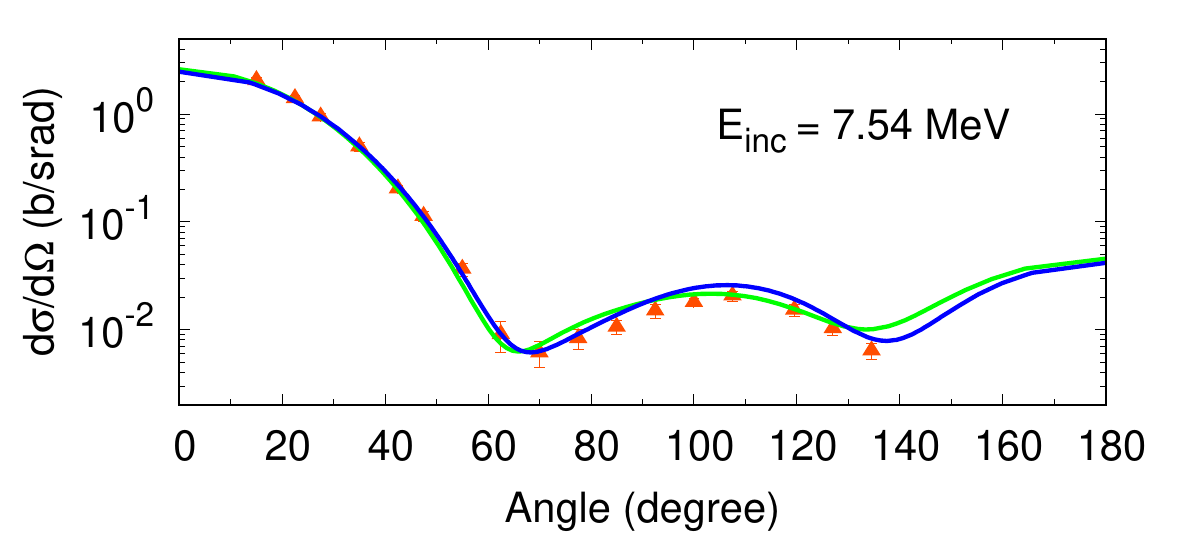} \\ \vspace{-3.4mm}
\includegraphics[scale=0.70,keepaspectratio=true,clip=true,trim=0mm 0mm 0mm 0mm]{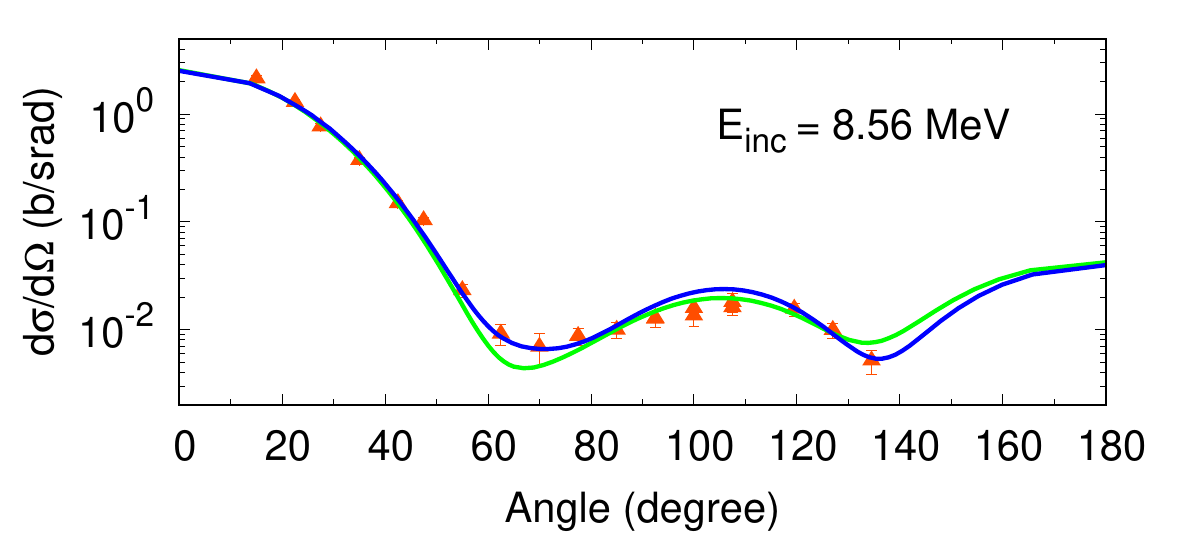} \\
\caption{(color online) \nuc{52}{Cr}(n,elas) elastic angular distribution in the fast region. Experimental data taken from Refs.~\cite{Korzh:1982,Kinney:1974}.}
\label{fig:cr52-angdist-elas}
\end{figure}

In Fig.~\ref{fig:cr52-angdist-elas} elastic angular distributions for \nuc{52}{Cr} are shown for select incident energies starting at 6~MeV, i.e., above the region where fluctuations in the elastic cross section are relevant. We see that even though our evaluation does not differ much from ENDF/B-VIII.0, where it does it is consistently in better agreement with experimental data. This is an indication in support of the optical model potential employed.

\begin{figure}
\includegraphics[scale=0.38,keepaspectratio=true,clip=true,trim=0mm 12mm 0mm 0mm]{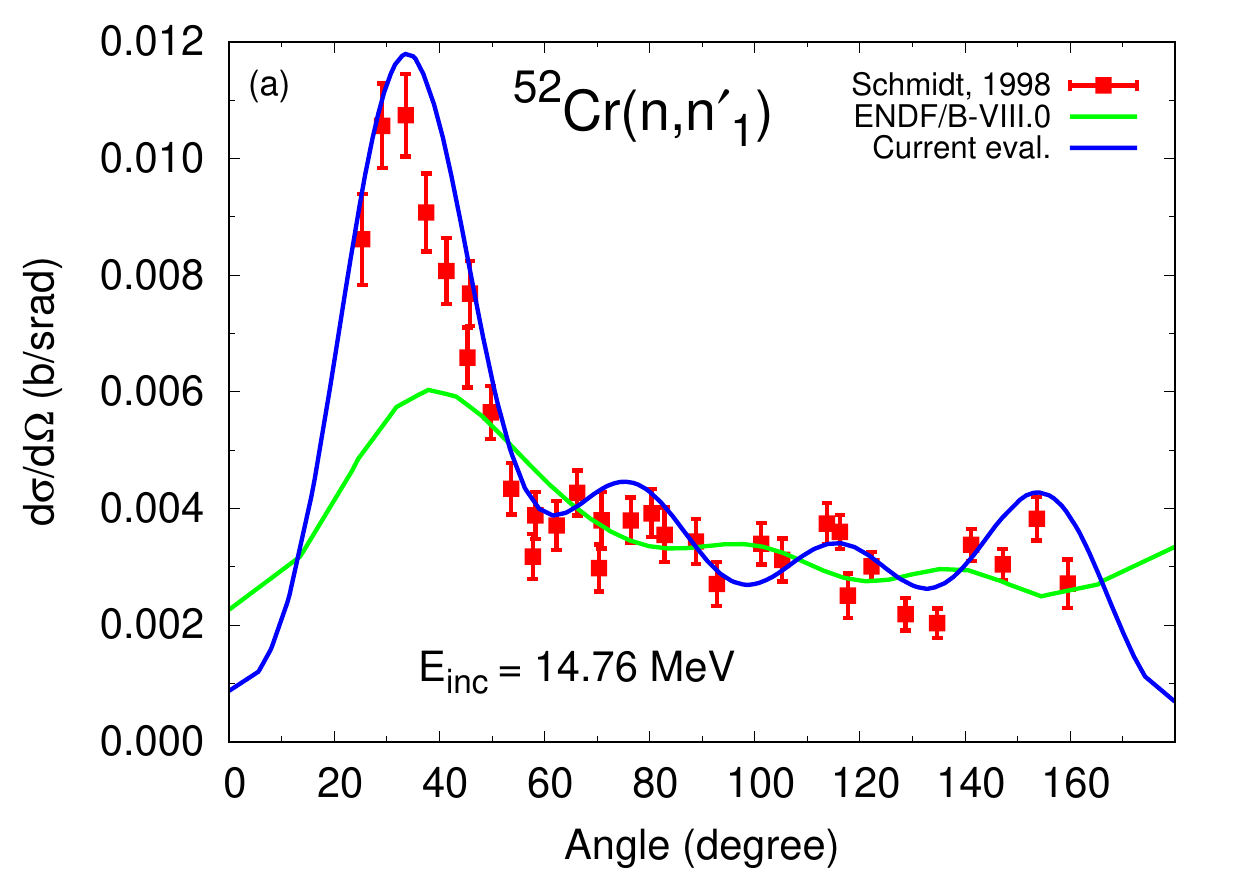} \hspace{-5.6mm}
\includegraphics[scale=0.38,keepaspectratio=true,clip=true,trim=22mm 12mm 0mm 0mm]{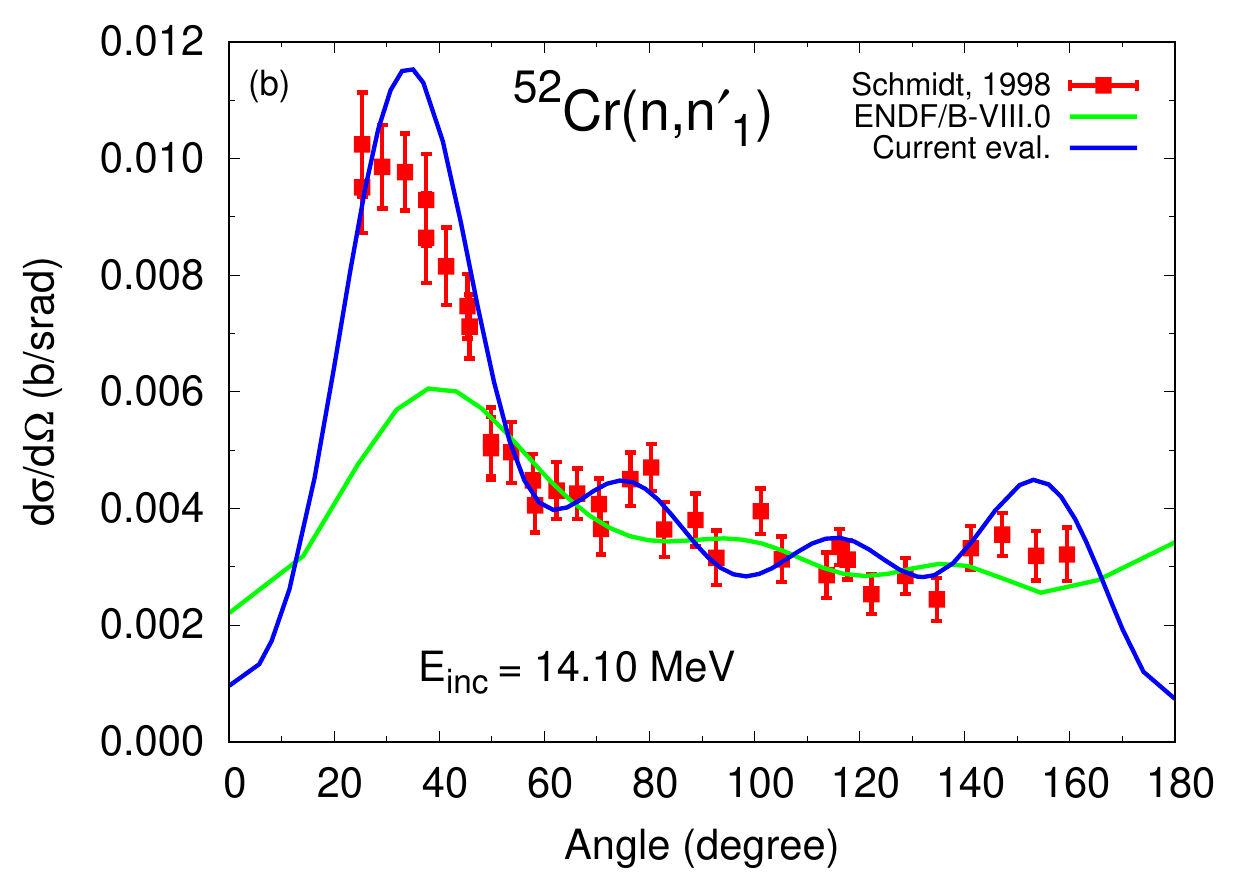} \\
\vspace{-2.6mm}
\includegraphics[scale=0.38,keepaspectratio=true,clip=true,trim=0mm 12mm 0mm 0mm]{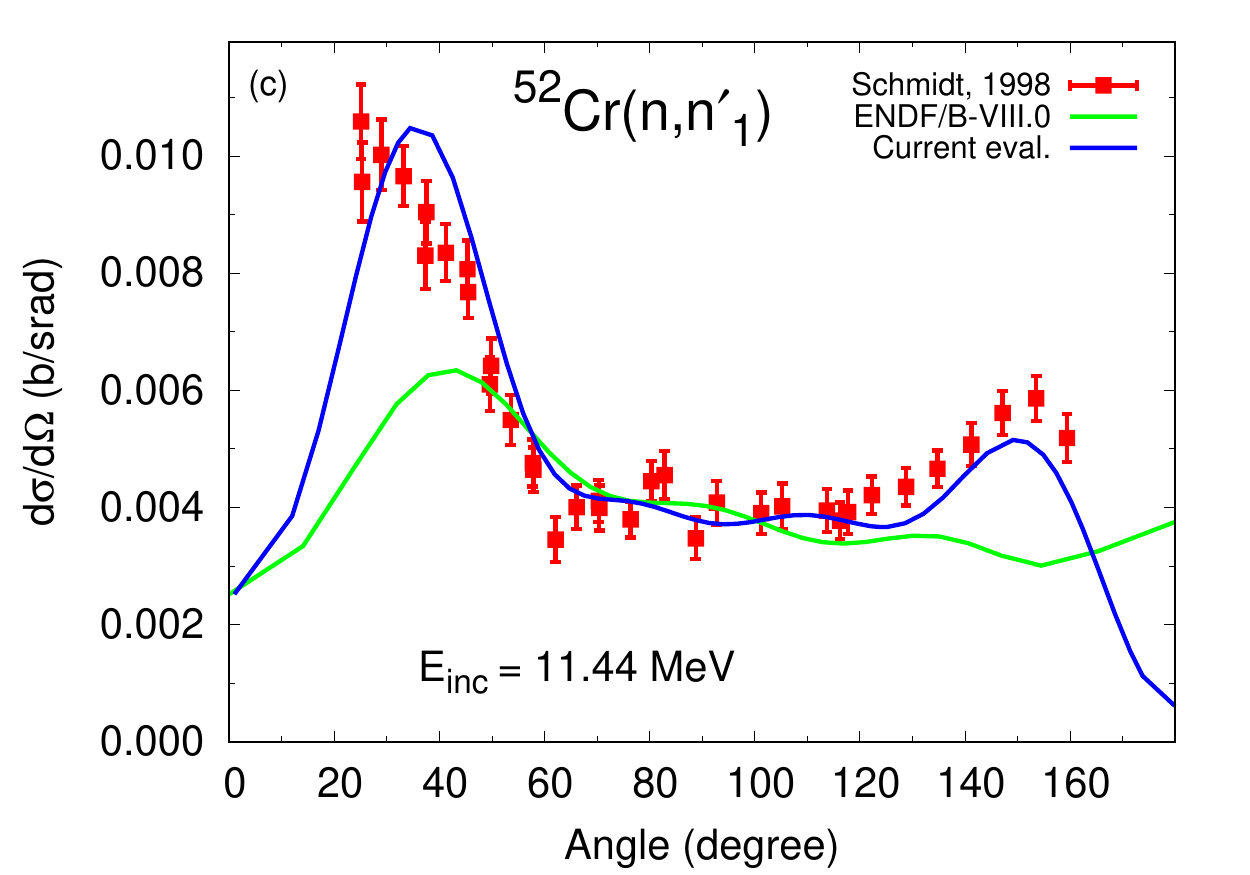} \hspace{-5.6mm}
\includegraphics[scale=0.38,keepaspectratio=true,clip=true,trim=22mm 12mm 0mm 0mm]{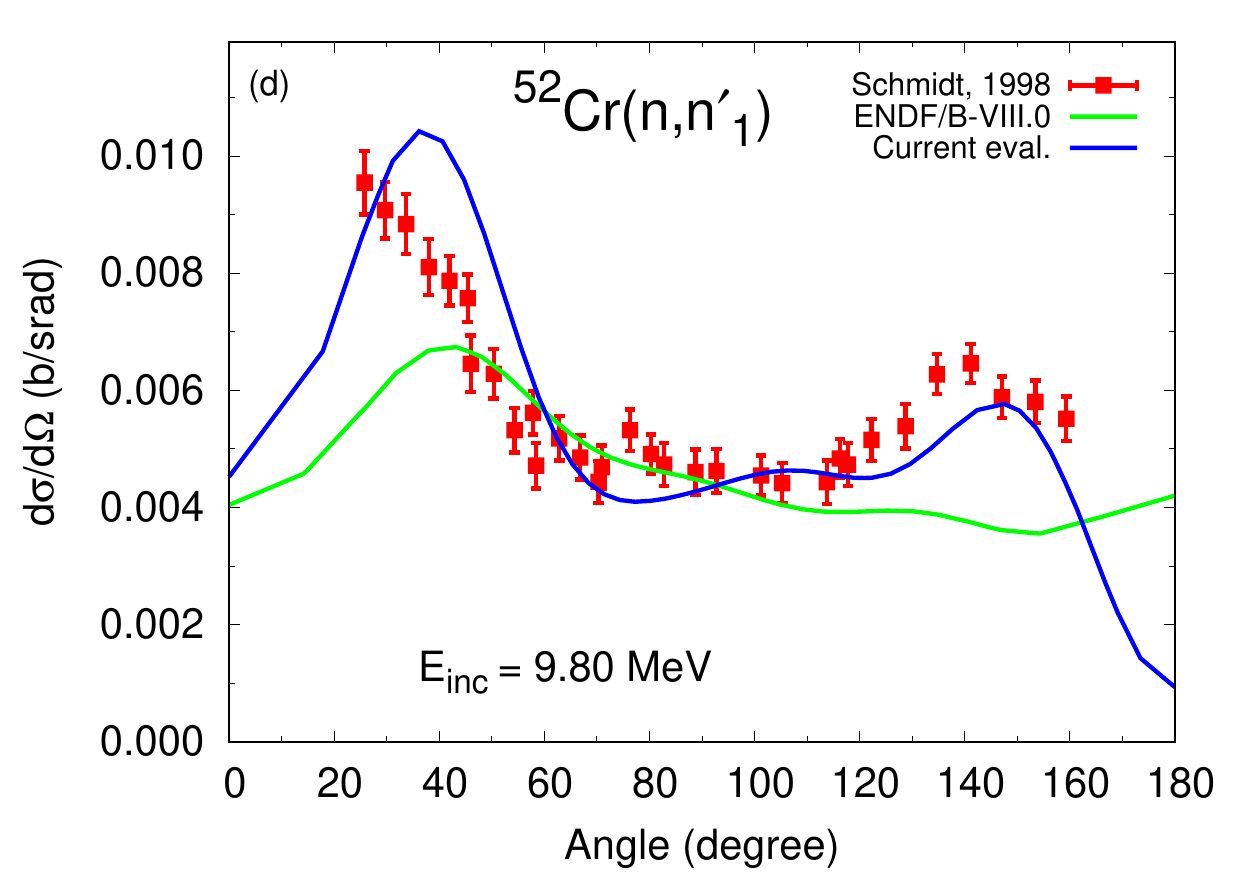} \\
\vspace{-2.6mm}
\includegraphics[scale=0.38,keepaspectratio=true,clip=true,trim=0mm 0mm 0mm 0mm]{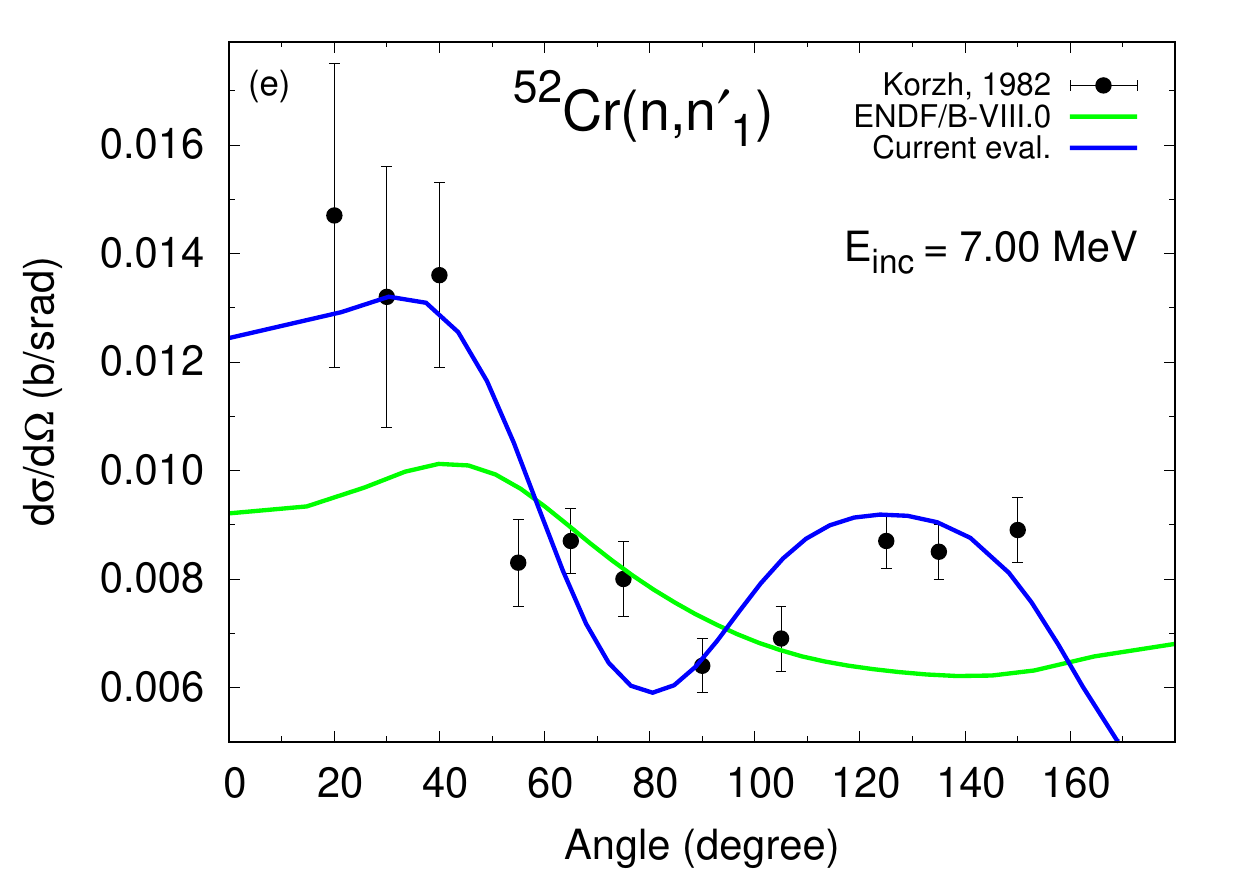}
\hspace{-5.6mm}
\includegraphics[scale=0.38,keepaspectratio=true,clip=true,trim=22mm 0mm 0mm 0mm]{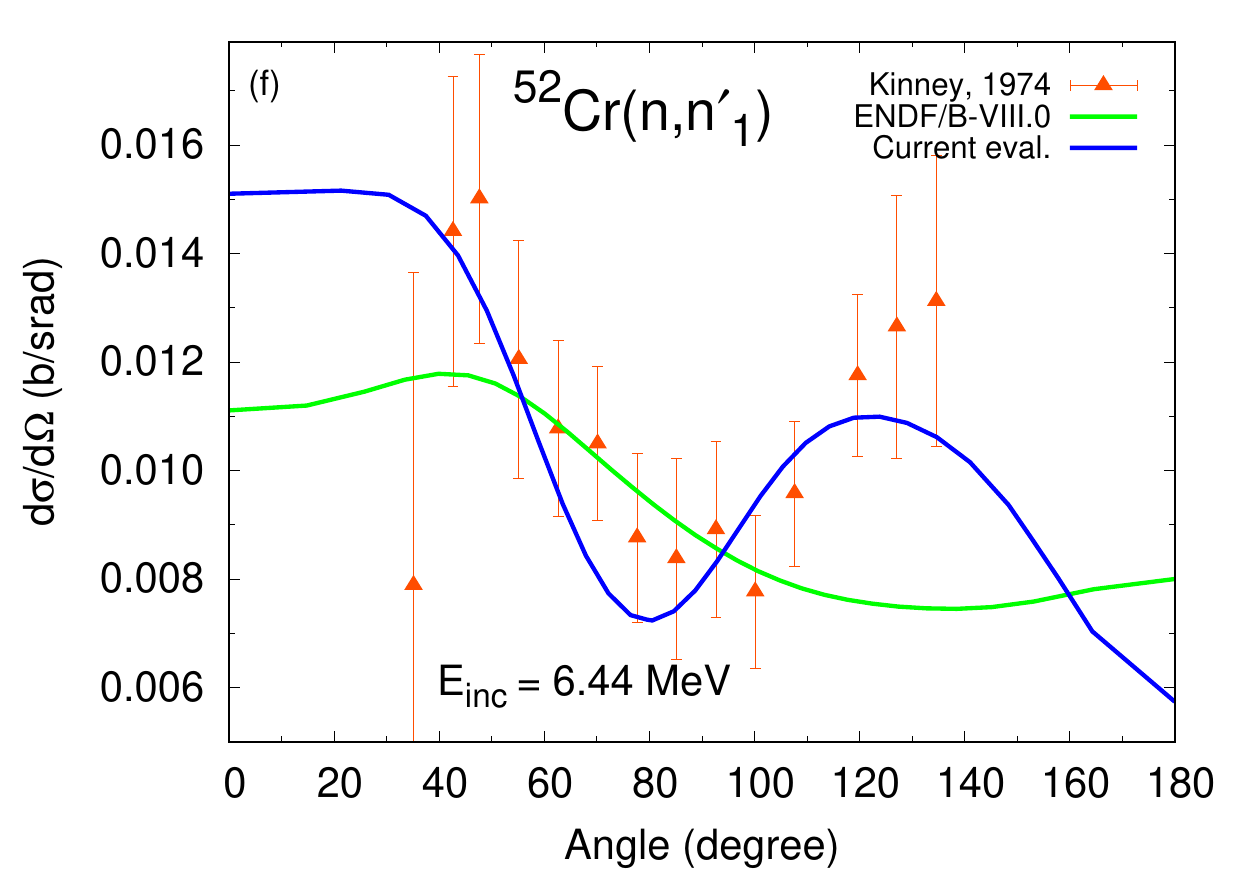}
\caption{(color online) \nuc{52}{Cr}(n,n$^{\prime}_{1}$) elastic angular cross section in the fast region. Data taken from Refs.~\cite{Schmidt:1998,Korzh:1982,Kinney:1974}.}
\label{fig:cr52-angdist-inel}
\end{figure}

Similarly, Fig.~\ref{fig:cr52-angdist-inel} shows evaluated inelastic angular distributions for the first inelastic state of \nuc{52}{Cr} compared to ENDF/B-VIII.0. Again, we focus on incident energies above 6~MeV, where the impact of cross-section fluctuations is much smaller. The most relevant experimental sets available are from Schmidt et al. \cite{Schmidt:1998}, Korzh et al. \cite{Korzh:1982}, and Kinney et al. \cite{Kinney:1974}. The Schmidt data set corresponds to a 12 m time-of-flight (TOF) experiment with neutron incident energies ranging from about 8 to 15 MeV, and is the most reliable of the three sets. From panels (a)-(d) of Fig.~\ref{fig:cr52-angdist-inel}, which show examples of the Schmidt set for 14.76, 14.10, 11.44, and 9.80~MeV respectively, we see that, while ENDF/B-VIII.0 is not able to fully describe the observed data, our evaluation agrees very well with data, even reproducing some details of the measured angular distributions. This behavior is observed for all incident energies of the Schmidt set, even though for the lower-energy ones the agreement at very low and very high scattering angles may not be as good. In panels (e) and (f) of Fig.~\ref{fig:cr52-angdist-inel} we show comparisons with the lower resolution experiments of Korzh and Kinney, at lower incident energies. Again we see that our current evaluation provides a much more realistic angular distributions for the first inelastic channel of \nuc{52}{Cr}. This is likely due to the fact that we use an optical model potential based on a significantly better structure model in our calculations than what was used in ENDF/B-VIII.0, namely the dispersive optical potential of Li \etal \cite{Li:2013}, which takes into account the soft-rotor degrees of freedom of chromium isotopes.

\begin{figure}
\includegraphics[scale=0.38,keepaspectratio=true,clip=true,trim=0mm 12mm 0mm 0mm]{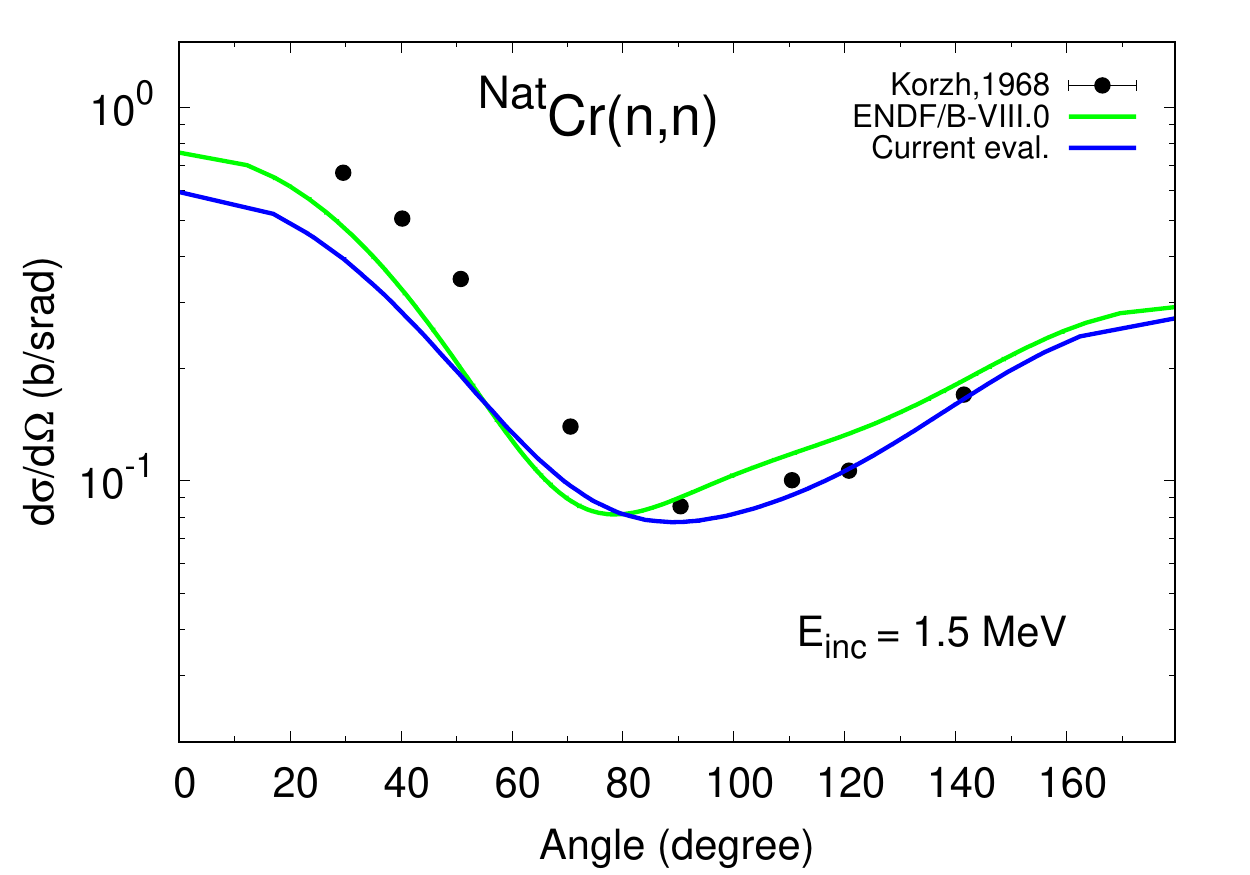} \hspace{-5.5mm}
\includegraphics[scale=0.38,keepaspectratio=true,clip=true,trim=17mm 12mm 0mm 0mm]{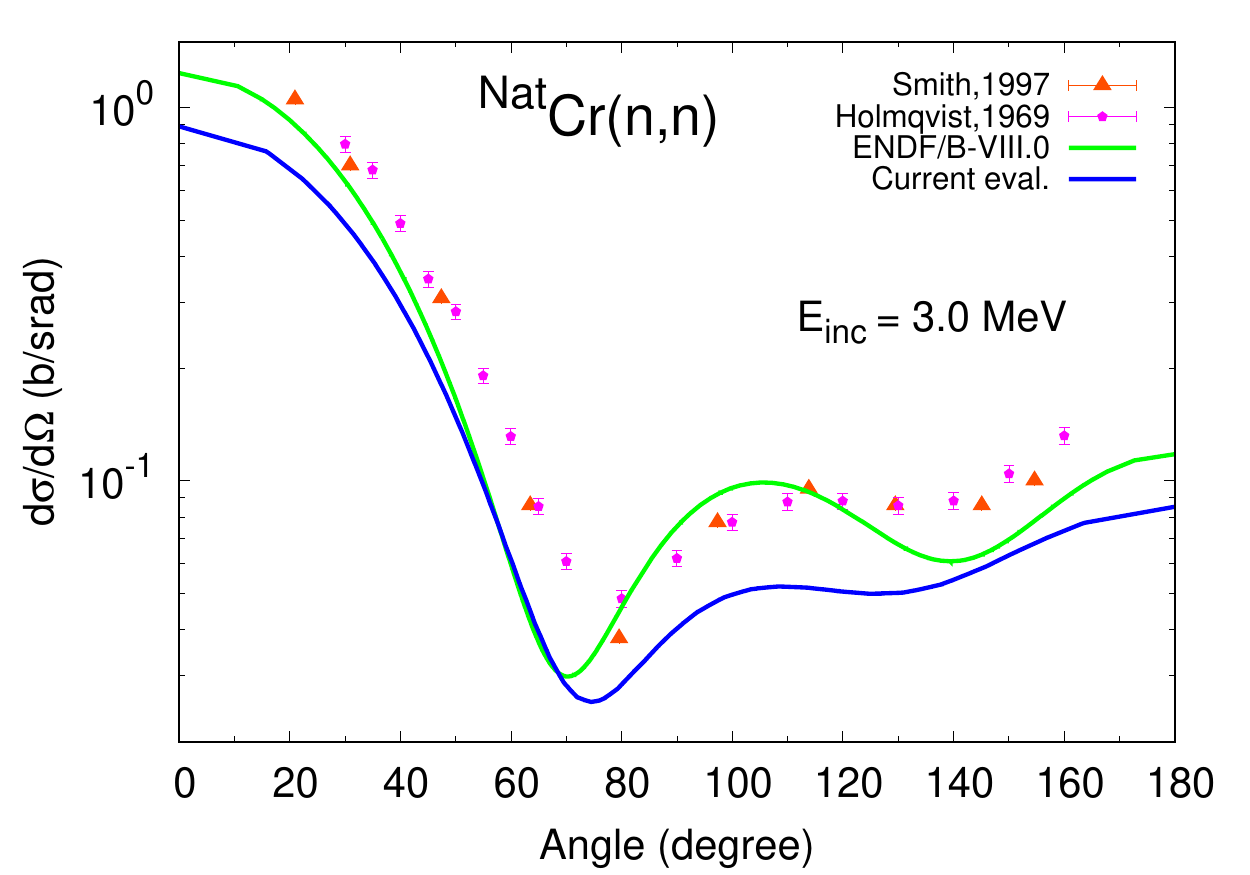} \\
\vspace{-2.55mm}
\includegraphics[scale=0.38,keepaspectratio=true,clip=true,trim=0mm 12mm 0mm 0mm]{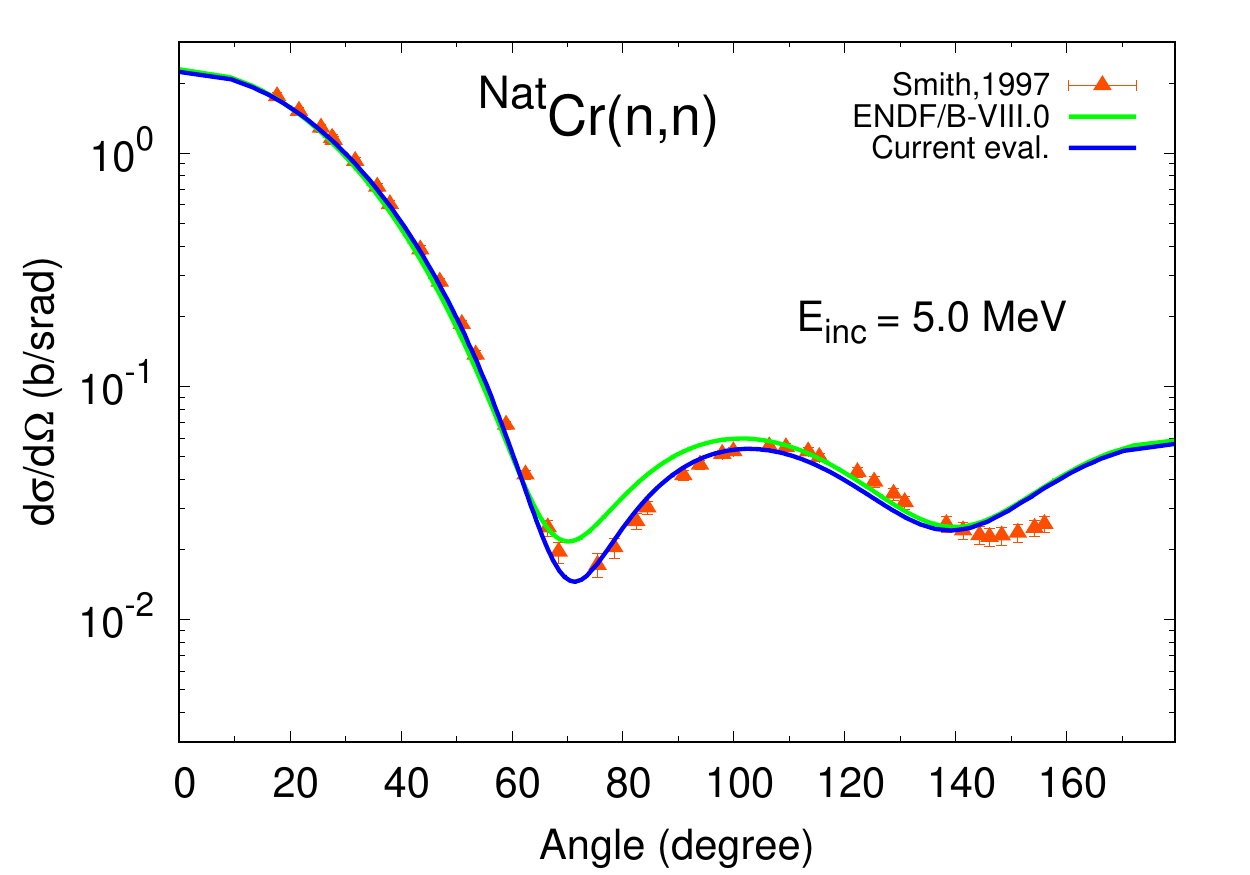} \hspace{-5.5mm}
\includegraphics[scale=0.38,keepaspectratio=true,clip=true,trim=17mm 12mm 0mm 0mm]{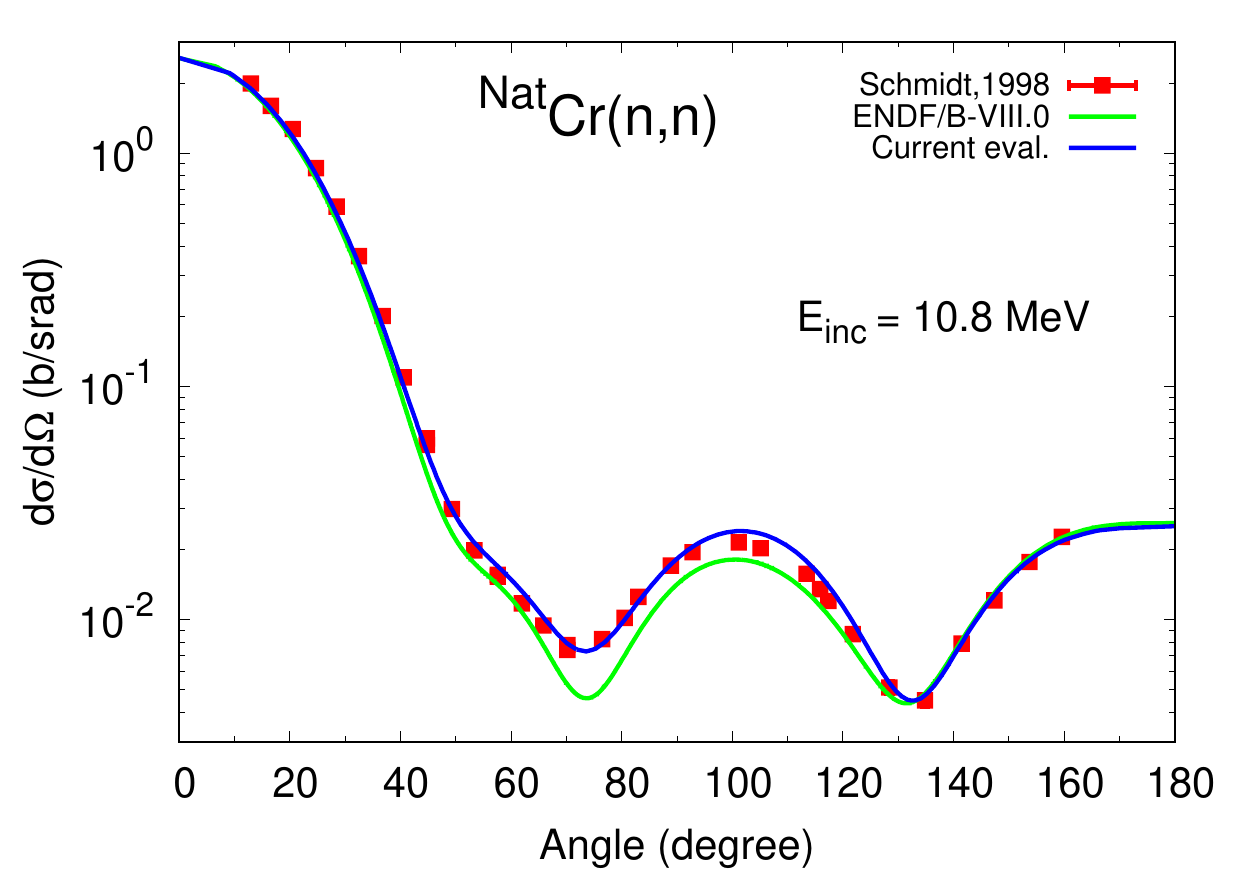} \\
\vspace{-2.55mm}
\includegraphics[scale=0.38,keepaspectratio=true,clip=true,trim=0mm 0mm 0mm 0mm]{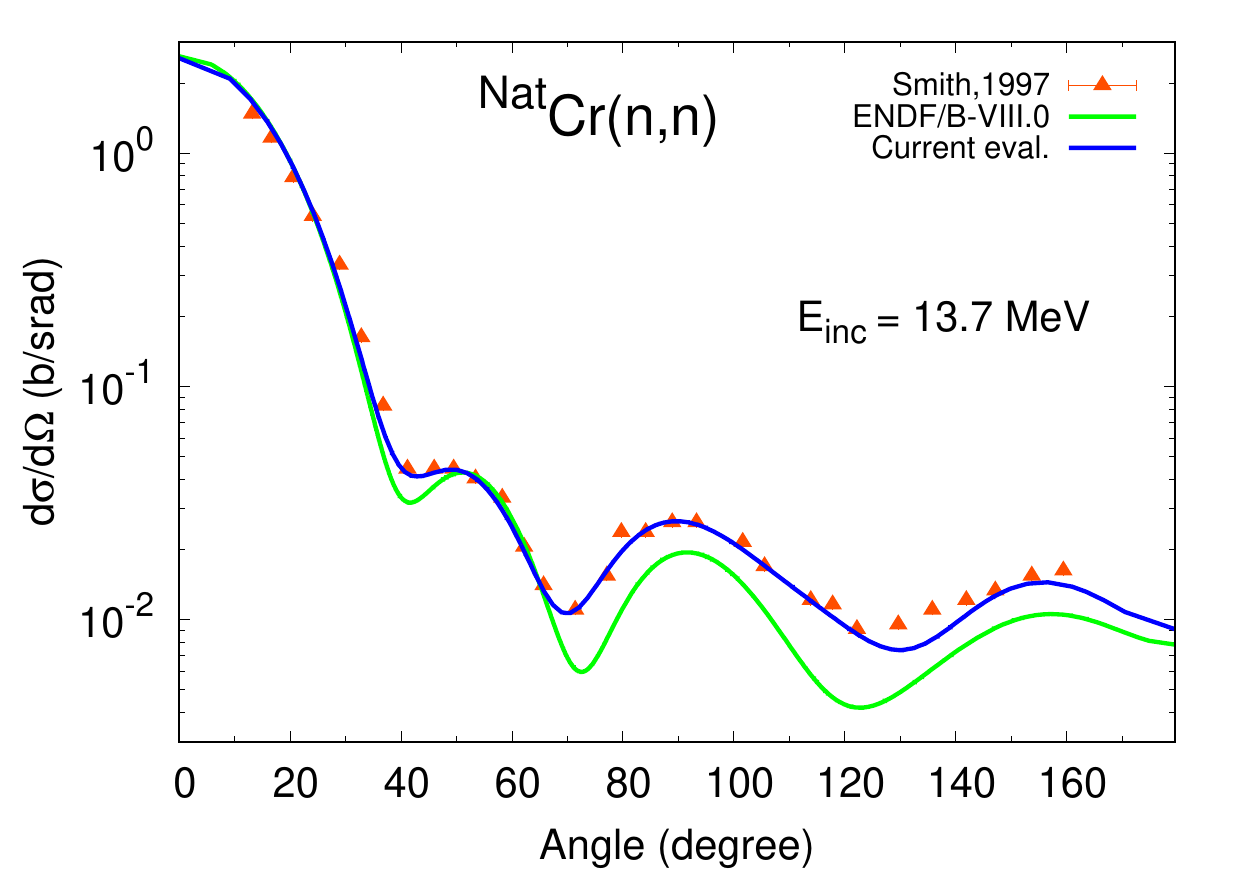} \hspace{-5.5mm}
\includegraphics[scale=0.38,keepaspectratio=true,clip=true,trim=17mm 0mm 0mm 0mm]{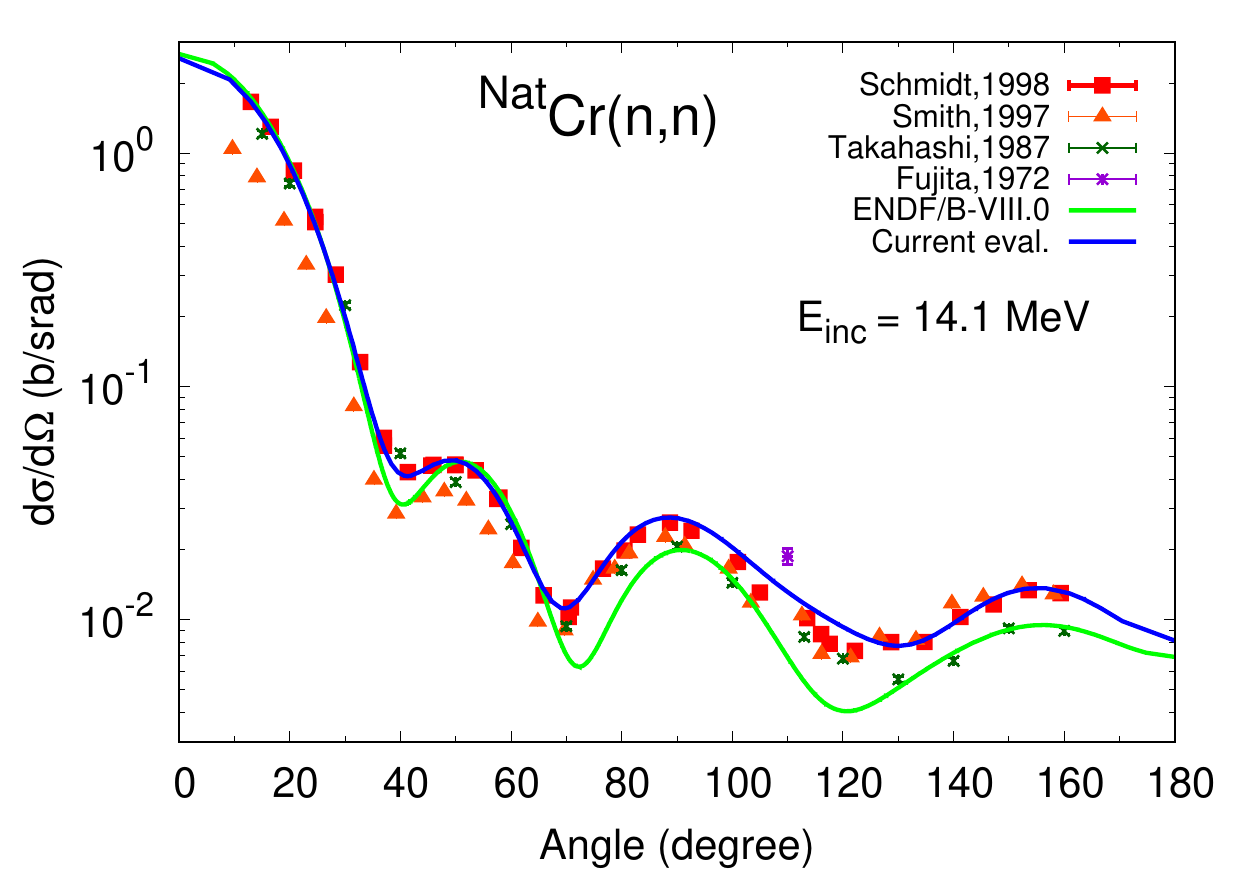} \\
\caption{(color online) \nuc{Nat}{Cr}(n,n) elastic angular distribution cross section in the fast region. Data taken from Refs.~\cite{Korzh:1968,Smith:1997,Holmqvist:1969,Schmidt:1998,Takahashi:1987Cr,Fujita:1972}.}
\label{fig:cr-nat-angdist-elas}
\end{figure}

There are also elastic and inelastic angular distribution data for natural samples of chromium in the fast region. The main data sets are the ones from Smith \etal~\cite{Smith:1997} and Schmidt \etal~\cite{Schmidt:1998}, in addition to other relevant experiments. Even though the inelastic angular distributions do not bring much more information than that of isotopic data, it is instructive to compare elastic angular distributions. Fig.~\ref{fig:cr-nat-angdist-elas} shows examples of such angular distributions for select different neutron incident energies. At higher energies there is a clearly better agreement of the current evaluation with data than ENDF/B-VIII.0. This become not so evident below $\sim$5 MeV where the optical model becomes not as reliable, and the agreement with data is poorer, both from the current evaluation and ENDF/B-VIII.0.

In contrast with Fig.~\ref{fig:cr52-angdist-elas} that shows that  \nuc{52}{Cr} isotopic elastic angular distributions from our evaluation is mildly in better agreement with data than ENDF/B-VIII.0, in the case of natural chromium our evaluation is in significantly much better agreement with measurements than previous ENDF/B-VIII.0 evaluation, except at 3 MeV of incident neutron energy as shown in Fig.~\ref{fig:cr-nat-angdist-elas}. This indicates that the angular distributions for minor isotopes have also been improved.

\subsection{Energy Spectra}
   \label{subSec:de}
There are several measurements of neutron energy spectra concentrated around 14 MeV.
At these incident energies, the high outgoing-energy end of the spectra is dominated by the elastic peak and direct transitions to the collective levels that are well modeled by CC and DWBA calculations. The middle part of the spectrum is governed by the PE emission,  while the low energy peak is the evaporation peak described by the Hauser-Feshbach model. Proper description of the neutron spectra is thus a very good overall test of the quality of the reaction modeling used in the evaluation.

\begin{figure}
\includegraphics[scale=0.70,keepaspectratio=true,clip=true,trim=0mm 9mm 0mm 0mm]{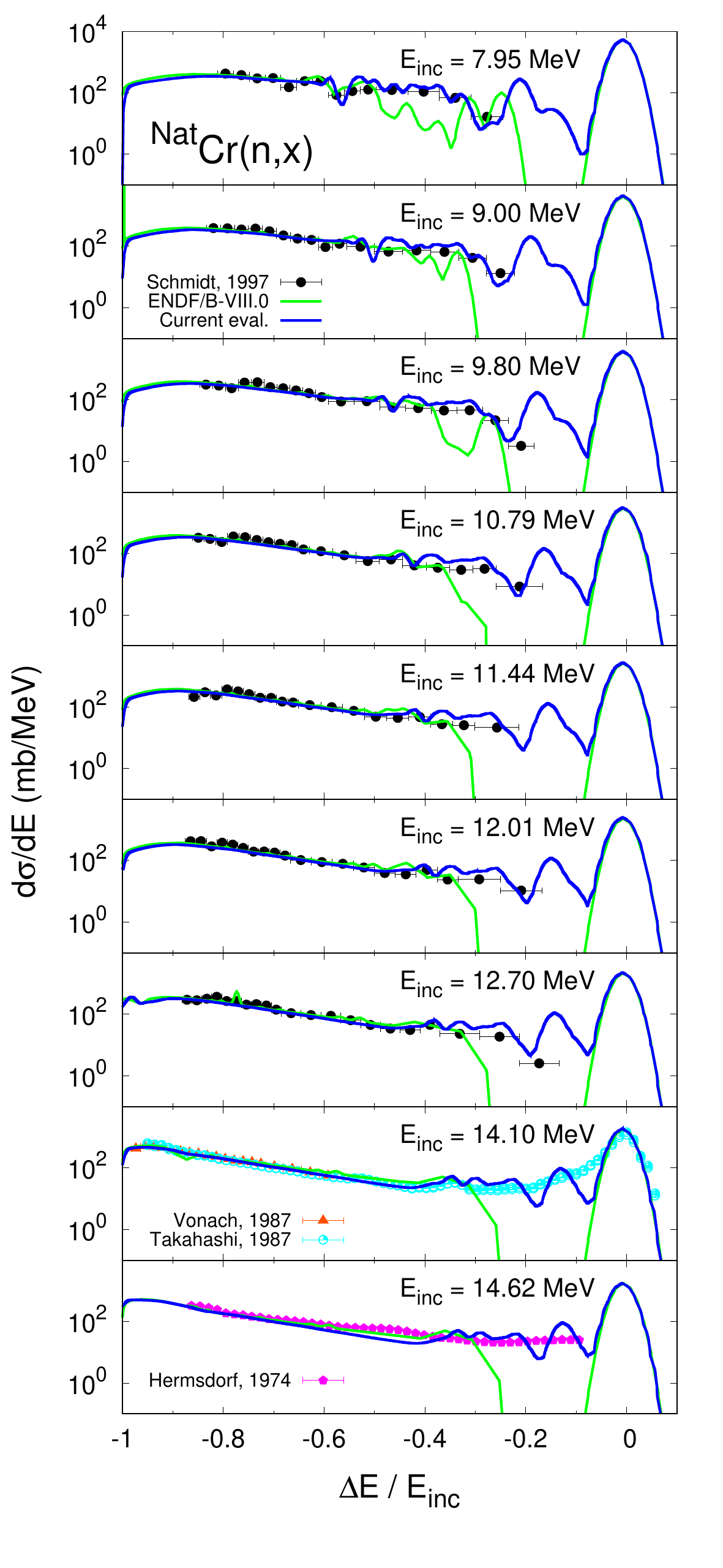}
\caption{(color online) Energy-differential spectra for \nuc{Nat}{Cr}(n,x), as a function of the normalized energy transfer $\Delta E / E_{\mathrm{inc}}$, where $ E_{\mathrm{inc}}$ is the incident neutron energy and $\Delta E = E_{\mathrm{out}} - E_{\mathrm{inc}}$ is the difference between the neutron outgoing energy and the incident energy. Experimental data taken from Refs.~\cite{Schmidt:1997:DD,Vonach:1980,Takahashi:1987Cr}.}
\label{fig:cr-nat-DE}
\end{figure}

\subsection{Double Differential Cross Sections}
   \label{subSec:ddx}

Energy-angle correlated cross sections add another dimension to the energy spectra, associated with the scattering angle.  These arguably correspond to an even more direct kind of measurement since energy spectra are usually obtained by integrating double-differential (DD) ones which unavoidably involves some approximations related to interpolation and extrapolation of angular distributions. 
However, employed models in EMPIRE rely on empirical Kalbach parameterization of double differential cross section \cite{Kalbach:81}.
Therefore, comparison of an evaluation with  double-differential cross sections provides for an additional stringent test of the modeling employed in the evaluation procedure.

\subsubsection{Neutron Double-Differential Spectra}
\label{subSec:nDDS}

There are double-differential (DD) neutron data available for \nuc{Nat}{Cr} for a variety of incident energies ranging from 7.95 MeV up to 14.8 MeV, each for a different set of scattering angles. Undoubtedly, the widest variety of measured angles is associated with experiments at neutron incident energies of 14.1 MeV due to the availability of almost monoenergetic D-T neutron sources. Because of that, and also due to the importance for fusion of 14--15~MeV D-T neutrons (e.g., as illustrated by the leakage experiment discussed in Section~\ref{Sec:validation}), we focus our DD discussion here on this incident energy. Conclusions can certainly be extrapolated to other incident energies where data are available.

\begin{figure*}
\includegraphics[scale=0.545,keepaspectratio=true,clip=true,trim=  0mm 12mm 3mm 2mm]{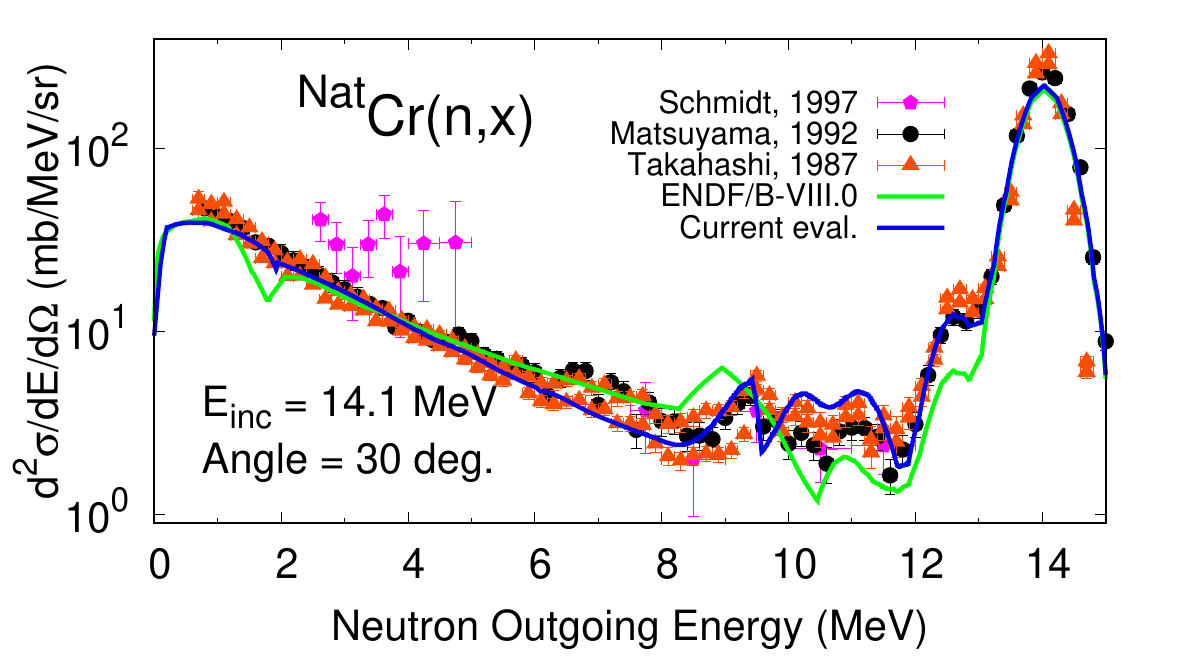} \hspace{-3.8mm}
\includegraphics[scale=0.545,keepaspectratio=true,clip=true,trim=7mm 12mm 3mm 2mm]{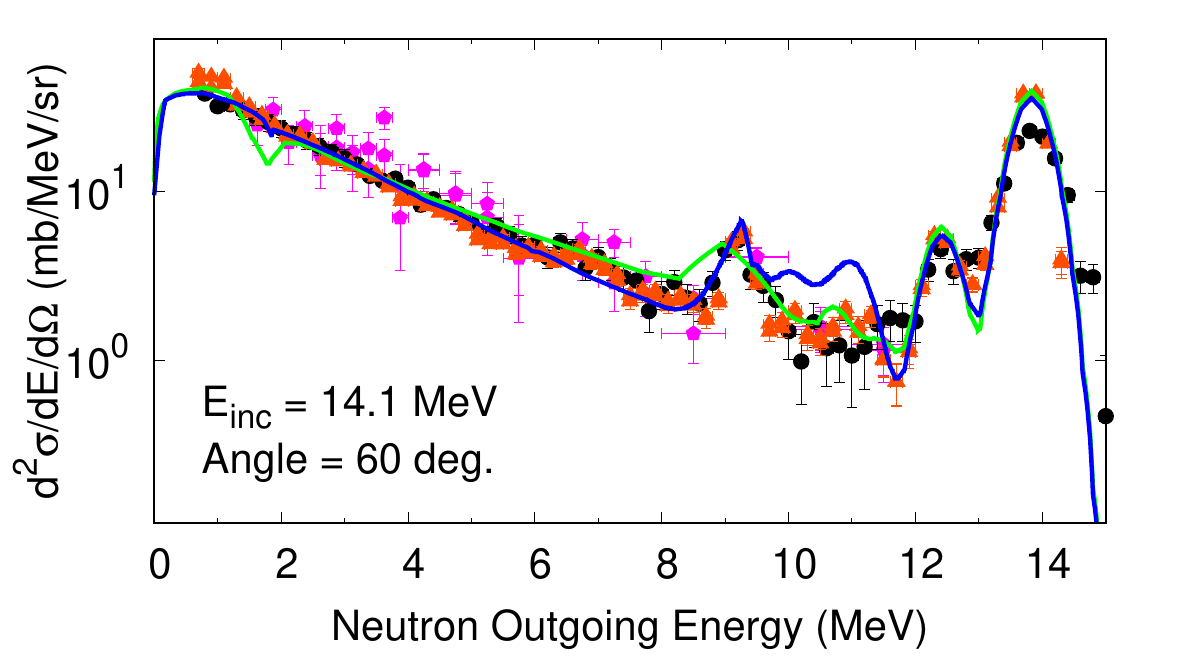} \hspace{-3.8mm}
\includegraphics[scale=0.545,keepaspectratio=true,clip=true,trim=7mm 12mm 3mm 2mm]{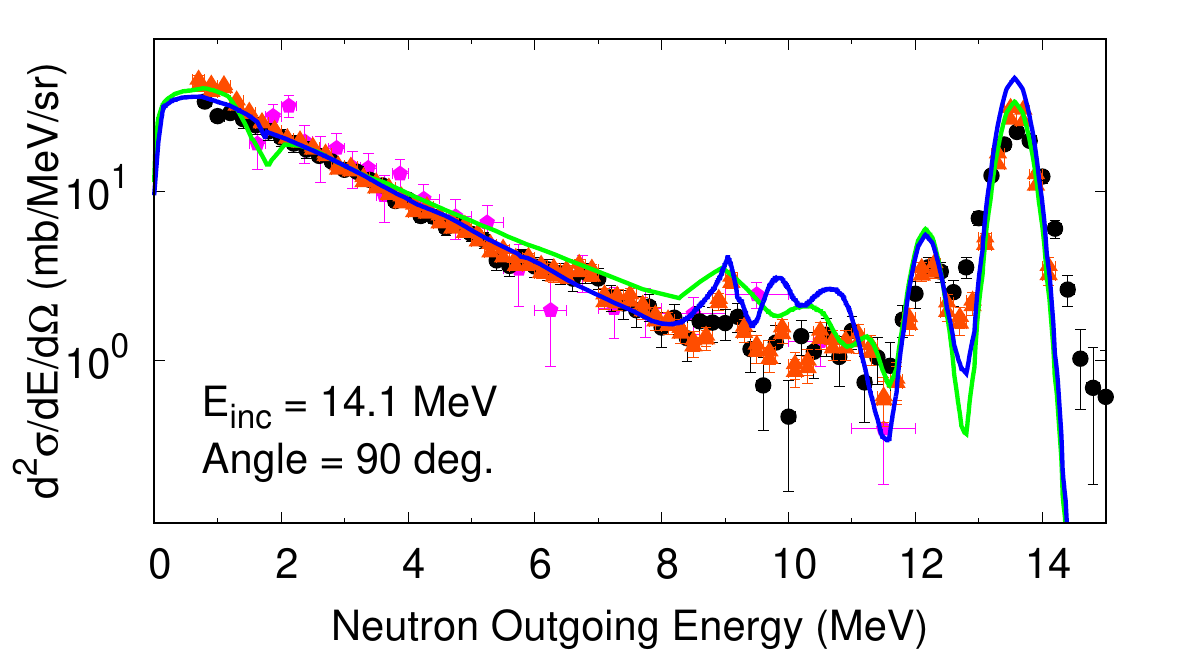}
\\ \vspace{-2.3mm}
\includegraphics[scale=0.545,keepaspectratio=true,clip=true,trim=  0mm 0mm 3mm 2mm]{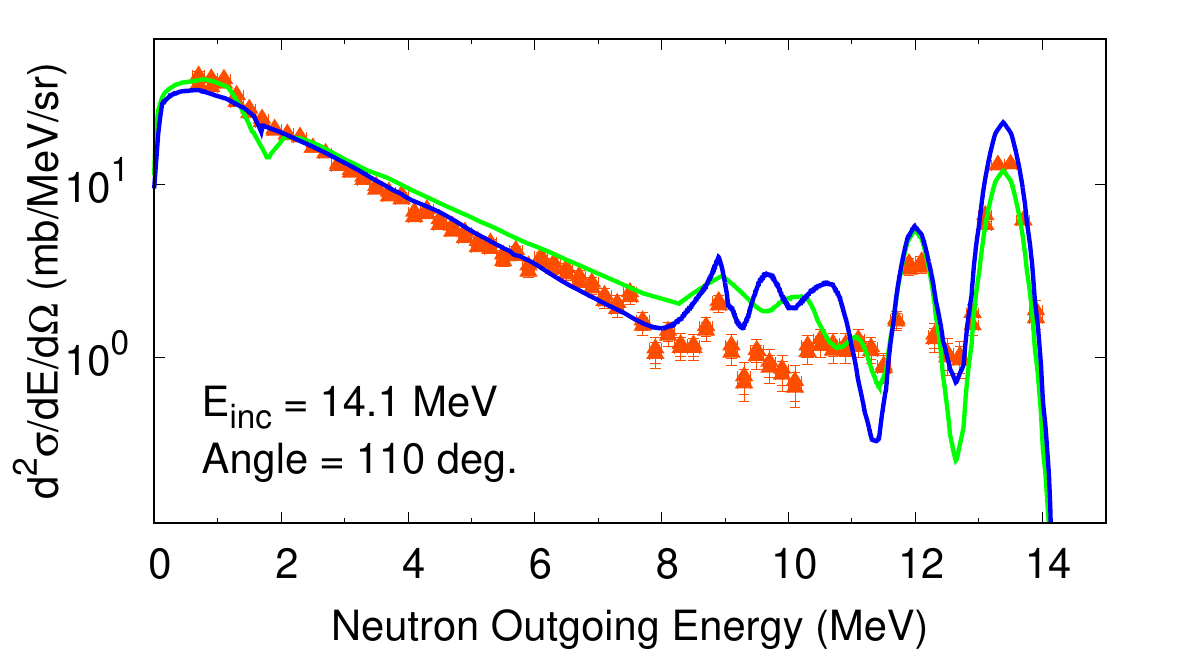} \hspace{-3.8mm}
\includegraphics[scale=0.545,keepaspectratio=true,clip=true,trim=7mm 0mm 3mm 2mm]{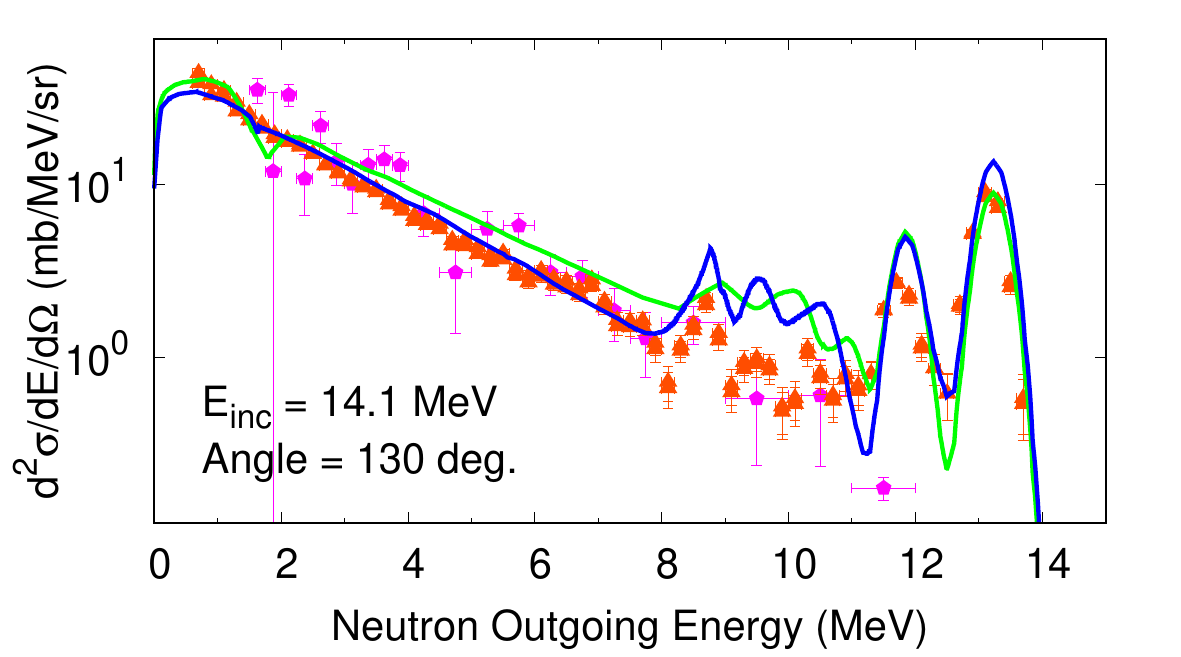} \hspace{-3.8mm}
\includegraphics[scale=0.545,keepaspectratio=true,clip=true,trim=7mm 0mm 3mm 2mm]{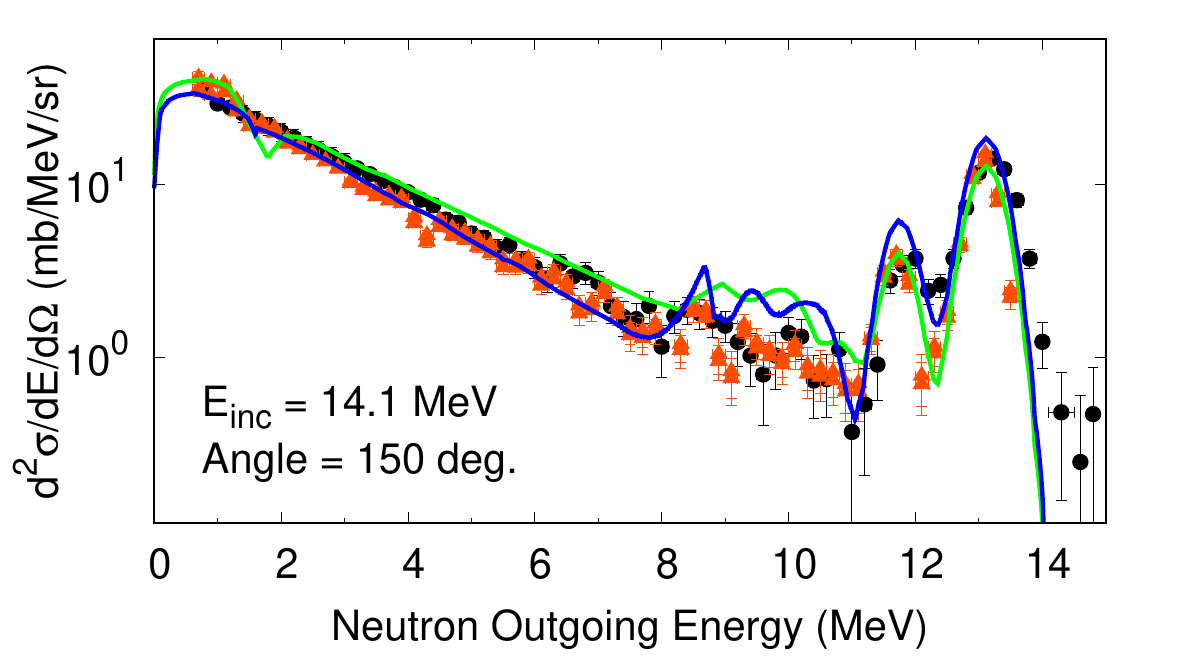}
\caption{(color online) Double-differential spectra cross sections for 14.1 MeV incident neutrons on \nuc{Nat}{Cr} as a function of the neutron outgoing energy at different scattering angles. Resolution broadening was applied to visually match resolution of the spectra data. Experimental data taken from Refs.~\cite{Schmidt:1997:DD,Matsuyama:1992Cr,Takahashi:1987Cr}.}
\label{fig:cr-nat-DD}
\end{figure*}

In Fig.~\ref{fig:cr-nat-DD} we show the DD spectra cross sections for 14.1 MeV incident neutrons on \nuc{Nat}{Cr} as a function of the neutron outgoing energy $E_{\mathrm{out}}$ at six different scattering angles, which were chosen as a general representation of the whole scattering solid angle: 30, 60, 90, 110, 130 and 150 degrees. The three main data sets available at this energy are from Schmidt \etal~\cite{Schmidt:1997:DD}, Matsuyama \etal~\cite{Matsuyama:1992Cr}, and Takahashi \etal~\cite{Takahashi:1987Cr}. Smith's data set is the most recent one and spans across more angles but often has no points around the elastic and first inelastic peak, while Takahashi and Matsuyama tend to have broader resolution. Even though Schmidt's data is generally consistent with the two other sets, there are discrepancies, especially at more forward angles,  as can be seen  in the 30 degree panel in Fig.~\ref{fig:cr-nat-DD}. Matsuyama and Takahashi data are normally smoother in the continuum/prequilibrium region ($E_{\mathrm{out}}=\sim2$  to $\sim$8 MeV) but present some structure in the discrete region $E_{\mathrm{out}}=\sim8$  to $\sim$13 MeV. By comparing the current evaluation with ENDF/B-VIII.0 in regards to agreement with data, it is generally observed that the current evaluation is better in the continuum and preequilibrium region with a smoother transition to the compound component. The present evaluation has a small peak in the outgoing neutron energy between 8 and 9 MeV, which seems to be supported by data. That is not completely clear only for very backward angles (e.g., 150 degrees). On the discrete-level part, there are scattering angles for which ENDF/B-VIII.0 seems better (e.g. 60 and 90 degrees), cases where the current evaluation seems to be in better agreement (e.g. 30 degrees), but also cases where there is not a clear way to determine which agrees better with data (e.g. 110 and 130 degrees). Noteworthy is that frequently there is a low cross-section point from Schmidt (eg. 90, 130 degrees) at around $E_{\mathrm{out}}\sim$11.5 MeV that our evaluations seems to agree to better. Regarding the first inelastic and elastic peaks, generally there is a small advantage for the current evaluation (e.g. 30 degrees) but both the evaluation and ENDF/B-VIII.0 agree well with data, especially if broader resolution is employed. Differences between ENDF/B-VIII.0 and current evaluation at the elastic peak vary with the scattering angle, which can be understood by the differences in angular distributions (see Fig.~\ref{fig:cr-nat-angdist-elas}).

\subsubsection{$\gamma$  Spectra}
\label{subSec:gDDS}

\begin{figure*}
\includegraphics[scale=0.615,keepaspectratio=true,clip=true,trim=  0mm 12mm 3mm 2mm]{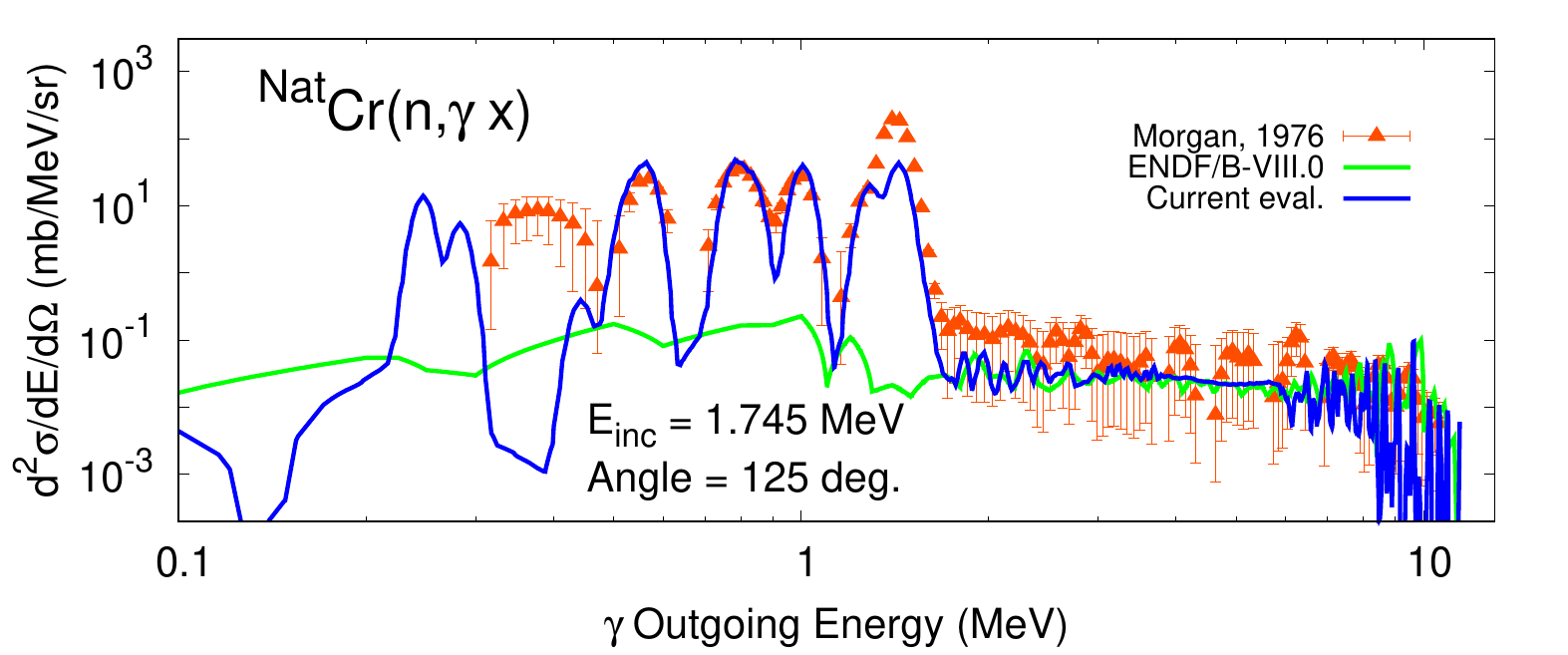} \hspace{-6.6mm}
\includegraphics[scale=0.615,keepaspectratio=true,clip=true,trim=15.5mm 12mm 3mm 2mm]{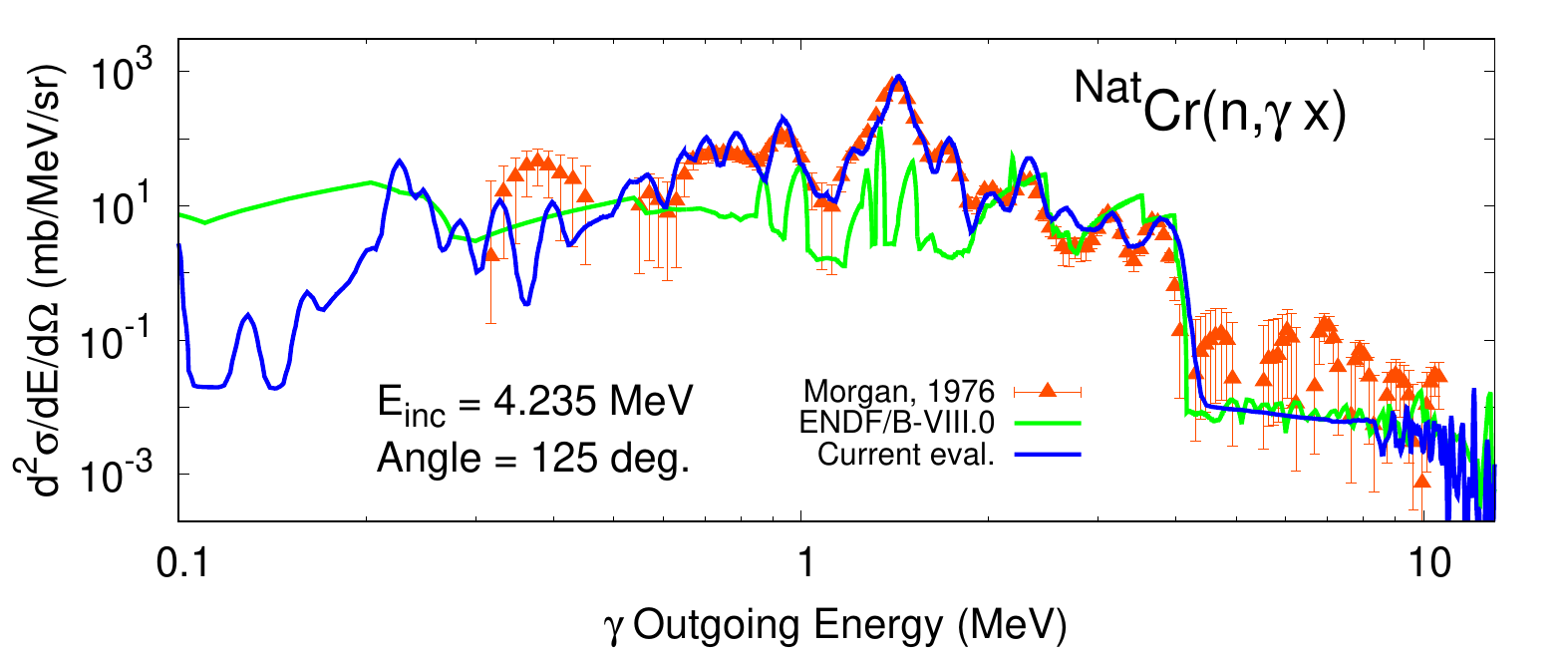}
\\ \vspace{-2.5mm}
\includegraphics[scale=0.615,keepaspectratio=true,clip=true,trim=  0mm 12mm 3mm 2mm]{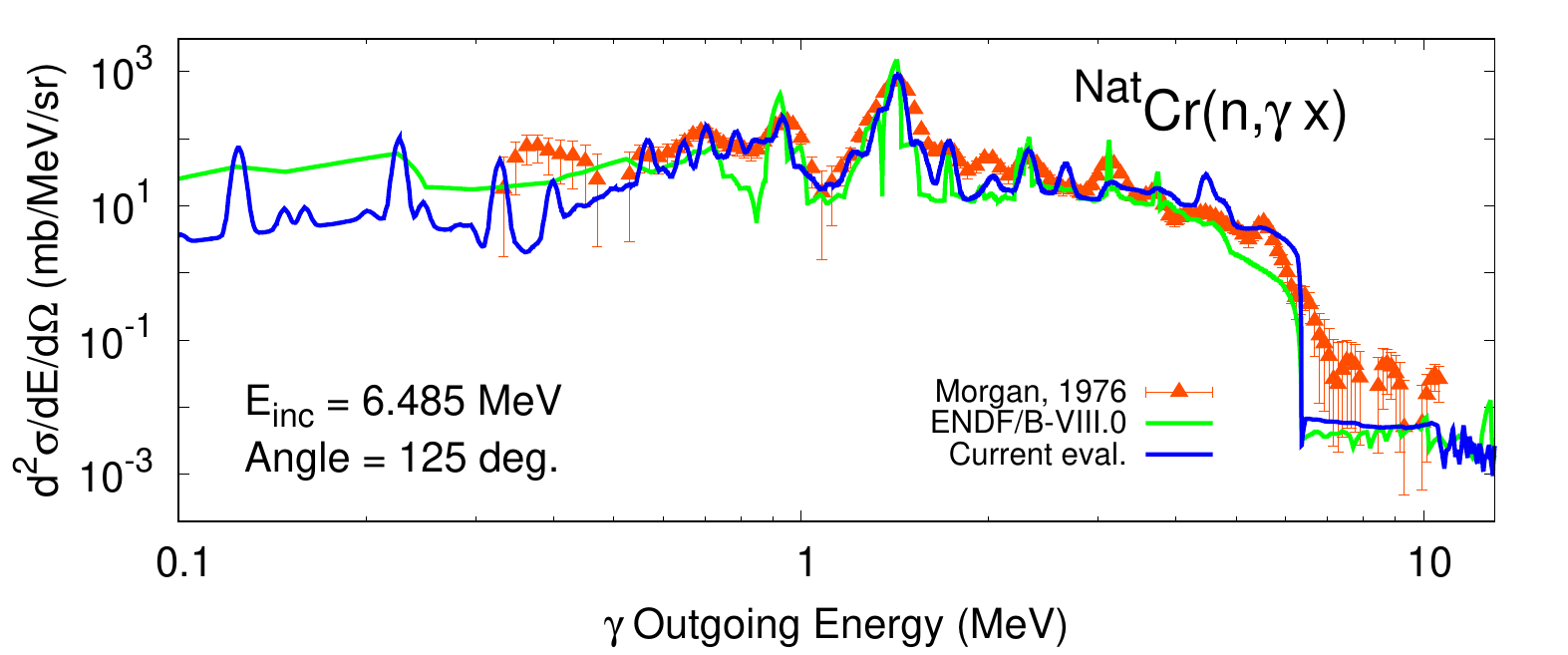}  \hspace{-6.6mm}
\includegraphics[scale=0.615,keepaspectratio=true,clip=true,trim=  15.5mm 12mm 3mm 2mm]{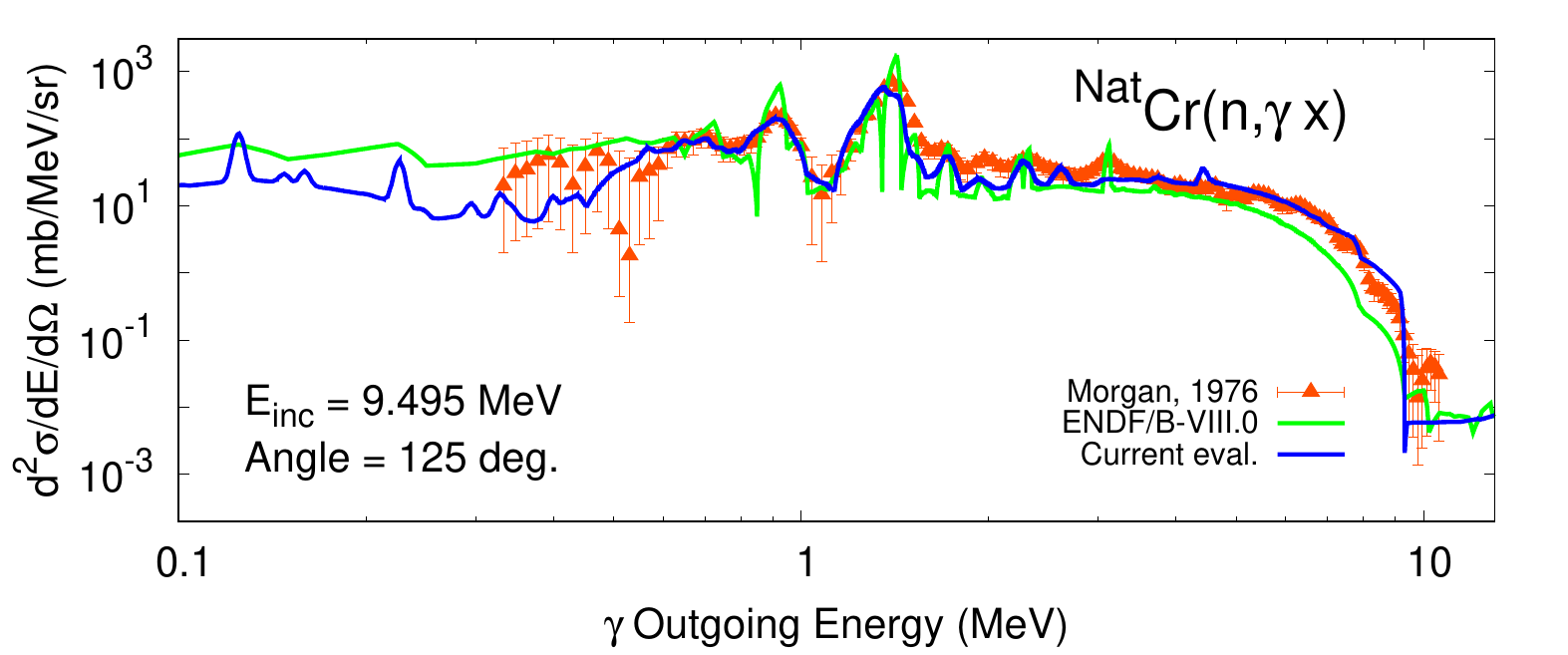}
\\ \vspace{-2.5mm}
\includegraphics[scale=0.615,keepaspectratio=true,clip=true,trim=0mm 0mm 3mm 2mm]{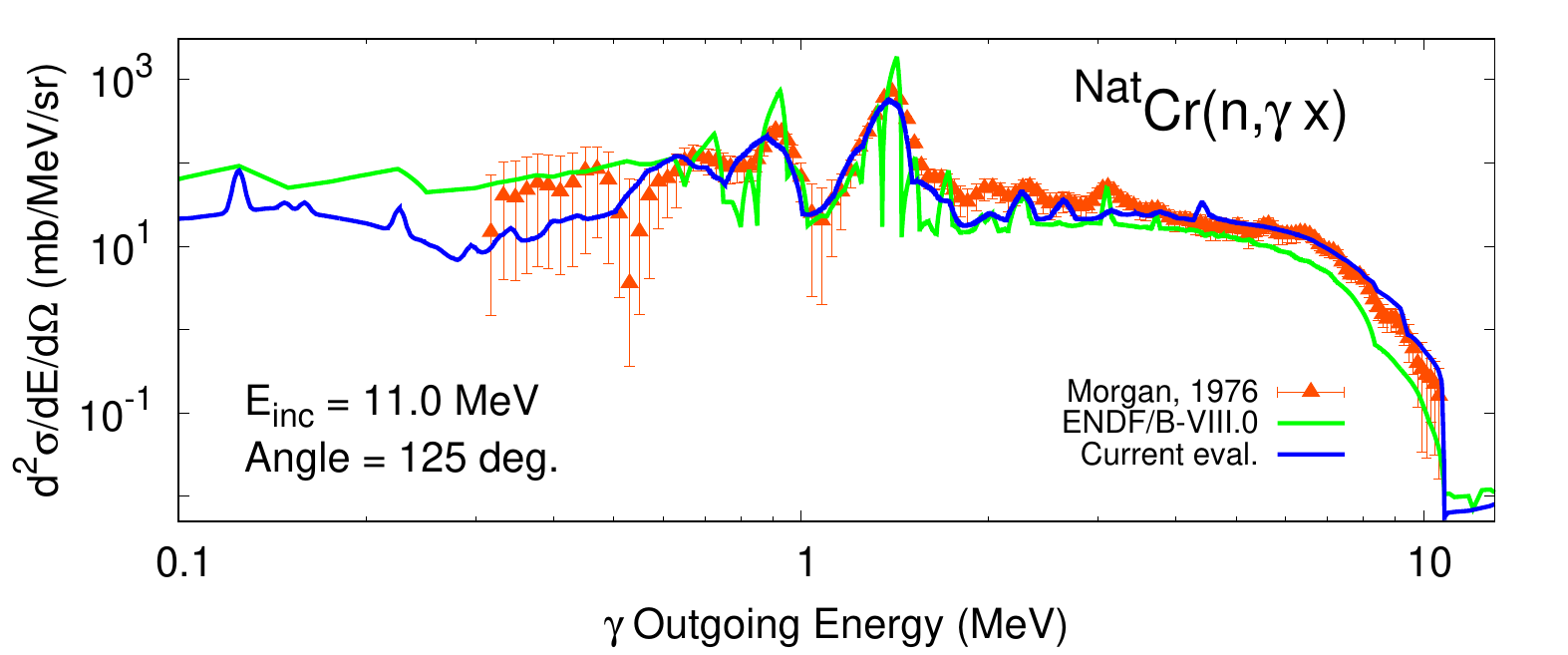} \hspace{-6.6mm}
\includegraphics[scale=0.615,keepaspectratio=true,clip=true,trim=15.5mm 0mm 3mm 2mm]{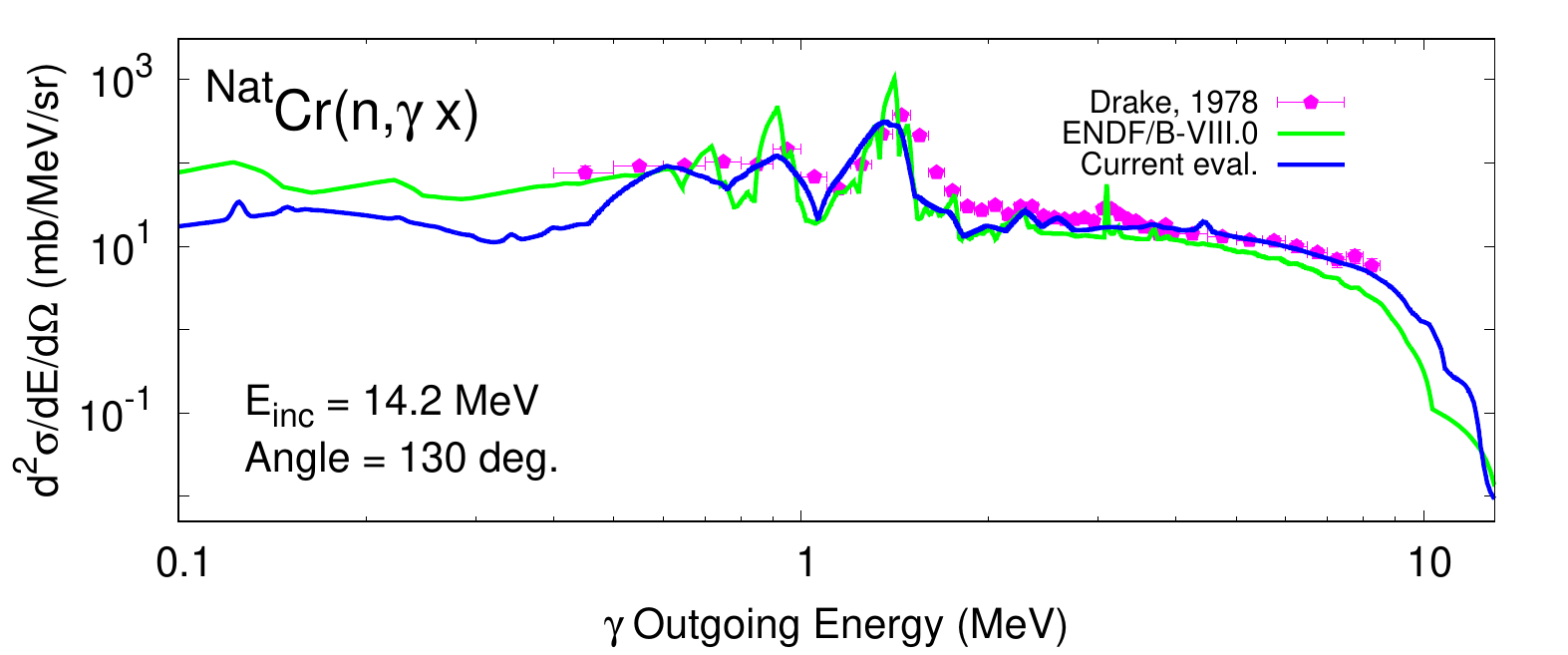}
\caption{(color online) Double-differential spectra for gamma production cross sections for a variety of neutron incident energies  on \nuc{Nat}{Cr} at the fixed scattering angle of 125 degrees, except for $E_{\mathrm{inc}} = 14.2$ MeV which is presented for 130 degrees, as a function of the outgoing $\gamma$ energy. Resolution broadening was applied to visually match resolution of the spectra data. Experimental data taken from Refs.~\cite{Morgan:1976,J.NSE.65.49.197801}.}
\label{fig:cr-nat-DD-gamma}
\end{figure*}

In addition to properly describing the cross section of the scattered neutrons for all angles and outgoing energies, for all incident energies, a fully consistent evaluation should also be able to accurately provide analogous information about the scattered gammas. In practice such task is not that simple, as the full evaluation is normally a complex combination of experimental data and models that may not be fully self-consistent. Nevertheless, it is still a very instructive way to assess strong points and deficiencies of the evaluation.  
In Fig.~\ref{fig:cr-nat-DD-gamma} we compare the evaluated double-differential spectra of gamma production, as a function of the outgoing $\gamma$ energy for select values of neutron incident energy and scattering angle, with experimental data.

We see that, in general, the peaks at gamma energies corresponding to the  inelastic transitions of the first few excited levels of \nuc{52}{Cr}, the major isotope in elemental chromium, are much better described by the current evaluation than ENDF/B-VIII.0. This is likely due to the effort done in the evaluation in consistently fitting inelastic gamma and neutron cross sections, as described in Sec.~\ref{sec:inel}. The consistently good agreement achieved and exemplified in Fig.~\ref{fig:inel_gammas} and Figs.~\ref{fig:cr52-MT4} and~\ref{fig:cr52-MT51-54} certainly contributed to the gamma spectra agreement in that energy range. A noteworthy exception is the gap in the spectra seen at $\sim$300 keV gamma energy. The cause of this may be due to transitions that have not been properly described but it has not been fully investigated yet. We leave this for a future work.

For  higher energy gammas, but below the neutron incident energy, the current evaluation also is in much better  agreement with the data than ENDF/B-VIII.0. Outgoing gammas with energies above the incident neutron energy, which correspond to transitions in the compound nucleus, are well described more or less equally well by both the current evaluation and ENDF/B-VIII.0, being slightly underestimated in a few cases. These are due to capture processes and the accurate description of capture cross section in the fast neutron region is hindered by the lack of experimental data (see Sec.~\ref{sec:capture_fast}).


%

\section{Validation}
\label{Sec:validation}

\subsection{Criticality Benchmarks}
   \label{subSec:crits}

Testing of chromium data on criticality benchmarks is important, since it has implications on criticality safety, as well as on design optimization of new types of reactors. In most cases chromium does not appear alone. Usually it is alloyed in stainless steel, which is a common structural material. Data testing therefore includes cases with a significant content of stainless steel.

The selection of benchmarks from the ICSBEP Handbook was based on previous work reported by various authors~\cite{CSEWG2010_ANL}, and especially on the sensitivity searches in the \href{https://www.oecd-nea.org/science/wpncs/icsbep/dice.html}{DICE system} of the NEA Data Bank~\cite{NEADB_DICE}. We converged on 22 cases, which are expected to have sensitivity to chromium. The list is given in Table~\ref{Table_bnch}. There is only one benchmark dedicated specifically to testing chromium data, namely the Russian $k_{\infty}$ measurement in the KBR facility (KBR-15). Other benchmarks mainly involve chromium is stainless steel (labeled as ``SS'' in Table~\ref{Table_bnch}) used either as cladding or a reflector.

\begin{table*}[htbp] \centering
  \caption{List of benchmarks from the ICSBEP Handbook selected for the validation of chromium data.}
  \label{Table_bnch}
  \begin{tabular}{r l l l l } 
  \toprule
  \toprule
  No. & ICSBEP Label       & Short name & Common name       & Comment      \\ 
  \midrule
    1 & HEU-COMP-INTER-005 & hci005-009 & KBR-09(SS)        & $k_\infty$       \\
    2 & HEU-COMP-INTER-005 & hci005-010 & KBR-10(Mo)        & $k_\infty$        \\
    3 & HEU-COMP-INTER-005 & hci005-015 & KBR-15(Cr)        & $k_\infty$        \\
    4 & HEU-COMP-THERM-011 & hct011-001 & RRC-KI-21x21-001  & SS$_{\mathrm{cladding }}$ \\
    5 & HEU-COMP-THERM-011 & hct011-002 & RRC-KI-21x21-002  & SS$_{\mathrm{cladding }}$ \\
    6 & HEU-COMP-THERM-011 & hct011-003 & RRC-KI-21x21-003  & SS$_{\mathrm{cladding }}$ \\
    7 & HEU-COMP-THERM-012 & hct012-001 & RRC-KI-18x18-001  & SS$_{\mathrm{cladding }}$ \\
    8 & HEU-COMP-THERM-012 & hct012-002 & RRC-KI-18x18-002  & SS$_{\mathrm{cladding }}$ \\
    9 & HEU-COMP-THERM-013 & hct013-001 & RRC-KI-14x14-001  & SS$_{\mathrm{cladding }}$ \\
   10 & HEU-COMP-THERM-013 & hct013-002 & RRC-KI-14x14-002  & SS$_{\mathrm{cladding }}$ \\
   11 & HEU-COMP-THERM-014 & hct014-001 & RRC-KI-10x10-001  & SS$_{\mathrm{cladding }}$ \\
   12 & HEU-COMP-THERM-014 & hct014-002 & RRC-KI-10x10-002  & SS$_{\mathrm{cladding }}$ \\
   13 & HEU-COMP-THERM-022 & hct022-001 & SPERT-III         & SS$_{\mathrm{cladding }}$ \\
   14 & HEU-MET-INTER-001  & hmi001     & ZPR-9/34          & SS$_{\mathrm{reflector}}$ \\
   15 & HEU-MET-INTER-001  & hmi001d    & ZPR-9/34          & SS$_{\mathrm{reflector}}$ \\
   16 & HEU-MET-THERM-016  & hmt016     & LACEF/Ni-Cr-Mo-Gd &              \\
   17 & IEU-COMP-THERM-005 & ict005     & KBR-21            & $k_\infty$       \\
   18 & LEU-SOL-THERM-012  & lst012-001 & TRACY-203c        &              \\
   19 & MIX-COMP-FAST-001  & mcf001     & ZPR-6/7           &              \\
   20 & MIX-MET-FAST-008   & mmf008-003 & ZEBRA-8C/2        & $k_\infty$      \\
   21 & PU-MET-INTER-002   & pmi002     & ZPR-6/10          & SS$_{\mathrm{reflector}}$ \\ 
 \bottomrule
 \bottomrule
  \end{tabular}
\end{table*}

One of the objectives in nuclear data evaluation is the condition that a new evaluation does not degrade the overall performance of a library. Of course it is possible that an inferior/older evaluation performs better in certain cases due to cancellation of errors in combination with other materials, but such cases need to be looked at very carefully. In particular, it is well known (and acknowledged in the main reference describing the \mbox{ENDF/B-VIII.0} library) that the iron evaluations have some deficiencies and improved evaluations developed within the INDEN collaboration of the IAEA~\cite{INDEN} are available. Similarly, improvements to the silicon evaluations have been made and are also available from the INDEN web page. Thus, the new chromium evaluations were tested together with the new iron and silicon evaluations. The results for libraries were compared and are identified by labels as follows:
\begin{itemize}
  \item {\bf \tt INDEN} Chromium, iron, silicon, and oxygen evaluations from INDEN, the rest from \mbox{ENDF/B-VIII.0}
  \item {\bf \tt Cr-BROND} Chromium evaluations from BROND-3.1~\cite{BROND-3.1}, iron, silicon, and oxygen evaluations from INDEN, the rest from \mbox{ENDF/B-VIII.0}
  \item {\bf \tt e80} All materials from \mbox{ENDF/B-VIII.0}
  \item {\bf \tt jeff33} All materials from \mbox{JEFF-3.3}
  \item {\bf \tt e71} All materials from \mbox{ENDF/B-VII.1}
\end{itemize}

\begin{figure}[htbp]
\begin{center}
\includegraphics[width=0.50\textwidth]{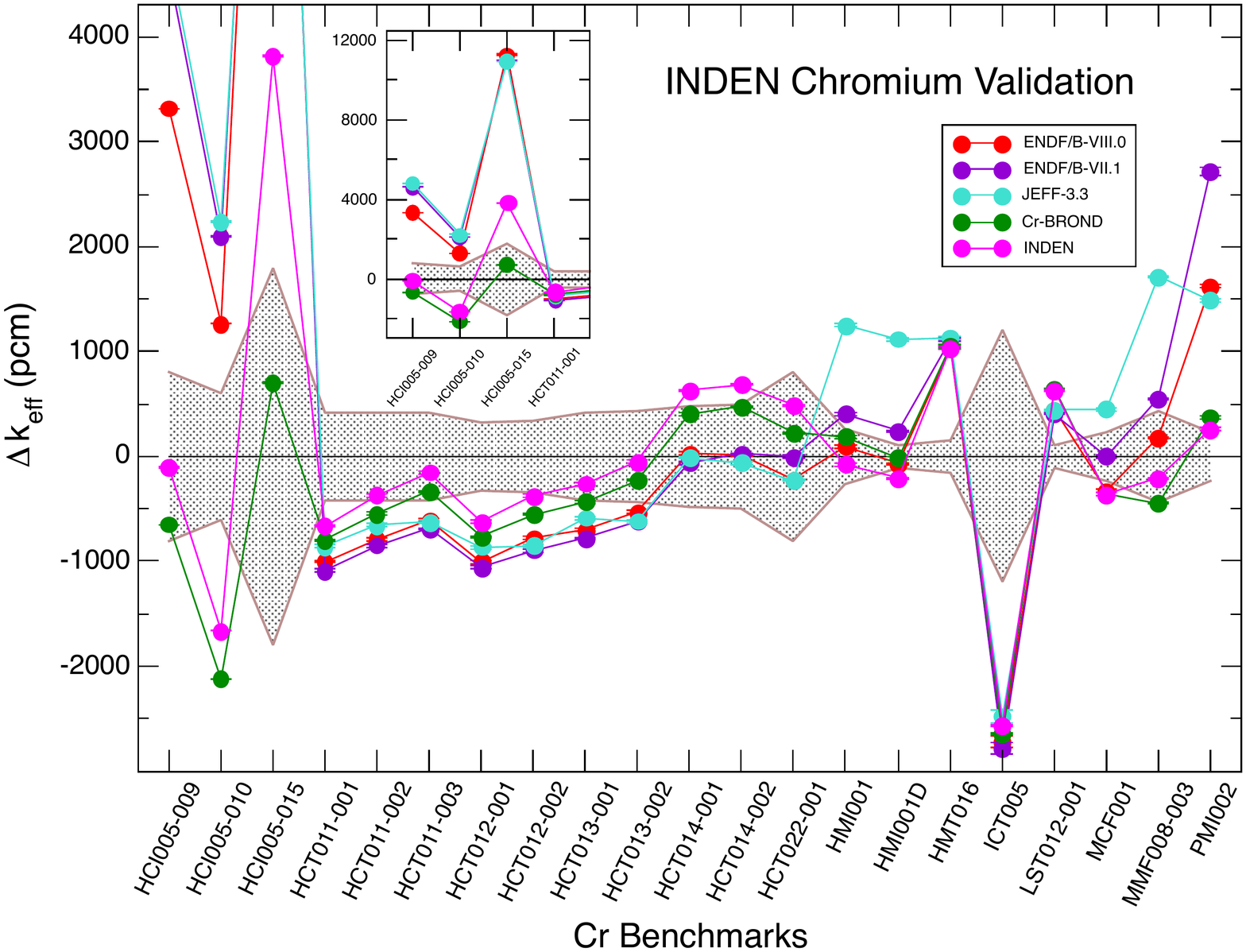}
\caption{(color online) Comparison of the differences from the reference benchmark values using different data libraries. Details of each calculation are discussed in text.}
\label{Fig_bnch}
\end{center}
\end{figure}

\begin{figure}[htbp]
\begin{center}
\includegraphics[width=0.52\textwidth]{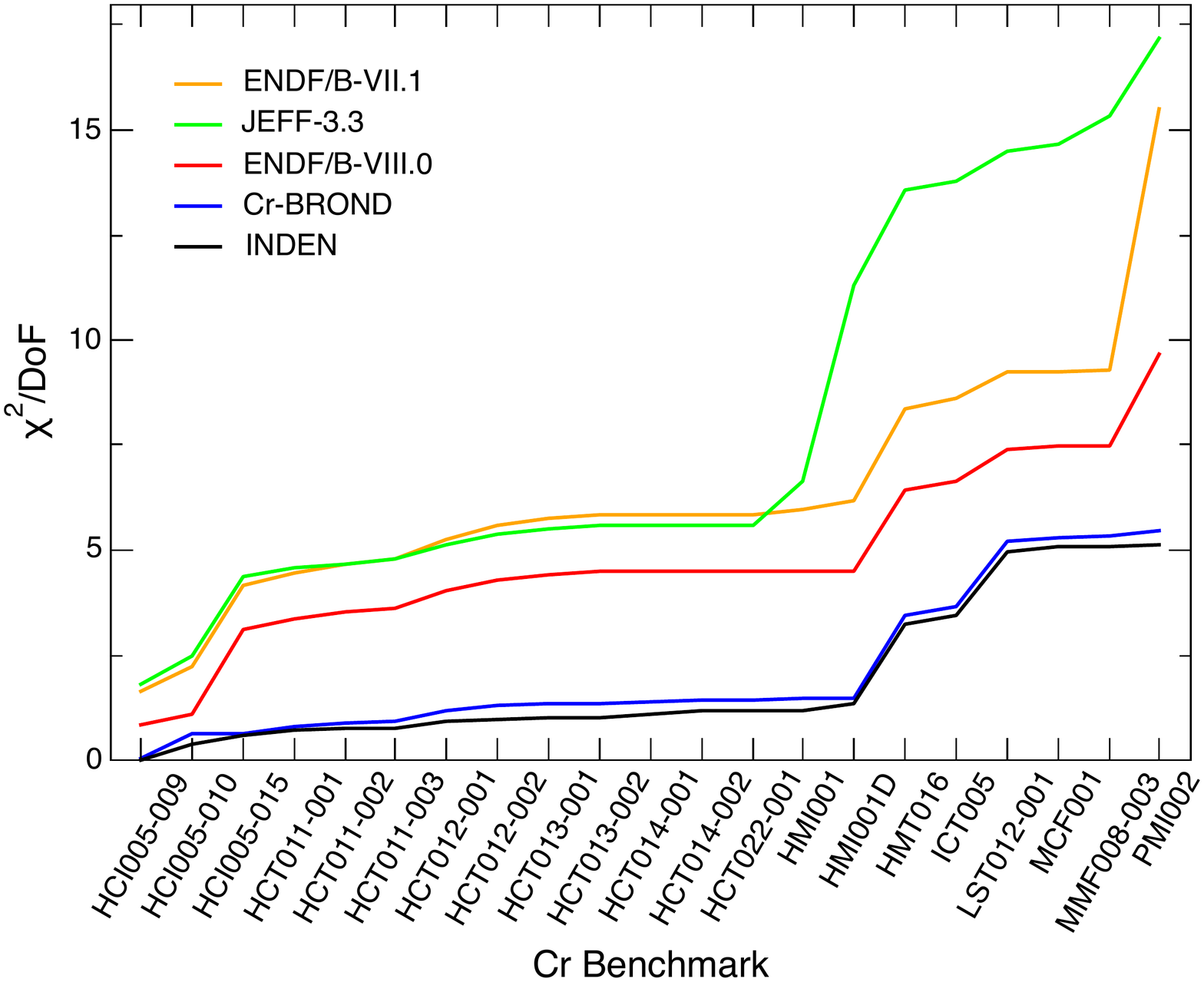}
\caption{(color online) Cumulative Chi-squared per degree of freedom for the selected list of benchmarks. Details of each calculation are discussed in text.}
\label{Fig_Chi}
\end{center}
\end{figure}

The BROND-3.1~\cite{BROND-3.1} evaluations of chromium isotopes have a significantly enhanced capture in the cluster of resonances near 5~keV, mainly from \nuc{50,53}{Cr}, linked to a much larger thermal capture as discussed in the Introduction. The evidence for the enhancement were the preliminary results of the measurements by Guber et al.~\cite{Guber:2011}. A lot of effort was invested to find a solution that would respect experimental data measured by several authors and provide good performance in the simulations of integral experiments. The results for the selected list of benchmarks are shown in Figure~\ref{Fig_bnch}. The Chi-squared per degree of freedom ($\chi^{2}$/DoF) is shown in Figure~\ref{Fig_Chi}. The figure shows that compared to \mbox{ENDF/B-VIII.0} the $\chi^{2}$/DoF is reduced by more than a factor of two when the INDEN evaluations are used and is similar to what achieved with chromium from \mbox{BROND-3.1}.

The \mbox{BROND-3.1} evaluation~\cite{BROND-3.1} showed the best performance for the KBR-15 benchmark, but in our opinion, for the wrong reason. In fact, \mbox{BROND-3.1} is not the best for the other two KBR benchmarks involving stainless steel and a molybdenum alloy. Additionally, the \mbox{BROND-3.1} evaluation is performing significantly worse than the current work for the HCT benchmarks from the Kurchatov institute (RRC-KI) which are highly sensitive to the chromium thermal capture, which is an evidence that the thermal capture of \mbox{BROND-3.1} is overestimated.

The current INDEN evaluation is off for the KBR-15 benchmark by 4000~pcm, but this is still a reduction of the discrepancy with pure \mbox{ENDF/B-VIII.0} data by a factor of three. Note that the results for \mbox{ENDF/B-VII.1, JEFF-3.3} and \mbox{ENDF/B-VIII.0} are out of scale for this benchmark. It must be borne in mind that $k_\infty$ measurements are extremely sensitive, but they also involve many corrections based on nuclear data, so it is possible that the uncertainty of the reference benchmark value is underestimated. The other big improvement is for the PMI002 (ZPR-6/10) benchmark, which has been a notorious outlier for a long time. With the current INDEN evaluations the predicted PMI002 reactivity is within the uncertainty interval of the measured value. Some improvement is observed also in the reactivity predictions of  the thermal benchmarks from the Kurchatov institute (RRC-KI) as discussed above. Other benchmarks do not show large sensitivity to the differences in the evaluations.


\subsection{D-T Neutron Leakage Experiments}
\label{subSec:leakage}

In 1980's a series of experiments was performed on the Oktavian facility for benchmarking nuclear data for fusion neutronics~\cite{Oktavian94}. One of the experiments involved a 40~cm diameter chromium sphere with a pulsed D beam incident of a tritiated Ti target that acted as the D-T (fusion) neutron source. Time-of-flight leakage spectrum was measured about 10~m from the target at an angle of 55 degrees relative to the beam. The source spectrum is given in Table~4.6 of the report. In a companion report~\cite{Oktavian98} the results with different data libraries are presented and a sample computational model for the MCNP Monte Carlo code is given. The source specifications are claimed to be the lethargy spectrum but the units were "1/MeV/n", which would imply the energy spectrum. By repeating the calculations with the \mbox{ENDF/B-IV} and \mbox{JENDL-3.2} libraries the plot in Fig.~4.11 of~\cite{Oktavian94} and in Fig.~5.3.9 of~\cite{Oktavian98} the interpretation of the spectra was affirmed.

The computational model for MCNP in~\cite{Oktavian98} is very simple, with the measured neutron spectrum for the source and the surface leakage current spectrum for the tally. A more detailed model was designed that took the sphere geometry accurately into account and the source was modeled from first principles by explicitly modeling the target, taking into account the angle and energy distribution of the neutrons from the D-T reaction. The approach followed closely the one applied by Milocco et al.~\cite{Milocco2010} for the analysis of other Oktavian benchmarks in the SINBAD data base. The detector was modeled as a point-detector about 9 m from the target and at an angle of 55~degrees. Gaussian resolution-broadening of the source was done within MCNP assuming 7~ns pulse width at half maximum to approximately take into account the finite pulse width and other effects in the measuring system.

Spectrum tally in the detector was collected in the time-domain and converted to energy spectrum externally with the ACEFLX code available locally. The conversion would be exact if the time offset and the flight path with which the measured ToF spectrum was converted into the energy spectrum. Since this information is not known exactly we took the library to use time offset of 4~ns and flight path of 9.1~m using relativistic kinematics.

\begin{figure}[htbp]
\begin{center}
\includegraphics[width=0.52\textwidth]{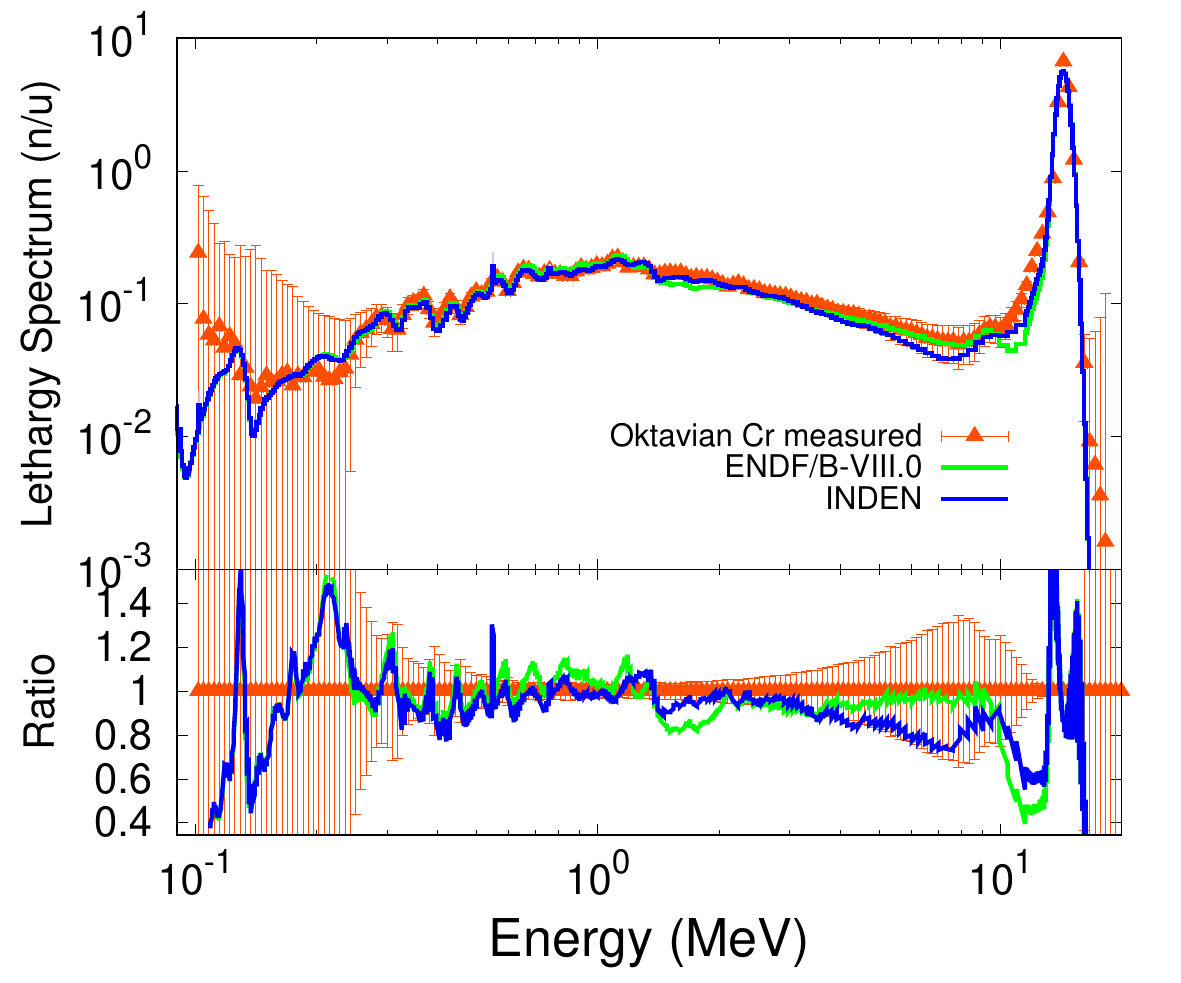}
\caption{(color online) Leakage spectrum from the Oktavian Cr sphere with a D-T source. Calculated spectra with \mbox{ENDF/B-VIII.0} and the new INDEN evaluation for chromium are compared to measured data.}
\label{Fig:Oktavian}
\end{center}
\end{figure}

Calculations were performed with the ENDF/B-VIII.0 library and with a modified library that included the new chromium evaluations. The results are presented in Fig.~\ref{Fig:Oktavian}. The figure shows excellent agreement in the energy range 0.25-0.6~MeV with both libraries. In the energy range 0.6-1.1~MeV the new INDEN evaluation is distinctly better. Between 1.1~MeV and 2~MeV some discrepancy is seen, which is much smaller with the INDEN evaluation. Above 2~MeV both evaluations lie within the experimental uncertainty, INDEN being slightly closer to the mean of the measured values. The converse is true between 4~MeV and 9~MeV. Above 9~MeV the discrepancies are larger, but these could be to the assumptions in modeling the target assembly, the resolution function, as well as the nuclear data.

\subsection{CEA Pile-oscillation Experiments in MINERVE Reactor}

The collaboration group at the CEA/DEN/CAD/DER/SPRC/Laboratoire d'\`Etudes de Physique in Cadarache, France, has developed an experimental program entitled MAESTRO~\cite{jef-1849:2017} for the validation of thermal capture cross sections of structural and moderating materials using pile-oscillation experiments in the MINERVE reactor. The experiment consisted of oscillating small samples in a critical MINERVE reactor core and measuring the response function (namely the reactivity worth) of the sample. Note that the  sensitivity was maximized for thermal neutrons and rapidly decreased for higher energy neutrons; a very low sensitivity to the keV region was determined. Among structural materials of interest pile-oscillation measurements were undertaken in MAESTRO Phase III (2013-2014) on natural chromium and iron samples. Pure rods of Fe and Cr materials were used in those experiments. Very accurate metrology of the dimensions($\pm 10\mu$m) and mass ($\pm 10$mg) were undertaken~\cite{jef-1849:2017}. Void (dummy) samples were used to cancel the reactivity worth due to the cladding. The measurement reactivity uncertainty of $\pm0.01$ pcm was achieved due to design improvements~\cite{jef-1849:2017}.

A 3D detailed full core model of the MINERVE reactor was implemented using TRIPOLI Monte Carlo model allowing to reach agreement within 3.4\% of the measured actinide fission and capture reaction rates 
\nuc{238}{U}(n,f)/\nuc{235}{U}(n,f), \nuc{237}{Np}(n,f)/\nuc{235}{U}(n,f), \nuc{239}{Pu}(n,f)/\nuc{235}{U}(n,f), \nuc{240}{Pu}(n,$\gamma$)/\nuc{239}{Pu}(n,f), and \nuc{242}{Pu}(n,$\gamma$)/\nuc{239}{Pu}(n,f) 
in the central position of the MINERVE core using the JEFF-3.2 transport library. This is the same position used to measure the samples of interest and these reaction rate measurements validate the calculated neutron flux at that location.

Results of measurements for chromium and iron samples 
gave (C/E -1) of -0.5($\pm1.1$)\% and +0.3($\pm1.2$)\%, respectively using JEFF-3.2 transport library as documented in Ref.~\cite{jef-1849:2017}. A sample of stainless steel 316L (SS316L) that contains about 18\% of chromium and 64\% of iron gave (C/E -1) of -0.7($\pm1.1$)\%~\cite{jef-1849:2017}, whilst a sample of stainless steel 304L (SS304L) that contains about 19\% of chromium and 68\% of iron gave (C/E -1) of +0.3($\pm1.1$)\%~\cite{jef-1849:2017}.

As can be seen in Table~\ref{table:natCrIntegralValues}, the thermal capture on natural chromium in this evaluation increased by 0.11 barn compared to JEFF-3.2\footnote{The resonance evaluation of chromium isotopes in ENDF/B-VIII.0 is identical to the evaluation in JEFF-3.2},
a 0.35\% increase driven by thermal capture increase in \nuc{53}{Cr}. Therefore, an estimated MAESTRO (C/E-1) value for natural chromium will improve benchmark result to -0.15($\pm1.1$)\%. There is an estimated reactivity worth for the SS316L sample of only +0.06\% due to the increased thermal capture of natural chromium resulting in an updated (C/E-1) value of -0.64($\pm1.1$)\% for SS316L, which is still acceptable within the quoted experimental benchmark uncertainties. The estimated reactivity worth for the SS304L sample is +0.07\% due to the increased thermal capture of natural chromium resulting in an updated (C/E-1) value of +0.37($\pm1.1$)\% for SS304L, which is also acceptable within the quoted experimental benchmark uncertainties.

The 7\% decrease of the thermal capture in $^{56}$Fe from 2.586 b in JEFF-3.2 to 2.396 b in the INDEN evaluation (as suggested by Firestone et al. \cite{PhysRevC.95.014328}) is estimated to decrease the (C/E-1) of the natural iron by -6.4\% resulting 
in an updated (C/E-1) value of -6.1($\pm1.1$)\% for the MAESTRO benchmark. Such large decrease in the (C/E-1) clearly rejects the proposed reduction of the thermal capture on natural iron. A similar rejection trend is observed for experimental values of SS316L and SS304L stainless-steel samples. 

In summary, MAESTRO benchmark results are fully compatible with new chromium evaluations presented in this work. The evaluation trend of increasing thermal capture is confirmed and improves the already good benchmark performance for thermal neutron capture on natural chromium. However, the proposed decrease of the thermal capture of \nuc{56}{Fe} as suggested by Firestone et al. \cite{PhysRevC.95.014328} cannot be supported by the MAESTRO benchmark as measurements on three independent samples (Fe, SS304L, and SS316L) rejected the proposed change.   

Future additional validation may be provided by a new benchmark experiment being developed by CEA/EPLP collaboration, the PETALE neutron transmission experiment \cite{PETALE} which data acquisition concluded in September 2020. PETALE is being performed at the EPFL/CROCUS facility (Lausanne,Switzerland), and first results are expected in 2021. This transmission experiment will, indeed, help to validate the elastic and inelastic cross sections for Cr, Fe and Ni elements, independently, as it consists of transmission through purely isotopic blocks of the aforementioned structural materials.

\section{Conclusions and Recommendations}\label{Sec:conclusions}

In the resolved resonance region, particularly in the cluster of
$s$-wave resonances of the \nuc{50,53}{Cr} isotopes, this work has
clarified and unraveled some of the most relevant issues related to the
measured and evaluated data with particular focus to the deficiencies
of the ENDF/B-VIII.0 library.

Regarding the measured data, this work showed that the normalization
of the neutron capture yield is of fundamental importance and in the
case of  data sets from Stieglitz \etal and Guber \etal reported in the EXFOR
library, the adopted normalization (i.e., by the inverse of the
thickness) is the incorrect choice as shown in
Figs.~\ref{fig:53cr-norm-stieg} and~\ref{fig:53cr-norm-guber}. In
general, the measured data should be both reported as neutron capture
yield data and checked for the correct normalization. Moreover, for
nuclei with large total cross section, the capture measurements should be
designed with the proper thickness sample to minimize the self
shielding effects and multiple-scattering corrections.

Although the large discrepancy between oxide
enriched data sets from  Stieglitz \etal and Guber \etal  was not directly resolved, a correction of about
3\% to the thickness of the  data from Guber \etal was found and the corrected value
used in the current evaluation. Also, the fit of capture yield
data from  Guber \etal measured on metal natural sample showed remarkable consistency
with  oxide enriched data from Stieglitz \etal as shown in
Figs.~\ref{fig:guber_metal_mltsc}--\ref{fig:stieglitz_oxide_mltsc}. This
implies that the \nuc{53}{Cr} oxide enriched data from Guber \etal are affected by a
normalization issue.

Regarding the evaluated data, a new set of resonance parameters was
derived by the fit of capture yield data measured on metal natural
sample and reported in Tab.~\ref{table:ResonanceWidths}. 
Direct capture was shown to play an important role in the resonance 
evaluation of near magic nuclei like chromium isotopes. 

Monte Carlo simulations revealed some inconsistency in describing the multiple
scattering corrections for the cases of oxide-enriched samples, for
which the multiple scattering corrections are large, and a careful
choice between the finite size and infinite slab approximation should
be performed. Additional work and numerical simulations as well as
possible corrections to the algorithm to compute the multiple
scattering corrections with defined approximations are planned as
future updates to the SAMMY code. Further verification of multiple 
scattering corrections used in SAMMY code and comparison to MCNP-calculated 
multiple scattered corrections are warranted. 

The fit of \nuc{50}{Cr} measured data was performed for the time being 
in finite size approximation; a revised set of resonance parameters
is needed as soon as the multiple scattering corrections can be described 
properly. These updates might also clarify the problem with alignment of
the measured data in the cluster of $s$-wave resonances.

The currently evaluated \nuc{53}{Cr} resonance integral value is
higher than the value reported in the ENDF/B-VIII.0 library, which is
consistent with Mughabghab Atlas 2017 evaluated value. However,
the current evaluated RI is still more than a barn lower than 
the value reported by BROND-3.1 (11.19~b) and Stieglitz's thesis (10.75~b).
The overestimated resonance integral value for BROND-3.1 is related to the
overestimated thermal capture cross  section and the assumptions on the 
fitted cross sections in the $s$-wave cluster of resonances. For Stieglitz, the value of the resonance integral can be calculated by
the infinite dilution formula as in Eq.~(119) of Stieglitz's thesis
together with the reported resonance parameters. However, in that
equation it is unclear the quantity ``M'' defined as ``reduced mass of
neutron'' contrary to the ANL report~\cite{ANL5800:1963} where the
equation was taken from and that quantity was defined as the neutron
mass.

Complete revaluations of the fast-region reaction for all stable chromium isotopes were performed taking into account new data (e.g., inelastic, (n,p) reactions) and new modeling (e.g. soft-rotor coupled-channel optical potential), leading to substantial improvement in the description of experimental data for angle-integrated cross sections, angular distributions, energy spectra,  and neutron and gamma double-differential spectra.

The new chromium evaluations combined with proposed INDEN iron evaluations were benchmarked with MCNP simulations
of a small suite of critical assemblies sensitive to natural chromium and iron data and with the Oktavian chromium shielding
benchmark.  Considerable performance improvements were observed both for critical assemblies and the shielding experiment.
Thermal capture values of chromium and iron INDEN evaluations were benchmarked with the CEA MAESTRO pile-oscillation
experiments. The chromium evaluations show small improvement in the observed C/E trend well within the quoted uncertainties. 
However, the proposed 7\% decrease of the thermal capture in natural iron\footnote{as suggested by Firestone et al. \cite{PhysRevC.95.014328}} is rejected by CEA MAESTRO pile-oscillation measurements.

Despite the significant advances in evaluated data and documented performance improvements of the proposed chromium evaluations, the authors would like to
remark that there are outstanding issues that still need to be addressed in a future work:
\begin{inparaenum}[a)]
\item A dispersive coupled-channel optical model needs to be improved below 4 MeV of neutron incident energy.
\item The modeling description of measured inelastic gammas as shown in Fig.~\ref{fig:inel_gammas} should be improved. Important work has been recently published in modeling of the \nuc{57}{Fe}(n,n$^{\prime}\gamma$) \cite{Negret:2017} that could serve as a template for such improvement.
\item The high energy part of the resolved resonance region and capture in particular can be improved. 
\item Additional work is needed on the evaluated resonance integrals of \nuc{50,53}{Cr} isotopes.
\item Capture gamma spectra of thermal neutrons in iron and chromium need to be reevaluated to address issues recently raised by Mauborgne \etal \cite{Mauborgne:2020}.
\end{inparaenum}

\acknowledgments{ 
Authors would like to acknowledge G. Noguere for very productive discussions and exchange of information. Work at Brookhaven National Laboratory was sponsored by the Office of Nuclear Physics, Office of Science of the U.S. Department of Energy under Contract No. DE-AC02-98CH10886 with Brookhaven Science Associates, LLC. ORNL is managed by UT-Battelle, LLC, for the U.S. Department of Energy under Contract No. DE-AC05-00OR22725. The U.S. Department of Energy Nuclear Criticality Safety Program sponsored the work presented in this paper. }

\bibliography{chromium_paper,nsr,exfor}

\end{document}